\documentclass[aps,groupedaddress,superscriptaddress,amsmath,amssymb,twocolumn,pra]{revtex4-2}
\usepackage{bbm}
\usepackage{graphicx}
\usepackage[normalem]{ulem}
\usepackage{epstopdf}
\usepackage{braket}
\usepackage{xcolor}
\usepackage[hidelinks]{hyperref}
\usepackage{esint}

\def\rb{{\bf r}}

\def\ub{{\bf u}}

\def\vb{{\bf v}}

\def\n{\boldsymbol{\nabla}}
\def\kb{{\bf k}}

\def\Rb{{\bf R}}

\def\sgn{{\rm sgn}}
\def\nl{^{\rm (nl)}}

\def\ep{\varepsilon}
\def\RR{\mathfrak{R}}
\newcommand{\sumprime}[1]{\sum_{#1}{\vphantom{\sum}}^{\!\!\prime}}

\begin{document}

\title{
Mesoscopic physics of nanomechanical systems}
\author{Adrian Bachtold}
\email{adrian.bachtold@icfo.es}
\affiliation{ICFO - Institut de Ciencies Fotoniques,
The Barcelona Institute of Science and Technology, 08860 Castelldefels, Barcelona,
Spain}
\author{Joel Moser}
\email{j.moser@suda.edu.cn}
\affiliation{School of Optoelectronic Science and Engineering \& Collaborative Innovation Center of Suzhou Nano Science and Technology, Soochow University, Suzhou, 215006, People's Republic of China}
\author{M. I. Dykman}
\email{dykman@pa.msu.edu}
\affiliation{Michigan State University, East Lansing, MI 48823, USA}

\date{\today}

\begin{abstract}

Nanomechanics has brought mesoscopic physics into the world of vibrations. Because nanomechanical systems are small, fluctuations  are significant, the vibrations become nonlinear already for comparatively small amplitudes, and new mechanisms of dissipation come into play. At the same time, the exquisite control of these systems makes them a platform for studying many problems of classical and quantum physics far from thermal equilibrium in a well-characterized setting. This review describes, at a conceptual level, basic theoretical ideas and explicative experiments pertaining to mesoscopic physics of nanomechanical systems. Major applications of nanomechanics in science and technology are also outlined. A broad range of phenomena related to the conservative as well as dissipative nonlinearity and fluctuations are discussed within a unifying framework. They include the linear response of single and coupled vibrational modes as well as nonlinear effects of periodic driving. Such driving breaks the continuous time-translation symmetry and the detailed balance, with conspicuous consequences for fluctuations, particularly in the presence of the driving-induced bi- and multistability. Mathematical techniques are described in the appendices to streamline the reading, but also to provide an introduction to the theory. The goal of the review is to show the richness of the physics at work. The continuous experimental and theoretical advances make nanomechanical systems a vibrant area of research, with many new phenomena to discover.
\end{abstract}

\maketitle

\tableofcontents


\begin{widetext}
\begin{center}
\begin{table*}[h]
\caption{List of Symbols}
 \begin{tabular}{|p{3cm}|p{15cm}|}
 \hline
Symbols & Description \\
 \hline\hline
  $q$ &   Mode coordinate; it is proportional to the displacement at the antinode of the mode\\
 \hline
  $p$ &  Momentum conjugate to the coordinate $q$ \\
 \hline
 $M$ & Effective mass of the mode \\
 \hline
 $u$ & Complex vibration amplitude\\
  \hline
   $S_{}(\omega)$ & Spectral density of fluctuations of $q$ at frequency $\omega$\\
 \hline
 $\chi(\omega)$ &  Susceptibility of the coordinate at frequency $\omega$ \\
 \hline
 $\omega_{0{}}= 2\pi f_{0{}}$ &  Eigenfrequency of the mode\\
 \hline
  $\gamma$ &  Parameter of the Duffing, or Kerr, nonlinearity\\
 \hline
  $\Gamma$&  Linear friction coefficient, decay rate of the vibration amplitude in the linear regime\\
 \hline
 $Q= \omega_{0{}}/2\Gamma$ &  Quality factor from energy decay measurements \\
 \hline
 $\Delta\omega$&  Spectral bandwidth of mechanical resonance, half width at half maximum of the resonant peak in $S_{}(\omega)$ \\
\hline
$Q_\mathrm{\omega}= \omega_{0{}}/2\Delta\omega$ &  Quality factor from spectral measurements \\
 \hline
  $\omega_F$ &  Angular frequency of a resonant driving force \\
 \hline
\end{tabular}
\label{table:symbol}
\end{table*}
\end{center}
\end{widetext}

\cleardoublepage


\section{Introduction}
\label{sec:introduction}

Studying vibrational motion has been one of the major areas of physics at least since Galileo. The advent of nanomechanical vibrational systems (NVSs) has opened a new direction in these studies.
NVSs are resonators for mechanical vibrations. To picture an example, one can think of a string of a musical instrument downscaled    to a diameter $\lesssim 100$~nm and a submicron length. The vibration frequencies range from kilohertz to gigahertz and can be tuned not only through the dimensions and the shape of a device, but also  {\it in situ} by electrostatic and optical means. In addition, the lifetime of vibrations has now been increased to hundreds of seconds and above thanks to progress in nanofabrication.

By their nature NVSs are mesoscopic. Because they are small, they display many features of microscopic systems. At the same time,  they are sufficiently large to enable studying an individual vibrational system rather than resorting to ensemble measurements, as in conventional molecular or solid-state vibrational spectroscopy.

NVSs   were developed in the 1990$^{\prime}$s~\cite{Cleland1996,Cleland1998} and quickly attracted  interest. Their vibrational eigenmodes display rich dynamics that involves a broad range of many-body effects stemming from the coupling to electrons, propagating phonons, photons, and two-level fluctuators. They are also of significant interest for various applications, which range from  ultrasensitive mass, charge, and force detection  to clocks. For example, the adsorption of mass onto an NVS can be detected with a resolution approaching 1~yg, while a force can be resolved with a sensitivity approaching 1~zN/Hz$^{1/2}$.
Over the years different aspects of the studies of NVSs along with their applications have been reviewed in a number of papers and books \cite{Cleland2003,Ekinci2005,Lifshitz2008,Poot2012,Aspelmeyer2014a,Schmid2016,Steeneken2021}.

In this review, we focus on the mesoscopic physics of NVSs, including dissipation, fluctuations, and nonlinear and far from thermal equilibrium phenomena in these systems. While these basic physical phenomena have been intensely investigated during the last few years, they have not yet been reviewed from a general perspective. Our aim is to provide a coherent and unifying description of the underlying concepts along with the experimental and theoretical results and to put them into a broad physics context. Details of the mathematical techniques are provided in Appendices.

We describe dissipation and thermal fluctuations of nanomechanical vibrations as resulting from the coupling to a thermal reservoir of a general form. Such description applies to both flexural and localized compression/shear-type modes. It allows us to analyze various specific dissipation mechanisms. They include the Landau-Rumer, Akhiezer, and thermoelastic relaxation due to scattering by propagating phonons and the phonon-induced clamping losses, as well as relaxation associated with the electrons in the nanoresonators and the leads and with the two-level systems.

A consequence of the small size of NVSs expected from the general arguments of statistical physics is the occurrence of comparatively strong quantum and classical fluctuations. These fluctuations play a significant role in the vibration dynamics. Another important aspect of the dynamics is the vibration nonlinearity. Because of the small system size, vibrations with even comparatively small amplitudes become nonlinear. Not only the restoring force displays nonlinearity, but also the rate of dissipative losses becomes amplitude dependent, which is associated with nonlinear friction.

The exquisite control of the NVSs and their versatility make them invaluable as a tool for studying the interplay of nonlinearity and fluctuations. This interplay leads to a broad range of  phenomena that manifest themselves in different settings, both in the classical and  quantum domains. Revealing and understanding them is an ongoing effort. We describe several of these phenomena studied with the NVSs, including the nondissipative broadening of the vibration spectra,  nonlinear inter-mode energy exchange, and self- and cross-modulation of the vibration frequencies.

We also describe how the nonlinearity makes NVSs vibrations a testing ground for exploring nonequilibrium phenomena. Several general types of such phenomena emerge where the vibrations are driven by a resonant field. Because the decay rates of the vibrations are usually small, even a weak field can lead to a significantly nonequilibrium behavior, such as the occurrence of bi- and multistability or chaos. Of particular interest, which goes beyond NVSs as such, are fluctuation effects away from thermal equilibrium. They range from noise-induced switching between coexisting metastable vibrational states to fluctuation squeezing. Since, on the one hand, driven nonlinear vibrations lack detailed balance while, on the other hand, the vibrations of NVSs are well-characterized, these vibrations provide a unique opportunity for addressing many generic problems of quantum and classical statistical physics far from thermal equilibrium.


\section{
Nanoresonators at a glance}
\label{sec:resonators_characteristics}

\subsection{Phenomenological description of the dynamics of a linear nanoresonator}
\label{subsec:phenomenology_linear}

We will describe the dynamics of an NVS mode in terms of the coordinate $q$ and momentum $p$ of an oscillator.  The mechanical displacement in the mode ${\bf u}_{}(\rb,t)$ as a function of the coordinate $\rb$ has a spatial profile $\boldsymbol{\varphi}(\rb)$, whereas $q(t)$ describes how the displacement varies in time,
\begin{align}
\label{eq:mode_displacement}
{\bf u}_{}(\rb,t) = q(t)\boldsymbol{\varphi}(\rb), \qquad \int \boldsymbol{\varphi}^2(\rb)d\rb =V,
\end{align}
where $V$ is the volume of the resonator. The momentum of the oscillator is $p=M{}\dot q$, where $M{}$ is the oscillator mass, $M{}=\int \rho(\rb)\boldsymbol{\varphi}^2(\rb)d\rb$ [$\rho(\rb)$ is the mass density]. Functions $\boldsymbol{\varphi}(\rb)$ for different modes are orthogonal. We note that in the analysis of the experimental data there is sometimes used a different normalization, i.e., it is set that $\max |\boldsymbol{\varphi}(\rb)| = 1$. With this normalization, the maximal value of $q(t)$ is the displacement amplitude.

The simplest theoretical model employed in the study of nanomechanical modes is a classical harmonic oscillator that performs Brownian motion \cite{Risken1996}, with a friction force proportional  to the velocity and with fluctuations due to thermal noise.
The noise is assumed to be Gaussian and $\delta$-correlated in time. If the oscillator coordinate is $q$ and the mass is $M$, the motion is described by the Langevin equation
\begin{align}
\label{eq:Brownian}
M{}\ddot q + 2M{}\Gamma\dot q + M{}\omega_0^2 q = f_T(t),\nonumber\\
\langle f_T(t)f_T(t')\rangle = 4M{}\Gamma k_BT \delta(t-t').
\end{align}
Here $\omega_0$ is the mode eigenfrequency and $\Gamma$ is the friction coefficient, which determines the decay rate of the vibrations in the absence of noise.

\begin{figure}[h]
\includegraphics[scale=0.8]{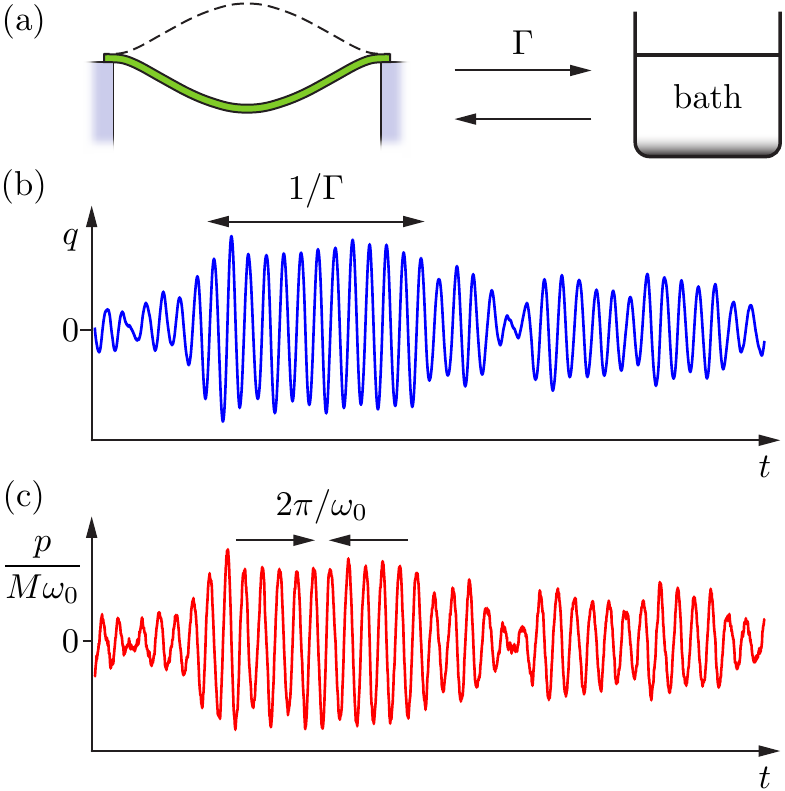} \hfill
\caption{(a) Mechanical oscillator coupled to a thermal bath. The coupling is quantified by the rate $\Gamma$ for an oscillator quantum to be transferred into the bath. The parameter $\Gamma$ enters both the friction force and the noise intensity in Eq.~(\ref{eq:Brownian}). (b,c) Time evolution of the displacement and the momentum of the oscillator performing Brownian motion. The characteristic correlation time of the amplitude fluctuations is given by $1/\Gamma$. One can measure $\Gamma$ using the spectral density of the displacement fluctuations; $\Gamma$ is the half width at half maximum of the spectral peak, see Eq.~(\ref{eq:Lorentz_classical}).}
\label{fig:thermal_bath}
\end{figure}

Without noise, the model of a damped oscillator  has been long used in physics; for example, it was used by Lorentz
in 1878 to describe the polarizability of matter. Later it was realized that, along with the friction, there comes noise; both of them result from the coupling of the oscillator to a thermal reservoir (thermal bath), see Fig.~\ref{fig:thermal_bath}.  The microscopic analysis was started by \textcite{Einstein1910b}. A detailed classical study was performed by  \textcite{Bogolyubov1945}, whereas the studies of the quantum dynamics were started in the late 50s - early 60s, see
\cite{Toda1958,Senitzky1961,Schwinger1961,Krivoglaz1961,Louisell1990,Ford1965,Ullersma1966a}; more references can be found in the paper by \textcite{Ford1988}. In many of these papers there was used a model in which the thermal bath was described by a set of harmonic oscillators and the coupling to the considered oscillator was linear in $q$. Over the years, such model of the bath, often called ``bosonic bath", has been one of the most frequently used in the study of quantum relaxation, cf.  \cite{Feynman1963,Caldeira1981,Grabert1988b}.

NVSs are one of the best examples of systems where dissipation described by the linear in $q$  coupling to a bosonic bath can play an important role. In this case, the bath is often formed by phonons in the resonator and in its support. Different mechanisms of the phonon-induced dissipation, such as clamping losses, thermoelastic, Landau-Rumer, and Akhiezer dissipation are discussed in Sec.\ref{subsec:phonon_scattering}. Nonlinear in $q$ coupling to a bosonic bath can also play a major role in the NVSs dynamics.

The most common way of characterizing the dynamics of NVSs is based on measuring either the spectral density of fluctuations of the displacement
\begin{align}
\label{eq:power_defined}
S_{}(\omega) =\int_{-\infty}^\infty dt \langle q(t)q(0)\rangle \exp(i\omega t)
\end{align}
or the susceptibility $\chi_{}(\omega)$, which characterizes the response to an external force at frequency $\omega$.
It is defined by the relation between the force-induced displacement $\delta q(t)$ and the force $F\exp(-i\omega t)$ added to the right-hand side of Eq.~(\ref{eq:Brownian}) as
\begin{align}
\label{eq:susceptibility_defined1}
\langle \delta q(t) \rangle = \chi_{}(\omega) F\exp(-i\omega t).
\end{align}

The susceptibility and the spectral density of fluctuations are related by the fluctuation-dissipation theorem, see Sec.~\ref{sec:FDT}, with Im~$\chi_{}(\omega)=(\omega/k_BT)S_{}(\omega)$ in the classical limit. In addition, the real and imaginary parts of the susceptibility,  Re~$\chi_{}(\omega)$  and Im~$\chi_{}(\omega)$, are related  by the Kramers-Kr\"onig relation~\cite{Landau1980}; Re~$\chi_{}(\omega)$  and Im~$\chi(\omega)$ determine, respectively, the in-phase and  out-of-phase components of the
force-induced displacement, whereas $|\chi(\omega)|$ determines the displacement amplitude.

Of particular interest for NVSs is the situation where the oscillator decay rate $\Gamma$ is small compared to the eigenfrequency $\omega_0$. In this case both $S_{}(\omega)$ and  Im~$\chi_{}(\omega)$ have sharp resonant peaks at frequency $\omega_0$. From Eq.~(\ref{eq:Brownian}) we find that near $\omega_0$ both functions have a Lorentzian peak,
\begin{align}
\label{eq:Lorentz_classical}
&S_{}(\omega) = \frac{k_BT}{M{}\omega_0^2}\frac{\Gamma}{\Gamma^2 + (\omega - \omega_0 )^2},\nonumber\\
&\chi_{}(\omega) = \frac{i}{2M{}\omega_0}[\Gamma - i(\omega-\omega_0)]^{-1}, \quad |\omega -\omega_0|\ll \omega_0.
\end{align}
Measuring the position and the halfwidth of the peak of $S_{}(\omega)$ and/or Im~$\chi_{}(\omega)$ is commonly used to determine  $\omega_0$ and $\Gamma$. It thus enables finding the  quality factor $Q$, which characterizes the energy relaxation and is conventionally defined as the ratio of the stored vibrational energy to the energy dissipated per cycle multiplied by $2\pi$,
\begin{align}
\label{eq:Q_definition}
Q=2\pi \frac{\text{stored energy}}{\text{dissipated energy per cycle}} \to \frac{\omega_0}{2\Gamma}.
\end{align}
This factor is independent of the vibration amplitude, for a linear mode.

A central role in the dynamics of NVS modes is played by the fluctuations of the mode eigenfrequencies. Such fluctuations can arise, for example, from the fluctuations of the resonator mass due to attachment or detachments of molecules, fluctuations of the spring constant due to charge and voltage fluctuations in the system, as well as from the interplay of the thermal fluctuations of the amplitude and the vibration nonlinearities. These frequency fluctuations increase the halfwidth of the spectral peak of $S_{}(\omega)$ and Im~$\chi_{}(\omega)$ and, generally, can change the shape of the spectrum. Therefore it is convenient to introduce an effective quality factor
\begin{align}
\label{eq:Q_omega}
Q_\omega= \omega_{0{}}/2\Delta\omega.
\end{align}
The parameter $\Delta\omega$ takes into account frequency fluctuations and replaces the decay rate $\Gamma$ in the standard expression (\ref{eq:Q_definition}). We use two different symbols, $Q$ and $Q_\mathrm{\omega}$, to emphasize this difference. Factor $Q_\omega$ does not describe energy relaxation but refers to the results of spectroscopic measurements. The physics behind frequency fluctuations is fairly rich and will be discussed at length in the review. However, to set the scene, in this section we will not consider them and will set $Q=Q_\omega$.  In  Tables \ref{table:ta}--\ref{table:tc} we provide the reported values of the quality factor in several types of nanomechanical systems.


\subsection{Most common types of nanomechanical resonators}
\label{subsec:types_of_resonators}

The first practical nanomechanical system was based on a silicon nanobeam \cite{Cleland1996,Cleland1998}. Since then a variety of nanomechanical resonators has been explored. Tables~\ref{table:ta} and \ref{table:tb} summarize the characteristics of resonators based on low-dimensional materials, such as carbon nanotubes, semiconductor nanowires, graphene, semiconductor layered membranes, and levitating particles. Table \ref{table:tc} summarizes those of resonators nanofabricated from bulk material using the top-down approach. In many of these examples the vibrational modes correspond to bending (flexural) modes of the resonator, a microscopic analog of the vibrations of a string or a membrane. Carbon nanotubes are the narrowest resonators; their diameters range typically between 1 and 3 nm. Graphene and semiconductor monolayers are the thinnest membranes as they are atomically thin. Other types of vibrational modes are investigated, such as torsional modes in mechanically based torque resonators~\cite{HAUER2013181}, the bulk (Lam\'e) modes in micromechanical resonators~\cite{Chandorkar2008} and layered nanomaterial resonators~\cite{Zalalutdinov2021}, and the localized modes in optomechanical crystals, see \cite{Eichenfield2009,MacCabe2020} and references therein.


Advances in fabrication have led to a steady improvement of the quality factor of the NVSs. Optimizing the surface of resonators turned out to be central. The microscopic nature of the dissipation due to surface defects is not clear, but it might be related to additional relaxation channels that open due to the surface  contamination and the degradation of crystallinity near the surface.
In particular, the quality factor of mechanical resonators based on nanotube and graphene becomes extremely large at cryogenic temperature \cite{Huttel2009}, reaching the range between one and ten million \cite{Moser2014,Guttinger2017,Urgell2020}. The demonstration of such large quality factors came as a surprise. For many years, researchers observed that quality factors would decrease with the volume of the resonator, and because of this trend it was unthinkable that nanotubes and graphene resonators could exhibit such giant quality factors. The large $Q$-factors reflect the high crystallinity of these nanoscale systems and show that surface contamination is reduced to a minimal amount.

Levitating particles feature large quality factors, up to 100 million, even at room temperature~\cite{Ricci2017}. In many experiments, the particles are trapped by a laser beam; the resonance frequency is given by the optical gradient force. Damping, which arises from collisions with the gas molecules in the sample chamber, becomes very low in high vacuum. Recently, particles have been levitated using new schemes, including ferromagnetic particles in Meissner traps made from superconducting materials~\cite{Gieseler2020,Vinante2020}, diamagnetic particles in magnetic traps~\cite{Hsu2016,OBrien2019}, and charged particles in Paul traps~\cite{Alda2016,Delord2020}.

High-stress silicon nitride is the material used by many groups to produce top-down resonators endowed with high quality factors~\cite{Kozinsky2006,Verbridge2006,Wilson-Rae2011,Unterreithmeier2009,Unterreithmeier2010a,Yu2012,Maillet2017,Zhou2019,Bothner2019}. Dissipation in the bulk of this high-stress material is remarkably low. The quality factor can be further enhanced by structuring high-stress silicon nitride films into trampoline geometries~\cite{Reinhardt2016,Norte2016} and nonuniform phononic crystal pattern~\cite{Tsaturyan2017, Ghadimi2018}. Engineering the shape of mechanical eigenmodes enables reducing losses near their supports. Quality factors as high as 800 million can be achieved at room temperature~\cite{Ghadimi2018} and $1.5$ billion at 30~mK \cite{Seis2021}.

An important role in studying nonlinear phenomena and the effects of mode coupling has been played by single- and coupled nanoresonators based on narrow nanowires grown by electrodeposition~\cite{Kozinsky2007}, thin nanofabricated structures~\cite{Defoort2015}, and suspended GaAs and AlN heterostructures that exploit piezoelectric effect to actuate and control the vibrations \cite{Masmanidis2007,Mahboob2008,Karabalin2009,Karabalin2009a,Karabalin2011,Yamaguchi2017}.

\begin{table*}[t]
\label{table:NVS_parameters}
\begin{center}
 \begin{tabular}{|p{2cm}|p{2cm}|p{2cm}|p{3cm}|p{4cm}|p{4cm}|}
 \hline
\multicolumn{6}{|c|}{{\bf Carbon nanotube}} \\
\hline
 $M$ (kg)& $k$ (N/m) & $\omega_0/2\pi$ (Hz) & $Q$ or $Q_\mathrm{\omega}$ & Description & Reference \\
 \hline\hline
   &   & $2.4\cdot10^{6}$ & $1.7\cdot10^{2}$ at 300~K & single-clamped multi-wall & \cite{Poncharal1999}\\
 \hline
  &  & $8.5\cdot10^{8}$ & 40 at 4~K & double-clamped single-wall bundle& \cite{Reulet2000} \\
 \hline
 $3.3\cdot 10^{-21}$  & $4\cdot10^{-4}$ & $5.5\cdot10^{7}$ & 80 at 300~K & double-clamped single-wall& \cite{Sazonova2004} \\
 \hline
 $4.8\cdot 10^{-22}$  & $1\cdot10^{-3}$ & $2.3\cdot10^{8}$ & $2\cdot10^{2}$ at 6~K & double-clamped single-wall& \cite{Chiu2008} \\
 \hline
  $5.3\cdot 10^{-21}$  & $2.7\cdot10^{-2}$ & $3.6\cdot10^{8}$ & $1.2\cdot10^{5}$ at 20~mK & double-clamped single-wall& \cite{Huttel2009} \\
\hline
     &  & $1.1\cdot10^{10}$ & $4.2\cdot10^{2}$ at 4~K & double-clamped single-wall& \cite{Chaste2011} \\
\hline
     &  & $3.9\cdot10^{10}$ & $3.3\cdot10^{4}$ at 0.1~K & double-clamped single-wall, device produced once& \cite{Laird2012} \\
 \hline
$5\cdot 10^{-19}$  & $5.1 \cdot10^{-6}$ & $5.1\cdot10^{5}$ & 250 at 300~K & double-clamped single-wall bundle & \cite{Stapfner2013}\\
\hline
$4.4\cdot 10^{-21}$ & $5.2\cdot10^{-4}$  & $5.5\cdot10^{7}$ & $4.8\cdot10^{6}$ at 30~mK & double-clamped single-wall & \cite{Moser2014}\\
 \hline
 $7.9\cdot 10^{-19}$ & $4.5\cdot10^{-8}$  & $3.8\cdot10^{4}$ & $2.2\cdot10^{3}$ at 300~K & single-clamped single-wall & \cite{Tavernarakis2018}\\
 \hline
 $2.7\cdot 10^{-21}$ & $8.9\cdot10^{-4}$  & $9.1\cdot10^{7}$ & $6.8\cdot10^{6}$ at 70~mK & double-clamped single-wall & \cite{Urgell2020}\\
 \hline\hline\hline

 \multicolumn{6}{|c|}{{\bf Semiconductor nanowire}} \\
\hline
 $M$ (kg)& $k$ (N/m) & $\omega_0/2\pi$ (Hz) & $Q$ or $Q_\mathrm{\omega}$ & Description & Reference \\
 \hline\hline
   $2.3\cdot 10^{-17}$  & 6.0 & $8.0\cdot10^{7}$ & $1.3\cdot10^{4}$ at 300~K &double-clamped Si wire& \cite{Feng2007} \\
   \hline
 $1.6\cdot 10^{-17}$  & $4.6\cdot10^{-2}$ & $8.5\cdot10^{6}$ & $1.0\cdot10^{3}$ at 300~K &single-clamped GaN wire& \cite{Henry2007} \\
 \hline
 $9.8\cdot 10^{-15}$  & $7.2\cdot10^{-4}$ & $4.3\cdot10^{4}$ & $1.6\cdot10^{5}$ at 300~K &single-clamped SiC wire& \cite{Perisanu2007} \\
 \hline
  $1.6\cdot 10^{-17}$  & $2.8\cdot10^{-5}$ & $2.1\cdot10^{5}$ & $1\cdot10^{4}$ at 300~K &single-clamped Si wire& \cite{Nichol2008} \\
 \hline
   $5.5\cdot 10^{-16}$ & 0.1 & $2.2\cdot10^{6}$ & $2\cdot10^{3}$ at 300~K &single-clamped Si wire& \cite{Gil-Santos2010} \\
 \hline
   $1.5\cdot 10^{-17}$ & 1 & $4.1\cdot10^{7}$ & $5\cdot10^{2}$ at 300~K &double-clamped Si wire& \cite{Sansa2012} \\
\hline
$3.5\cdot 10^{-16}$ & $1.5\cdot10^{-4}$& $1.1\cdot10^{5}$ & $2.9\cdot10^{3}$ at 300~K &single-clamped SiC wire& \cite{Gloppe2014} \\
\hline
$3.5\cdot 10^{-15}$ & $8.6\cdot10^{-2}$& $7.9\cdot10^{5}$ & $5.8\cdot10^{3}$ at 4.2~K &single-clamped GaAs/AlGaAs wire& \cite{Montinaro2014} \\
\hline
$1\cdot 10^{-16}$ & $14$& $5.9\cdot10^{7}$ & $2.8\cdot10^{3}$ at 16~K &double-clamped InAs wire& \cite{Mathew2015} \\
\hline
$5.9\cdot 10^{-17}$ & $6.8\cdot10^{-5}$& $1.7\cdot10^{5}$ & $5.9\cdot10^{4}$ at 4~K &single-clamped Si wire& \cite{Sahafi2020}\\
\hline

\end{tabular}
\caption{Figures of merit of mechanical resonators based on nanoscale systems}
\label{table:ta}
\end{center}
\end{table*}

\begin{table*}[t]
\begin{center}
 \begin{tabular}{|p{2cm}|p{2cm}|p{2cm}|p{3cm}|p{4cm}|p{4cm}|}
 \hline
\multicolumn{6}{|c|}{{\bf Graphene}} \\
\hline
 $M$ (kg)& $k$ (N/m) & $\omega_0/2\pi$ (Hz) & $Q$ or $Q_\mathrm{\omega}$ & Description & Reference \\
 \hline\hline
 $1.4\cdot 10^{-18}$  & $0.3$ & $7.0\cdot10^{7}$ & 78 at 300~K & double-clamped monolayer& \cite{Bunch2007} \\
 \hline
 $7.8\cdot 10^{-17}$  & $10$ & $5.7\cdot10^{7}$ & $3\cdot10^{3}$ at 300~K & multilayer graphene oxide drum& \cite{Robinson2008} \\
  \hline
 $2.2\cdot 10^{-18}$  & $1.4$ & $1.3\cdot10^{8}$ & $1.4\cdot10^{4}$ at 5~K & double-clamped monolayer& \cite{Chen2009} \\
  \hline
   & & $7.5\cdot10^{7}$ & $9\cdot10^{3}$ at 10~K & double-clamped monolayer& \cite{Zande2010}\\
   \hline
 $1.9\cdot 10^{-17}$  & $3$ & $6.4\cdot10^{7}$ & $2.5\cdot10^{2}$ at 300~K & double-clamped monolayer& \cite{Singh2010}\\
  \hline
 $3.9\cdot 10^{-19}$ &0.35  & $1.5\cdot10^{8}$ & $1.0\cdot10^{5}$ at 90~mK & double-clamped monolayer& \cite{Eichler2011a}\\
   \hline
   $2.2\cdot 10^{-17}$ & 2.8 & $5.7\cdot10^{7}$ & $1.4\cdot10^{3}$ at 4.2~K & double-clamped monolayer & \cite{Song2012}\\
 \hline
  $2.7\cdot 10^{-16}$  & $14$ & $3.6\cdot10^{7}$ & $2.2\cdot10^{5}$ at 14~mK & multilayer drum& \cite{Singh2014} \\
   \hline
  $7.9\cdot 10^{-16}$  & $6.5\cdot10^{-2}$ & $1.4\cdot10^{6}$ & $8.2\cdot10^{2}$ at 300~K & monolayer drum& \cite{Cole2015} \\
 \hline
  $9.6\cdot 10^{-18}$  & $0.8$ & $4.6\cdot10^{7}$ & $1.0\cdot10^{6}$ at 15~mK & multilayer drum& \cite{Guettinger2017} \\
 \hline\hline\hline

 \multicolumn{6}{|c|}{{\bf Semiconductor layer}} \\
\hline
 $M$ (kg)& $k$ (N/m) & $\omega_0/2\pi$ (Hz) & $Q$ or $Q_\mathrm{\omega}$ & Description & Reference \\
 \hline\hline
   &  & $2.6\cdot10^{7}$ & $1.1\cdot10^{2}$ at 300~K & monolayer MoS$_2$ drum& \cite{Castellanos-Gomez2013} \\
   \hline
     & & $2.0\cdot10^{7}$ & $7.1\cdot10^{2}$ at 300~K & multilayer MoS$_2$ drum& \cite{Lee2013a}\\
 \hline
     & & $2.2\cdot10^{7}$ & 41 at 300~K & monolayer MoS$_2$ drum & \cite{Leeuwen2014}\\
 \hline
     & & $4.1\cdot10^{7}$ & $6.9\cdot10^{2}$ at 300~K & double-clamped MoS$_2$ bilayer & \cite{Samanta2015}\\
 \hline
    $2.3\cdot 10^{-17}$ & 2.9 & $5.7\cdot10^{7}$ & $4.7\cdot10^{4}$ at 3.5~K & monolayer WSe$_2$ drum& \cite{Morell2016}\\
     \hline
  &  & $1.8\cdot10^{7}$ & 48 at 300~K & multilayer MoS$_2$ drum& \cite{Davidovikj2017} \\
  \hline
  &  & $8.4\cdot10^{6}$ & $2.1\cdot10^{2}$ at 300~K & multilayer black phosphorus drum& \cite{Islam2018} \\
   \hline
  $4.3\cdot 10^{-17}$& $2.2$ & $3.6\cdot10^{7}$ & $3.7\cdot10^{4}$ at 3~K & monolayer MoSe$_2$ drum& \cite{Morell2019} \\
\hline\hline\hline

 \multicolumn{6}{|c|}{{\bf Other layered crystals}} \\
\hline
 $M$ (kg)& $k$ (N/m) & $\omega_0/2\pi$ (Hz) & $Q$ or $Q_\mathrm{\omega}$ & Description & Reference \\
 \hline\hline
    $1.8\cdot 10^{-15}$  & $41$ & $2.4\cdot10^{7}$ & $2.1\cdot10^{2}$ at 10~K & double-clamped NbSe$_2$ multilayer & \cite{Sengupta2010} \\
  \hline
    $3.7\cdot 10^{-17}$  & $4.0$ & $5.2\cdot10^{7}$ & $2.4\cdot10^{5}$ at 15~mK & NbSe$_2$-graphene-NbSe$_2$ drum& \cite{Will2017} \\
  \hline
    $1.3\cdot 10^{-16}$  & $0.65$ & $11\cdot10^{7}$ &  & graphene-MoSe$_2$ drum& \cite{Kim2018} \\
\hline\hline\hline

 \multicolumn{6}{|c|}{{\bf Levitating particles}} \\
\hline
 $M$ (kg)& $k$ (N/m) & $\omega_0/2\pi$ (Hz) & $Q$ or $Q_\mathrm{\omega}$  & Description & Reference \\
 \hline\hline
    $3.1\cdot 10^{-14}$  & $1.1\cdot10^{-4}$  & $9.7\cdot10^{3}$ & $2.1\cdot10^{4}$ at 300~K & SiO$_2$ particle in optical trap & \cite{Li2011} \\
  \hline
    $3.1\cdot 10^{-18}$  & $1.7\cdot10^{-7}$  & $3.7\cdot10^{4}$ & $1\cdot10^{7}$ at 300~K & SiO$_2$ particle in optical trap & \cite{Gieseler2012} \\
  \hline
    $8.3\cdot 10^{-17}$  & $1.1\cdot10^{-6}$ & $1.8\cdot10^{4}$ & $1.8\cdot10^{4}$ at 300~K & SiO$_2$ particle in optical trap& \cite{Millen2015} \\
  \hline
  $\sim 1\cdot 10^{-18}$  & $\sim 2\cdot10^{-10}$ & $2.1\cdot10^{3}$ & $1.1 \cdot10^{3}$ at 300~K & charged particle in Paul trap& \cite{Conangla2018} \\
  \hline
    $\sim 1\cdot 10^{-13}$  & $\sim 7\cdot10^{-7}$ & $4.1\cdot10^{2}$ & $9.2 \cdot10^{1}$ at 542~K & diamagnetic particle in magnetic trap& \cite{OBrien2019} \\
  \hline
    $\sim 6\cdot 10^{-10}$  & $\sim 7\cdot10^{-5}$ & $5.6\cdot10^{1}$ & $2.1 \cdot10^{6}$ at 4.2~K & ferromagnetic particle in Meissner trap& \cite{Vinante2020} \\
  \hline

\end{tabular}

\caption{Figures of merit of mechanical resonators based on nanoscale systems}
\label{table:tb}
\end{center}
\end{table*}

\begin{table*}[t]
\begin{center}
 \begin{tabular}{|p{2cm}|p{2cm}|p{2cm}|p{3cm}|p{4cm}|p{4cm}|}
 \hline
\multicolumn{6}{|c|}{{\bf Top-down nanofabricated resonators}} \\
\hline
 $M$ (kg)& $k$ (N/m) & $\omega_0/2\pi$ (Hz) & $Q$ or $Q_\mathrm{\omega}$ & Description & Reference \\
 \hline\hline
 $5.6\cdot 10^{-14}$  & $6.5\cdot 10^{-6}$ & $1.7\cdot10^{3}$ & $6.7\cdot10^{3}$ at 4.8~K & single-clamped Si beam& \cite{Stowe1997} \\
 \hline
 $3.4\cdot 10^{-17}$  & $1.3\cdot 10^{3}$ & $1.0\cdot10^{9}$ & $5\cdot10^{2}$ at 4.2~K & double-clamped SiC beam& \cite{Huang2003} \\
  \hline
 $1.9\cdot 10^{-15}$  & $1.5$ & $4.5\cdot10^{6}$ & $2.1\cdot10^{5}$ at 300~K & double-clamped Si$_3$N$_4$ beam&    \cite{Verbridge2006} \\
   \hline
 $1.4\cdot 10^{-10}$  & $1\cdot 10^{2}$ & $1.4\cdot10^{5}$ & $1.1\cdot10^{5}$ at 2.5~K & double-clamped GaAs/AlGaAs resonator&   \cite{Mahboob2008} \\
   \hline
 $3.3\cdot 10^{-16}$  & $6.3\cdot 10^{4}$ & $2.2\cdot10^{9}$ & $2.7\cdot10^{3}$ at 300~K & optomechanical Si crystal& \cite{Eichenfield2009} \\
 \hline
 $7.7\cdot 10^{-15}$  & $1.2\cdot 10^{8}$ & $2\cdot10^{10}$ & $\sim 1\cdot10^{3}$ at 300~K & GaAs/AlAs microcavity& \cite{Fainstein2013} \\
 \hline
 $1.6\cdot 10^{-12}$  & $6.7\cdot 10^{-2}$ & $3.2\cdot10^{4}$ & $1.5\cdot10^{6}$ at 3~K & single-clamped diamond beam&  \cite{Tao2014} \\
 \hline
 $4\cdot 10^{-12}$ & $3\cdot10^{-1}$  & $4.1\cdot10^{4}$ & $4.5\cdot10^{7}$ at 300~K & Si$_3$N$_4$ trampoline&  \cite{Reinhardt2016} \\
  \hline
 &  & $1.4\cdot10^{5}$ & $1\cdot10^{8}$ at 300~K & Si$_3$N$_4$ tethered membrane&   \cite{Norte2016} \\
  \hline
 &  & $1.4\cdot10^{7}$ & $4.3\cdot10^{4}$ at 25~mK & torque Si resonator& \cite{Kim2016} \\
 \hline
 $1.6\cdot 10^{-11}$ & $3.7\cdot10^{2}$  & $7.7\cdot10^{5}$ & $2.1\cdot10^{8}$ at 300~K & Si$_3$N$_4$ membrane with engineered mode& \cite{Tsaturyan2017} \\
 \hline
 $4.1\cdot 10^{-15}$  & $1.1\cdot 10^{-4}$ & $2.5\cdot10^{4}$ & $1.6\cdot10^{5}$ at 0.14~K & single-clamped diamond ladder&   \cite{Heritier2018} \\
 \hline
  $\sim 5\cdot 10^{-15}$  & $\sim 1$ & $\sim 2.5\cdot10^{6}$ & $\sim 1\cdot10^{8}$ at 300~K &Si$_3$N$_4$ nanobeam with engineered mode&    \cite{Ghadimi2018} \\
 \hline
  &  & $1.0\cdot10^{7}$ & $2.6\cdot10^{6}$ at 300~K & Lam\'e-mode Si resonator&   \cite{Rodriguez2019} \\
 \hline
$1.3\cdot 10^{-16}$  & $1.5\cdot 10^{5}$ & $5.3\cdot10^{9}$ & $4.9\cdot10^{10}$ at 7~mK & optomechanical Si crystal&   \cite{MacCabe2020} \\
 \hline

\end{tabular}
\caption{Figures of merit of mechanical resonators fabricated from bulk material with the top-down approach. }
\label{table:tc}
\end{center}
\end{table*}


\subsection{Driving resonators}
\label{subsec:driving}

The most straightforward way to excite nanomechanical resonators is by driving them with a directly applied force at an angular frequency $\omega_F$ close to the eigenfrequency  $\omega_{0{}}$. Due to nonlinearity
of the system, such a direct drive can also efficiently excite vibrations when $\omega_F\simeq \omega_{0{}} \cdot N$ or $\omega_F\simeq \omega_{0{}}/N$ with an integer $N>1$, but
their amplitude is usually smaller for the same drive amplitude~\cite{Nayfeh2004}. Another method that is often employed consists in modulating the spring constant~\cite{Rugar1991,Turner1998}. Such parametric driving is equivalent to the modulation of the resonant frequency. It is most efficient when the drive frequency is close to $2\omega_{0{}}$. The vibration amplitude is very nonlinear in the amplitude of the parametric drive; the amplitude remains small until the driving amplitude reaches a threshold value, after which  the resonator vibrates at half the drive frequency

\begin{figure}[h]
	 \includegraphics[scale=0.8]{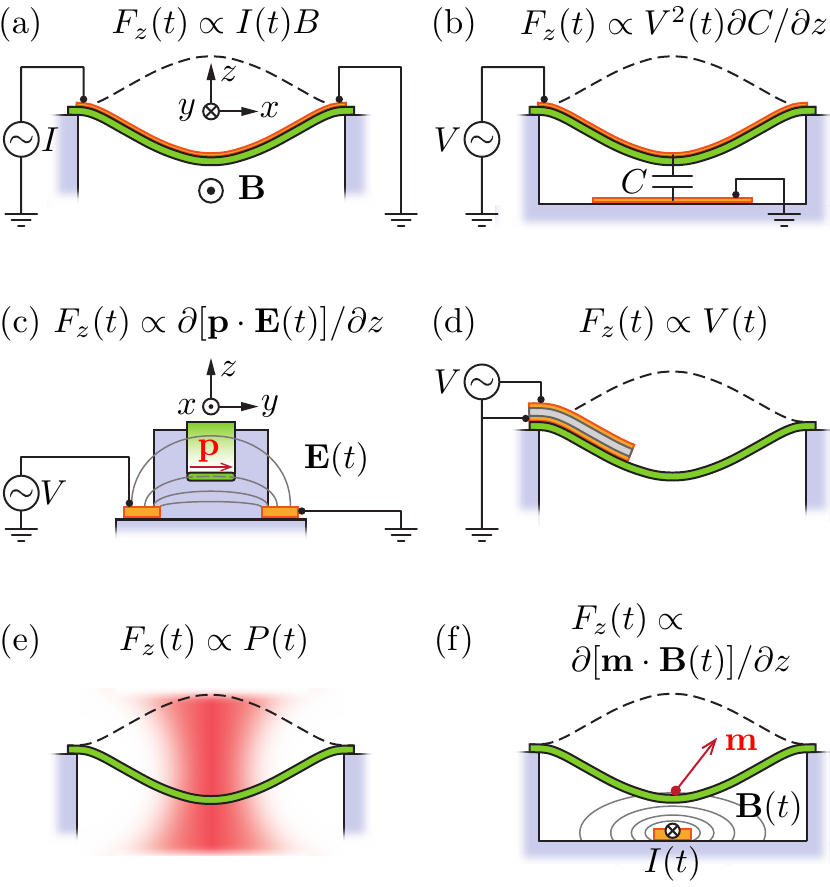}
	\caption{Excitation of flexural vibrations. (a) Magnetomotive force, which is generated by a static magnetic field and an alternating current along the mechanical beam (shown in green). (b) Capacitive force, which is created by applying an oscillating voltage between the bottom gate electrode and the mechanical beam (green) that is conducting. (c) Dielectric force, which is acting on a mechanical beam (green) made from a dielectric material using a time-modulated electric field gradient. The figure is a cross section of the device (see the coordinate frame). (d) Piezo-electric force, where the mechanical beam (green) is actuated with a piezo-electric film (grey) sandwiched between two metal electrodes. (e) Optical force, which is obtained by modulating the intensity of the laser beam (red) focused on the mechanical beam (green). (f) Magnetic force, where either the magnetic moment of the mechanical beam (green) or the magnetic field gradient is modulated (or both).}
	\label{fig:differentforces}
\end{figure}

The actuation of a nanoscale resonator was achieved for the first time with the magnetomotive drive (Fig.~\ref{fig:differentforces}a) \cite{Cleland1996,Cleland1998}. In this method, flexural vibrations of a conducting nanobeam are excited by the Lorentz force, which emerges when an alternating current is applied through the nanowire placed in a perpendicular magnetic field. The vibrations are detected by measuring the electromotive force generated along the length of the nanobeam using a network analyzer.

A widely used drive is the capacitive force, see Fig.~\ref{fig:differentforces}b~\cite{Rugar1991}. It is implemented by applying a static voltage $V_\mathrm{g}^\mathrm{DC}$ and an oscillating voltage with a comparatively low amplitude $\delta V_\mathrm{g}^\mathrm{AC}$ between the resonator and a nearby gate electrode. The force is
\begin{equation}
	\label{eq:capacitiveforce}
	F^\mathrm{AC}= C^{\prime}_\mathrm{g}V_\mathrm{g}^\mathrm{DC}\delta V_\mathrm{g}^\mathrm{AC}
\end{equation}\\
where $C^{\prime}_\mathrm{g}\equiv \partial C_\mathrm{g}/\partial q$ is the derivative of the gate-resonator capacitance with respect to the mode displacement $q$. As a rough estimate, $C'_{\rm g}$ is given by the ratio of the capacitance to the characteristic distance between the resonator and the gate. For simplicity, we omit the work function difference between the resonator and the gate electrode in Eq.~(\ref{eq:capacitiveforce}), which leads to an offset in $V_\mathrm{g}^\mathrm{DC}$. This driving force is effective when the resonator is an electrical conductor or is covered with a metal plate. A resonator made of a dielectric material can be driven by applying an electric field gradient between two electrodes structured near the resonator (Fig.~\ref{fig:differentforces}c) \cite{Unterreithmeier2009}. A dielectric material in vacuum moves towards the region with highest electric field. This dielectric force has the same origin as the capacitive force, since they are both related to the displacement-induced electrostatic energy gain of a capacitor. When the resonator is made of a piezo-electric material, a  piezo-electric force is created by applying an electric field between two electrodes, which are usually patterned on the resonator itself (Fig.~\ref{fig:differentforces}d) \cite{Masmanidis2007,Mahboob2008}.

Optical drive is also frequently used (Fig.~\ref{fig:differentforces}e). Photothermal forces are straightforward to apply to nanoscale resonators. One needs simply to modulate the intensity of the laser focused on the resonator. Absorption-induced heating displaces the resonator through thermal expansion \cite{Bunch2007}. Care has to be taken if one wants to keep the resonator temperature low. This requires weak absorption. When the absorption is sufficiently suppressed, the force from the heating is overcome by the force traditionally associated with radiation pressure that comes from the reflection of photons by the resonator. This force is often small, but it has the advantage of not causing heating. It is used in quantum optomechanics experiments where losses are detrimental to the manipulation of quantum states~\cite{Aspelmeyer2014a}. Various optical excitation protocols developed by now enable driving different types of NVS vibrations, including the vibrations localized at defects in phononic optomechanical crystals, see \cite{MacCabe2020} and references therein.

Magnetic forces are usually used in magnetic resonance force microscopy experiments to detect electron and nuclear spins \cite{Poggio2010}. In some of these experiments, the spins located on the mechanical resonators are periodically flipped using magnetic resonance techniques; the associated time-modulated magnetic moment together with the magnetic field gradient of a nearby magnet results in an oscillating force (Fig.~\ref{fig:differentforces}f). Alternatively, the magnet is placed on the mechanical resonator and the spins on the surface of a chip.


\subsection{Frequency control}
\label{subsec:freq_tunability}

The eigenfrequency $\omega_0$ of a flexural mode of an NVS can be tuned by direct forces in two different ways. The frequency depends on the static gradient of the force. In the case of the capacitive force, the shift of the spring constant is given by
\begin{equation}
	\label{eq:capacitivesoftening}
	\Delta k= -\frac{\partial F}{\partial q} =-\frac{1}{2}C^{\prime\prime}_\mathrm{g}(V_\mathrm{g}^\mathrm{DC})^2.
\end{equation}\\
Usually $C^{\prime\prime}_\mathrm{g}\equiv \partial^2 C_\mathrm{g}/\partial q^2 > 0$ and a dc gate voltage directly leads to an electrostatic softening of a resonator \cite{Kozinsky2006,Solanki2010,Eichler2011a,Wu2011a,Stiller2013}.

In addition, the eigenfrequency can be tuned by the dc force via the change of the equilibrium position and the associated elongation of the resonator. Through nonlinear elasticity, such a change modifies the mechanical tension in the resonator. This effect leads to an increase of the frequency. The mechanism is broadly used with the capacitive force for frequency control in soft resonators, such as carbon nanotubes~\cite{Purcell2002,Sazonova2004,Rechnitz2021}. Piezo-actuators can also be used to tune $\omega_{0{}}$ by enhancing the separation between the supports of a doubly clamped resonator; using this technique, the  eigenfrequency of  a carbon nanotube could  be increased more than 20 times \cite{Ning2014}.

Time-dependent frequency control is important for various applications, such as parametric drive. In the case of frequency control with the capacitive force, the eigenfrequency can be modulated in time using a combination of dc and ac gate voltages {\cite{Rugar1991}}. An alternative and highly efficient approach is based on time-dependent piezo-electric actuation \cite{Yamaguchi2017}.


\subsection{Detection of displacement}
\label{subsec:detection}

The detection of motion becomes increasingly difficult when resonators get smaller. A variety of methods has been developed to address the problem. The most broadly used methods are based on magnetomotive, capacitive, and optical or microwave measurements. Other methods include piezo-electric~\cite{Masmanidis2007,Mahboob2008}, piezo-resistive \cite{He2008,Lee2013b}, scanning probe microscopy~\cite{Garcia-Sanchez2007}, scanning and transmission electron microscopy~\cite{Buks2001a,Nigues2015,Tsioutsios2017}, and field-emission~\cite{Purcell2002} measurements.

The capacitive method relies on the motion-induced modulation of the capacitance $\delta C_\mathrm{g}$ between a conducting resonator and a nearby gate.
Modulation of the capacitance in the presence of a gate voltage $V_\mathrm{g}^\mathrm{DC}$ leads to the charge modulation $\delta Q=\delta C_\mathrm{g} \cdot V_\mathrm{g}^\mathrm{DC}$. Where the vibrating resonator acts as a field-effect transistor, its conductance oscillates as $\delta G=\frac{\partial G }{\partial Q }\delta Q$. This causes a change of the current through the resonator $\delta I=\delta G\cdot \delta V_\mathrm{sd}$ in response to a voltage $\delta V_\mathrm{sd}$ applied across it. When $\delta V_\mathrm{sd}$ is  oscillating at frequency $\omega_\mathrm{sd}$ close to $\omega_{0{}}$, the current can be conveniently measured at a low-frequency $\approx \mid \omega_\mathrm{sd}-\omega_{0{}}\mid$. This method was first applied to a GaAs resonator supporting a single-electron transistor made of aluminium~\cite{Knobel2003} and to nanotube resonators~\cite{Sazonova2004,Gouttenoire2010}. Yet another way to measure the oscillating capacitance is based on embedding a conducting nanoresonator into a superconducting cavity and measuring its radio-frequency reflection and transmission~\cite{Teufel2008,Rocheleau2010,Weber2014,Singh2014,Song2014a,Blien2020}. A similar method can be used with a dielectric resonator by integrating it with a nearby electric cavity~\cite{Faust2012}.

In the optical frequency range, high-precision vibration detection is often based on interferometry. Vibrations of a nanoresonator embedded in an optical cavity can be detected from the modulation of the resonant frequency of the cavity measured by its transmission or reflection \cite{Aspelmeyer2014a}.
Another method is based on focusing a laser beam onto the resonator and detecting the modulation of the reflected or scattered light \cite{Carr1997,Yeo2013,Gloppe2014}. Here we describe in more details  optical detection of the vibrations of graphene-based resonators, a method used by many groups to detect vibrations of monolayer- and few-layer systems \cite{Barton2012}. The laser beam illuminates a graphene membrane suspended over a metal gate. When the graphene monolayer is covered by an adsorbed contamination layer, an optical cavity is created between the monolayer and the gate, and the vibrations modulate this cavity. In contrast, in the case of a clean monolayer, a standing wave is formed by the interfering incident and reflected laser beams with the reflection from the metal gate. A clean graphene monolayer does not affect the standing wave much, since its reflection coefficient is small. In contrast, the absorption coefficient of graphene is about 0.02 for visible light, which is a remarkably large value considering that the material is only one atom thick. The displacement of the graphene layer changes the absorbed intensity. Therefore, the motion of the graphene resonator can be measured by recording the reflected light intensity.

Optical detection of micro-mechanical systems has attracted much interest in the context of improving the sensitivity of LIGO detectors of gravitational waves, as these detectors are also based on vibrational systems. One of the important challenges is reducing the photon counting noise and the photon radiation pressure noise. It can be addressed using squeezed states of light, and the experimental results on measuring a micromechanical membrane with squeezed light have been recently reported \cite{Kleybolte2020}.

\subsubsection{Quantum regime}
\label{subsubsec:q_detection}

Significant progress has been made on detecting small displacements that occur where a nano- or micro-scale vibrational system is in the quantum regime, in which case the vibration amplitude is $\propto \hbar^{1/2}$. Detecting vibrations and characterizing their energy distribution in this regime is one of the central problems of quantum optomechanics. An important approach is based on illuminating an optical or a microwave cavity that contains a nanoresonator and measuring the spectrum of the emitted radiation. This spectrum contains lines shifted from the incident light frequency by the frequency of the vibrational mode. The ratio of the intensities of the lower- and higher-frequency lines, i.e., of the Stokes and anti-Stokes components, is determined by the effective temperature of the mode \cite{Kippenberg2008,Teufel2011,Aspelmeyer2014a,Riedinger2016,Pfeifer2016,Clark2017,Reed2017,Tebbenjohanns2020,Delic2020}. A detailed information about the vibrational quantum state of mechanical modes can be obtained by connecting these modes to a qubit \cite{O'Connell2010,Satzinger2018,Chu2018,Arrangoiz-Arriola2019}.

Mechanical resonators in the quantum regime open new directions of research, including quantum squeezing of the mechanical motion \cite{Wollman2015,Pirkkalainen2015a,Lecocq2015}, measurement-based quantum control of mechanical motion \cite{Wilson2015,Rossi2018a}, quantum backaction evading measurements \cite{Suh2014,OckeloenKorppi2016,Moeller2017}, entanglement between mechanical resonators \cite{Riedinger2018,Ockeloen-Korppi2018,LepinayMercier2021,Kotler2021,Wollack2021a}, and fundamental measurements with levitating particles cooled to the ground state \cite{Delic2020,Magrini2021,Tebbenjohanns2021}.


\subsection{Measurement of the spectral response}
\label{subsec:measuring_spectra}

The measurement of the spectral response is the most common method to study mechanical resonators.  This response provides the important characteristics of a resonator. They include the resonant frequencies of the eigenmodes, their spectral bandwidth $\Delta\omega$, and the quality factor $Q_\omega$ introduced in Eq.~(\ref{eq:Q_omega}). There are two major approaches to measuring the spectral response. One of them is based on  applying an oscillating force to the resonator, sweeping the frequency of the force, and measuring the displacement of the resonator with a lock-in amplifier. This gives the mechanical susceptibility, $\chi_{}(\omega)$, as discussed in Sec.~\ref{subsec:phenomenology_linear}.

\begin{figure}[t]
\includegraphics[scale=0.8]{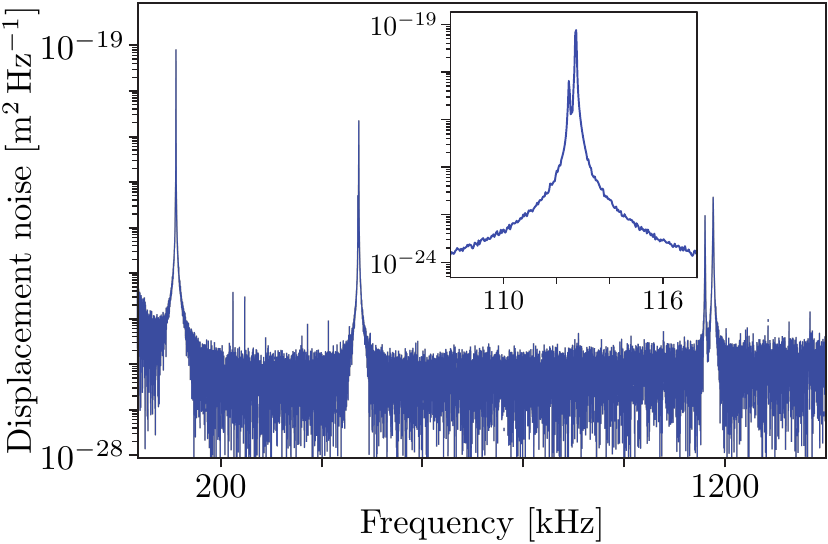}
	\caption{Power spectrum $S_{}(\omega)$ of a single-clampeded resonator based on a SiC nanowire measured optically at 300~K. The spectrum formed by the three lowest-frequency eigenmode ``families'' are shown. Each mode family is composed of two peaks (see inset), which correspond to the two eigenmodes that vibrate in perpendicular directions. The peaks allow determining the mode eigenfrequencies and the $Q_\omega$ factors. The noise floor of the spectrum quantifies the displacement sensitivity of the detection. Adapted from \cite{Gloppe2014}.}
	\label{fig:NoiseSpectrumSiCResonator}
\end{figure}

The other  approach is to measure the power spectrum $S_{}(\omega)$, i.e., to measure the  spectrum of the displacement that results from thermal and quantum fluctuations with no regular external force applied. The power spectrum is measured by simply feeding the output signal of the resonator detector into a spectrum analyser. In practice, the power spectrum is often more difficult to measure than the spectrum of response to a periodic drive, especially when no care is taken to reduce the technical noise in the measurement circuit. Figure~\ref{fig:NoiseSpectrumSiCResonator} shows the measured power spectrum of a cantilever based on a SiC nanowire featuring different eigenmodes.

\section{Sensing and clocks}
\label{sec:sensing}
Nanomechanical resonators have attracted considerable interest due to remarkable sensing capabilities. Because of their small mass, such resonators are exquisite sensors of external forces and mass deposition. We describe here the basics of sensing. Nanomechanical resonators are also used as clocks in commercial products.

\subsection{Force sensing}
\label{sec:forcesensing}
Force sensing consists in converting a weak force into a displacement that is then measured by electrical or optical means. The linear response of the displacement $q(t)$ to an external force $F\cos\omega t$ is given by the mechanical susceptibility $\chi_{}(\omega)$. If the noise is disregarded, the force-induced displacement increment is $\delta q(t) = \mathrm{Re} [\chi(\omega)F\exp(-i\omega t)]$. The detection of an oscillating force is optimized by matching its frequency to the mechanical eigenfrequency, where $|\chi(\omega)|$ is maximal.

\begin{figure}[t]
\includegraphics[scale=0.8]{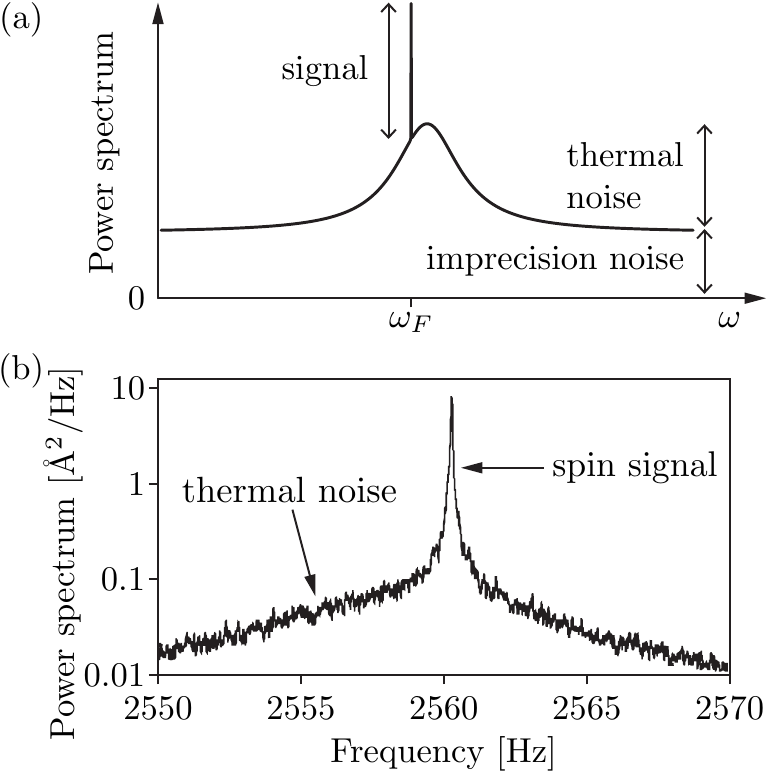}
	\caption{(a) Sensing of a coherent external force using spectral measurements. The power spectrum has contributions from driving-induced displacement, thermal vibrations, and the imprecision in the vibration detection. The signal to be detected is the narrow peak at the frequency of the external force. The integration time $\tau$ required to resolve this peak can be obtained by setting the resolution bandwidth $\Delta f_\mathrm{RBW}$ of the spectrum analyzer so that the signal-to-noise ratio is one, $\tau=1/\Delta f_\mathrm{RBW}$. The peak height of the coherent external force in the spectrum gets larger as $\Delta f_\mathrm{RBW}$ is reduced. The effect of frequency noise is disregarded. (b) Power spectrum of the vibrations of a microcantilever in the force sensing of nuclear spin experiments. The signal induced by the periodically driven spins is the narrow peak. The imprecision in the vibration detection is negligible in this experiment. Figure adapted from \cite{Poggio2007}.
	}
	\label{fig:sensing}
\end{figure}

The measured displacement is a superposition of the displacement induced by the force and thermal vibrations, i.e., a superposition of the signal and  noise, respectively (Fig.~\ref{fig:sensing}a). The fundamental limit of force sensing is set by the condition that the signal exceeds the thermal noise. Quantitatively, this limit can be characterized by noting that, by its meaning, the power spectrum of the mode $S(\omega)$ is the mean-square thermal displacement at frequency $\omega$ per unit frequency. At the same time, it follows from the definition of the susceptibility $\chi(\omega)$, Eq.~(\ref{eq:susceptibility_defined1}), that the squared displacement induced by a regular force $F\cos\omega t$ and averaged over the force period is $\overline{(\delta q)^2} = |\chi(\omega)|^2F^2/2$. The ratio of these displacements  $S_{F}(\omega) =2S_{}(\omega)/|\chi_{}(\omega)|^2$ shows how strong should the force be so that it  be can be detected with a signal-to-noise ratio equal to 1.

The ratio $S_F(\omega)$ takes a particularly simple form in the important case where the power spectrum of the vibrations has a Lorentzian peak at the vibration eigenfrequency $\omega_0$ with halfwidth $\Delta\omega \ll \omega_0$, see Sec.~\ref{subsec:eigenfrequency_fluctuations}. For such a spectrum, to the leading order in $\Delta\omega/\omega_0$ the resonant susceptibility is
\begin{align}
\label{eq:simple_Lorentzian}
\chi_{}(\omega) =  \frac{i}{2M{}\omega_0}[\Delta\omega - i(\omega-\omega_0)]^{-1}, \quad \Delta\omega, |\omega-\omega_0| \ll \omega_0,
\end{align}
cf. Eq.~(\ref{eq:Lorentz_classical}). Then in  the same range $|\omega-\omega_0|\ll \omega_0$
\begin{equation}
S_{F}(\omega) \approx 8k_\mathrm{B}TM \Delta\omega\equiv \frac{ 4k_\mathrm{B}TM{}\omega_0}{Q_\omega}. 
\label{eq:forcesensitivity}
\end{equation}
The function $S_{F}(\omega)$ is independent of frequency over the whole range of the peak in the spectrum of the resonator, that is, beyond the region $|\omega-\omega_0|\lesssim \Delta\omega$. Such broad-band force sensing is somewhat non-intuitive, given that the force response of the displacement is strongly frequency-dependent. We emphasize that force sensing can be comparatively broadband even for high-$Q$ resonators. Forces with frequencies far away from other resonances, can be detected provided that thermal noise and imprecision noise in the displacement detection are low enough.

Equation (\ref{eq:forcesensitivity}) suggests a strategy for detecting tiny forces. The best force sensitivity has been achieved with carbon nanotubes, which are the operating resonators with the smallest mass, $\sqrt{S_{F}}=(4.3\pm 2.9) \times 10^{-21} \mathrm{N/\sqrt{Hz}}$ \cite{Bonis2018}. Resonators based on silicon carbide nanowires can reach $\sqrt{S_{F}}=4.0 \times 10^{-20}\mathrm{N/\sqrt{Hz}}$ \cite{Fogliano2021}. Resonators microfabricated from bulk material are often more handy to use as sensors; their sensitivity
has reached $\sqrt{S_{F}}=(1.9\pm 0.6) \times 10^{-19} \mathrm{N/\sqrt{Hz}}$ \cite{Heritier2018}. Fluctuations of the nanoresonator frequency worsen the force sensitivity in Eq.~\ref{eq:forcesensitivity}, as they widen the mechanical linewidth and decrease the associated quality factor $Q_{\omega}$ \cite{Moser2013}.

The transduction of the mechanical vibrations into a measurable electrical or optical output signal is challenging with small resonators. The transduction can deteriorate the force sensitivity by adding technical noise. For instance, it contributes $\sim$14\% to the force sensitivity of the nanotube studied by \textcite{Bonis2018}. This so-called imprecision noise is quantified by the noise floor in the spectral measurement of a resonator (Fig.~\ref{fig:sensing}a).

Figure~\ref{fig:sensing}b shows a force sensing experiment of nuclear spins \cite{Poggio2007}. The nuclear spins of $^{19}$F atoms in a CaF$_2$ crystal are detected by attaching the crystal to a microcantilever. The nuclear spins are flipped back and forth at the mechanical resonance frequency.  The associated time-modulated magnetic moment together with an applied magnetic field gradient creates a force that drives the microcantilever. The measured power spectrum of the vibrations in Fig.~\ref{fig:sensing}b shows a narrow peak associated with the spins on top of the broad resonance of the thermal vibrations of the microcantilever. The goal of such experiments is to achieve magnetic resonance imaging with atomic resolution~\cite{Degen2009,Nichol2013}.

Force sensing has been used with great success in recent advances in various fields. These include the Casimir force \cite{Chan2001,Chan2008,Klimchitskaya2009,Liu2021b,Gong2021}, nanomagnetism~\cite{Forstner2012,Losby2015,Rossi2019}, scanning probe microscopy imaging of vectorial forces~\cite{Li2007,Lepinay2016,Rossi2016}, light-matter interaction~\cite{Gloppe2014}, persistent currents in normal metal rings~\cite{Bleszynski-Jayich2009}, and detecting a phonon flux in superfluid $^{4}\mathrm{He}$~\cite{Guenault2020}.

\subsection{Mass sensing}
\label{sec:masssensing}
\begin{figure}[t]
\includegraphics[scale=0.9]{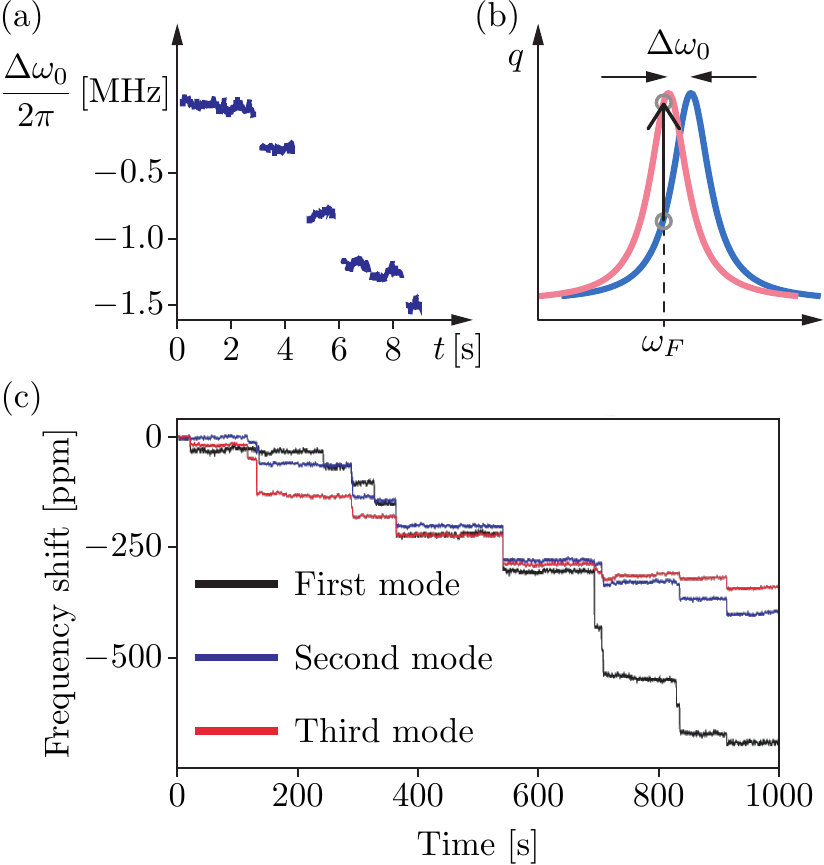}
	\caption{(a) Sensing of the adsorption of naphthalene molecules onto a nanotube resonator by monitoring the resonance frequency as a function of time at 4.3~K. Each shift in the resonance frequency is associated with adsorption of one molecule. The suspended nanotube is $\sim 150$~nm long and its fundamental mode vibrates at 1.8~GHz. The figure is adapted from ~\cite{Chaste2012}. (b) Mass sensing associated with one adsorption event. The resonator is driven at the frequency $\omega_F$. The change in the resonance frequency due to the adsorption event results in the increase of the driven vibration amplitude. (c) Real-time record of the eigenfrequencies of the first three flexural modes of the microcantilever exposed to a flux of gold nanoparticles. Adapted from  \cite{Malvar2016}.  Similar measurements were carried out with  protein macromolecules \cite{Hanay2012} and \textit{E. coli} bacteria \cite{Malvar2016}.
}
	\label{fig:masssensing}
\end{figure}

Mass sensing relies on monitoring how the eigenfrequency of a nanomechanical resonator changes when an additional mass is adsorbed onto its surface. Most experiments are done with flexural modes \cite{Ekinci2004,Yang2006}. If one assumes that this mass $m_{\rm add}$ is uniformly distributed over the resonator, the resonator itself is uniform, and $m_{\rm add}$ is small compared to the resonator mass $M_\mathrm{NVS}{}$,  then the relative change of the resonator eigenfrequency is
\begin{equation}
\delta \omega_0 /\omega_0 =-m_{\rm add}/2M_\mathrm{NVS}.
\label{eq:masssensing}
\end{equation}
As an example, Fig.~\ref{fig:masssensing}a shows a series of downward shifts in $\omega_0$ consistent with single adsorption events of naphthalene molecules onto a nanotube resonator.

There are different methods of monitoring the eigenfrequency. The simplest one relies on driving the resonator slightly off resonance while recording the vibration amplitude with a lock-in amplifier. A change in the eigenfrequency results in a change of the displacement amplitude $A$  (Fig.~\ref{fig:masssensing}b). The resonator settles to new amplitude and phase over the decay time. The frequency shift can be quantified from the measured change of $A$ using the slope of the mechanical susceptibility at the drive frequency. Implied in the analysis is that the adsorbed mass does not change over the duration of the measurement, which itself exceeds the decay time; however, the analysis can be also extended to the case where adsorbates attach and detach with a rate comparable to the relaxation rate  \cite{Dykman2010a}.

Phase-locked loop measurements are also often used to track the resonance frequency in mass sensing experiments. The method was first developed for high-$Q$ cantilevers in atomic force microscopy \cite{Albrecht1991}. It was used for faster detection of the frequency shift compared to the method discussed in the previous paragraph. However, a phase-locked loop is efficient when the imprecision noise in the detection of the displacement is small compared to the driven vibration amplitude; this is often hard to achieve for small resonators, such as nanotube resonators measured capacitively.

Mass spectrometry of molecules and nanoparticles requires overcoming the assumption of the uniform distribution of the adsorbed mass over the resonator. This can be achieved by tracking the resonance frequency of several flexural eigenmodes~\cite{Hanay2012,Hanay2015}. The method utilizes the fact that the resonance frequency shift induced by an adsorbed molecule depends both on its mass and its position on the resonator. The frequency shift of an $n$th eigenmode due to a particle attached to a point $\Rb$  on the surface of a nanoresonator is
\begin{align}
\label{eq:Hanay_shift}
\delta\omega_n(\Rb)/\omega_n =-m_{\rm add}\boldsymbol{\varphi}_n^2(\Rb)/2M_\mathrm{NVS}.
\end{align}
Here, $\boldsymbol{\varphi}_n(\Rb)$ is the dimensionless displacement of the resonator at the point $\Rb$ due to the $n$th eigenmode, see Eq.\ref{eq:mode_displacement}.

The frequency shift is the largest when the molecule is located at the antinode of the eigenmode, where its vibration amplitude is maximal, while it is equal to zero when the molecule sits at a node. The more modes can be measured, the better is the resolution of the mass of the absorbed molecule and of its location.

Figure~\ref{fig:masssensing}c shows how the measured  eigenfrequencies of a microcantilever get reduced due to adsorption of 100~nm diameter gold nanoparticles \cite{Malvar2016}. Quasi-instantaneous jumps are simultaneously observed in the frequencies of the first three flexural modes when a single nanoparticle is adsorbed. The frequency shift is different for the three modes, since it depends on the nanoparticle location. The data analysis gives an average nanoparticle mass of $11.6 \pm 3.8$~fg.

The exceptional sensing capabilities of mechanical resonators enabled advances in many directions. These include characterizing analytes using mass spectrometry~\cite{Hanay2015,Malvar2016,Sage2018}, the detection of large-mass biological particles that cannot be probed with commercial mass spectrometers based on mass-to-charge ratio measurements~\cite{Dominguez-Medina2018}, weighting biomolecules and single cells in fluid using microchannels integrated in resonators~\cite{Burg2007}, gas chromatography~\cite{Li2010a,Venkatasubramanian2016}, probing density and viscosity of liquids~\cite{Gil-Santos2015}, the diffusion of adsorbed atoms on the surface of a resonator~\cite{Yang2011}, the formation and transitions between solid- and liquid-phase monolayers of adsorbed atoms~\cite{Wang2010,Tavernarakis2014}, properties of helium superfluid thin films~\cite{Noury2019,Sachkou2019,Guenault2019}, and in-situ nanofabrication~\cite{Gruber2019}.

The mass resolution of a resonator is limited by frequency fluctuations.
As we will see in Sec.~\ref{sec:frequency_fluctuations}, there is a fundamental limit on how small frequency fluctuations and the mass resolution can be. This fundamental limit in  the linear regime is related to the thermal displacement noise. It was found that, for some resonators, to improve mass sensing in a linear regime one may want to decrease the $Q$ factor \cite{RoySwapan2018}. It was proposed to further increase mass sensitivity by driving a nanoresonator into a nonlinear regime and using the squeezing of thermomechanical fluctuations \cite{Buks2006a}. Mass sensing measurements in the nonlinear regime has recently been performed by \textcite{Yuksel2019}. 

On the practical side, of particular importance are comparatively slow frequency fluctuations, with the correlation time comparable or exceeding the decay time. Identifying the origin of such fluctuations is usually difficult. For instance, they can arise from the diffusion of atoms over the resonator surface~\cite{Atalaya2011,Atalaya2011a,Yang2011}, the electrostatic interaction between the resonator and trapped charges in the substrate~\cite{Siria2012}, and temperature drifts that modify the resonator stress. Slow frequency fluctuations are usually quantified using the Allan deviation. Their mechanisms and the Allan deviation are discussed in Sec.~\ref{sec:frequency_fluctuations}. These frequency fluctuations can be reduced, which has enabled reaching mass resolution $(1.7 \pm 0.5) \times 10^{-27} \mathrm{kg}$ with a 2~s integration time; it has been achieved with a nanotube thanks to its extremely small mass~\cite{Chaste2012}. This mass resolution is comparable to the mass of one proton.

\subsection{Photothermal-based sensing}
\label{sec:photothermalsensing}

Nanoresonators can be used for several kinds of sensing based on light absorption. The experiments are done using resonators with tensile stress. The underlying idea is that photon absorption causes heating; the associated thermal expansion reduces the stress, changing the mechanical frequency of the resonator. Because the frequency can be detected with high accuracy, these experiments are exquisitely sensitive, with the sensitivity limited by frequency fluctuations, as in the case of mass sensing. Such photothermal sensing enables optical absorption spectroscopy of single particles and molecules located on the resonator~\cite{Larsen2013}, high-precision single-molecule imaging~\cite{Chien2018}, high-speed detection of electromagnetic radiation with graphene nanomechanical bolometers~\cite{Blaikie2019}, and thermal transport measurements of phonons in MoSe$_2$ monolayer drums through the detection of the thermal gradient over these suspended devices~\cite{Morell2019}.

\subsection{Clocks and clock-based systems}
\label{sec:clocks}
Mechanical resonators are used with great success as ultra-stable timing references. Although these resonators are mostly produced at the micro-scale, we would like to mention them as they have an important technological impact on our society. The current generation of these devices utilizes silicon technology with a wafer-encapsulation process to keep the self-sustaining oscillators in vacuum ~\cite{Kim2007}. They can be now found in nearly all mobile phones, for example. One of the leading manufacturers, SiTime, report clocks for mobile and other battery powered devices with $\pm 3$~ppm frequency stability over the industrial temperature range between $-40$~C and $+85$~C, while the power consumption is kept low, below 5~$\mu$W~\cite{Zaliasl2015}.
Telecom applications require still better clocks, and there has been achieved frequency stability $<\pm 0.1$~ppm and the Allan deviation about $8 \times 10^{-11}$ at 1~s integration time in the temperature range $-45$~C and $105$~C~\cite{Roshan2016}.

Nanoscale electromechanical resonators have also been used as self-sustained oscillators. The challenge is to integrate the device in an electrical circuit, so that it produces a continuous high-frequency signals when powered by a DC input in the feedback loop. The vibrations are transduced into an electrical signal, that is amplified with an adjustable gain and phase before being fed back to the resonator. The difficulty is to obtain a sizeable electrical signal with minimum imprecision noise added by the transduction of the vibrations. A frequency stability of $\pm 2$~ppm is demonstrated in a single-crystal SiC electromechanical resonator, albeit over a short test time, less than 1000~s~\cite{Feng2008}. Interesting approaches to suppressing the noise which are based on using a resonator in the nonlinear regime near bifurcation points of the response to an external drive have been proposed by \textcite{Greywall1994,Yurke1995,Kenig2012a,Kenig2013} and the possibility to go beyond the limit imposed by the thermomechanical noise was demonstrated \cite{Villanueva2013}.


\section{Fluctuation-dissipation theorem and the reaction force from the thermal bath}
\label{sec:FDT}

A simple general model that leads to the Brownian  dynamics (\ref{eq:Brownian}) is based on the assumption that the nanoresonator is an oscillator coupled  to a thermal bath and that the coupling  is linear in the oscillator coordinate. In this section we describe the connection between the phenomenological theory (\ref{eq:Brownian}) and a microscopic theory. We specify the conditions of applicability of the phenomenological description, where the friction force is determined by the instantaneous value of the velocity and where the noise can be described as $\delta$-correlated in time (the Markovian approximation). We then show that, for resonators with a high $Q$ factor, a significantly  less restrictive formulation can be developed, if one is interested in the dynamics of the vibration amplitude and the slow part of its phase. Such dynamics is Markovian on times that largely exceed the vibration period even where the model (\ref{eq:Brownian}) is inapplicable. The analysis provides an expression for the decay rate $\Gamma$ in terms of the parameters of the thermal reservoir, and also shows that the frequency of the mode is changed as a result of the coupling and can become temperature-dependent. The formulation immediately extends to the case where the mode dynamics is quantum.


\subsection{Coupling of the oscillator to a thermal bath}
\label{subsec:classical}

The description of dissipation of the oscillator is based on the picture in which the oscillator is coupled to a system with many degrees of freedom. In particular, a nanomechanical mode is coupled to phonons in the nanoresonator and in the substrate that supports  the nanoresonator, to the electronic degrees of freedom, to an extended set of two-level fluctuators, and so on. The many-degrees-of-freedom system coupled to the oscillator can be usually thought of as a thermal bath.  The leading term in the expansion of the coupling energy in the oscillator coordinate $q$ is linear in $q$ and can be written as
\begin{align}
\label{eq:H_int}
H_i = q h_{\rm b},
\end{align}
where $h_\mathrm{b}$ depends on the dynamical variables of the bath only.
It gives the force $F_b\equiv -h_{\rm b}$ that the bath exerts on the oscillator.
Because of the large number of the dynamical variables, the excitation spectrum of the bath is (quasi)continuous. The form of the function $h_\mathrm{b}$ depends on the particular type of the bath. For example, if the bath is formed by phonons in the resonator or the substrate, typically $h_{\rm b}$ is a series in the phonon coordinates. The nonlinear terms in this series are behind such familiar mechanisms of dissipation of a nanomechanical mode as the thermoelastic, Landau-Rumer, and Akhiezer relaxation, see Sec.~\ref{subsec:phonon_scattering}.

To formally describe the dynamics of the mode and the bath we assume that for $t\to -\infty$ they are uncoupled and the bath is in thermal equilibrium. Because the bath is ``big'', it is only weakly perturbed by the coupling once the latter is turned on.  The response of the bath can be then described by the linear response theory \cite{Landau1980},
\begin{align}
\label{eq:bath_suscept1}
h_{\rm b}(t) \approx h_{\rm b}^{(0)}(t) + \delta h_{\rm b}(t).
\end{align}
Here, $h_{\rm b}^{(0)}(t)$ is the function $h_{\rm b}$ in the absence of the coupling. The force $-h_{\rm b}^{(0)}(t)$ describes the effect of  thermal fluctuations of the unperturbed bath on the oscillator. It is a random force, and it can be chosen to have zero mean, $\langle h_{\rm b}^{(0)}(t)\rangle=0$.

The term $\delta h_b(t)$ is the perturbation caused by the coupling.  When averaged over the fluctuations of the bath, it can be written as
\begin{align}
\label{eq:bath_suscept2}
\langle \delta h_{\rm b}(t)\rangle=- \int_0^\infty dt'{\cal X}_{\rm b}{}(t')q(t-t').
\end{align}
Function ${\cal X}_{\rm b}{}(t)$ is the generalized susceptibility of the bath with respect to the oscillator coordinate $q$. Equation (\ref{eq:bath_suscept2}) is just an expression of the causality principle: the response of the bath at time $t$ depends on the values of $q$ at earlier times.

Because the bath is in thermal equilibrium, its susceptibility and fluctuations are related by the fluctuation-dissipation theorem (FDT). This relation has the form
\begin{align}
\label{eq:bath_LRT}
S_{\rm b}{}(\omega) =2\hbar [\bar n(\omega)+1]\,{\rm Im}~\chi_{\rm b}{}(\omega),
\end{align}
where $\chi_\mathrm{b}(\omega)$ is the Fourier transforms of the susceptibility $\mathcal{X}_\mathrm{b}(t)$ [see Eq.~(\ref{eq:bath_suscept_defined})], $S_\mathrm{b}(\omega)$ is the power spectrum of the bath fluctuations,
\begin{align}
\label{eq:bath_power_spectrum}
&S_{\rm b}{}(\omega)=\int_{-\infty}^\infty dt e^{i\omega t} s_{\rm b}{}(t), \; s_{\rm b}{}(t)=\langle h_{\rm b}^{(0)}(t)h_{\rm b}^{(0)}(0)\rangle,
\end{align}
and $\bar n(\omega)=[\exp(\hbar\omega/k_BT)-1]^{-1}$ is the thermal occupation number of vibrations at frequency $\omega$. The real and imaginary parts of $\chi_{\rm b}(\omega)$ are related by the Kramers-Kr\"onig relation. Therefore Eq.~(\ref{eq:bath_LRT}) fully defines the bath susceptibility in terms of the power spectrum $S_\mathrm{b}(\omega)$.

To the leading order in the coupling, one is tempted to replace $\delta h_{\rm b}(t)$ in Eq.~(\ref{eq:bath_suscept1}) with $\langle \delta h_{\rm b}(t)\rangle$. Then the force from the bath on the mode takes the form
\begin{align}
\label{eq:force_decoupled}
F_\mathrm{b}(t) =F_\mathrm{b}^{\rm (r){}}(t) -h_{\rm b}^{(0)}(t), \quad F_\mathrm{b}^{\rm (r){}}(t)\equiv  -\langle\delta h_{\rm b}(t)\rangle.
\end{align}
The term $F_\mathrm{b}^{\rm (r){}}(t)$ is the reaction force from the bath. It results, as seen from Eq.~(\ref{eq:bath_suscept2}), from the perturbation of the bath by the oscillator and depends on the oscillator coordinate. In the optomechanics literature the reaction force is often called the dynamical backaction \cite{Kippenberg2008}. It is reminiscent of the backaction in the theory of quantum measurements, which describes the effect of the measuring apparatus on the measured quantum system and comes from the interaction between the apparatus and the system, cf. the Heisenberg microscope \cite{Heisenberg1927}. We note that the force  $F_\mathrm{b}^{\rm (r){}}(t)$ emerges not only in the quantum, but also in the classical description of the dynamics.

From Eq.~(\ref{eq:bath_suscept2}), the force $F_\mathrm{b}^{\rm (r){}}(t)$ is retarded: it depends on $q(t')$ with $t'\leq t$. One should keep in mind that $F_\mathrm{b}^{\rm (r){}}(t)$ is an approximation of the reaction force. It relies on the perturbation theory, and one should make sure that the perturbation theory holds for long times on the order of the lifetime of the considered vibrational mode.

In the classical theory, the above approximation for the reaction force applies in the important case where the power spectrum $S_{\rm b}(\omega)$ of the thermal noise   is almost constant for frequencies ranging from much smaller to much larger than $\omega_0$. When the power spectrum is flat, the noise correlation function $-h_\mathrm{b}^{(0)}(t)$ is approximately a $\delta$-function, $h_\mathrm{b}^{(0)}(t)\propto \delta(t)$. This means that the noise part of the force from the bath $F_\mathrm{b}$ is a white noise. One can show that the backaction part of $F_\mathrm{b}(t)$ has then a term $\propto -dq(t)/dt$, see Appendix~\ref{subsec:Ohmic_dissipation}. This maps the dynamics of the mode onto the phenomenological equation of Brownian motion (\ref{eq:Brownian}).   In the quantum theory, the case was studied by \textcite{Caldeira1981} for the bath being a set of harmonic oscillators and $h_{\rm b}$ being linear in the coordinates of these oscillators. It was called Ohmic dissipation.


\subsection{Brownian motion of the complex vibration amplitude}
\label{subsubsec:slow_variables}

For an NVS mode, the analysis of the effect of coupling to a thermal bath can be extended to a significantly broader situation using that the mode decay rate $\Gamma$ is small compared to the eigenfrequency $\omega_0$ or, equivalently, the $Q$-factor is large. The approach we will describe allows one also to study nonlinear effects. In addition, it allows avoiding the assumptions of the Ohmic dissipation model. The underlying physics is that a mode with a large $Q{}$ factor is a ``filter". It is sensitive primarily to perturbations in a narrow frequency range around the eigenfrequency $\omega_0$. In particular, it is sensitive to the bath fluctuations  in this frequency range. Of primary interest therefore is the amplitude and phase of the mode.

A natural approach to the analysis of the dynamics is offered by the Krylov-Bogoliubov method of averaging \cite{Kryloff1947,Bogoliubov1961}.
Here, the first step is to switch from the fast-oscillating coordinate $q(t)$ and momentum $p(t)$ of the mode to two new variables, the complex amplitude $u\equiv u(t)$ and its conjugate,
\begin{align}
\label{eq:u(t)}
&q(t)=u(t)\exp(i\omega_0t) + u^*(t)\exp(-i\omega_0t),\nonumber\\
&p(t) = iM\omega_0[u(t)\exp(i\omega_0t) - u^*(t)\exp(-i\omega_0t)].
\end{align}
The real and imaginary parts of $u(t)$ have a simple physical meaning. They are just the quadratures of the vibrations: if we write the mode displacement as $q(t)=A(t)\cos[\omega_0 t+\phi(t)]$, then Re~$u(t) = \frac{1}{2}A(t)\cos \phi(t)$ and Im~$u(t) = \frac{1}{2}A(t)\sin\phi(t)$.

In the absence of coupling to a thermal bath, the mode oscillates at frequency $\omega_0$ with constant amplitude and phase, and then $u(t) =$const. Because of the coupling,  $u(t)$ will vary in time, but for weak coupling the change over the vibration period $2\pi/\omega_0$ will be small. This can be used to show that, in ``slow'' time compared to $2\pi/\omega_0$,  one can disregard retardation of the reaction force. This means that the value of  $F_\mathrm{b}^{\rm (r){}}(t)$ at time $t$ is determined by the value of $u(t)$ at the same time $t$, see Eq.~(\ref{eq:coupling_expansion}).  Then taking into account the explicit expressions for the reaction force (\ref{eq:coupling_expansion}) and the thermal noise $h_{\rm b}^{(0)}(t)$, we obtain, see Appendix~\ref{subsec:averagin_cplng}
\begin{align}
\label{eq:eom_classical}
&\dot u = -(\Gamma - iP_{})u + \xi(t).
\end{align}
Here the term $\xi(t)=(i/2M{}\omega_0)h_{\rm b}^{(0)}(t)\exp(-i\omega_0 t)$ describes the noise that comes from the fluctuations in the thermal bath. The meaning of the parameters $\Gamma$ and $P$ is seen from the solution of
Eq.~(\ref{eq:eom_classical}) for the regular part of the complex amplitude, $ \langle u(t)\rangle  = \langle u(0)\rangle \exp[-(\Gamma -iP_{})t]$. When combined with Eq.~(\ref{eq:u(t)}), it shows that $\Gamma$ is the  decay rate of the vibrations, whereas $P$ is the change of their frequency due to the coupling to the bath, $\omega_0\to \omega_0+P$, i.e., $P$ is an analog of the polaronic effect for a vibrational system. Both $\Gamma$ and $P$ are expressed in terms of the bath susceptibility $\chi_\mathrm{b}(\omega)$ at the mode eigenfrequency (Appendix~\ref{subsec:averagin_cplng}),
\begin{align}
\label{eq:Gamma_and_P}
\Gamma =\frac{{\rm Im}\,\chi_{\rm b}{}(\omega_0)}{2M{}\omega_0},\quad P_{}=-\frac{{\rm Re}\,\chi_{\rm b}{}(\omega_0)}{2M{}\omega_0}.
\end{align}
With the account taken of Eq.~(\ref{eq:bath_LRT}), the decay rate $\Gamma$  is also simply expressed in terms of the power spectrum $S_{\rm b}(\omega_0)$ of the random force exerted by the bath. In the classical limit $k_BT\gg \hbar\omega_0$ we have
\begin{align}
\label{eq:Gamma_classical}
\Gamma = (4M{}k_BT)^{-1}S_{\rm b}(\omega_0)
\end{align}
We will use this expression below in the analysis of the decay of nanomechanical modes due to their coupling to phonons, two-level fluctuators, and electrons.

As seen from Eq.~(\ref{eq:Gamma_and_P}), both the decay rate $\Gamma$ and the frequency shift $P_{}$ depend on temperature. The temperature dependence of $P_{}$ is a simple microscopic mechanism of the temperature dependence of the measured eigenfrequency of NVSs, and this dependence can be determined from the experimentally measured power spectrum or the response curves. For a carbon nanotube, such measurements were reported by \textcite{Tepsic2021}. In what follows, we incorporate $P_{}$ into the definition of $\omega_0$, i.e., replace $\omega_0\to \omega_0+P_{}$.

The noise $\xi(t)$ in Eq.~(\ref{eq:eom_classical}) is Gaussian and zero-mean. It is $\delta$-correlated in the ``slow" time compared to the vibration period $2\pi/\omega_0$ and the time over which bath correlations decay,
\begin{align}
\label{eq:noise_correlator}
\langle \xi^*(t)\xi(t')\rangle \approx (\Gamma k_BT/M{}\omega_0^2)\delta(t-t'),
\end{align}
whereas the correlator $\langle \xi(t)\xi(t')\rangle$ and its complex conjugate can be disregarded (Appendix~\ref{subsec:averagin_cplng}). It is seen from the simple relations (\ref{eq:eom_classical}) - (\ref{eq:noise_correlator}) between the decay rate of the mode and the noise from the thermal bath that the stationary distribution of the mode is the Boltzmann distribution, and  $\langle |u|^2\rangle = k_BT/2M{}\omega_0^2$.

The time evolution of the complex oscillator amplitude described by Eq.~(\ref{eq:eom_classical}) is Markovian. There is no delay, the response of the bath is instantaneous in  slow time. The corrections disregarded in Eqs.~(\ref{eq:eom_classical}) and (\ref{eq:noise_correlator}) are small if the bath susceptibility $\chi_{\rm b}{}(\omega) $ weakly varies with $\omega$ in a comparatively narrow band centered at $\omega_0$. The width of this band is determined by the time dependence of $u(t)$ and is $\sim \Gamma, |P|$. It is small compared to $\omega_0$ and to the reciprocal correlation time of fluctuations in the bath at frequency $\sim \omega_0$. This is in contrast with the model of the Brownian motion described by Eq.~(\ref{eq:Brownian}), which requires near constancy of $\chi_{\rm b}(\omega)$ in the frequency band broader than $\omega_0$.

We reiterate that the Markovian equation of motion (\ref{eq:eom_classical}) is an approximation. Its applicability has to be checked and the expressions for the decay rate and the polaronic frequency shift have to be derived using a microscopic model of the bath and the coupling.

\subsection{Quantum description of the mode dynamics}
\label{subsec:quantum_aspects}

The above analysis can be immediately extended to the quantum regime. In quantum terms, functions $u(t), u^*(t)$ become operators, with $u^*$ understood as $u^\dagger$. They are just the operators $\frac{1}{2}[q(t)\mp ip(t)/M{}\omega_0]\exp[\mp i\omega_0 t]$ of a harmonic oscillator in the Heisenberg representation and are simply related to the ladder operators $a^\dagger$ and $a$, with
\begin{align}
\label{eq:u_and_a_dagger}
&u(t) = (\hbar/2M{}\omega_0)^{1/2}a^\dagger(t)e^{-i\omega_0t},\nonumber\\
&[u(t),u^\dagger(t)] = -\hbar/2M{}\omega_0.
\end{align}

The equation of motion for $u(t)$ (\ref{eq:eom_classical}) is linear. It applies not only in the classical case, but also in the case where $u(t)$ is an operator, becoming a {\it quantum Langevin equation}. Importantly, the noise $\xi(t)$ in this formulation is an operator, too, and the operators $\xi(t)$ and $\xi^\dagger(t)$ do not commute ($\xi^\dagger$ replaces $\xi^*$). In particular, with the account taken of Eq.~(\ref{eq:bath_LRT}) we have instead of Eq.~(\ref{eq:noise_correlator})
\begin{align}
\label{eq:noise_correlator_quantum}
\langle \xi^\dagger(t)\xi(t')\rangle &\approx e^{\hbar\omega_0/k_BT}\langle \xi(t)\xi^\dagger(t')\rangle\nonumber\\
&\approx [\hbar (\bar n+1)\Gamma/M{}\omega_0]\delta(t-t')
\end{align}
where $\bar n \equiv \bar n(\omega_0)$ is the thermal occupation number of the mode.

The non-commutativity of $\xi(t)$ and $\xi^\dagger(t)$ is important. Indeed, the mean values $\langle u(t)\rangle $ and $\langle u^\dagger(t)\rangle$ decay in time. However, the commutation relation between $u(t)$ and $u^\dagger(t)$ should be independent of time. Using Eq.~(\ref{eq:noise_correlator_quantum}) for the noise correlators, one immediately finds from Eq.~(\ref{eq:eom_classical}) and the conjugate equation for $u^\dagger$ that $\langle [u(t),u^\dagger(t)]\rangle$ remains constant.

On the formal side, the approximations made in deriving the equations of motion (\ref{eq:eom_classical}) - (\ref{eq:noise_correlator_quantum}) and the equivalent master equation (see Appendix~\ref{subsubsec:Born}) correspond to the familiar ladder approximation in the diagrammatic technique \cite{Abrikosov1975}. We note that a non-Markovian quantum Langevin equation has been also discussed in the literature; it has been consistently  derived microscopically for the case of the coupling to a bath of harmonic oscillators, with the coupling being effectively bilinear in the dynamical variables of the mode and the bath oscillators, see \cite{Ford1988} and references therein.

The decay rate $\Gamma$ in the quantum theory is simply related to the rate of  transitions between the oscillator energy levels due to the coupling to the bath. As shown in Appendix~\ref{subsubsec:Born}, $\Gamma$ is a half of the rate $W_{1\to 0}$ of transitions from the first excited to the ground state of the mode due to the coupling to the bath for $T=0$.

\subsubsection{The power spectrum and the susceptibility of a weakly damped mode}
\label{subsubsec:linear_susceptibility}

The picture of the dynamics of the oscillator as vibrations at frequency $\omega_0$ with slowly varying amplitude and phase provides a physical insight into why the oscillator power spectrum $S_{}(\omega) = \int_{-\infty}^\infty dt \langle q(t) q(0)\rangle \exp(i\omega t)$ has a peak at frequency  $\omega$ close to $\omega_0$. Expressing the coordinate $q(t)$ in terms  the complex amplitude $u(t)$ and its complex conjugate in the classical limit or the operators $u(t), u^\dagger(t)$, to extend to the quantum description, we find from Eq.~(\ref{eq:u(t)}) that, for $|\omega-\omega_0|\ll\omega_0$,
\begin{align}
\label{eq:power_spectrum_defined}
S_{}(\omega)\approx 2{\rm Re}\int_0^\infty dt \langle u^\dagger (t) u(0)\rangle
e^{i\omega t}. \quad
\end{align}
From Eqs.~(\ref{eq:eom_classical}) and (\ref{eq:noise_correlator_quantum}), for the considered linear oscillator
\begin{align}
\label{eq:Lorentz}
S_{}(\omega) = \frac{\hbar}{M{}\omega_0}(\bar n +1)\frac{\Gamma}{\Gamma^2 + (\omega - \omega_0)^2}.
\end{align}
This expression extends to the quantum domain the result (\ref{eq:Lorentz_classical}). Importantly, the frequency $\omega_0$ here incorporates the polaronic shift $P_{}$ , and therefore the position of the maximum of  $S_{}(\omega)$ is temperature-dependent. We note that the peak of the  power spectrum of the radiation {\it emitted} by a quantum oscillator is described by Eq.~(\ref{eq:Lorentz}) in which the factor $\bar n +1$ is replaced by the thermal occupation number $\bar n$.

The general expression for the oscillator susceptibility $\chi_{}(\omega)$ in response to a weak drive at frequency $\omega$ near resonance, $ |\omega-\omega_0|\ll\omega_0$,  reads
\begin{align}
\label{eq:susceptibility_general}
\chi_{}(\omega) =\frac{ i}{\hbar (\bar n +1)}\int_0^\infty dt \langle u^\dagger (t) u(0)\rangle
e^{i\omega t}.
\end{align}
For a linear oscillator the susceptibility is given by Eq.~(\ref{eq:Lorentz_classical}).

It should be emphasized that the general expressions (\ref{eq:power_spectrum_defined}) and (\ref{eq:susceptibility_general}) for the power spectrum and the susceptibility are not limited to the case of a linear oscillator. They describe the power spectrum of a vibrational mode even where the vibrations are nonlinear as long as the spectrum has a narrow peak with width much smaller than $\omega_0$ and the reciprocal correlation time of the reservoir.

The quantum description becomes relevant for the experiments where the mechanical mode is close to the quantum ground sate. This regime was first achieved in nanomechanics by \textcite{O'Connell2010}. In this experiment a mechanical breathing mode vibrating at 6~GHz was cryogenically cooled using a dilution fridge at 25~mK, and a non-classical state of motion was created. Recently cryogenic cooling down to the quantum regime was accomplished for a nanomembrane flexural mode with $\omega_0=2\pi\times 15.1$~MHz by lowering its temperature to $500~\mu$K \cite{Cattiaux2021}.   Mechanical modes endowed with long lifetimes have been cooled down into the quantum regime by coupling them to photon cavities and using parametric drive \cite{Chan2011,Teufel2011,Verhagen2012}. Such cooling is discussed in the review by \textcite{Aspelmeyer2014a}.


\section{Relaxation mechanisms of  nanomechanical resonators}
\label{sec:mechanisms_linear}

The general expression for the decay rate allows us to consider various microscopic mechanisms of energy dissipation of low-frequency modes in NVSs. The major types of the relevant thermal reservoirs are (i) phonons in the substrate, (ii) thermal phonons in the nanoresonator, (iii) electrons in the nanoresonator and in the leads, and (iv) two-level systems. We use the term ``phonon'' somewhat loosely, as the systems we are discussing do not necessarily have translational symmetry or even spatial uniformity. However, the relevant eigenmodes, even though not plane waves, are spatially extended and have a quasi-continuous frequency spectrum, resembling phonons. In thin nanobeams and membranes the spectrum consists of bands of modes extended along the nanobeams or membranes and having different transverse spatial structures. Generally, in disordered systems there are also localized vibrational excitations, which we will not discuss here.


\subsection{Scattering by phonons}
\label{subsec:phonon_scattering}

The problem of linear damping of low-frequency eigenmodes in nanoresonators due to phonon scattering overlaps with the problem of sound absorption in dielectrics, which has been intensely studied since 1930s \cite{Landau1937,Akhiezer1938}, see \cite{Woodruff1961,Maris1966,Gurevich1988,Garanin1992,Collins2013,Lindenfeld2013,Feng2015} and references therein. It is also closely related to the problem of decay of low-frequency resonant modes and gap modes in crystals with defects \cite{Brout1962,Kagan1962,Krivoglaz1961,Krivoglaz1964a}.

\subsubsection{Clamping losses}
\label{subsub:clamping}

The simplest decay mechanism of nanomechanical vibrations is radiation of vibrational excitations (phonons) into the supporting structure. This mechanism is an important contributor to the so-called clamping losses. Vibrations of the NVS create time-dependent strain and stress in the area where the resonator is clamped. They serve as phonon-radiating antennas. The supporting structure is much larger than the resonator, and its phonons are a thermal bath for the resonator. This picture of decay via phonon emission is common for nano- and micro-mechanical systems. In micromechanical systems the corresponding losses are often called anchor losses.

The simplest model of the coupling to the bath of phonons in the support is where the coupling is linear both in the coordinate $q$ of the NVS mode and in the phonon coordinates $q_\kappa$. It is described by the Hamiltonian $H_i=qh_{\rm b}$,  Eq.~(\ref{eq:H_int}), with
\begin{align}
\label{eq:linear_phonon_cplng}
h_{\rm b}=\sum_\kappa  V_\kappa
(b_\kappa+b_\kappa^\dagger).
\end{align}
Here, $b_\kappa$ and $b_\kappa^\dagger$ are the annihilation and creation operators of the $\kappa$th phonon, $\omega_\kappa$ is the phonon frequency; index $\kappa$ includes the wave vector and the branch of the phonon. The full Hamiltonian of the coupled NVS mode and the phononic bath is
\begin{align}
\label{eq:linear_H}
&H=H_0+ H_i +H_b, \qquad H_0=\hbar\omega_0a^\dagger a , \nonumber\\
& H_b=\sum_k\hbar\omega_\kappa b_\kappa^\dagger b_\kappa
\end{align}
[we remind that the mode coordinate is $q=(\hbar/2M{}\omega_0)^{1/2}(a+a^\dagger)$].

The analysis of Sec.~\ref{subsubsec:slow_variables} shows that, for the model (\ref{eq:linear_phonon_cplng}), the decay rate $\Gamma$ and the shift of the eigenfrequency $P_{}$ of the NVS mode have the form \cite{Bogolyubov1945}
\begin{align}
\label{eq:decay_clamping}
\Gamma=\frac{\pi}{4M{}\omega_0^2}g_{\rm cl}(\omega_0), \quad P_{}=\frac{1}{2M{}\omega_0}\fint d\omega\frac{g_{\rm cl}(\omega)}{\omega^2-\omega_0^2},
\end{align}
where $g_{\rm cl}(\omega) = (2\omega/\hbar)\sum_\kappa |V_\kappa|^2\delta(\omega-\omega_\kappa)$ is the phonon  density of states weighted with the interaction. The decay rate (\ref{eq:decay_clamping}) is independent of temperature, it cannot be eliminated by cooling down the nanoresonator. However, it is small for low-frequency NVS modes, because the density of phonons at frequency $\omega_0$ in a 3D support is $\propto (\omega_0/\omega_D)^2\ll 1$ ($\omega_D$ is the Debye frequency). From Eq.~(\ref{eq:decay_clamping}), the decay rate can be reduced by either decreasing the density of states of phonons in the support or the coupling to these phonons.

The phonon-radiation decay of nano- or microresonators has been attracting much attention both on the theory side, see \cite{Angelescu1998,Cross2001,Park2004,Photiadis2004,Bindel2005,Wilson-Rae2008}, and on the experimental side, see \cite{Yasumura2000,Judge2007,Anetsberger2008,Eichenfield2009,Pandey2009,Unterreithmeier2010a,Schmid2011,Cole2011,Wilson-Rae2011,Rieger2014,Villanueva2014,Chakram2014,Meenehan2014,Ghaffari2015,Pfeifer2016,Norte2016,Tsaturyan2017,Ghadimi2018} and references therein. Approximate expressions for the decay in cantilevers and membranes are summarized by \textcite{Schmid2016}.

Separating radiation decay from other decay mechanisms is not necessarily simple in the experiment, cf. \cite{Unterreithmeier2010a,Yu2012,Ghaffari2015}. On the theory side, to find the coupling parameters $V_\kappa$ one has to find the force that the resonator mode exerts on the phonons. For atomic displacements in the contact area, a separation of the contributions from vibrational modes of the resonator and of the support is nontrivial. Formally, the resonator and the support are a single system with a single set of eigenmodes. The considered resonator ``mode'' is a superposition of the exact eigenmodes, which have frequencies lying within a band centered at $\omega_0$, with a width $\sim \Gamma$,  and have comparatively large amplitudes inside the resonator. Such a picture is reminiscent of the theory of tunneling decay in quantum mechanics, with the NVS mode being an analog of the state localized in a potential well and decaying into extended states outside the well. The extended states are the analogs of the phonons of the support. Both formulations have a counterpart in the analysis of resonant vibrations in solids that are mostly localized on defects \cite{Brout1962,Kagan1962}; see \cite{Barker1975}.

One of the approaches to the problem of clamping losses is based on calculating the transmission ${\cal T}$  of elastic waves through the contact (the clamping area). \textcite{Cross2001} performed detailed calculations of the transmission for the support and the resonator being of the same thickness. They used a scalar model of elastic waves and the  stress-free boundary conditions at the edges. The displacement field in the contact area was expanded into a superposition of incident and reflected waves on the resonator side and a propagating wave in the support, while keeping the field continuous on the both sides of the contact. Once ${\cal T}$ has been calculated, it is easy to find the decay rate. For example, for an eigenmode of a cantilever, which is a standing wave formed as a superposition of the two waves propagating in the opposite directions, $\Gamma\sim v_g{\cal T}/L$, where $v_g$ is the group velocity of the wave and $L$ is the length of the cantilever.

Extensive calculations of the decay rate were done for several types of modes in nanoresonators by conditionally separating the displacement at the edge into that from the NVS mode and the phonons in the support \cite{Wilson-Rae2008}. This separation allowed finding the stress from the resonator mode in the contact area and then  the force this mode exerts on the support modes, which gives the parameters $V_\kappa$. The model was compared to measurements on high-stress Si$_3$N$_4$ membranes with circular and square geometries \cite{Wilson-Rae2011} and AlGaAs suspended-plate resonators \cite{Cole2011}.

\begin{figure}[h]
\includegraphics[scale=0.8]{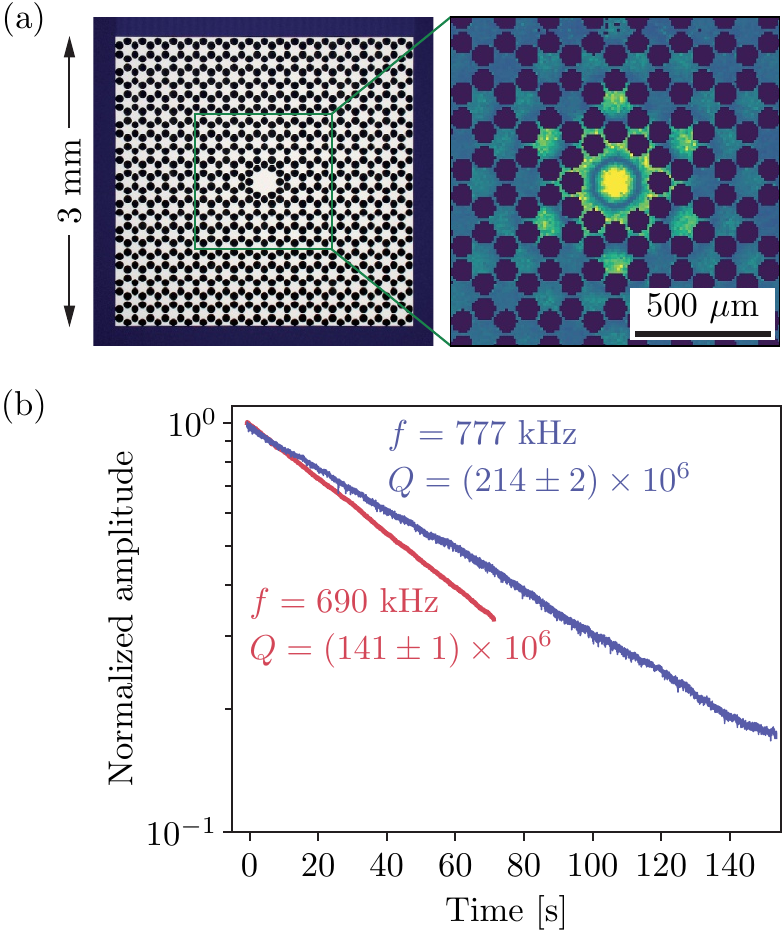} \hfill
\caption{(a) Engineering the shape and the support structure of the mechanical eigenmode. The measured shape of the localized flexural mode is indicated by the intensity of the yellow colour in the right panel. The silicon nitride membrane under tension is patterned with holes (shown in black) to (\textit{i}) decrease clamping losses thanks to the quasi-bandgap created by the phononic crystal in the membrane, and (\textit{ii}) reduce dissipation with ``soft clamping'' by avoiding areas with large curvature. The membrane, which is about 100~nm thick, is supported by a silicon frame (left panel). (b) Energy decay measurements of two different modes at room temperature. Figure adapted from \cite{Tsaturyan2017}.
}
\label{fig:softclamping}
\end{figure}

Significant progress in reducing clamping losses in micromechanical resonators has been made by creating gaps in the phonon density of states in the contact area. If phonons with frequencies close to $\omega_0$ are decoupled from the resonator mode, the decay (\ref{eq:decay_clamping}) is eliminated (in practice, suppressed).  Different means of creating spectral gaps have been developed for micromechanical resonators, cf.  \cite{Pandey2009,Mohammadi2009,Ghaffari2013,Bahr2014,Gokhale2017,Yu2014,MacCabe2020}. One of the most commonly used methods is based on using phononic crystals.  It allows one to create a mode localized mostly within an interior of a membrane or a nanobeam near a ``defect'' of the phononic crystal, with the frequency in the gap of the crystal excitation spectrum, see Figs.~\ref{fig:softclamping}a,b. This significantly reduces the coupling to the support phonons. Such localized modes are  counterparts of the modes localized at defects in solids \cite{Lifshitz1942a,Lifshitz1942,Lifshitz1942b}, which have frequencies that lie outside the phonon bands. Most of the studies of localized modes in crystals with defects have been done using  ensemble measurements \cite{Barker1975}, in contrast to NVSs, where the modes are accessed individually.

The importance of clamping (anchor) losses has led to a development of various numerical algorithms, some of which have been incorporated into the standard software (COMSOL). An example is  the method of a perfectly matched layer (PML), which was first applied to the problem of anchor losses by \textcite{Bindel2005}. In this numerical method, to avoid reflection of the irradiated phonons back into the resonator there is introduced a layer adjacent to the boundary that truncates the unbounded support. The elasticity equations are artificially modified, so that inside the layer the solution decays exponentially, which is technically accomplished using a complex-valued coordinate change in combination with a finite element implementation. At the same time, the solution in the region that includes the resonator is not changed (perfect matching).

\subsubsection{Anelastic relaxation and dissipation dilution for flexural modes}
\label{subsubsec:anelastic_relaxation}

At a phenomenological level, decay of vibrational modes in micro- and nano-resonators is often described by anelastic relaxation. The term was introduced by \textcite{Zener1948}. Anelastic relaxation comes from thermal relaxation, motion of interstitial atoms, interstate transitions in two-level systems, grain boundary relaxation, and other processes \cite{Zener1937,Zener1958,Nowick1972}. In many cases the losses are described by assuming that the Young modulus has an imaginary part. One can think of this imaginary part as a result of the delay of the elastic response, hence the term ``anelasticity''. Generally, the losses depend on the mode frequency and often display a characteristic peak as a function of frequency \cite{Zener1948}. In this approach, a complex Young modulus is a property of a material, cf. \cite{Saulson1990}. However, for  the low-frequency modes studied in nanomechanical resonators the frequency dispersion of the complex Young modulus is often disregarded, with an exception of thermoelastic relaxation, see below. The decay rate and the quality factor $Q$, which is defined by Eq.~(\ref{eq:Q_definition}), are then expressed in terms of the imaginary part of the Young modulus.

To extend applications of nano- and micromechanical systems it is important  to increase the quality factor. It was shown, in the context of interferometric detectors of gravitational waves \cite{Gonzalez1994} that, for flexural modes of a suspended loaded wire, the quality factor may become  higher than the intrinsic $Q$ of the material. This effect has become known as dissipation dilution. The physics of dissipation dilution is based on the fact that the elastic energy of a flexural mode in a stretched wire or, more generally, any stressed  nano- or microresonator comes not only from the internal stress associated with the strain, but also from the externally applied tension, whereas the losses are often due to the intrinsic properties of the material.

Detailed understanding of dissipation dilution requires a careful analysis of the interplay between the tension and the losses. It was first shown numerically by \textcite{Unterreithmeier2010a} that, in a multimode resonator, the effect can be accounted for by taking into account the actual shape of the flexural modes while assuming the complex Young modulus to be frequency-independent. The results were successfully compared with their experimental data on up to 9 modes in  SiN nanostrings. In this approximation, the losses can be analyzed \cite{Unterreithmeier2010a,Schmid2011,Yu2012} by writing the energy loss per vibrational cycle in the same form as the stress-related elastic energy and replacing the real Young modulus with its imaginary part $E_2$. For example,  in a plate of thickness $h$ with a transverse displacement $\zeta(x,y)$  in the $z$ direction, the density of losses per unit area is
\begin{align}
\label{eq:Regal}
W_{\rm anelast}=&\frac{\pi h^3}{12\omega_0(1-\nu_\mathrm{P}^2)} E_2\left\{(\partial^2_x\zeta + \partial^2_y \zeta)^2 \right.
\nonumber\\
&\left.+2(1-\nu_\mathrm{P})\left[(\partial_x\partial_y\zeta)^2 -\partial_x^2\zeta \,\partial^2_y\zeta\right]\right\}
\end{align}
where $\nu_\mathrm{P}$ is the Poisson ratio.  \cite{Yu2012} extended this expression to spatially nonuniform systems, $E_2 \to E_2(x,y)$. An explicit calculation for a nanobeam was done by \textcite{Schmid2011}.

As seen from Eq.~(\ref{eq:Regal}), a major contribution to losses comes from the areas of the largest curvature. For clamped membranes and nanobeams under tension, these areas are close to the clamps. Therefore one may expect that localizing flexural modes to the central part of nanobeams  or membranes, i.e., away from the clamps, may reduce the losses. Such ``soft clamping'' has been successfully implemented using phononic crystals in multi-mode nano-membranes (Figs.~\ref{fig:softclamping}a,b) \cite{Tsaturyan2017} as well as in multimode nanobeams \cite{Ghadimi2018}. An alternative strategy consists in clamping a vibrating structure at the  anti-node of a perimeter mode~\cite{Bereyhi2021}. Record-high Q factors have been obtained along with very high $Q\times f$ factors, which in long nanobeams were as large as $1.1\times 10^{15}$~Hz for $f=1.33$~MHz \cite{Ghadimi2018} at room temperature. Thus isolating the modes inside suspended structures can be advantageous not only for reducing phonon emission into the substrate, but also for reducing intrinsic material losses.



\subsubsection{Landau-Rumer relaxation}
\label{subsubsec:Landau_Rumer}

At the microscopic level, a major intrinsic contribution to decay of low-frequency micro- and nanomechanical modes comes from their nonlinear coupling to other vibrational modes. The usually studied NVS modes have frequencies small compared to the Debye frequency $\omega_D$. Therefore the density of states of higher-frequency modes is usually much larger than the density of states at $\omega_0$, and it is the coupling to such modes that leads to decay of the low-frequency modes. This decay is somewhat reminiscent of sound absorption in dielectrics, which has been intensely studied since 1930s \cite{Landau1937,Akhiezer1938}, see \cite{Maris1966,Gurevich1988,Garanin1992,Collins2013,Lindenfeld2013,Feng2015} and references therein. The relevant lowest order nonlinearity is cubic. It leads to processes in which three vibrational modes are involved.

For cubic nonlinearity, the coupling (\ref{eq:H_int}) is described by the Hamiltonian $H_i=qh_{\rm b} $ with
\begin{align}
\label{eq:anharmonic_linear}
 h_{\rm b}\equiv h_b^{(3)}=\sumprime{\kappa\kappa'}V_{\kappa\kappa'}b_\kappa^\dagger b_{\kappa'} + \sum_{\kappa\kappa'}( V'_{\kappa\kappa'}b_\kappa^\dagger b_{\kappa'}^\dagger + {\rm H.c.})
\end{align}
(the prime over the sum indicates that $\kappa\neq \kappa'$). In this expression $b_\kappa$ and $b_\kappa^\dagger$ are the annihilation and creation operators, with $\kappa$ now enumerating the modes localized mostly inside the resonator. Coupling to such modes is often  stronger than to the modes in the support, and they serve as a thermal bath for the considered low-frequency mode. To simplify the language and to distinguish them from the considered mode, we will call these high-frequency modes ``phonons''.

Generally, because of the possible ripples and other inhomogeneites of the nanoresonator, the modes  of the quasicontinuous frequency spectrum are not standard plane waves. This is why we use $\kappa$ rather than the wave vector to enumerate them.  The nanomechanical modes we are interested in are also not plane waves, whether these are flexural modes or modes localized near defects of a phononic crystal. Therefore, in distinction from the sound absorption problem, in the scattering described by the coupling $qh_\mathrm{b}$ the momentum is not conserved. This makes the problem similar to that of dissipation of modes localized on defects in disordered solids. Such problem for the coupling Hamiltonian (\ref{eq:anharmonic_linear}) was considered by \textcite{Krivoglaz1961,Krivoglaz1964a}.

The coupling parameters $V_{\kappa\kappa'}$ have to be calculated with the account taken of the actual structure of the involved modes, cf. \cite{Iyer2016,Atalaya2016,MacCabe2020}. We note that the nonlinear coupling of the considered NVS modes to the modes in the support and the corresponding nonlinear clamping losses may be also important because of the same density of states argument. This coupling is described by the Hamiltonian (\ref{eq:anharmonic_linear}) with $\kappa$ referring to the modes primarily localized in the support. To the best of our knowledge, such coupling has not yet been studied for nanoresonators either theoretically or in the experiment.

\begin{figure}[h]
\includegraphics{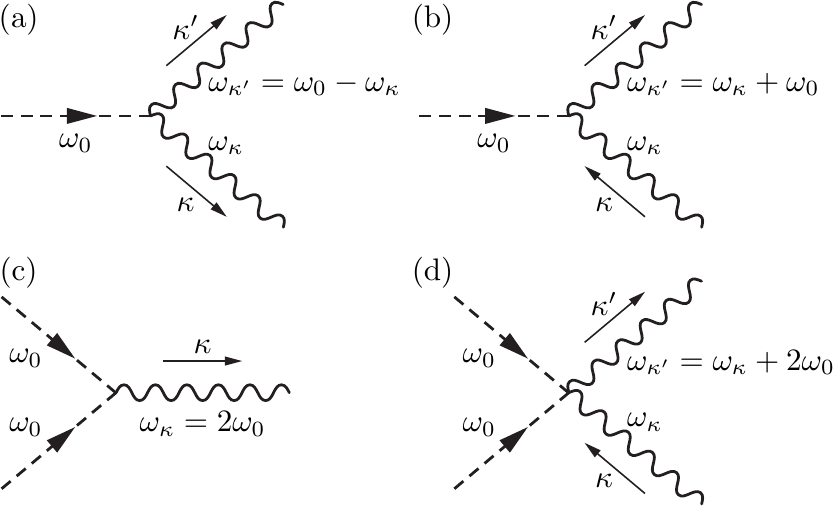} \hfill
		\caption{Scattering of a considered NVS mode due to its nonlinear coupling to phonons; $\omega_0$ is the NVS mode frequency, and $\omega_\kappa, \omega_{\kappa'}$ are the frequencies of the phonons with quantum numbers $\kappa, \kappa'$. (a) Decay into two phonons. (b) The Landau-Rumer mechanism: anti-Stokes scattering of a phonon $\kappa$ into a phonon $\kappa'$ by the NVS mode. (c) Scattering of two quanta of the NVS mode into a phonon, a process leading to nonlinear friction. (d) Analog of the Landau-Rumer scattering that leads to nonlinear friction.
}
\label{fig:scattering_diagram}
\end{figure}

The term $\propto V'_{\kappa\kappa'}$ in Eq.~(\ref{eq:anharmonic_linear}) describes a decay process where one quantum of the NVS mode disappears and there emerge two quanta of the thermal bath (two phonons) with total energy $\hbar\omega_0$, see Fig.~\ref{fig:scattering_diagram}(a). The rate of such scattering is very small for low-frequency NVS modes because of the low density of states of the relevant phonons.
For $\hbar\omega_0\ll k_BT$ of a much higher probability is the process where a phonon is inelastically scattered off the considered NVS mode into another phonon;  for example, a phonon $\kappa$ is scattered into a phonon $\kappa'$ and the energy difference $\hbar\omega_{\kappa'} - \hbar\omega_\kappa $ is equal to $\hbar\omega_0$. This Raman-type scattering is sketched in Fig.~\ref{fig:scattering_diagram}(b). The coupling leading to such a process is given by $V_{\kappa\kappa'}$  in Eq.~(\ref{eq:anharmonic_linear}). The scattering probability is comparatively large when the thermal occupation  numbers of the modes $\kappa, \, \kappa'$ are large and their density of states is large, too. This means $k_BT \gg \hbar\omega_0$, a condition met in most experiments, with a few  exceptions, see \cite{O'Connell2010,Chu2017,Satzinger2018,Chu2018,Arrangoiz-Arriola2019,MacCabe2020,Cattiaux2021,Wollack2021a} and references therein.

The phonons $\kappa,\, \kappa'$ involved in the scattering shown in Fig.~\ref{fig:scattering_diagram}(b) are themselves experiencing decay. The rate of the decay of the considered low-frequency mode $\Gamma$ strongly depends on the interrelation between the phonon relaxation time $\tau_\kappa$ and the vibration period of the considered  mode $2\pi/\omega_0$.
If $ \omega_0\tau_\kappa\gg 1$,  we can disregard decay of  high-frequency phonons. Then from Eqs.~(\ref{eq:Gamma_classical}) and (\ref{eq:anharmonic_linear}) we find
\begin{align}
\label{eq:Landau_Rumer}
\Gamma=\Gamma^{\rm LR}=&\frac{\pi}{2M{}\hbar\omega_0}\sum_{\kappa,\kappa'}|V_{\kappa\kappa'}|^2[\bar n(\omega_\kappa) - \bar n(\omega_{\kappa'})] \nonumber \\
&\times\delta(\omega_\kappa+\omega_0-\omega_{\kappa'}).
\end{align}
This expression has the same form as the expression for the decay rate of a mode localized on a defect \cite{Krivoglaz1961,Krivoglaz1964a} and is similar to the expression by \textcite{Landau1937} for the sound absorption coefficient in solids. An analysis of this expression in the case of a breathing mode in a nanobeam phononic crystal \cite{MacCabe2020} showed that the corresponding decay rate is much smaller than the observed value (which itself was extremely small, with the $Q$-factor reaching $ 5\times 10^{10}$.) It should be noted that the density of states arguments and the symmetry arguments may lead to a four-quantum decay process having a higher rate than the three-quantum one \cite{Landau1949,Landau1949a,DeMartino2009}.


\subsubsection{Thermoelastic and Akhiezer relaxation}
\label{subsubsec:TER}

Of primary interest for nano- and micromechanical resonators is the situation where the decay rate of high-frequency phonons exceeds the frequencies of the considered modes. Phonon decay significantly complicates the calculation of the power spectrum of $h_{\rm b}^{(3)}$, which gives the decay rate $\Gamma$, see Eq.~(\ref{eq:Gamma_classical}). In such a calculation the interaction between the phonons should be explicitly taken into account. This interaction comes from the nonlinearity of the crystalline lattice, which is described by nonlinear terms in the phonon Hamiltonian, i.e., the bath Hamiltonian. To the lowest order, in the bath nonlinearity, one has to replace the bath Hamiltonian $H_b$ given by  Eq.~(\ref{eq:linear_H}) with $H_b + H_b^{\rm (3)}$,
\begin{align}
\label{eq:lattice_nonlinearity}
H_b^{\rm (3)} = \frac{1}{2}\sum_{\kappa_1\kappa_2\kappa_3}V_{\kappa_1\kappa_2\kappa_3}b_{\kappa_1}^\dagger b_{\kappa_2}^\dagger b_{\kappa_3} +{\rm H.c.}
\end{align}
The Hamiltonian (\ref{eq:lattice_nonlinearity}) describes processes where one phonon decays into two other phonons or, vice versa, two phonons annihilate and one phonon emerges, so that the overall phonon energy is conserved, $\omega_{\kappa_1} + \omega_{\kappa_2} = \omega_{\kappa_3}$. Other cubic in $b_\kappa, b_\kappa^\dagger$ terms have been dropped in Eq.~(\ref{eq:lattice_nonlinearity}), as they do not describe phonon decay, to the leading order.

In Appendix~\ref{sec:app_Akhiezer} we outline a way to calculate the decay rate of the  low-frequency NVS mode using the general formulation of Sec.~\ref{sec:FDT}, with the account taken of the nonlinear mode couplings (\ref{eq:anharmonic_linear}) and (\ref{eq:lattice_nonlinearity}). The calculation is somewhat involved. Here we give a qualitative phenomenological picture for two important limiting cases. We note, however, that  both cases follow from the same general analysis.

\underline{\it The thermoelastic relaxation} (TER) case is where the eigenfrequency $\omega_0$ is so small that, because of the coupling (\ref{eq:lattice_nonlinearity}), the high-frequency vibrations have time to come to thermal equilibrium locally in different parts of the vibrating nanoresonator. One can then introduce a local position-dependent temperature $T(\rb)$ inside the nanoresonator.  For flexural modes, this implies that the mean free path of thermal phonons $l_T$, which is determined by the coupling (\ref{eq:lattice_nonlinearity}), is small compared to all dimensions, including the transverse dimension of the resonator.

The TER mechanism was proposed by \textcite{Zener1938}. A detailed analysis of the TER for flexural modes is given by \textcite{Lifshitz2000}. The underlying physics can be understood \cite{Landau1986} if one thinks of thermal expansion and of generating heat by bending a beam or a membrane. For a small temperature change $\delta T$, thermal expansion is proportional to $\delta T$, that is, the relative change of the volume is $\delta V/V = 3\alpha_T\,\delta T$, where $\alpha_T$ is the linear thermal expansion coefficient [it is determined by the coupling parameters (\ref{eq:anharmonic_linear}), see the discussion below Eq.~(\ref{eq:Gruneisen})]. On the other hand, it follows from the thermodynamics that there is an inverse process. Changing the volume leads to heating or cooling. The heat produced by a small volume increment $\delta V$ is $T\partial_T F_{\rm {V-T}}$, where $F_{\rm V-T}$ is the free energy density associated with thermal expansion, $F_{\rm V-T}=
-E\alpha_T \delta T\,\delta V/(1-2\nu_\mathrm{P})V$.

In flexural vibrations one part of the nanoresonator is periodically squeezed ($\delta V<0$) whereas the other is stretched in the counterphase. Therefore there emerges a temperature gradient across the resonator. It dissipates via thermal conductivity, which leads to vibration decay.

If the thickness of the nanoresonator in the bending direction is $l_\perp$, the characteristic time of thermal diffusion across it is
\begin{align}
\label{eq:Zener_time}
\tau_Z = C\rho l_\perp^2/\pi^2\kappa_T,
\end{align}
where $C$ is the specific heat per unit mass and $\kappa_T$ is the thermal conductivity \cite{Zener1938}.
If this time is small compared to the vibration period, the nanoresonator is essentially isothermal and TER is not efficient. In the opposite limit $\omega_0 \tau_Z \gg 1$ TER is not efficient either, since the heat does not have time to propagate across the resonator over the vibration period and is locally averaged out over the compression-extension cycle. An intuitive approximation for the decay rate is provided by the Zener expression \cite{Zener1938}
\begin{align}
\label{eq:TER}
\Gamma^{\rm TER} =\frac{E\alpha_T^2T\omega_0}{2C\rho}\frac{\omega_0\tau_Z}{1+(\omega_0\tau_Z)^2}.
\end{align}
In agreement with the above qualitative arguments, the decay rate becomes small both where $\omega_0\tau_Z$ is large or small.

A quantitative analysis of the dynamics of a flexural mode in a nanobeam can be done by writing the equation of motion for the displacement in the bending direction which, along with the elastic force, takes into account the force from the thermal expansion. This equation and the thermal diffusion equation form a system of two coupled linear equations. The complex eigenvalues of these equations   describe the decay rate and the frequency shift of the flexural mode due to the thermoelastic effect \cite{Lifshitz2000}.

On the experimental side, the TER has been seen in both micro- and nanomechanical resonators at room temperature, see \cite{Roszhart1990,Yasumura2000,Verbridge2006,Chandorkar2009,Ghaffari2015}. As the temperature is decreased, the decay rate (\ref{eq:TER}) also decreases and other decay mechanisms come into play. In addition,  the mean free path of thermal phonons in nanoresonators can become larger than the resonator thickness, so that the system is no longer in the TER regime.

\underline{\it Akhiezer damping.} The expression for the decay rate changes if $\omega_0$  largely exceeds the rate of heat diffusion across the resonator, $\omega_0\gg \tau_Z^{-1}$, even though it can still be small or comparable to the relaxation rate of thermal phonons. Then the phonons do not have time to equilibrate locally to different temperatures in different parts of the resonator.  The decay mechanism in this case was discussed by \textcite{Akhiezer1938} in the context of ultrasound absorption in solids. The corresponding mechanism of ultrasound absorption is called Akhiezer damping. It has been attracting much attention and has been studied for various models of the phonon-phonon coupling, cf. \cite{Woodruff1961,Maris1966,Gurevich1968,Maris1968,Garanin1992} and references therein.

The idea of the Akhiezer damping directly extends to the decay of low-frequency vibrational modes in nano- and micromechanical systems. In this context it  was analyzed in several papers \cite{Kiselev2008,Kunal2014,Iyer2016,Atalaya2016,Hamoumi2018}; a detailed experimental study of the temperature dependence of Akhiezer damping in Si micromechanical resonator is described by \textcite{Rodriguez2019}, see also \cite{Ghaffari2013} for a review of the earlier work.

To give an idea of the mechanism we consider the coupling of a low-frequency mode to high-frequency phonons in the deformation potential approximation \cite{Gurevich1988}. This approximation corresponds to the choice of the coupling parameters $V_{\kappa\kappa'}$ in Eq.~(\ref{eq:anharmonic_linear})   based on the picture \cite{Akhiezer1938} in which a low-frequency mode, with a spatially smooth displacement field ${\bf u}_{}(\rb)$, weakly locally distorts the crystal. The distortion leads to coordinate-dependent changes $\delta\omega_\kappa$  of the frequencies $\omega_\kappa$ of high-frequency modes,
\begin{align}
\label{eq:Gruneisen}
\delta\omega_\kappa =- \omega_\kappa\gamma_\kappa ^{\rm (G)}\n \ub{}, \qquad {\bf u}_{}(\rb) = q\boldsymbol{\varphi}(\rb)
\end{align}
(we remind that $\boldsymbol{\varphi}(\rb)$ is the dimensionless local displacement due to the considered low-frequency mode).

The parameter  $\gamma_\kappa^{\rm (G)}$ determines the coupling constants $V_{\kappa_1\kappa_2}$ in Eq.~(\ref{eq:anharmonic_linear}).  The average of $\gamma_\kappa ^{\rm (G)}$ over the phonons with the weight given by the phonon heat capacities  gives the Gr\"uneisen parameter $\gamma^{\rm (G)}$. This parameter is immediately related to the thermal expansion coefficient $\alpha_T$, i.e., $ \gamma^{\rm (G)} = E\alpha_T /C\rho(1-2\nu_\mathrm{P})$. The approximation (\ref{eq:Gruneisen}) is often generalized by replacing $\n \ub{}$ with the strain tensor associated with $q\boldsymbol{\varphi}(\rb)$, in which case $\gamma_\kappa ^{\rm (G)}$ also becomes a tensor.

Finding the Akhiezer damping rate in nanomechanics requires, as a first step, solving the full quantum kinetic equation for the two-phonon correlation function of thermal phonons, see Appendix~\ref{sec:app_Akhiezer}. This equation goes beyond the conventional kinetic equation for the occupation numbers of phonons \cite{Atalaya2016}.
However, to see the Akhiezer effect qualitatively, one can start with Eq.~(\ref{eq:Landau_Rumer}) that describes phonon scattering off the low-frequency mode. Since high-frequency phonons have finite lifetimes, their energies are uncertain, and in  Eq.~(\ref{eq:Landau_Rumer}) the $\delta$-function of the energy conservation law,  $\delta(\omega_\kappa-\omega_{\kappa'} + \omega_0)$,  can be replaced  by a Lorentzian with a halfwidth given by a phonon decay rate $1/\tau_{\rm ph}$. This rate is the characteristic value of the decay rate $\tau_\kappa^{-1}$ of thermal phonons, which is quadratic in the parameters of the phonon-phonon coupling $V_{\kappa_1\kappa_2\kappa_3}$. Then from Eqs.~(\ref{eq:Landau_Rumer}) and (\ref{eq:Gruneisen})
\begin{align}
\label{eq:Akhiezer}
\Gamma^{\rm Akh} =a^{\rm Akh} C(\gamma^{\rm (G)})^2\frac{\omega_0\tau_{\rm ph}}{1+(\omega_0\tau_{\rm ph})^2}.
\end{align}

The lifting of the energy conservation constraint described by Eq.~(\ref{eq:Akhiezer}) leads to an increase  of the decay rates of low-frequency NVS modes compared to the Landau-Rumer theory (\ref{eq:Landau_Rumer}). It also leads to a specific temperature dependence of the decay rate. If the mean free path of thermal phonons is small compared to the size of the resonator, this dependence is similar to the temperature dependence of ultrasound absorption. The parameter $a^{\rm Akh}$ in the latter case was found by \textcite{Woodruff1961} for a simple model of the coupling to acoustic phonons and for the ultrasound frequency $\omega_0\ll\tau_{\rm ph}^{-1}$, in which case $a^{\rm Akh} =\omega_0 T /3\rho v_s^2$, where $v_s$ is the average speed of sound.
In this regime $\Gamma^{\rm Akh}$ weakly depends on temperature provided $T$ exceeds the Debye temperature. This is because the phonon scattering rate is proportional to the phonon occupation number, i.e., $ \tau_{\rm ph}^{-1} \propto T$ \cite{Gurevich1988}, whereas $C$ is independent of $T$. For temperatures low compared to the Debye temperature, on the other hand, $\Gamma^{\rm Akh}\propto T^{-1}$ in clean systems, where $\tau_{\rm ph}\propto T^{-5}$.

The decay rate (\ref{eq:Akhiezer}) displays a pronounced dependence on the mode eigenfrequency $\omega_0$. It is small both for  $\omega_0\tau_{\rm ph}\ll 1$ and in the opposite limit $\omega_0\tau_{\rm ph}\gg 1$, where the decay is described by the Landau-Rumer-type theory.

The low-temperature behavior of $\Gamma^{\rm Akh}$ changes in thin nanoresonators. When the wavelength of thermal phonons becomes comparable to one of the dimensions of a resonator, the phonon spectrum is quantized and the density of states of phonons is changed. This leads to a change of  both the specific heat and the phonon decay rate. Moreover,  nanoresonators can be, and often are inhomogeneous on the scale of the phonon mean-free path, because of bending, twisting, ripples, etc., which requires a modification of the theory \cite{Atalaya2016}. On the experimental side, in contrast to micromechanical resonators \cite{Rodriguez2019}, the study of Akhiezer damping and the accompanying frequency shift in nanoresonators is at an early stage \cite{Tepsic2021}.

To relate to the previous discussion, we note that the decay of thermal phonons, that underlies the Akhiezer relaxation, is one of the microscopic mechanisms of the anelastic relaxation described by a complex Young modulus, see Sec.~\ref{subsubsec:anelastic_relaxation}. 

\subsection{Losses due to surface effects and two-level systems}
\label{subsec:surface_losses}

Nanomechanical resonators are characterized by a large surface-to-volume ratio. Therefore surface scattering and the defects associated with surfaces may be an important source of mode relaxation, cf. \cite{Yasumura2000,Ekinci2005,Unterreithmeier2010a,Yu2012,Faust2014,Villanueva2014,Hamoumi2018} and references therein. In particular, \cite{Villanueva2014} performed detailed measurements of surface losses in SiN membranes at room temperature as a function of the thickness and also compared different data in the literature; they concluded that the $Q$-factor linearly increases with the increasing thickness. This is expected for surface losses if one thinks of the $Q$ factor as the ratio of the energy stored, which is proportional to the volume, to the energy losses, which linearly increase with the surface area.

Generally, one can think of the surface losses as resulting from ``static'' and ``dynamical'' effects. A simple static effect comes from the static disorder, which leads to scattering  of thermal phonons. The effect is particularly strong  where the phonon mean free path becomes comparable to the thickness. The disorder-induced scattering relaxes the momentum conservation law in phonon-phonon scattering and thus decreases the lifetime of thermal phonons. This leads to the decrease of the relaxation rate of the low-frequency NVS modes in the Akhiezer regime for $\omega_0\tau_{\rm ph} <1$, as seen from Eq.~(\ref{eq:Akhiezer})  \cite{De2016}. However, as mentioned above, one can also think of thermal ``phonons'' somewhat differently, by associating them with the exact vibrational excitations of the disordered system in the harmonic approximation. The coupling of such thermal excitations to the low-frequency NVS modes is different compared to a system with no disorder, cf. \cite{Atalaya2016}. This can increase the Landau-Rumer and Akhiezer relaxation rates compared to those calculated in the absence of disorder.

The dynamical effects of surface disorder come from the defects with internal degrees of freedom, which can absorb energy from the low-frequency NVS modes. The best-known type of such defects are two-level systems (TLSs),  which were introduced by \textcite{Anderson1972,Phillips1972} to explain the anomalous heat capacity and thermal conductivity of glasses at low temperatures. TLSs exist not only on surfaces, but also in the bulk. Their density of states may be higher than the density of states of thermal phonons for low temperatures. For higher temperatures, where the density of states of thermal phonons is higher, the TLSs can ``mediate'' energy transfer from the low-frequency NVS modes to thermal phonons.

It is believed that of utmost importance for relaxation of low-frequency modes in nanoresonators are TLSs with level spacing that significantly exceeds $\hbar\omega_0$. The relaxation is due not  to resonant interlevel transitions of the TLSs, but to nonresonant processes. It has been discussed for a broad range of nanoresonators, such as gold \cite{Venkatesan2010}, polycrystalline aluminum \cite{Hoehne2010}, silica \cite{Riviere2011}, aluminum covered silicon \cite{Lulla2013}, SiN \cite{Faust2014} and GaAs \cite{Hamoumi2018} nanobeams, as well as graphene-based heterostructure membranes \cite{Will2017}, half-ring crystalline Si resonator \cite{Hauer2018}, and phononic crystals \cite{MacCabe2020}; see also \cite{Imboden2014} for a review of the early work.

The dominant mechanism of nonresonant  coupling to a TLS is the modulation of the level spacing by the strain from the vibrational mode. Such coupling is often called  dispersive. It is easy to visualize if one thinks of the TLS states as intrawell states of a particle in a double-well potential, which are hybridized by interwell tunneling. The strain modulates the wells differently, that leads to the change of their relative depths and thus the level spacing. The Hamiltonian of the coupling reads
\begin{align}
\label{eq:H_TLS}
H_{\rm TLS} = C_{\rm TLS} q (\hat n_2 - \hat n_1),
\end{align}
where $\hat n_i$ is the operator of the occupation of the $i$th state of the TLS ($i=1,2$), $q$ is the NVS mode coordinate, and $C_{\rm TLS}$ is the coupling constant.

Relaxation of the mode results from the finite lifetime of the TLS states. By periodically modulating the TLS level spacing, the mode modulates the state populations with a delay determined by the interrelation between $\omega_0$ and the interstate switching rate $\tau_{\rm TLS}^{-1}$. It is this delay that leads to the absorption of the mode energy, i.e., to the mode relaxation. If $\omega_0\ll \tau_{\rm TLS}^{-1}$, the TLS adiabatically follows the mode-induced strain, with essentially no absorption. In the opposite limit,  $\omega_0\gg \tau_{\rm TLS}^{-1}$ the TLS averages out the mode-induced strain, again, with very little absorption. Overall, the absorption coefficient $\Gamma^{\rm TLS}$ as a function of the mode eigenfrequency $\omega_0$ is described by the so-called Debye peak, which was found by \textcite{Debye1929} in the analysis of the dielectric response due to reorientation of polar molecules in crystals,
\begin{align}
\label{eq:Debye_decay}
\Gamma^{\rm TLS}\propto C^2_{\rm TLS}(k_BT)^{-1}\frac{\omega_0\tau_{\rm TLS}}{1+\omega_0^2\tau_{\rm TLS}^2}.
\end{align}
This expression can be immediately derived from the general formulation (\ref{eq:bath_LRT}).

The TLS  relaxation rate $\tau_{\rm TLS}^{-1} $ is determined by the coupling to phonons (or electrons). Generally, this rate depends on the geometry of the nanoresonator and the associated change of the phonon spectrum \cite{Behunin2016}.  At low temperatures, it is dominated by interstate tunneling and single-phonon processes, with $\tau_{\rm TLS}^{-1}\propto \coth(E/2k_BT)$, where $E$ is the level spacing. At higher temperatures phonon scattering off TLSs comes into play. To extend the results to still higher temperatures, TLSs are often thought of as particles in a double-well potential, with $\tau_{\rm TLS}^{-1}$ being determined by the rate of activated interwell switching, $\tau_{\rm TLS}^{-1}\propto \exp(-\Delta U/k_BT)$, where $\Delta U$ is the barrier height, see \cite{Enss2005}.


The overall temperature dependence of the decay rate of the low-frequency NVS modes is obtained by summing the contributions $\Gamma^{\rm TLS}$ for different TLSs. For low temperatures, $T\lesssim 1-3$~K, it is often described by a power law \cite{Venkatesan2010,Lulla2013}, as expected for some models of the TLSs  \cite{Seoanez2008}. For higher temperatures, because of the exponential falloff of  $\tau_{\rm TLS}$ with the increasing temperature, $\Gamma^{\rm TLS}$ may display a peak as a function of temperature where $\omega_0\tau_{\rm TLS} =1$ \cite{Faust2014}. The overall behavior of the decay rate with temperature depends on whether there are various types of TLSs or, as in the case of certain surface defects, the distribution of the TLS parameters is narrow, cf. \cite{Faust2014,Hamoumi2018}.

Generally, resonant absorption by low-energy TLSs with the level spacing $\hbar\omega_0$ can also contribute to the mode decay \cite{Remus2009}. It would be characterized by absorption saturation and the associated decrease of the decay rate with the increasing mode amplitude, similar to ultrasound absorption \cite{Golding1973} and the absorption of microwave radiation in superconducting cavities \cite{Gao2007}. However, this behavior is most clearly manifested for
$\hbar\omega_0\gtrsim k_BT$, a demanding condition in nanomechanics.


\subsection{Electronic relaxation}
\label{sub:electronrelaxation}

The thermal bath can be the electrons flowing through the resonator. It can also be the conducting electrons in a device capacitively coupled to the resonator. Both layouts are similar. Electron transport is used in many resonators to transduce mechanical vibrations into a measurable signal and to drive the motion via the capacitive force, i.e., by modulating the potential $V_{\rm g}$ between the gate electrode and the nanoresonator, as discussed in Sec.~\ref{sec:resonators_characteristics}. Conversely, the coupling to the electron system leads to the relaxation of mechanical vibrations via electrical dissipation.

\begin{figure}[h]
\includegraphics{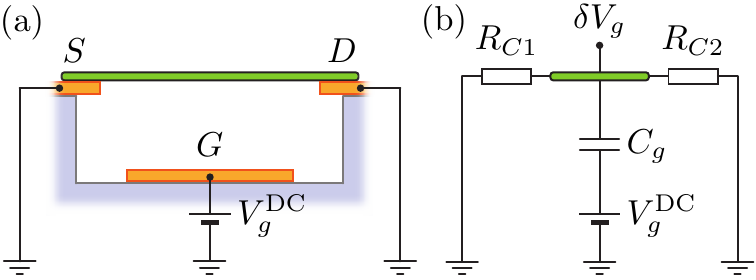} \hfill
\caption{(a) Schematic and (b) equivalent circuit of a suspended wire capacitively coupled to the gate electrode $G$ and electrically contacted to the electrodes $S$ and $D$. The resistances at the interface between the wire and the electrodes are $R_\mathrm{c1}$ and $R_\mathrm{c2}$. }
\label{fig:electricallosses}
\end{figure}

The simplest relaxation mechanism is Ohmic losses in the electronic circuit.
To illustrate this mechanism, we consider a mechanical resonator based on a suspended wire coupled to the gate electrode which, along with the elements of the electronic circuit, serve as the thermal reservoir, see Fig.~\ref{fig:electricallosses}a. The coupling to the electronic degrees of freedom comes from the potential $V_{\rm g}$ in Fig.~\ref{fig:electricallosses}b that fluctuates due to the noise in the electron system. The coupling is described by the Hamiltonian
\begin{align}
\label{eq:fluct_potential}
H_i = qh_{\rm b}, \quad h_{\rm b} = -C'_{\rm g}\overline{V}_{\rm g} \delta V_{\rm g}
\end{align}
We assume here that the fluctuations of the potential $\delta V_{\rm g}$ are small compared to its mean value $\overline{V}_{\rm g} \equiv V_{\rm g}^{\rm DC}$, cf. Eq.~(\ref{eq:capacitiveforce}).
From the general expression (\ref{eq:Gamma_and_P}), the contribution $\Gamma_e$ of the coupling (\ref{eq:fluct_potential}) to the decay rate $\Gamma$ of the resonator in the classical case $k_BT\gg \hbar\omega_0$ is
\begin{align}
\label{eq:electronic_decay}
\Gamma_e = (4M{}k_BT)^{-1}  (C'_{\rm g}V_{\rm g}^{\rm DC})^2\int_{-\infty}^\infty dt e^{i\omega_0 t}\langle \delta V_{\rm g}(t)\delta V_{\rm g}(0)\rangle.
\end{align}
The power spectrum of the voltage fluctuations has a simple form for the Ohmic resistance in the device shown in Fig.~\ref{fig:electricallosses}a, where we disregard the Coulomb interaction between the electrons and phase-coherent effects. For simplicity, we assume that the impedance of the circuit at the frequency $\omega_0$ is given by large resistances $R_\mathrm{c1}$ and $R_\mathrm{c2}$ at the interface between the wire and the electrical leads. The corresponding resistors are connected in parallel to the ground, see Fig.~\ref{fig:electricallosses}b, so that the total resistance is  $R_{\rm eff}=R_\mathrm{c1}R_\mathrm{c2}/(R_\mathrm{c1} + R_\mathrm{c2})$. Then the correlator of $\delta V_{\rm g}$ is determined just by the Johnson$-$Nyquist noise, resulting in
\begin{align}
\Gamma_e= (C'_{\rm g}V_{\rm g}^{\rm DC})^2R_{\rm eff}/2M.
\label{eq:electronicdissipation}
\end{align}
We note that this expression for the decay rate can be obtained directly from Eq.~(\ref{eq:fluct_potential}) by noting that the force on the resonator is $C'_{\rm g}\overline{V}_{\rm g} \delta V_{\rm g}$. The motion of the resonator modulates the charge on it, leading to a current $\approx C'_{\rm g}V_{\rm g}^\mathrm{DC}\dot q$. As a result, the potential $\delta V_{\rm g}$ is modulated. If the vibration frequency $\omega_0$ is small compared to the relaxation rate of the circuit $(R_{\rm eff}C_{\rm g})^{-1}$, then $\delta V_{\rm g}$ follows $\dot q(t)$ adiabatically, so that the corresponding part of $\delta V_{\rm g}$ has the form $[\delta V_{\rm g}]_q = -R_{\rm eff} C'_{\rm g}V_{\rm g}^{\rm DC} \dot q$. By substituting this expression into the force on the resonator, we obtain the friction force $-2M{}\Gamma_e \dot q$. An alternative derivation based on calculating the energy loss due to the resistance of the nanoresonator is given by \textcite{Song2012}.

The electronic relaxation has been measured in resonators based on two-dimensional systems, such as graphene, WSe$_2$ monolayers, and van der Waals stacks~\cite{Song2012,Morell2016,Will2017}. The measured mechanical dissipation rate increases quadratically with $V_\mathrm{g}^\mathrm{DC}$, in agreement with Eq.~(\ref{eq:electronicdissipation}). The electrical resistance obtained from the mechanical dissipation is in a reasonable agreement with the resistance of the device. A quantitative comparison is often challenging, especially when the spatial flow of the vibrations-induced current is not known precisely due to the geometry of the device.

The electron-electron interaction can strongly modify the relaxation rate of a nanoresonator. The effect has been studied in several theoretical papers for low-resistive nanoresonators capacitively coupled to stationary normal and superconducting single-electron transistors (SETs) and in the layout where the nanoresonator itself is an SET that is capacitively coupled to an immobile gate electrode \cite{Mozyrsky2004,Armour2004,Blencowe2005,Clerk2005,Bennett2006,Pistolesi2007,Usmani2007,Micchi2015}. The underlying physics is related to the dependence of the potential of the SET island, and thus the tunneling rate, on the position of the nanoresonator. Experiments were carried out on SETs~\cite{Woodside2002,Knobel2003,Stomp2005,Zhu2005,Lassagne2009,Steele2009,Bennett2010,Ganzhorn2012,Meerwaldt2012,Benyamini2014,Okazaki2016,Ares2016,Deng2016,Willick2017,Blien2020,Wen2020,Urgell2020,Vigneau2021}, superconducting SETs~\cite{LaHaye2004,Naik2006,LaHaye2009,Pirkkalainen2015}, and double-quantum dots~\cite{Benyamini2014,Khivrich2019}.

The theory takes advantage of the fact that the response of a SET to the position is usually fast on the time scale of the vibration period, i.e., $\omega_0$ is small compared to the tunneling rate. The analysis can be formulated in terms of the linear response of the SET to the vibrations, cf. Sec.~\ref{sec:FDT}. Measurements have shown that the coupling can dramatically increase the mechanical dissipation~\cite{Naik2006,Lassagne2009,Steele2009}.
When the voltage applied between the source and the drain electrodes of the SET is larger than $k_BT/e$, the electronic bath is no longer in equilibrium, and it can cool the thermal vibrations \cite{Clerk2005}. In addition, such an out-of-equilibium electronic bath in an SET can suppress the total dissipation rate of the mechanical resonator to zero, leading to self-oscillation \cite{Usmani2007,Wen2020}. Cooling and self-oscillations can also be produced by an electrothermal reaction force associated with the electrical power dissipated in SETs \cite{Urgell2020}.

The so-called electron shuttles~\cite{Erbe2001,Gorelik1998,Fedorets2004,Koenig2012} can be operated in the self-oscillation regime, too. These are devices where the metal island of the SET placed on a nano- or microcantilever is oscillating between the source and drain leads -- in each oscillation period, the island mechanically transfers a quantized number of electrons from one lead to the other.

There has been also investigated the interplay of the NVS dynamics with other many-electron effects and the effects of the topology and coherence of the electron system. Those include the Kondo~\cite{Goetz2018} and the quantum Hall effects~\cite{Singh2012,Chen2016b}, the electronic Fabry-P\'{e}rot interference in a nanoresonator~\cite{Moser2014},  and the effect of Aharonov-Bohm oscillations in a topologically nontrivial nanowire~\cite{Kim2019}. Yet other manifestations of the coupling of the NVS modes and the electron subsystems were studied in  ballistic p–n junctions~\cite{Jung2019}, field-effect transistors~\cite{Sazonova2004}, and quantum-point contact devices~\cite{Poggio2008}.


\section{Conservative and dissipative nonlinearity}
\label{sec:nonlinearity_general}

Vibration nonlinearity is one of the most important and interesting features of the NVS modes. As mentioned before, since nanomechanical resonators are small, it comes into play already for small vibration amplitudes. Sometimes even thermal fluctuations can be sufficiently large to take the vibrations to a nonlinear regime. This regime is also reached with a modest resonant driving if the quality factor is high; in fact, for high-Q modes care must be taken of staying in the linear regime.

Usually nonlinear effects are separated into  conservative and dissipative. Conservative nonlinearity corresponds to the restoring force being a nonlinear function of the mode coordinate $q$ or, equivalently, to the potential energy of the mode $U(q)$ being different from $M{}\omega_0^2q^2/2$.   If $U(q)$ has inversion symmetry, $U(q) = U(-q)$, as for bending modes in a straight nanotube or a flat membrane, the leading nonparabolic term in $U(q)$  is quartic in $q$. The potential energy of the mode then has the form
\begin{align}
\label{eq:Duffing_potential}
U(q) = \frac{1}{2}M{}\omega_0^2 q^2 + \frac{1}{4}M\gamma q^4.
\end{align}
Such nonlinearity is often called the Duffing nonlinearity or, in terms of nonlinear optics, the Kerr nonlinearity. It has been seen in the majority of NVSs.

Conventionally, in nonlinear dynamics the nonlinearity is considered to be strong where the nonparabolic part of the potential becomes of the same order of magnitude as the parabolic part \cite{Arnold1989}. However, in nanomechanics, in most studies the conservative nonlinearity of NVSs is weak in this sense,
\begin{align}
\label{eq:weak_nonlinearity}
|\gamma| \langle q^4\rangle \ll \omega_0^2\langle q^2\rangle.
\end{align}
However, even where the condition (\ref{eq:weak_nonlinearity}) holds, the effect of the nonlinearity on the dynamics can be strong, provided the decay rate of the mode is small, i.e., $Q\gg 1$. This is clear from the following argument. The nonlinear part of the restoring force $-M\gamma q^3$ shows that the effective ``spring constant'' gets effectively either stronger or softer with the increasing vibration amplitude, depending on whether $\gamma>0$ or $\gamma<0$, respectively. Therefore the resonance frequency becomes dependent on the vibration amplitude $A_{}$. For weak nonlinearity, the shift in $\omega_0$ is quadratic in $A_{}$, see Appendix~\ref{subsec:nonlin_averaging},
\begin{equation}
\delta \omega_0=\frac{3}{8}\frac{\gamma}{\omega_0}A_{}^2.
\label{eq:Duffing_frequency_shift}
\end{equation}
The Duffing nonlinearity becomes important once this frequency shift becomes comparable to the frequency uncertainty associated with the decay rate $\Gamma$. In quantum terms, the energy levels of the mode become nonequidistant, and the nonlinearity becomes important once this nonequidistance becomes comparable to the level width $\propto \hbar\Gamma$, see Appendix~\ref{sec:Duffing_spectra_Appendix}.

Dissipative nonlinearity corresponds, in the phenomenological description, to the friction force being a nonlinear function of the velocity and coordinate. This function changes sign upon time reversal, analogous to the linear friction force $-2M{}\Gamma\dot q$.  In its simplest form, the nonlinear friction force is $\propto q^2\dot q$ [\cite{vanderPol1926}] or $\propto \dot q^3$ [\cite{Rayleigh1894}].
Similar to the conservative nonlinearity, for weakly damped modes the nonlinear dissipative force is important where it is comparable with the linear friction force $-2M{}\Gamma \dot q$, which is much weaker than the restoring harmonic force $-M{}\omega_0^2 q$. In this section we discuss the mechanisms of nonlinearity and some of the key manifestations of nonlinearities in mechanical resonators. Nonlinear resonant phenomena are discussed in more detail in Sec.~\ref{sec:nonlinear_phenomena}.



\subsection{Mechanisms of conservative nonlinearity}
\label{subsec:conserv_nonlin_mechanism}

There are several mechanisms of nonlinearity of the restoring force in nanomechanics. The simplest of them is the nonlinear dependence of the stress (tension) on the displacement field of the mode. A familiar example is provided by a doubly-clamped pre-stressed thin beam \cite{Landau1986}. The change $\Delta L$ of the length $L$ of the beam due to  the transverse displacement $X(z,t)$ in a flexural vibrational mode  is $\Delta L \approx \int_0^L dz (\partial X/\partial z)^2/2$ for small $|dX/dz|$ ($z$ is the coordinate along the beam). The elongation leads to the tension $ES\Delta L/L$, where $S$ is the area of the beam cross-section. This tension adds to the tension $T$ inside the beam, so that the overall restoring force due to the tension is
\begin{align}
\label{eq:tension}
F_T(z) \approx \left[T + (ES/2L)\int dz (\partial X/\partial z)^2\right]\partial^2 X/\partial z^2.
\end{align}
If one substitutes $X(z,t) = q(t)\varphi (z)$, where $\varphi(z)$ is the normalized spatial profile of the mode (which is sinusoidal for strong tension), one can see that the cubic in $X(z,t)$  term leads to the force $-\gamma M q^3$ in the equation of motion for $q(t)$ [Eq.~(\ref{eq:Brownian})], with $\gamma = (E/2L\rho)[\int d\rb (d\varphi/dz)^2]^2$ \cite{Lifshitz2008}.

Nonlinear terms in the restoring  force  can come also from other sources. In particular, they can come from the nonlinear dependence of the resonator-to-gate capacitance $C_{\rm g}$ on the displacement of the resonator. As discussed in Sec.~\ref{subsec:freq_tunability}, for a given mode, the second derivative of $C_{\rm g}$ over the displacement associated with the mode leads to the change of the mode frequency. The higher-order derivatives of $C_{\rm g}$ lead to a force which is quadratic or cubic in the displacement, i.e., has the form $-M\beta q^2 - M\gamma q^3$, cf. \cite{Kozinsky2006,Chan2008a,Eichler2011a,Meerwaldt2012a,Eichler2013}.
The parameters $\beta$ and $\gamma$, which are proportional to the third and fourth derivatives of $C_{\rm g}$, respectively, are quadratic in the gate voltage, cf. Eq.~(\ref{eq:capacitivesoftening}). They scale approximately as the cube and the fourth power of the ratio of the displacement amplitude to the distance between the resonator and the relevant electrodes.

The nonlinearity can also result from electron-vibrational coupling \cite{Steele2009,Lassagne2009,Meerwaldt2012,Yang2016,Moskovtsev2017}. 
Measurements on nano- and micromechanical systems showed that the dependence of the vibration frequency on the amplitude, the so-called backbone curve, can be more complicated than Eq.~(\ref{eq:Duffing_frequency_shift}) \cite{Kacem2009,Polunin2016,Samanta2018,Huang2019,Ochs2021}. This indicates that, in some cases, the restoring force can be proportional to higher powers of the displacement. The backbone curve can become nonmonotonic. At the extrema the vibration frequency is independent of the amplitude, leading to the so called zero-dispersion phenomena \cite{Soskin2003}.

\begin{figure}[h]
\includegraphics[scale=0.8]{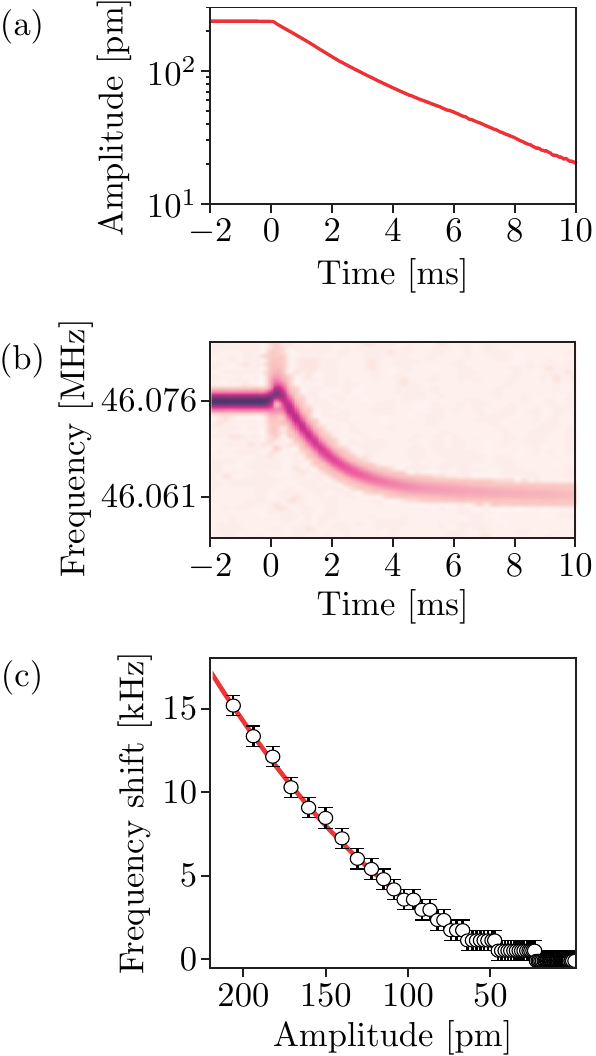} \hfill
\caption{Duffing backbone curve measured in a multi-layer graphene drum. (a) Ring-down measurement. At time $t=0$, the mechanical driving force is
switched off and the vibrational amplitude starts to decay. (b) Time dependence of the short-time Fourier transform of the vibrations during the ring-down measurement. (c) Frequency shift as a function of vibrational amplitude. The red line is the quadratic dependence expected from Eq.~\ref{eq:Duffing_frequency_shift}. Figure adapted from \cite{Guttinger2017}.
}
\label{fig:DuffingFrequancyShiftmeasured}
\end{figure}

The backbone curve of weakly damped modes can be measured directly in the ring-down measurement. It is based on exciting vibrations to a comparatively large  amplitude and measuring their frequency and amplitude as functions of time as the amplitude decays, see Figs.~\ref{fig:DuffingFrequancyShiftmeasured}(a-c) \cite{Londono2015,Polunin2016,Guttinger2017}. This method applies where the nonlinearity is comparatively strong, so that the overall frequency change is much larger than the decay rate.

Usually nanomechanical resonators have several well-resolved low-frequency eigenmodes at a time. These can be flexural or torsional modes, or standing acoustic waves \cite{Barnard2012,Eichler2012,Westra2010, Castellanos-Gomez2012,Mahboob2013,Matheny2013,Yamaguchi2013,Miao2014,Hanay2015,Mathew2016}. They are nonlinearly coupled. The nonlinearity of the elasticity and of the capacitance are the leading sources of this coupling. For flexural modes this is seen already from Eq.~(\ref{eq:tension}) if one takes into account that $X(z)$ is a sum of the displacements of different modes. In the case of nanobeams one should also take into account that nonlinear tension comes from modes that involve displacements in different directions. This is done by adding $(\partial Y/\partial z)^2$ to $(\partial X/\partial z)^2$ in Eq.~(\ref{eq:tension}) \cite{Landau1986}. Similarly, the capacitance of the circuit that incorporates a resonator depends on different contributions to the  displacement, which come from different modes, leading to the mode coupling. A broad range of experiments have been done also on the modes in coupled nanoresonators \cite{Buks2002a,Mahboob2008a,Karabalin2009,Okamoto2009a,Karabalin2011,Mahboob2014,Sun2016,Deng2016,Dong2018}.

The nonlinear part of the energy of the multimode nanoresonator is
\begin{align}
\label{eq:nonlin_multimode}
&U_{\rm nl}(q_1,q_2,...) = \frac{1}{3}M\sum \beta_{n_1 n_2 n_3}q_{n_1}q_{n_2}q_{n_3} \nonumber\\
&+ \frac{1}{4}M\sum \gamma_{n_1n_2n_3n_4}q_{n_1}q_{n_2}q_{n_3}q_{n_4}+\ldots ,
\end{align}
where the subscripts $n_{1,2,3,4}$ enumerate the modes and summation over these subscripts is implied. The mass $M$ here is chosen as an effective mass of one of the modes, cf. the discussion below Eq.~(\ref{eq:mode_displacement}).

For comparatively weak nonlinearity, the major effects of the nonlinear mode coupling can be conditionally  separated into resonant and dispersive. Nonlinear resonant effects occur where a linear combination of the frequencies of several modes or their overtones is equal or close to the frequency of another mode or its overtone \cite{Eichler2012,Antonio2012,Samanta2015,DeAlba2016,Guttinger2017,Czaplewski2018,Houri2019a}. The manifestations of the resonant couplings are discussed in Sec.~\ref{sec:coupling_modes}.

The dispersive coupling is important, on the other hand, where the mode frequencies are different and resonant conditions do not hold. The primary consequence of such coupling is the dependence of the vibration frequency of one mode on the amplitudes of other modes or, in quantum terms, the dependence of  the spacing between the energy levels of one mode on the occupation numbers of the other modes \cite{Ivanov1965,Dykman1971,Dykman1973a,Santamore2004,Westra2010,Venstra2012,Matheny2013,Vinante2014,Miao2014,Sun2016,Maillet2017,Ari2018}.     Similar to Eq.~(\ref{eq:Duffing_frequency_shift}), if the amplitudes of the modes are $A_m$, the change $\delta\omega_n$ of the frequency of the mode $n$ due to the dispersive coupling described by the  quartic in $q_{n_i}$ terms in Eq.~(\ref{eq:nonlin_multimode}) is
\begin{align}
\label{eq:dispersive_shift}
\delta\omega_n = \frac{3}{4\omega_n}\sumprime{m}\gamma_{nnmm}A_m^2,
\end{align}
where the prime indicates that $m\neq n$ in the sum over $m$. We note that the parameters $\gamma_{nnmm}$ here have corrections $\propto \beta_{nmk}^2$ due to the cubic in $q_{n_i}$ terms in Eq.~(\ref{eq:nonlin_multimode}), cf. \cite{Dykman1973a}.

\begin{figure}[h]
\includegraphics{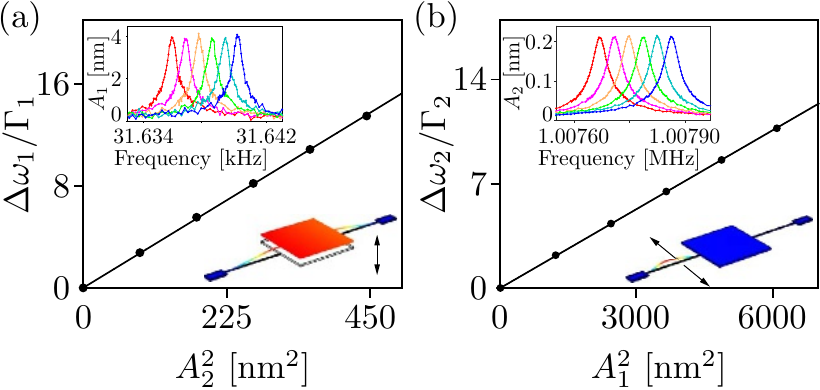} \hfill
\caption{Quadratic dependence of the shift of the mode frequency on the amplitude of another mode. The quadratic dependence indicates the dispersive mode coupling expected in Eq.~\ref{eq:dispersive_shift}. The resonator made of polycrystalline silicon consists of a plate supported on its opposite sides by two beams. Mode~1 involves the translational motion of the plate with both beams bending in the same direction, whereas only one of the beams vibrates in mode~2. The insets in the panels (a) and (b) show the spectra of the linear response of the modes 1 and 2, respectively. Note the strong difference in the mode eigenfrequencies and decay rates. Figure adapted from \cite{Sun2016}.
}
\label{fig:Sun2016freq_shift}
\end{figure}

Figure~\ref{fig:Sun2016freq_shift} shows measurements on two coupled NVS modes featuring the characteristic quadratic dependence of the shift of the  eigenfrequency of one mode on the amplitude of the other mode. Dispersive coupling can also be measured between modes of different nature, such as the flexural mode of a graphene drum and its optical phonon modes~\cite{Zhang2020}.

\subsection{Mechanisms of nonlinear friction}
\label{subsec:nonlin_friction_mechanisms}

A basic microscopic mechanism of nonlinear friction in mesoscopic vibrational systems is a decay process in which two quanta of the considered mode scatter into excitations of a thermal reservoir \cite{Dykman1975a}. The corresponding process is sketched in Fig.~\ref{fig:scattering_diagram}(c), (d), see also Fig.~\ref{fig:interlevel_decay}(b).
An important contribution to nonlinear friction of low-frequency nanomechanical modes can come from the processes described by the quartic nonlinearity. In such processes, a thermal phonon with frequency $\omega_\kappa\gg 2\omega_0$ is scattered off the considered nanomechanical mode with frequency $\omega_0$ into another high-frequency phonon, see Fig.~\ref{fig:scattering_diagram}(d). Since the density of states of thermal phonons is often much higher than of the phonons with frequencies $\approx \omega_0$, these processes may be the leading  cause of nonlinear friction. The frequency difference $2\omega_0$ between the involved thermal phonons can be smaller than their decay rate. This complicates the analysis and makes it similar, to some extent, to the analysis of the thermoelastic or Akhiezer relaxation for linear friction \cite{Atalaya2016}. Another contribution to nonlinear friction was discussed for nonlinear leakage of the flexural modes into bulk acoustic modes \cite{Croy2012}, the nonlinear-friction analog of the standard clamping losses.

A general formulation of the theory of nonlinear friction is similar to that of the linear friction in Sec.~\ref{sec:FDT} \cite{Dykman1975a}. The relevant Hamiltonian of the coupling of the mode to a thermal bath is quadratic in the mode coordinate to allow for the processes where two vibrational quanta of the mode are created or annihilated,
\begin{align}
\label{eq:H_nonlin_friction}
H_i\nl = q^2 h_{\rm b}\nl.
\end{align}
Here $h_{\rm b}\nl$ depends on the dynamical variables of the bath. Similar to the analysis of linear friction, one can express the coefficient of nonlinear friction $\Gamma\nl$ in terms of the susceptibility of the bath with respect to the mode $\chi_{\rm b}\nl(\omega) $,
\begin{align}
\label{eq:coeff_nonlin_fric}
\Gamma\nl =\hbar^{-1}q_0^4\,{\rm Im}~\chi_{\rm b}\nl(2\omega_0), \quad q_0=(\hbar/2M{}\omega_0)^{1/2},
\end{align}
The susceptibility $\chi_{\rm b}\nl(\omega) $ is related by the fluctuation-dissipation relation (\ref{eq:bath_LRT})  to the power spectrum $S_{\rm b}\nl(\omega)$ of  the coupling $h_{\rm b}\nl$ calculated disregarding the effect of the mode on the bath,
\[S_{\rm b}\nl(\omega)=\int_{-\infty}^\infty dt e^{i\omega t} \langle h_{\rm b}^{\rm (nl)}(t)h_{\rm b}^{\rm (nl)}(0)\rangle.
\]
If the fluctuation spectrum $S_{\rm b}\nl(\omega)$ is flat over a broad frequency range that significantly exceeds $2\omega_0$, the classical dynamics of the mode is the Brownian motion described by Eq.~(\ref{eq:Brownian}) with an additional term of the van der Pol friction force
\begin{align}
\label{eq:nonlin_VdPforce}
f_{\rm VdP} = -4M{}\Gamma\nl (q/q_0)^2\dot q.
\end{align}
In the classical temperature range $k_BT\gg \hbar\omega_0$, from Eq.~(\ref{eq:coeff_nonlin_fric}), $\Gamma\nl \propto q_0^4/\hbar \propto \hbar$,  and therefore $\hbar$ drops out of the nonlinear friction coefficient $\Gamma\nl/q_0^2$.
Important for a phenomenological description of nonlinear friction is that the thermal noise that comes along with the friction force $f_{\rm VdP}$ in Eq.~(\ref{eq:Brownian})  depends on the mode coordinate.

In the case of weakly damped modes, which is of interest for nanomechanics, the dynamics can be understood without requiring that  $S_{\rm b}\nl(\omega)$ be flat over the range from $\omega=0$ to $\omega \gg 2\omega_0$. It suffices that  $S_{\rm b}\nl(\omega)$ is smooth in the range centered at $2\omega_0$ with a width that largely exceeds $ \Gamma, \Gamma\nl$ . The classical motion can be conveniently described using the slowly varying complex amplitude $u(t)$, Eq.~(\ref{eq:u(t)}); again, here we assume that the polaronic frequency shift has been incorporated into $\omega_0$. Instead of Eq.~(\ref{eq:eom_classical}), the equation for $u(t)$ now reads \cite{Dykman1984}
\begin{align}
\label{eq:eom_nonlin}
\dot u =& -\Gamma u  - \left(\frac{2\Gamma\nl}{q_0^2} - i\frac{3\gamma}{2\omega_0}\right)u|u|^2 \nonumber\\
 &+\xi(t) + u^*\xi\nl(t)
\end{align}
where $\xi\nl(t)$ is white Gaussian noise, $\langle \xi\nl (t)[\xi\nl (t')]^*\rangle =( 4\Gamma\nl k_BT/\hbar\omega_0)\delta(t-t')$ [we use the Stratonovich convention \cite{Risken1996} for the multiplicative noise $u^*(t)\xi\nl(t)$]. The nonlinear friction term in Eq.~(\ref{eq:eom_nonlin}) has the same form both for the van der Pol and Rayleigh phenomenological nonlinear friction forces. We emphasize that, if the nonlinear friction comes from the coupling to a thermal reservoir, the associated noise term necessarily depends on the dynamical variables of the mode, as seen from the last term in Eq.~(\ref{eq:eom_nonlin}).

Along with the nonlinear friction, we have included into Eq.~(\ref{eq:eom_nonlin}) the Duffing nonlinearity, cf.~Appendix~\ref{subsec:nonlin_averaging}. The coupling to the bath of the form (\ref{eq:H_nonlin_friction}) leads to a renormalization of the Duffing parameter $\gamma$ \cite{Dykman1975a}, which we assume to have been done.

Nonlinear friction in nanomechanical resonators was first measured in driven spectra by exciting the system to a comparatively large vibration amplitude \cite{Eichler2011a,Zaitsev2012}. Such friction results notably in the mechanical linewidth that changes as the vibration amplitude is increased. It can be also observed in ring-down measurements \cite{Polunin2016}, where the decay rate varies as the vibration amplitude gets lower. The systems featuring nonlinear friction include carbon nanotubes as well as single- and multi-layer graphene resonators \cite{Eichler2011a,Miao2014,Singh2016,Guttinger2017,Dolleman2018,Keskekler2021}, PdAu nanobeams  \cite{Zaitsev2012}, silicon MEMS \cite{Nabholz2018}, and  Si$_3$N$_4$ membranes with engineered modes \cite{Catalini2021}. A strong nonlinear friction was observed in a micromechanical resonator submerged into liquid helium at ultralow temperatures; it was related to the amplitude-dependent attachment of vortices and provided a probe of quantized vorticity \cite{Barquist2020}.  Nonlinear friction plays an important role in the dynamics of coupled resonators when they are driven into the regime of self-sustained vibrations \cite{Mahboob2015,Mahboob2016}. It also strongly affects  parametric resonance \cite{Lin2015}, and the related features can be used to develop new types of phase-locked loops \cite{Miller2019}. Another application demonstrated by \textcite{Chen2016} is self-sustained micromechanical  vibrations in a system with linear feedback.

As seen from Eq.~(\ref{eq:coeff_nonlin_fric}), the nonlinear friction is comparatively large if the power spectrum of the thermal bath, and thus the susceptibility Im~$\chi_\mathrm{b}^{\mathrm{(nl)}}(\omega)\propto S_\mathrm{b}^{\mathrm{(nl)}}(\omega)$ are large for $\omega\approx 2\omega_0$. This happens if the considered mode is coupled to a mode with frequency close to $2\omega_0$ and a relaxation rate much higher than $\Gamma, \Gamma^{\mathrm{(nl)}}$. The nonlinear friction in this case corresponds to the process where two quanta of the considered mode scatter into a quantum of the second mode.

\begin{figure}[h]
\includegraphics{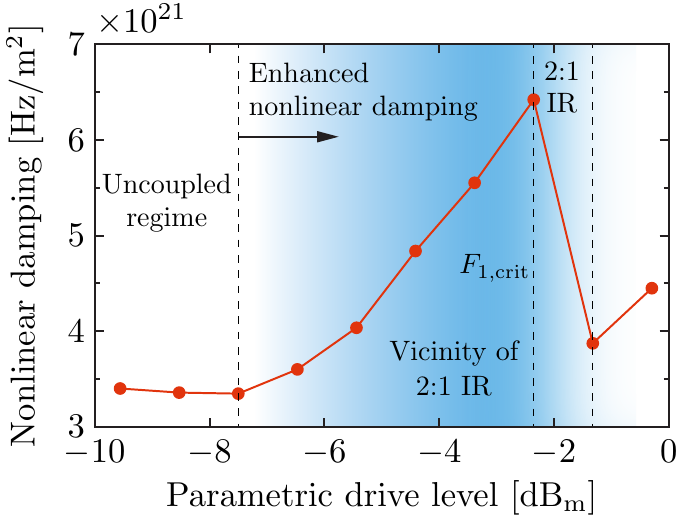} \hfill
\caption{The effective nonlinear damping coefficient of a parametrically driven mode when its vibration frequency is close to half the eigenfrequency of a second mode, i.e., near the $2:1$ internal resonance (IR). Adapted from \cite{Keskekler2021}.
}
\label{fig:Nonlinearfriction21}
\end{figure}

Figure \ref{fig:Nonlinearfriction21} shows an experiment demonstrating the effect \cite{Keskekler2021}. The effective nonlinear friction of the parametrically driven lowest mode of a graphene nanodrum is seen to strongly increase with the drive amplitude, and thus with the vibration amplitude, where the drive frequency $\omega_p\approx 2\omega_0$ was close to the eigenfrequency of the next lowest drum modes. In this system the inequality between the relaxation rates was not sufficiently strong and to describe the observations it was necessary to go beyond the approximation of linear response of the higher-frequency mode underlying Eq.~(\ref{eq:coeff_nonlin_fric}). This explains the behavior of the effective nonlinear damping in Fig.~\ref{fig:Nonlinearfriction21}.

Nonlinear friction can be engineered by parametrically driving coupled mechanical modes at proper combination frequencies. The friction coefficient can be made not only positive, but also negative depending on the drive frequency \cite{Dong2018}.  In microwave electromagnetic cavities, a positive nonlinear friction was engineered in order to create long-lived coherent quantum states, including cat states that can encode qubit states \cite{Leghtas2015,Touzard2018}.




\subsection{Effect of the nonlinearity on the spectra}
\label{subsec:spectra_nonlinear_conservative}

As indicated above, for underdamped vibrational systems, the effects of the vibration nonlinearity become pronounced where the nonlinear part of the restoring or dissipative force is comparable to the linear friction force, while all these forces are much smaller than the linear restoring force $-M\omega_0^2q$. Respectively, the vibrations  remain close to sinusoidal. However, the change of the dynamics can be dramatic. It is determined by the nontrivial interplay of the nonlinearity and linear damping.

\subsubsection{Spectral effects of the Duffing nonlinearity}
\label{subsubsec:nonlinearity_mode}

For not too large  vibration amplitudes, the primary effect of the conservative nonlinearity is the dependence of the vibration frequency on the mode amplitude, Eq.~(\ref{eq:Duffing_frequency_shift}), or mode amplitudes, for coupled modes, Eq.~(\ref{eq:dispersive_shift}). Due to this dependence, the  amplitude fluctuations of the mode (or, in quantum terms, fluctuations of the state populations) are converted into eigenfrequency fluctuations. This leads to spectral broadening and the overall change of the shape of the spectra, making the spectra profoundly non-Lorentzian. The theory of the spectra was first considered in several limiting cases in the quantum regime by \textcite{Ivanov1966}. Later a full classical and quantum theory was developed by \textcite{Dykman1971,Dykman1973a}. Experimentally, the  Duffing nonlinearity-induced evolution of the spectra with the varying fluctuation intensity  was explored in different types of systems, such as silicon nitride nanobeams \cite{Maillet2017}, levitated silica nanoparticles \cite{Gieseler2013,Amarouchene2019}, and a micromechanical trampoline resonator \cite{Huang2019}. \cite{Maillet2017} studied not only the evolution of the power spectrum of NEMS with the varying noise intensity, but also the in-phase and quadrature components of forced vibrations, which give the imaginary and real parts of the susceptibility of a mode with Duffing nonlinearity.

The classical physics of the effect can be readily understood \cite{Dykman1971} by noting that, in thermal equilibrium, the vibration amplitude fluctuates. In the harmonic approximation, the energy of the mode is $M\omega_0^2A_{}^2/2$, where $A$ is the vibration amplitude. Respectively,  the mean square amplitude is $\langle A_{}^2\rangle = 2k_BT/M\omega_0^2$. The amplitude fluctuations lead to the spread of the vibration frequency. From Eq.~(\ref{eq:Duffing_frequency_shift}), this spread, i.e., the characteristic magnitude of the frequency fluctuations is $\overline{\delta\omega_0} =3|\gamma|\langle A_{}^2\rangle/8\omega_0$. A critically important parameter is the relation between $\overline{\delta\omega_0}$ and the decay rate $\Gamma$, i.e, the parameter
\begin{align}
\label{eq:motional_narrowing}
\alpha_0  =\overline{\delta\omega_0}/2\Gamma = 3\gamma k_BT/8M\omega_0^3\Gamma.
\end{align}
It can be called the motional narrowing parameter, to draw the similarity (although somewhat indirect) with the motional narrowing effect in nuclear magnetic resonance \cite{Anderson1954,Kubo1954}.

Indeed, the parameter $\Gamma^{-1}$ is the correlation time of the amplitude fluctuations in the absence
of nonlinear friction, cf.~Eqs.~(\ref{eq:eom_classical}) and (\ref{eq:eom_nonlin}), and thus the correlation time of the frequency fluctuations due to the Duffing nonlinearity. For $|\alpha_0|\ll 1$ the correlation time is small compared to the time $(\overline{\delta\omega_0})^{-1}$. Then the frequency fluctuations are averaged out and $\overline{\delta\omega_0}$ gives just a characteristic shift of the mode frequency while the width of the spectrum is determined primarily by the decay rate $\Gamma$. This is similar to the fast averaging of fluctuations that underlies nuclear magnetic resonance in liquids.

On the other hand, for $|\alpha_0|\gg 1$, the spectrum of the oscillator can be thought of as a superposition of ``partial spectra''  centered at the frequencies $\omega_0+\delta\omega_0$ that depend on the instantaneous value of the amplitude $A_{}$.
The multitude of the frequencies is an analog of the inhomogeneous broadening in solid state spectroscopy.
The contribution of a partial spectrum at frequency $\omega_0+\delta\omega_0$ to the whole spectrum is determined by the probability density
of having a given $\delta\omega_0$, which is determined by the Boltzmann distribution over $\delta\omega_0\propto A^2$.
The overall width of the spectrum is $\sim \overline{\delta\omega_0}\gg \Gamma$, i.e., the spectral width is determined by the nonlinearity, not the decay, and the spectrum is strongly non-Lorentzian and asymmetric. The quantum picture is discussed in Appendix~\ref{sec:Duffing_spectra_Appendix}.

The intermediate range $|\alpha_0|\sim 1$ is most interesting theoretically, as it shows how the partial spectra of the limit $|\alpha_0|\gg 1$ shrink and deform with the decreasing nonlinearity or fluctuation intensity. Somewhat surprisingly, the spectrum is nevertheless described by a simple explicit expression both in the classical and quantum domains \cite{Dykman1971,Dykman1973a}, see Appendix~\ref{sec:Duffing_spectra_Appendix}. Figure~\ref{fig:classic_Duffing} shows the evolution of the spectrum with the varying $\alpha_0$.

Dispersive nonlinear coupling between the modes affects the spectrum in a similar way   \cite{Dykman1971,Dykman1973a}. As seen from Eq.~(\ref{eq:dispersive_shift}), the typical  spread of the frequency of mode $n=0$ due to fluctuations of the amplitude of mode $m>0$ is   $\overline{\delta\omega_0} = 3|\gamma_{00mm}| \langle A_m^2\rangle/4\omega_m$. The correlation time of the relevant fluctuations is the reciprocal decay rate $\Gamma_m^{-1}$ of the mode $m$. Therefore for $\overline{\delta\omega_0}\gtrsim \Gamma_m$ the spectrum is broadened and becomes non-Lorentzian, see  Appendix~\ref{sec:Duffing_spectra_Appendix}.
The corresponding spectral broadening  has been suggested as a major broadening mechanism for flexural modes in carbon nanotubes \cite{Barnard2012}, graphene sheets \cite{Miao2014}, doubly clamped beams \cite{Venstra2012,Matheny2013} as well as microcantilevers\cite{Vinante2014}.

\begin{figure}[h]
\includegraphics[scale=0.8]{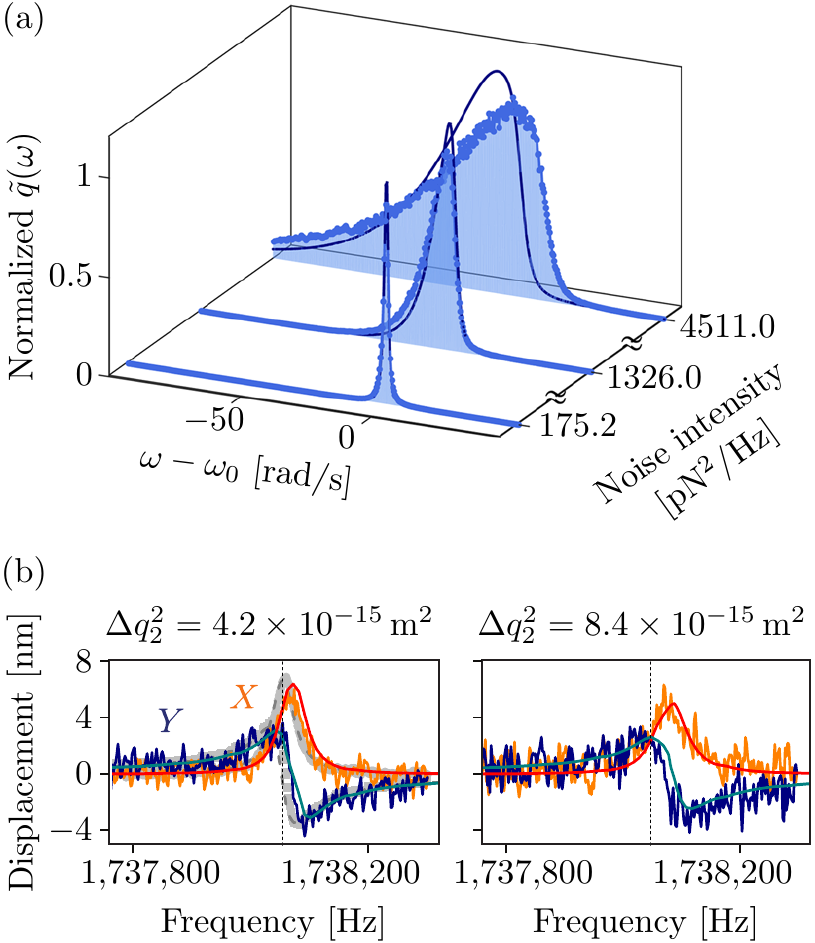} \hfill
\caption{(a) The evolution of the spectrum of a micromechanical trampoline resonator with the Duffing nonlinearity (\ref{eq:Duffing_potential}) upon increasing the effective temperature that is determined by the applied broad-band noise. When the noise intensity is small, the
spectrum is Lorentzian. At large noise intensity, the spectrum becomes highly asymmetric due to the interplay between the noise and the Duffing nonlinearity. At still larger noise intensities it was necessary to take into account higher-order nonlinearity. The solid lines correspond to the spectra expected without any free parameters. Adapted from  \cite{Huang2019}. (b) The in-phase and quadrature component of a weakly driven nanomechanical mode dispersively coupled to a fluctuating  mode $m=2$ for different values of the mean-square displacement $\Delta q_2$. The displacement fluctuations of the $m=2$-mode are driven by electric noise. When these fluctuations are small, the response of the measured mode (gray data points) is symmetric with respect to the resonance frequency (vertical dashed line). For large fluctuations, the response becomes asymmetric. Adapted from \cite{Maillet2017}. The solid lines in (a) and (b) are calculations with no free parameters.
}
\label{fig:broadening_Duffing}
\end{figure}

Figure~\ref{fig:broadening_Duffing} presents experimental results, which show how the power spectrum and the real and imaginary parts of the susceptibility are changed with the varying noise intensity. Panel (a) refers to the case where the dispersive coupling to other modes is small. The spectrum is determined by the internal mode nonlinearity and its shape
evolves from a Lorentzian to a strongly asymmetric peak with the increasing $\alpha_0$.
Panel (b), on the other hand, illustrates the effect of nonlinear dispersive mode coupling. The internal Duffing nonlinearity could be disregarded. The shape of the spectrum depends on the scaled coupling parameter  $\alpha_1=3\gamma_{0011}k_BT/4M\omega_0\omega_1^2\Gamma_1$.

An interesting behavior occurs where the number of the modes dispersively coupled to the considered mode is large, even though the coupling to each mode taken separately is small. In this case the power spectrum $S_{}(\omega)$ may become Gaussian in its central part \cite{Zhang2015a}, as first found numerically by \textcite{Barnard2012}, see Appendix~\ref{subsec:classical_spectrum}.

Dispersive coupling could be used for quantum non-demolition measurements of phonons~\cite{Santamore2004a}. In such a measurement, the number of phonons of one mode would be continuously measured by recording the resonance frequency of the second mode. A realization of such an experiment in nanomechanics is challenging, since it requires strong dispersive coupling compared to the decay rates of the modes.

\subsubsection{Dispersive coupling of a nanomechanical mode to a qubit}
\label{subsubsec:qubit_coupling}

A promising direction of nanomechanics is the study of the effects of coupling of nanomechanical vibrations to controllable two-level systems, qubits. The involved physics is closely related to the physics of the electronic states of defects coupled to phonons in solids \cite{Stoneham2001}, NV centers being one of the defects of utmost current interest. Coupling nanomechanical modes to qubits opens a path toward studying the electron-phonon effects with an unprecedented control. It also provides new means to manipulate and measure the quantum states of mechanical resonators operating in the GHz range \cite{O'Connell2010,Gustafsson2014,Satzinger2018,Chu2018,Bienfait2019}. Compared to superconducting resonators, GHz mechanical resonators can have longer lifetimes \cite{MacCabe2020} and are usually much more compact \cite{Safavi-Naeini2019}. These systems hold promise for scalable qubit architectures where quantum information is stored in bosonic systems \cite{Hann2019,Ofek2016,Lescanne2020}.

In the systems studied so far, the qubit-mode coupling energy was smaller than the mode level spacing $\hbar\omega_0$. In this case, if the level spacing of the qubit $\hbar\omega_{\mathrm{qbt}}$ is significantly different from $\hbar\omega_0$, of primary interest is the study of the dispersive coupling. Such coupling has attracted much attention in solid-state physics \cite{Sild1988}.  If the qubit is described as a spin-1/2 system with the Hamiltonian $H_{\rm qbt} = \hbar\omega_{\mathrm{qbt}}\sigma_z/2$, where $\sigma_z$ is the Pauli matrix, the dispersive coupling to the nanomechanical mode has the form
\begin{align}
\label{eq:qubit_dispersive}
H_i^{\mathrm{qbt}} = \frac{1}{2}\hbar \gamma_{\mathrm{qbt}}\sigma_z q^2,
\end{align}
where $\gamma_{\mathrm{qbt}}$ determines the coupling energy.

The ensuing physics is very similar to the physics of dispersively coupled vibrational modes. If one thinks of the mode vibrations classically, one can see that the coupling-induced change of the qubit transition frequency is determined by the vibration amplitude $A_{}$,
\[\delta\omega_{\mathrm{qbt}} = \frac{1}{2}\gamma_{\mathrm{qbt}}A_{}^2.\]
In the classical limit, $k_BT\gg \hbar\omega_0$, the shape of the qubit spectrum is determined by the ratio of the characteristic value of the frequency shift $\overline{\delta\omega}_{\mathrm{qbt}} =  \gamma_{\mathrm{qbt}}k_BT/M\omega_0^2$ to the reciprocal correlation time of the frequency fluctuations $\Gamma$, going from the ``motional narrowing'' regime where this ratio is small to the ``inhomogeneous broadening'' regime where it is large.

In the quantum regime, $\delta\omega_{\mathrm{qbt}}$ takes on discrete values that correspond to different occupation numbers $n_0$ of the vibrational mode,
\begin{align}
\label{eq:qubit_frequecy_shift}
\delta\omega_{\mathrm{qbt}}(n_0) = (\hbar\gamma_{\mathrm{qbt}}/M\omega_0)(n_0+1/2).
\end{align}
If $(\hbar\gamma_{\mathrm{qbt}}/M\omega_0)\gg \Gamma(2\bar n+1)$, the lines at $\omega_{\mathrm{qbt}}+\delta\omega_{\mathrm{qbt}}(n_0)$ with different $n_0$ weakly overlap. This leads to a fine structure of the qubit spectrum and  enables identifying the population of the mode Fock states $\Ket{n_0}$. The shape of the lines of the fine structure is close to Lorentzian, with halfwidth $\Gamma_{\mathrm{qbt}}+ 2\Gamma[\bar n(2n_0+1)+n_0]$ that linearly increases with $n_0$, similar to the case of dispersively coupled modes, cf. Eqs.~(\ref{eq:suscept_time_explicit}) - (\ref{eq:cpld_modes_parameters}); here $\Gamma_{\mathrm{qbt}}$ is the halfwidth of the qubit spectral line in the absence of coupling to the mode. Such a regime is an analog of the inhomogeneous broadening.

If $(\hbar\gamma_{\mathrm{qbt}}/M\omega_0)\lesssim \Gamma(2\bar n+1)$, the lines with different $n_0$ overlap. In this case quantum mechanics does not allow one to identify individual Fock states from the spectrum. As explained in Appendix~\ref{sec:Duffing_spectra_Appendix}, the amplitudes of the transitions with different $n_0$ are coupled. The overall qubit spectrum shrinks down with the decreasing $\hbar\gamma_{\mathrm{qbt}}/M\omega_0\Gamma$. Where this parameter is small the contribution of the dispersive coupling to the halfwidth of the qubit spectrum is $(\hbar\gamma_{\mathrm{qbt}}/M\omega_0\Gamma)^2 \Gamma \bar n(\bar n+1)/2$ \cite{Krivoglaz1965}. For systems in thermal equilibrium, the spectrum for an arbitrary $\hbar\gamma_{\mathrm{qbt}}/M\omega_0\Gamma$ was described by \textcite{Dykman1987}. Using a qubit to study the vibration-number statistics in the case of a driven mode was considered by \textcite{Clerk2007}.

\begin{figure}[h]
\includegraphics[width=7cm]{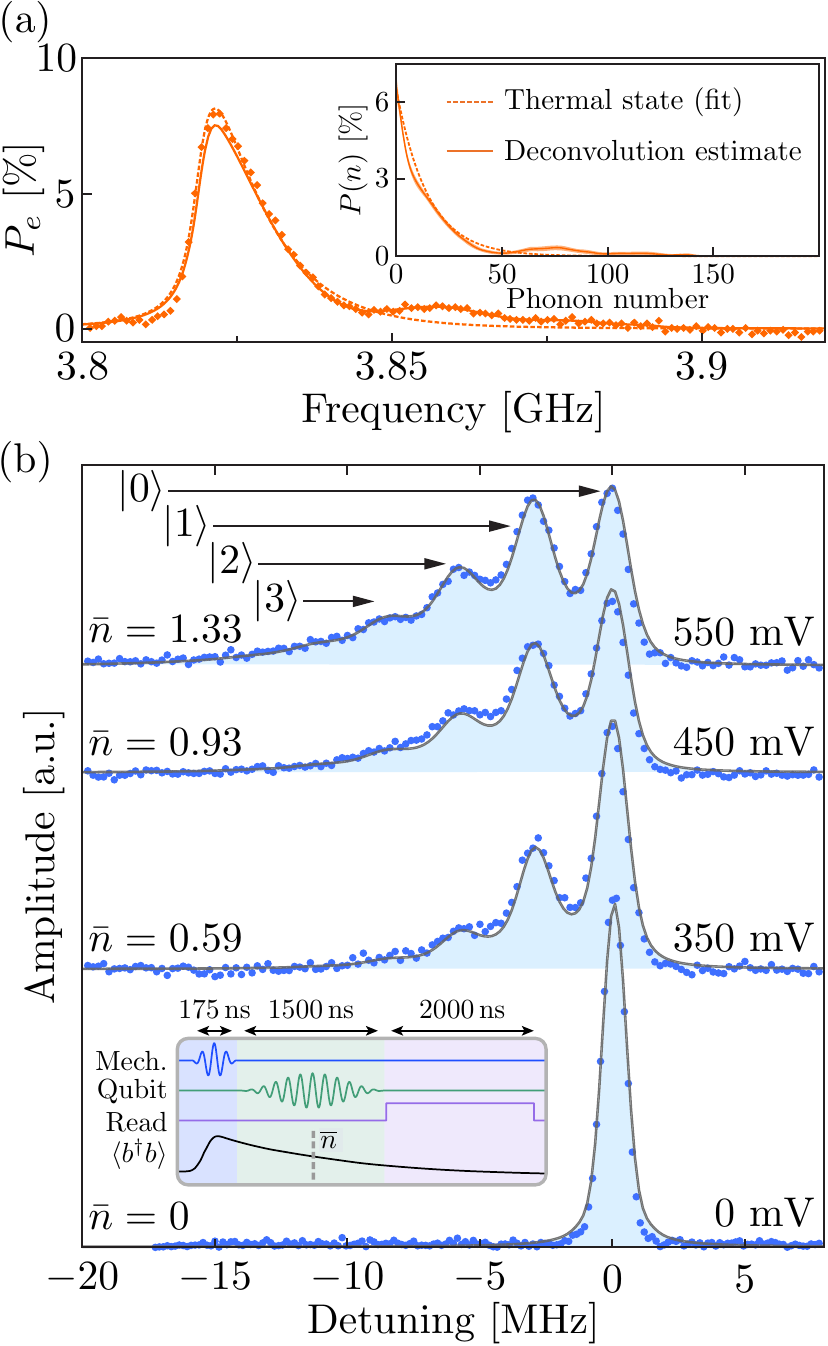} \hfill
\caption{(a) Spectroscopy of the qubit coupled to the micromechanical oscillator at thermal equilibrium in a dilution fridge. The asymmetry of the peak indicates an oscillator occupation number that is low but significantly larger than one. The oscillator vibrates at 25~MHz, whereas the qubit frequency is 3.82~GHz. The coupling parameter $\gamma_{\mathrm{qbt}}$ is positive. Inset: Phonon populations extracted from the spectrum with a fit assuming a thermal distribution (dashed line) or with a
Bayesian-based deconvolution algorithm (full line). Adapted from  \cite{Viennot2018}. (b) Pump–probe measurement consisting of a short phonon excitation pulse followed by a longer qubit spectroscopy pulse. The detuning on the horizontal axis is relative to the qubit frequency $\omega_{\mathrm{qbt}}$ = 2.317 GHz in the absence of a phonon excitation pulse; the mechanical oscillator vibrates at 2.405~GHz. The coupling parameter $\gamma_{\mathrm{qbt}}$ is negative. The initial phonon populations prepared by the pulse decay over the course of the measurement but are nevertheless visible as individual peaks separated
by $\hbar\gamma_{\mathrm{qbt}}/M\omega_0$. The blue points is a fit using numerical master-equation simulations. Adapted from \cite{Arrangoiz-Arriola2019}.
}
\label{fig:qubit_spectrum}
\end{figure}

In Fig.~\ref{fig:qubit_spectrum} we present the results of the measurements of the spectrum of Josephson junction qubits coupled to  a membrane resonator  \cite{Viennot2018} and to a high-frequency phononic crystal defect \cite{Arrangoiz-Arriola2019}. In the first system there was studied the regime where a large number of vibrational states was occupied, whereas in the second system it was possible to reach small occupation numbers and to resolve the fine structure of the spectrum.

It should be noted that the experiments by \textcite{Arrangoiz-Arriola2019} and a part of the experiments by \textcite{Viennot2018} were performed where the vibrational system was away from thermal equilibrium. The full analysis of the spectrum in this case requires taking into account the coupling of the complex amplitudes of transitions between the states of the mode in the transient regime \cite{Dykman1975,Zhang2017b}.


\subsubsection{Broadening of the power spectrum due to nonlinear friction}
\label{subsubsec:nonlin_friction_broadening}

Similar to the conservative nonlinearity, nonlinear friction also leads to a broadening and a change of the shape of the oscillator power spectrum \cite{Dykman1975a}. The effect depends on temperature even where the nonlinear friction coefficient $\Gamma^\mathrm{(nl)}$ defined by Eq.~(\ref{eq:coeff_nonlin_fric}) is temperature-independent. In the classical limit, this is clear already from Eq.~(\ref{eq:eom_nonlin}). Indeed, the nonlinear friction force is increasing with the increasing vibration amplitude. Therefore, as the mean squared amplitude is increasing with the increasing temperature, so is the nonlinear friction force.

It is seen from Eq.~(\ref{eq:eom_nonlin}) that, since in thermal equilibrium $\langle |u|^2\rangle = k_BT/2M\omega_0^2$, the characteristic parameter of nonlinear friction in the classical theory is $2\Gamma^\mathrm{(nl)}k_BT/\hbar\omega_0$. The evolution of the spectrum with the varying ratio $\gamma^\mathrm{(nl)}\equiv 2\Gamma^\mathrm{(nl)} k_BT/\hbar\omega_0\Gamma$ is shown in Fig.~\ref{fig:nonlin_fric_spectrum}. If the vibration frequency does not depend on the amplitude, the spectrum remains symmetric, but it becomes profoundly non-Lorentzian. Its width increases with the increasing nonlinear friction, and thus with the increasing temperature.

Generally, both nonlinear friction and conservative nonlinearity are present in NEMS and MEMS. Their interplay leads to characteristic features in the spectrum which should allow one to identify the presence of the both mechanisms \cite{Dykman1975a}.
An important indicator of the effect of the conservative nonlinearity, which has been seen in experiments, is an asymmetry of the spectrum.
We are not aware of experiments where a change of the spectrum due to nonlinear friction would have been established. This is to be contrasted with the observation of nonlinear friction in strongly driven systems described in Sec.~\ref{subsec:nonlin_friction_mechanisms}.


\section{Nonlinear resonant phenomena: a laboratory for studying physics far from thermal equilibrium}
\label{sec:nonlinear_phenomena}

Nonlinear resonant response of vibrational modes to an external drive leads to several groups of phenomena, which are of interest for nanomechanics but also makes nanomechanics a testing ground for such diverse areas as  statistical physics far from thermal equilibrium, nonlinear dynamics, and quantum cavity/circuit electrodynamics. This is because, for the characteristic small damping, the response becomes nonlinear already for comparatively weak resonant driving. This allows one to study nonlinear effects in a well-controlled fashion over a broad parameter range.


\subsection{Bistability of resonantly excited vibrations}
\label{subsec:bistability}

One of the simplest and yet very rich nonlinear effects studied in nanomechanics is hysteresis of the vibrations excited by applying a close to resonant driving force.  The hysteresis emerges when the frequency $\omega_F$ or the amplitude $F$ of a moderately strong drive is swept across a certain range. The onset of the hysteresis is a consequence of  coexistence of two stable vibrational states of the nonlinear mode, i.e., of the mode bistability \cite{Landau2004a}. Such hysteresis was observed in nanomechanical systems early on \cite{Husain2003}. In most cases it is well described already by the simplest model of the vibration  nonlinearity, the Duffing model introduced in Sec.~\ref{sec:nonlinearity_general}. In this section we will concentrate on the nonlinear resonant response described by this model.

\begin{figure}
\includegraphics[width=4.2 cm]{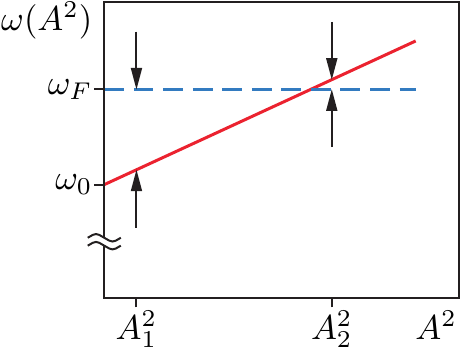}
\caption{A sketch of the dependence of the vibration frequency on the vibration amplitude $A$ for a comparatively small nonlinearity. From Eq.~(\ref{eq:Duffing_frequency_shift}),  the frequency linearly depends on  $A^2$, and for the Duffing model $\omega(A^2)-\omega(0)\approx 3\gamma A^2/8\omega(0)$ (note that $\omega(0)\equiv \omega_0$). When the mode is driven by a field at frequency $\omega_F$, it can have stable vibrational states with comparatively small and comparatively large amplitudes $A_1$ and $A_2$, for which $\omega(A^2)$ is further away or closer to $\omega_F$.
}
\label{fig:sketch_bistability}
\end{figure}

A phenomenological  equation of motion (\ref{eq:Brownian}) extended to include the close to resonance driving $F\cos\omega_F t$ and the Duffing nonlinearity reads
\begin{align}
\label{eq:driven_Brownian}
&M\ddot q + 2M\Gamma\dot q +M\omega_0^2 q + M\gamma q^3 \nonumber\\
&=  F\cos\omega_F t + f_T(t), \quad |\omega_F -\omega_0|\ll \omega_0.
\end{align}
Qualitatively, the occurrence of two stable vibrational states in the absence of noise can be inferred in the following way, cf. Fig.~\ref{fig:sketch_bistability}. Suppose the drive frequency $\omega_F$ is close to the eigenfrequency $\omega_0$, but the detuning $|\omega_F -\omega_0|$ considerably exceeds the decay rate. Then one may expect that the amplitude of the forced vibrations is comparatively small. It is given by the value $A_1$ in Fig.~\ref{fig:sketch_bistability}. However, the nonlinearity leads to a change of the mode frequency $\delta\omega_0$ with the mode amplitude, as described by Eq.~(\ref{eq:Duffing_frequency_shift}). If the vibration amplitude is sufficiently large and takes on the value $A_2$ in Fig.~\ref{fig:sketch_bistability}, the shifted frequency $\omega_0+\delta\omega_0$ can become very close to $\omega_F$, so that there is strong resonance. Such resonance will make vibrations with the corresponding large amplitude self-consistent. This leads to two stable states: Forced vibrations can either have large amplitude and be at good resonance or have small amplitude and be detuned from the resonance.

\begin{figure}[h]
\includegraphics[scale=0.9]{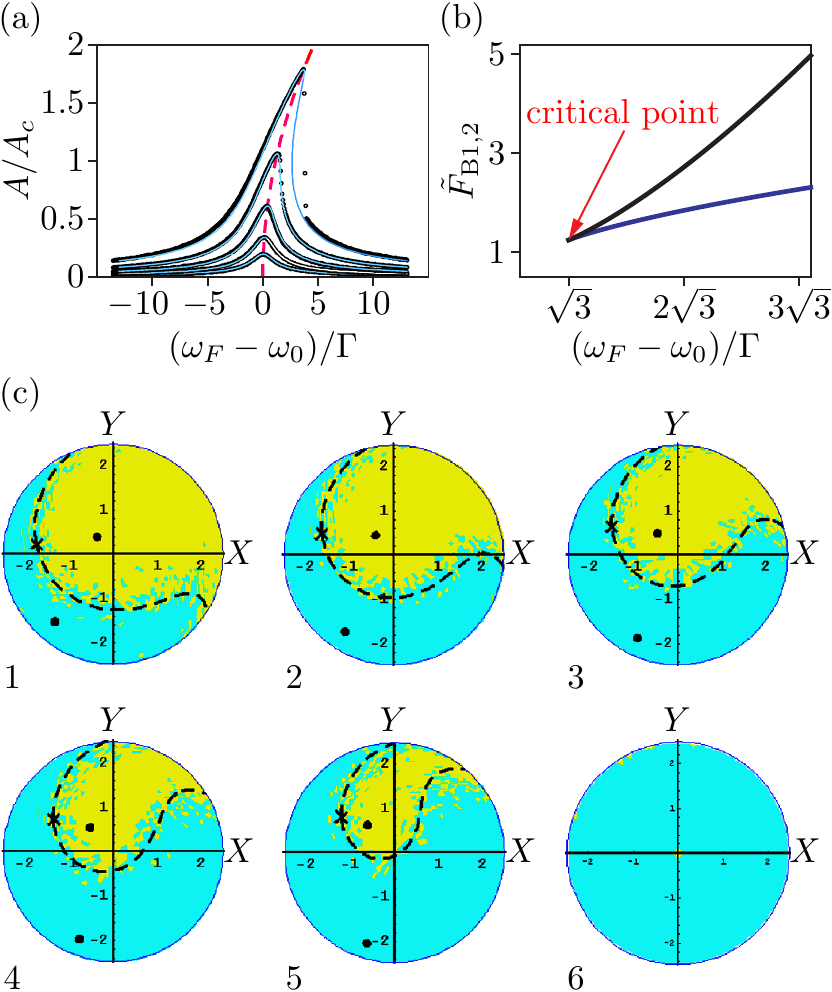} \hfill
\caption{(a) Vibration amplitude $A$ of a resonantly driven platinum nanowire versus frequency, for various driving powers showing the onset of bistability \cite{Kozinsky2007}. The curves refer to the drive amplitude $F/F_c = 0.249, 0.443, 0.788, 1.401, 2.492$. The amplitudes $A$ and $F$ are  normalized by their values $A_c$ and $F_c$ at the critical point in (b). (b) The bifurcation diagram of a resonantly driven Duffing oscillator; the scaled amplitude of the drive is $\tilde F= (3\gamma/32\omega_F^3\Gamma^3)^{1/2}F$. The bistability occurs in the interior of the region limited by the bifurcation curves $\tilde F_{B1,2}(\omega_F)$ given by Eq.~(\ref{eq:bif_resonant}). At the critical point $\tilde F_{B1}=\tilde F_{B2}=\tilde F_c$ the three stationary vibrational states (two stable and one unstable) merge, $\tilde F_c =(8/\sqrt{27})^{1/2} $. (c) Basins of attraction of the platinum nanowire resonator at $(\omega_F-\omega_0)/\Gamma= 4.26$ for increasing drive values, $F/F_c=(1)1.867, (2)2.049, (3)2.237, (4)2.434, (5)2.640, (6)2.741$. The variables $X$ and $Y$ are the scaled in-phase and quadrature components of the vibrations $Q$ and $P$, Eq.~(\ref{eq:quadratures}). Blue and yellow colors indicate the initial states in which the system is prepared by driving it at frequency $\omega_F$ with a certain amplitude $F$. The amplitude is then switched to a value above $F/F_c$ and the drive phase is adjusted. After that the system goes to the final high- or low-amplitude stable state depending on where it was prepared.  The theoretical positions of the stable states and the saddle point, and the separatrix curve are indicated by the black points, the black cross, and the dashed black curve, respectively \cite{Kozinsky2007}.
}
\label{fig:kozinsky_hysteresis}
\end{figure}

Semi-quantitatively, the onset of bistability can be understood already using the susceptibility of the oscillator $\chi_{}(\omega)$. By the definition of the susceptibility, the squared amplitude of forced vibrations is $A^2 = |\chi_{}(\omega_F)|^2 F^2$.  If we now use for $\chi_{}(\omega)$ the familiar expression (\ref{eq:Lorentz_classical}) for the susceptibility of a harmonic oscillator, but replace the eigenfrequency $\omega_0\to \omega_0+\delta\omega_0$ with $\delta\omega_0$ given by Eq.~(\ref{eq:Duffing_frequency_shift}),  the resulting equation for the squared amplitude reads
\begin{align}
\label{eq:bistability_naive}
A^2 = \frac{F^2}{(2M\omega_0)^2}\left[\Gamma^2 + \left(\omega_F-\omega_0- \frac{3\gamma A^2}{8\omega_0}\right)^2\right]^{-1}.
\end{align}
This equation describes the above qualitative arguments of ``tuning'' the mode in and out of good resonance, i.e., varying the detuning $\omega_F-\omega_0- 3\gamma A^2/8\omega_0$ by changing $A$. Formally, Eq.~(\ref{eq:bistability_naive}) is a cubic equation for $A^2$, and it can have three solutions. The solutions with the smallest and the largest $A^2$ can be shown to be stable \cite{Landau2004a}. The characteristic dependence of the amplitude $A$ on the drive frequency detuning $\omega_F-\omega_0$ for a nanowire measured by \textcite{Kozinsky2007}
is shown in Fig.~\ref{fig:kozinsky_hysteresis}(a).

The dynamics of a resonantly driven mode is conventionally described in terms of the in-phase and quadrature components $Q$ and $P$, respectively. They correspond to the coordinate and momentum of the mode in the rotating frame,
\begin{align}
\label{eq:quadratures}
&q(t)=Q\cos\omega_F t + P\sin\omega_Ft, \nonumber\\
&p(t)= M\omega_F(-Q\sin\omega_F t + P\cos\omega_F t).
\end{align}
The dynamical variables  $Q$ and $P$  are advantageous from the point of view of the experiment, as they  can be directly measured with a lock-in amplifier by setting its frequency to $\omega_F$.
We note that $(Q-iP)/2$ has the same form as the complex amplitude $u(t)$ introduced in Eq.~(\ref{eq:u(t)}) except that the frequency $\omega_0$ in Eq.~(\ref{eq:u(t)}) is replaced by $\omega_F$.  In terms of $Q$ and $P$, the vibration amplitude is $(Q^2+P^2)^{1/2}$ and the phase is $-\arctan(P/Q)$. In  quantum description, $Q$ and $P$ are operators, with $[Q,P]=i\hbar M\omega_F$.

The transient dynamics of the mode can be pictured in terms of the motion on the $(Q,P)$-plane, which is the phase plane in the rotating frame, see Appendix~\ref{subsec:resonant_forced}. The time evolution of $Q$ and $P$ is slow compared to the fast oscillations at frequency $\omega_F$. The stationary values of $Q$ and $P$ describe the stationary states of vibrations at frequency $\omega_F$ in the laboratory frame, as seen from Eq.~(\ref{eq:quadratures}). These values correspond to points on the $(Q,P)$-plane. If the mode has one stable vibrational state, there is one such point. In the range of bistability, there are three stationary states on the $(Q,P)$-plane. Two of them correspond to the stable vibrational states with different amplitudes and phases, and the third one corresponds to an unstable state (a saddle point). The whole $(Q,P)$-plane is divided into two regions: if prepared initially in one region, the mode evolves toward one stable state, whereas from the other region it evolves to the other stable state.

\cite{Kozinsky2007} managed to directly map the $(Q,P)$-plane  of a driven nonlinear mode and identify the basins of attraction to different stable states using the vibrations of a doubly clamped platinum nanowire, which were actuated and detected magnetomotively. The results are shown in Fig.~\ref{fig:kozinsky_hysteresis}~(a) and (c).

The number of stable vibrational states changes at the bifurcation parameter values, i.e., at the bifurcation points. These points lie on the lines in the space of the parameters of the driving force $(F,\omega_F)$, as illustrated in Fig.~\ref{fig:kozinsky_hysteresis}(b). On the upper (black) line the small-amplitude state merges with the unstable state. The experimental results \cite{Kozinsky2007} in Fig.~\ref{fig:kozinsky_hysteresis}~(c) show how, with the increasing driving amplitude, the small-amplitude state (state 1) approaches  the unstable state (panels 1 - 5). Ultimately both these states disappear (panel 6) and the mode has only one stable vibrational state.

A resonantly driven mode can display bistability if, in addition to conservative nonlinearity, the friction force is also nonlinear.  In the presence of nonlinear friction the decay rate is amplitude-dependent. This can be described by replacing $\Gamma\to \Gamma+ \Gamma^\mathrm{(nl)}A^2/2q_0^2$ in Eq.~(\ref{eq:bistability_naive}). The dynamics of the mode in this case was studied by \textcite{Buks2006a}.


\subsection{Parametric excitation}
\label{subsec:parametric_main}

Vibrations of nanomechanical systems can be also resonantly excited using parametric pumping. The corresponding pumping can be described as the modulation of the vibration eigenfrequency $\omega_0$ at a frequency $\omega_p$ close to $2\omega_0$. If the modulation is weak, it does not excite vibrations, the equilibrium state $q=p=0$ is stable. However, a sufficiently strong modulation can make this state unstable. It leads to an onset of two stable vibrational states. In each of these states the system vibrates at frequency $\omega_p/2$. The phases of the vibrations are fixed by the modulation parameters and differ by $\pi$ in the different states.  In the optics terms, the system is sometimes called a degenerate parametric oscillator to emphasize that the vibration frequencies are the same in the both states. Arguably, such parametrically excited vibrations  provide the simplest example of the onset of period doubling in a nonlinear system, since their period is twice the period of the driving.

The condition for the onset of period-2 vibrations and their amplitude can be obtained using the same naive arguments that led to Eq.~(\ref{eq:bistability_naive}).  Indeed, the simplest phenomenological equation of motion  that describes  parametric resonance  reads
\begin{align}
\label{eq:driven_parametric}
&M\ddot q + 2M\Gamma\dot q +M\omega_0^2 q + M\gamma q^3 \nonumber\\
&= q F_p\cos\omega_p t + f_T(t), \quad |\omega_p -2\omega_0|\ll \omega_0.
\end{align}
One can again relate the vibration amplitude to the force using the resonant susceptibility $\chi(\omega_p/2)$. However, the resonant force has now to be written with the account taken of the fact that we are seeking vibrations of the form $q(t) = A\cos[(\omega_pt/2)+\phi]$, and therefore the resonant part of the force $qF_p\cos\omega_pt$ is $(AF_p/2) \cos[[(\omega_pt/2)-\phi]$ (the phase $\phi$ has to be found separately). Substituting  this expression for the force into the equation for the amplitude gives $A^2 = |\chi(\omega_p/2)|^2 (AF_p/2)^2$. Using for the susceptibility the same expression as in Eq.~(\ref{eq:bistability_naive}) we obtain that either $A=0$, i.e., there are no vibrations, or
\begin{align}
\label{eq:amplitude_parametric}
& \frac{F_p{}^2}{(4M\omega_0)^2}\left[\Gamma^2 + \left(\delta\omega_p- \frac{3\gamma A^2}{8\omega_0}\right)^2\right]^{-1} =1,\nonumber\\
&\delta\omega_p = (\omega_p/2)-\omega_0.
\end{align}
Equation (\ref{eq:amplitude_parametric}) shows in particular that vibrations at frequency $\omega_p/2$ are excited provided the modulation is sufficiently strong to overcome the dissipation, $|F_p|/4M\omega_0 >\Gamma$. As in the case of the resonant driving discussed in Sec.~\ref{subsec:bistability} above, in the presence of nonlinear friction one should replace in Eq.~(\ref{eq:amplitude_parametric}) $\Gamma\to \Gamma+ \Gamma^\mathrm{(nl)}A^2/2q_0^2$.

Equation (\ref{eq:amplitude_parametric}) is a quadratic equation for $A^2$. It shows that, besides the zero-amplitude state where $A=0$, the modulated mode can have either one or two pairs of period-2 states, depending on whether Eq.~(\ref{eq:amplitude_parametric}) has one or two positive roots $A^2>0$.  The stability of these states
and the overall dynamics of a parametrically modulated mode can be conveniently described by changing  from $q(t), p(t)$ to the quadratures $Q(t), P(t)$. The corresponding transformation
is similar to Eq.~(\ref{eq:quadratures}),
\[q+ip/(M\omega_p/2) = (Q+ iP)\exp(-i\omega_pt/2).\]
The equations for the quadratures $Q,P$ are given in Appendix~\ref{subsec:parametric}. The stable period-2 states correspond to the symmetrically located fixed points on the phase plane $(Q,P)$.

\begin{figure}
\includegraphics[scale=0.9]{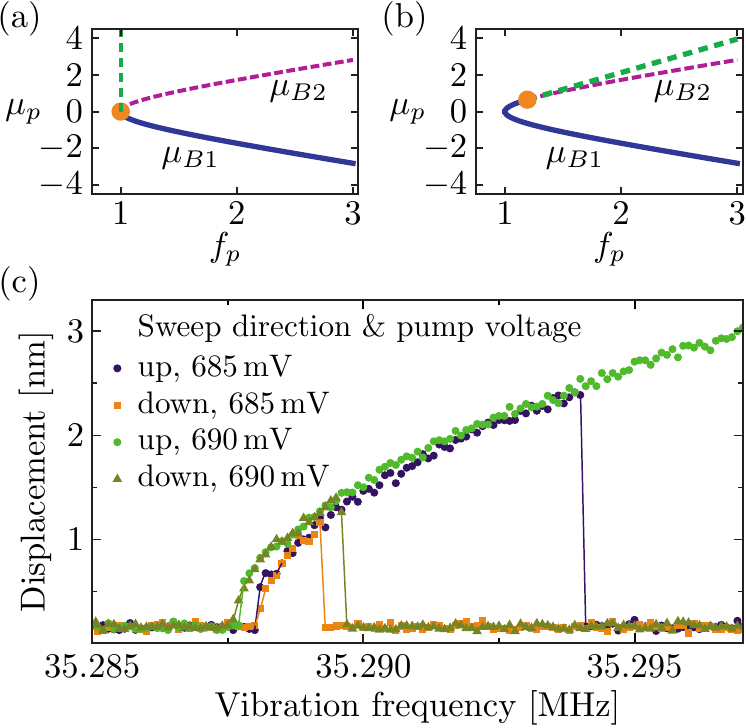}
\caption{The bifurcation diagrams of a parametrically modulated mode without (a) and with (b) nonlinear friction. The scaled modulation parameters are $\mu_p =(\delta\omega_p/\Gamma) \sgn\gamma, \; f_p= F_p/2M\Gamma\omega_p$. (a) With no nonlinear friction, the period-2 states are stable for $f_p>1$ and for $\mu_p>\mu_{B1} = -(f_p^2-1)^{1/2}$. For $\mu_p>\mu_{B2}=-\mu_{B1}$ the zero-amplitude state also becomes stable, leading to three stable states in the region delimited by the magenta and green dashed lines. (b) In the presence of nonlinear friction the three stable states coexist in the region bound by the line $\mu_{B2}(f_p)$ and the green dashed line, cf.~\cite{Lin2015}.
(c) Hysteresis of the response of a parametrically modulated nanoresonator \cite{Karabalin2010}.
}
\label{fig:parametric_bifurcation}
\end{figure}

The parameter ranges where a parametrically modulated mode has different numbers of coexisting states are separated by the bifurcation lines shown in Fig.~\ref{fig:parametric_bifurcation}~(a,b).
An example of the hysteretic behavior of parametrically modulated nanoresonators is shown in Fig.~\ref{fig:parametric_bifurcation}~(c). The experimental data by \textcite{Karabalin2010} shown in this figure were obtained using doubly clamped piezoelectrically controlled nanobeams. In the simplest case, the occurrence of the hysteresis is a consequence of nonlinear friction. As seen from Fig.~\ref{fig:parametric_bifurcation}~(b), if one increases the frequency (the parameter $\mu_p$) starting from the negative value, there are excited period-2 vibrations once the blue line $\mu_{B1}$ is crossed. These vibrational states disappear once the green dashed line is crossed. On the other hand, if one starts from above the green dashed line and decreases the frequency, the zero-amplitude state remains stable until the magenta line $\mu_{B2}$ is crossed. It is possible to break the symmetry between the two parametrically driven states that differ by $\pi$ in the phase by applying a force at $\omega_p/2$ \cite{Ryvkine2006a}, as measured by \textcite{Mahboob2010} and \cite{Leuch2016}.


\subsection{Fluctuations of driven modes}
\label{subsec:driven_fluctuations}

An important aspect of researching resonantly driven or parametrically modulated nanomechanical systems is the possibility to use them for studying fluctuation phenomena far from thermal equilibrium. Because nanomechanical systems are well-characterized, they are well suited for such studies. 
We will discuss several features of  fluctuations in driven NVSs.

\subsubsection{Fluctuation squeezing}
\label{subsubsec:squeezing}

An important generic feature of  fluctuations  of a periodically driven mode follows from the very fact  that the driving breaks  the continuous time-translation symmetry of the mode dynamics. An immediate consequence of the symmetry breaking is the possibility of fluctuation squeezing. In squeezing,  fluctuations of one of the vibrational components (quadratures) are reduced below their level in the absence of driving, whereas fluctuations of the other component are increased. In the absence of driving, the quadratures are the vibration components that oscillate as $\cos\omega_0t$ and $\sin\omega_0t$. If the system has a continuous symmetry, the origin of time can be shifted.  A shift in time by $\pi/2\omega_0$ results in the interchange of the quadratures. This shows that the variances of their fluctuations should be equal.  A periodically driven system, in contrast, has a discrete time-translation symmetry. It is symmetric only with respect to changing time by the period of the drive.  Therefore the quadratures may no longer be interchanged and their variances are generally different.

Historically, squeezing was first detected in quantum optics \cite{Slusher1985}. It attracted significant attention, since it can reduce fluctuations of a quadrature below their level in the quantum ground state of the mode and thus enable high-precision measurements \cite{Caves1981}. The technique has been implemented in laser interferometers for gravitational wave detection \cite{Tse2019,Acernese2019}. In nano- and micro-mechanical systems, squeezing in the quantum regime was demonstrated using the techniques of cavity optomechanics,   \cite{Wollman2015,Lecocq2015,Pirkkalainen2015}.

However, the concept of squeezing of fluctuations in vibrational systems equally applies to the classical regime.
Squeezing of thermal fluctuations of nanomechanical systems has been  achieved in several experiments using parametric pumping \cite{Rugar1991,Suh2010,Mahboob2010a}. The squeezing is obtained with a comparatively weak pumping, below the threshold for exciting period-2 vibrations. For weak noise, it can be described disregarding the mode nonlinearity, see Appendix~\ref{subsec:parametric}.

Figure~\ref{fig:squeezetang}a shows the measurements of the thermal vibration noise represented in the $(Q,P)$-phase plane \cite{Poot2015}.  The quadratures are measured with a lock-in amplifier by setting its frequency to $\omega_0$. Without pump, the variances of both quadratures are the same, as expected for the continuous time symmetry of the problem. When the resonator is parametrically driven at $2\omega_0$, the variances are no longer equal, pointing to time-translation symmetry breaking. The variance of the $Q$-quadrature is squeezed by the parametric drive. The strongest squeezing corresponds to 3 dB variance suppression, which is reached when the $P$ quadrature diverges (Fig.~\ref{fig:squeezetang}b).

With feedback control \cite{Vinante2013,Szorkovszky2013,Poot2014,Sonar2018} it was possible to achieve squeezing by 15.1dB \cite{Poot2015}, well above the conventional 3~dB limit for parametric pumping of a linear mode, cf. Eq.~(\ref{eq:squeezed_param}). Classical two-mode squeezing in mechanical resonators by non-degenerate parametric amplification has been also
reported \cite{Mahboob2014,Patil2015,Pontin2016} as well as the possibility to obtain squeezing in a two-mode system using post-selection \cite{Asano2019}. Classical squeezing was proposed as a way to reduce heating in computers \cite{Klaers2019}; it also represents an important asset for high-precision sensing \cite{DiFilippo1992,Natarajan1995,Mahboob2010a,Szorkovszky2013} and thus  may lead to a new generation of nanomechanical detectors.

\begin{figure}
\includegraphics[scale=0.8]{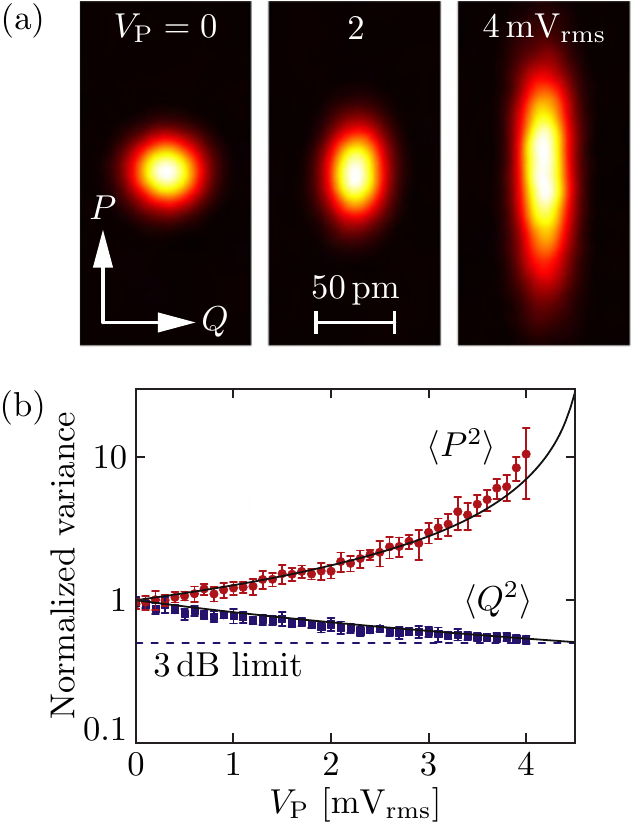}
\caption{(a) Thermal vibration noise plotted in the $(Q,P)$ phase plane and measured with the pump turned off (thermal equilibrium, pump power $V_P=0$~mV) and the pump turned on (nonzero $V_P$). (b) Variance of the $Q$ and $P$ quadratures for increasing pump power normalized by the variance at zero pump power. The dashed line indicates the 3 dB limit for parametric squeezing \cite{Poot2015}.
}
\label{fig:squeezetang}
\end{figure}

Much less attention has  been paid to squeezing of mode fluctuations due to a resonant coordinate-independent force, cf. Eq.~(\ref{eq:driven_Brownian}).
The general argument regarding the broken time-translation symmetry applies in this case, too. However, for a driven linear mode, forced vibrations are just linearly superimposed on thermal vibrations, and therefore there is no squeezing. The situation changes in the nonlinear regime. Here the fluctuations are affected by the driving and the effect can be resonantly strong.

For weak noise, a resonantly driven mode primarily fluctuates about its state of forced vibrations. These are the fluctuations of the deviations of the quadratures $\delta Q, \delta P$ from their values in the stable state that become squeezed in the nonlinear regime.
The occurrence of the squeezing could be inferred from the strongly asymmetric phase trajectories in the rotating frame in the neglect of dissipation in Fig.~\ref{fig:phase_portraits}~(b), which were discussed already in the early work on a resonantly driven Duffing oscillator \cite{Dykman1979b,Dmitriev1986}, cf. also \cite{Siddiqi2004}. A theory of squeezing was developed by \textcite{Buks2006a}. A strong suppression of a spectral component of a quadrature was observed in a  nanomechanical Duffing resonator by \textcite{Almog2007a} in a narrow parameter range near the critical point in Fig.~\ref{fig:kozinsky_hysteresis}~(b) using conventional homodyne detection.

Homodyne measurements are strongly impeded by frequency fluctuations, which play an important role in nanomechanical systems, see Sec.~\ref{sec:frequency_fluctuations}. The limitations are particularly pronounced in systems with small damping, where the noise in the in-phase component increases with the  increasing drive strength \cite{Fong2012}, see Fig.~\ref{fig:hong_freq_fluct}. However, it appears that, for underdamped vibrational modes, squeezing can be found by measuring the spectrum of their response to an additional weak probe field \cite{Ochs2021a} or, for classical fluctuations, by measuring the power spectrum \cite{Huber2020}.

\begin{figure}
\includegraphics{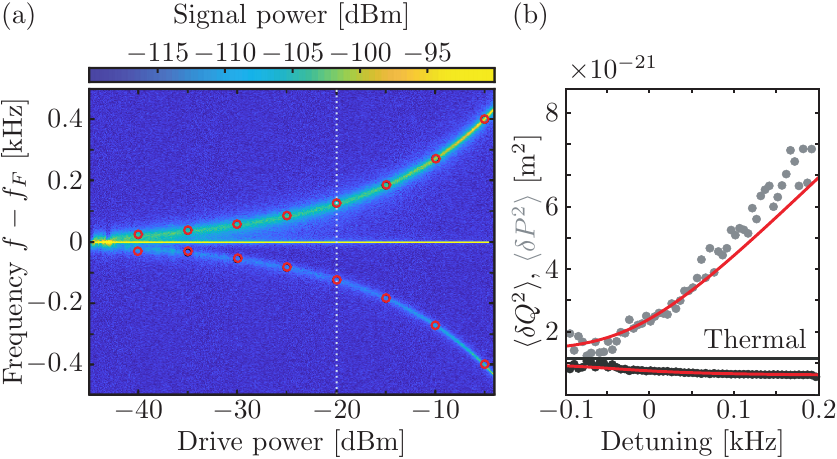}
\caption{(a) Power spectrum of a resonantly driven nanomechanical mode as a function of frequency $f$ for $\omega_F\equiv 2\pi f_F=\omega_0$. The line at $f-f_F=0$ is plotted with reduced brightness to improve the visibility of the satellites, which are due to fluctuations about the stable state of forced vibrations. (b) Variance of the scaled in-phase and quadrature fluctuations $\langle\delta Q^2\rangle$ and   $\langle\delta P^2\rangle$ as a function of the detuning $(\omega_F-\omega_0)/2\pi$. Black line indicates thermomechanical fluctuations at 293~K, and red lines indicate the theory.  Adapted from \cite{Huber2020}.
}
\label{fig:squeeze_resonant}
\end{figure}

The spectral measurements of squeezing exploits the nature of the dynamics of a  driven strongly underdamped mode. When viewed in the rotating frame, this dynamics involves weakly damped oscillations about a stable state of forced vibrations, as described in Appendix~\ref{subsubsec:fluctuations_driven}. The frequency of these oscillations $\omega_\mathrm{rot}$ is much smaller than the strong-drive frequency $\omega_F$. When viewed in the laboratory frame, they lead to peaks in the power spectrum at frequencies $\omega_F\pm\omega_\mathrm{rot}$.

Figure~\ref{fig:squeeze_resonant} shows the power spectrum of fluctuations of a  resonantly driven NVS mode with clearly resolved peaks at $\omega_F\pm\omega_\mathrm{rot}$ \cite{Huber2020}. The peaks have profoundly different intensities, a direct consequence of squeezing.

The peaks at $\omega_F\pm\omega_\mathrm{rot}$ emerge also  in the spectrum of the response to an additional weak probe field. The ratio of the areas $\mathcal{A}^{\pm}$ of these spectral peaks gives the squeezing  parameter $\phi$ \cite{Dykman2011,Dykman2012a,Ochs2021},
\begin{align}
\label{eq:squeeze_parameter}
\frac{\langle \delta Q^2\rangle} {\langle\delta P^2\rangle} = e^{-4\phi}, \quad \coth^4 \phi=\frac{\mathcal{A}^+}{\mathcal{A}^-}.
\end{align}
This expression holds for thermal and quantum fluctuations about the larger-amplitude vibrational state in the range of bistability, cf.  Fig.~\ref{fig:kozinsky_hysteresis}~(b); for the smaller-amplitude state one should replace  $\coth \phi \to \tanh \phi$.

One of the peaks of the response spectrum corresponds to amplification of the probe field by the strong driving field \cite{Dykman1979b,Ochs2021}.  A distinctly double-peak structure of the response spectrum in a suspended nanomembrane was seen by \textcite{Antoni2012}.


\subsection{Rare large fluctuations far from thermal equilibrium}
\label{subsec:rare_events}

Classical and quantum fluctuations in driven nanomechanical systems are not described by the statistical physics of systems in thermal equilibrium, and their probabilities are not determined by the thermodynamic potentials. Revealing general features of such fluctuations is both challenging and important. To a large extent, these features are related to the absence of detailed balance.
Detailed balance requires that, in the stationary regime, the probabilities of transitions between the states of a system be balanced pairwise. This means that the probability of a transition $A\to B$ between arbitrary  states $A$ and $B$ is equal to the probability of the transition $\bar B \to \bar A$. Here the overline indicates that, in the corresponding states, the signs of odd in time variables have been reversed \cite{Lifshitz1981a}; we assume that there is no magnetic field. For a classical vibrational mode, one can associate the mode states with small areas in phase space; for states $A$ and $B$ these areas are centered at points $q_A,p_A$ and $q_B,p_B$, whereas $\bar A$ and $\bar B$ are centered at $q_A,-p_A$ and $q_B,-p_B$, respectively.

Detailed balance is a requisite of thermal equilibrium. It differs from the stationarity condition that the probability to go from $A$ to $B$, $C$, etc is equal to the total probability to come to $A$ from $B$, $C$, etc.
Driven nonlinear modes do not have detailed balance, as a rule, cf.~\cite{Roberts2021}. They have proved to be invaluable as a means to study statistical physics with no detailed balance because, on the one hand, they are comparatively simple while, on the other hand, they display a nontrivial behavior.


\subsubsection{Dynamics of nonequilibrium systems in rare large fluctuations}
\label{subsubsec:optimal_paths}

Of special  interest in terms of their generic features are nonequilibrium phenomena related to comparatively rare large fluctuations away from a stable state and switching between coexisting stable states. These phenomena encompass chemical and biological reactions as well as switching in lasers, driven nano-magnets, and other systems of current interest. Even though much work has been done on the theory of switching in systems lacking detailed balance, to the best of our knowledge, vibrational systems have been the only ones where the theory could be quantitatively tested in the experiment. Moreover, with these systems the qualitative features of large rare fluctuations, including the involved scaling behavior, have been studied.

For weak on average fluctuations, most of the time the system performs small-amplitude fluctuations about its stable state (or one of its dynamically stable states). Still occasionally there occur large fluctuations, in which the system moves far away from this state in phase space. They may result in switching to another stable state.

A key idea behind the understanding of large rare fluctuations was put forward by \textcite{Onsager1953, Machlup1953} in the analysis of linear thermal equilibrium systems. They showed that, very counter-intuitively, in a large fluctuation to a given point in phase space, even though the motion is random, a system most likely moves along a certain trajectory. Moreover, for an overdamped system this trajectory is the time-reversed trajectory of moving back to the stable state from this point in the absence of fluctuations.

The concept of the corresponding most probable trajectory of a rare fluctuation extends to nonlinear systems and to nonequilibrium systems. However, in systems lacking detailed balance finding such a trajectory is a far from trivial problem, and the topology of such trajectories is also far from trivial \cite{Dykman1994d}. This problem is fundamental in the theory of nonequilibrium systems, from physics to biology. It attracted  much attention over the years \cite{Kamenev2011}.

\begin{figure}
\includegraphics[width=7.5cm]{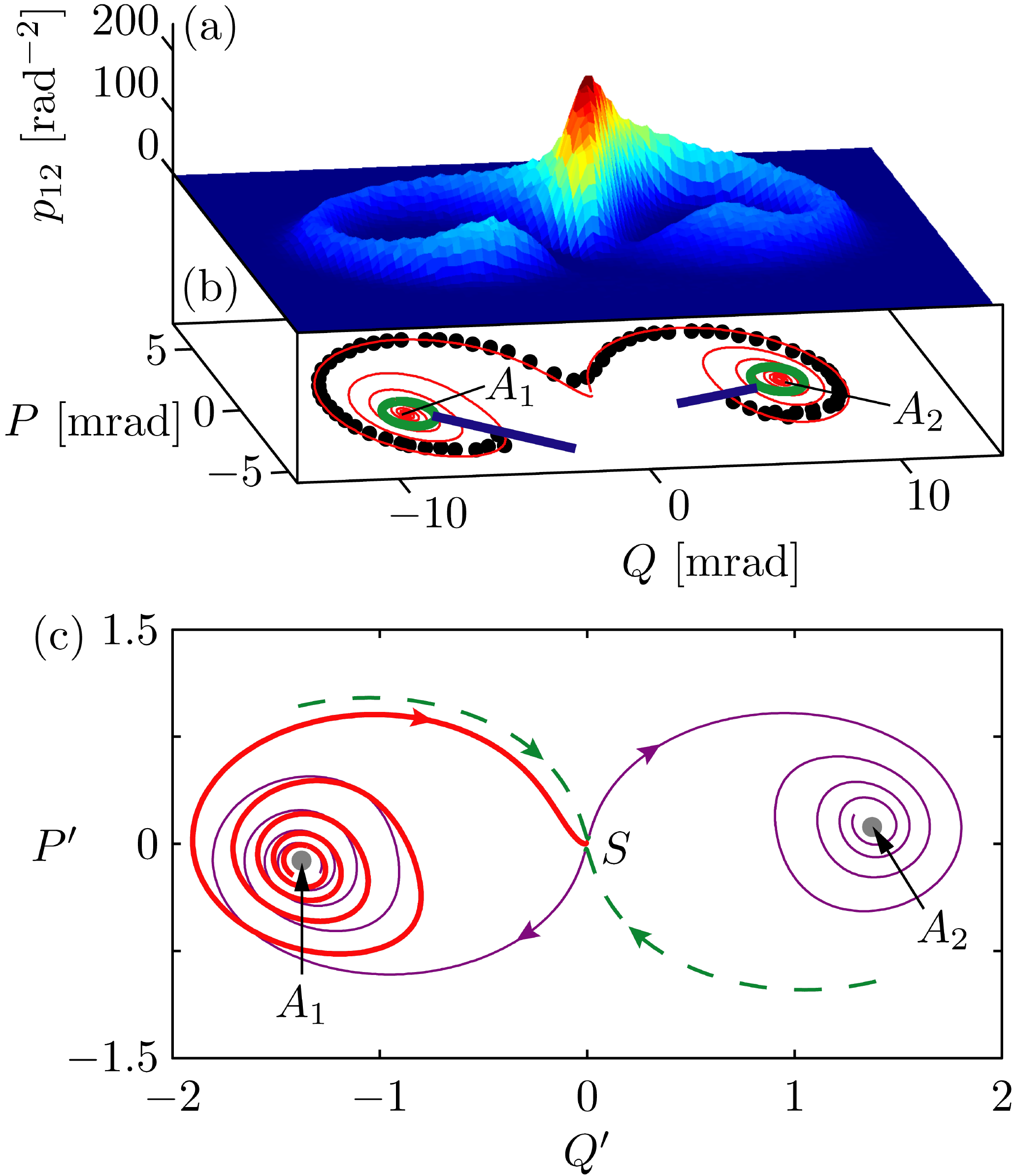}
\caption{(a) Probability distribution of the switching paths of a parametrically driven torsional microelectromechanical resonator. The distribution  $p_{12}(Q,P)$  is measured for switching out of stable state $A_1$  into state  $A_2$ in the rotating frame; $Q$ and $P$ are the quadratures, which are measured in the units of the rotation angle of the resonator. (b) The peak locations of the distribution are plotted as black circles and the theoretical most probable switching path is indicated by the red line. All trajectories originate from within the green circle in the vicinity of  $A_1$ and later arrive at the green circle around  $A_2$. The radii of the green circles give the typical fluctuation amplitude. The portion of the distribution outside the blue lines is omitted. (c) Comparison of the most probable switching path (MPSP) and the noise-free trajectory; $Q'$ and $P'$ are the rescaled dimensionless quadratures $Q$ and $P$. The noise-free trajectories from the saddle point ${\cal S}$ to the stable states are shown by thin solid lines (magenta). The thick red line shows the first portion of the MPSP  from $A_1$ to $S$, where the system is driven by noise. The MPSP as a whole is comprised by this trajectory and the noise-free trajectory from ${\cal S}$ to $A_2$.  The green dashed line shows the separatrix. Adapted from \cite{Chan2008a,Chan2008b} }
\label{fig:switching_trajectory}
\end{figure}

A direct experimental observation of the most probable trajectory in a system lacking detailed balance was done with a micromechanical system \cite{Chan2008a,Chan2008b}, see Fig.~\ref{fig:switching_trajectory}(a). The system was parametrically driven into the range where it had two stable vibrational states $A_1$ and $A_2$,
see Sec.~\ref{subsec:parametric}. A comparatively weak noise  caused mostly small-amplitude fluctuations about the states, but occasionally also led to interstate switching. In the experiment, the system was  prepared in the state $A_1$ and its trajectory was recorded. After it was found in the vicinity of the state $A_2$, a portion of the trajectory in the region between the blue lines in Fig.~\ref{fig:switching_trajectory}~(b) was saved and the experiment was repeated. The distribution $p_{12}$ of the paths followed in switching was obtained by superposing the saved portions of the trajectories.

The sharp peak of the distribution in Fig.~\ref{fig:switching_trajectory}~(a) shows that indeed, in switching the system is most likely to move along a certain path, the most probable switching path (MPSP). Except for the vicinity of the saddle point $Q=P=0$, the  distribution of the trajectories about the MPSP was Gaussian and was in full agreement with the theory, which extended the previously developed approach   \cite{Dykman1992d,Luchinsky1997b} to the problem of switching.   For switching from the zero-amplitude state of a parametrically excited microcantilever, observation of switching paths was reported by \textcite{Requa2007}.

Figure~\ref{fig:switching_trajectory}~(c) compares the MPSP (red line) with the noise-free trajectory (magenta line) along which the system would move from the saddle point to state $A_1$. It clearly shows that these trajectories do not have time-reversal symmetry, which would be the symmetry $(Q,P)\to (Q,-P)$. This is an unambiguous indication of the lack of detailed balance.

The form of the MPSP depends not only on the dynamics of the system in the absence of fluctuations, but also on the source of the fluctuations, that is, on the properties of the noise, including its  spectrum and  statistics. The results in Fig.~\ref{fig:switching_trajectory} refer to the case where the noise is thermal, cf. Eq.~(\ref{eq:driven_parametric}), and therefore is Gaussian and white in slow time, see Sec.~\ref{sec:FDT}.


\subsubsection{Scaling behavior of the rates of switching between stable vibrational states}
\label{subsubsec:scaling_of_switching}

Quite generally, for a Gaussian and not necessarily white noise the rate of switching between coexisting stable states scales with the noise intensity $D$ as \cite{Dykman1990}
\begin{align}
\label{eq:activation_general}
W_\mathrm{sw} = C_\mathrm{sw}\exp(-R/D).
\end{align}
This expression is similar to the expression by \textcite{Kramers1940} for the escape rate of a Brownian particle from a potential well. For the Brownian particle the noise intensity is $k_BT$ and $R=\Delta U$, where $\Delta U$ is the height of the potential barrier to be overcome in escape. In analogy with the Arrhenius law, $R$ is often called the effective activation energy. To the best of our knowledge, a  vibrational mode resonantly driven into bistability was the first physical system with no detail balance where $R$ was calculated \cite{Dykman1979b}. Both $R$ and the prefactor $C_\mathrm{sw}$ in Eq.~(\ref{eq:activation_general}) are independent of the noise intensity.

\begin{figure}
\includegraphics[scale=0.8]{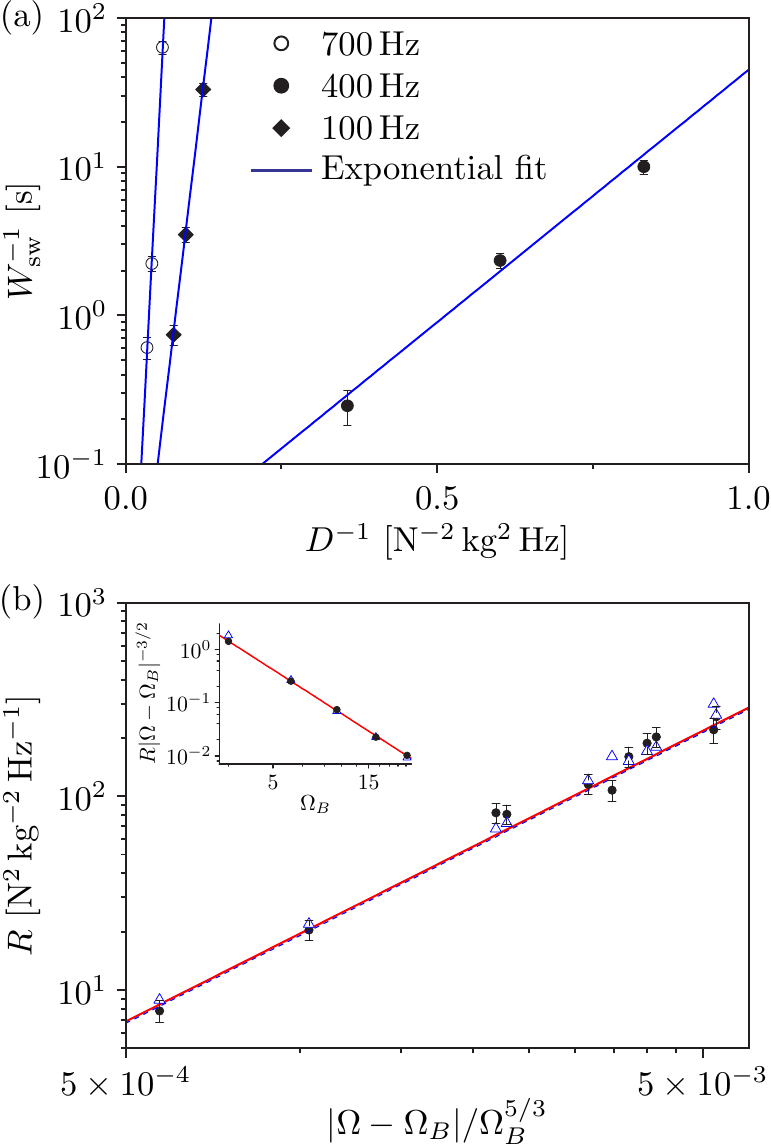}
\caption{(a) The activation dependence of the switching rate of a resonantly driven nanomechanical mode $W_\mathrm{sw}$ on the noise intensity $D$ for different detuning of the drive frequency from its bifurcational value. (b) The effective activation energy $R$  of the switching rate  as a function of the distance from the bifurcational value of the drive frequency; $\Omega =| \omega_F-\omega_0|/\Gamma$, and $\Omega_B$ is the value of $\Omega$ at the bifurcation point. The full circles indicate the experimental points, the open (blue) triangles the prediction of the full numerical simulation, the (red) full lines the linear fit to the data, and the dashed (blue) lines the prediction of the asymptotic theory \cite{Dykman1980}. Inset: Scaling with the parameter $\Omega_B$. Adapted from \cite{Defoort2015}.
}
\label{fig:collin_switching}
\end{figure}

The activation dependence of the switching rate on the noise intensity  was demonstrated for a  parametrically excited vibrational mode of a single electron in a Penning trap \cite{Lapidus1999}. It has been observed in various types of nano- and micromechanical resonators driven into bistability by either resonant driving force or parametric driving close to twice the mode eigenfrequency, see  \cite{Aldridge2005,Stambaugh2006,Chan2007,Venstra2013,Defoort2015,Dolleman2019} and references therein. An example is shown in Fig.~\ref{fig:collin_switching}~(a).

Another feature of switching in nonequilibrium systems is the scaling behavior of the switching rates, where the activation energy $R\propto |\ln W_\mathrm{sw}|$ scales as a power law of the parameters.
For equilibrium systems, where $R$ corresponds to the barrier height $\Delta U$,  such scaling was predicted for Josephson junctions \cite{Kurkijarvi1972} for the regime where a junction is not driven by an ac field and is in a quasiequilibrium state. This scaling has been broadly used to determine the junction parameters \cite{Fulton1974}.

Generally,  the scaling emerges close to bifurcation points where there disappears the stable state  from which the system is switching \cite{Dykman1979b,Dykman1980,Knobloch1983,Graham1987a,Dykman1998}. For resonantly and parametrically driven nonlinear modes these bifurcation points lie on the lines shown in Figs.~\ref{fig:kozinsky_hysteresis}(b) and \ref{fig:parametric_bifurcation}(a), (b). The scaling of $W_\mathrm{sw}$  near bifurcation points was seen in a number of experiments, both for resonantly driven \cite{Stambaugh2006,Defoort2015,Dolleman2019} and parametrically modulated \cite{Chan2007} micro- and nanomechanical modes, parametrically modulated atomic vibrations in a magneto-optical trap \cite{Kim2005}, and resonantly driven vibrations of Josephson junctions \cite{Siddiqi2006a,Vijay2009}.
The scaling exponents are different in the cases of resonant and parametric driving. Remarkably, not only the effective activation energy $R$ scales as the power law of the distance to the bifurcation point (see Appendix~\ref{subsubsec:bif_fluctuations}), but so does also the prefactor $C_\mathrm{sw}$ in Eq.~(\ref{eq:activation_general}), with a different exponent \cite{Dykman1980}.

The high degree of control of nanomechanical systems has enabled measuring the scaling exponents not only of $R$ but also of $C_\mathrm{sw}$. Figure~\ref{fig:collin_switching}(b) shows the  scaling behavior of $R$ for a resonantly driven nanomechanical system \cite{Defoort2015}. The data are obtained for different points on the bifurcation line $F_{B1}$ in Fig.~\ref{fig:kozinsky_hysteresis}(b). It was demonstrated that
\[R\propto |\omega_F-(\omega_F)_B|^{3/2},\quad
C_\mathrm{sw}\propto |\omega_F - (\omega_F)_B|^{1/2}\]
where $(\omega_F)_B$ is the bifurcational value of the drive frequency $\omega_F$. Interestingly, in agreement with the numerical analysis, \cite{Defoort2015,Kogan2008a} the scaling holds in a comparatively broad range.

\begin{figure}
\includegraphics{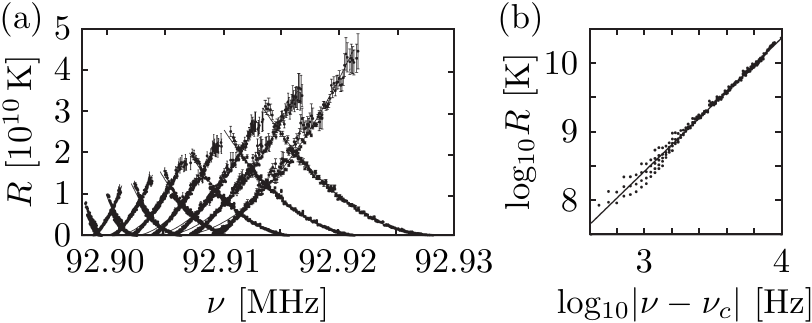}
\caption{Measured activation energies of switching between the coexisting large- and small-amplitude states of a resonantly driven nanoresonator. The switching is induced by an externally applied noise that mimics thermal noise. (a) The activation energies as functions of the drive frequency $\nu=\omega_F/2\pi$ for different values of the resonant  force amplitude $F$.  The activation energy for switching from the large-amplitude state monotonically decreases with the increasing drive frequency, whereas that from the small-amplitude state monotonically increases. (b) Log-log plot showing $R$ as a function of the difference between $\nu$ and its  critical value $\nu_c$. The amplitude of the force $F$ is close to the critical amplitude and is adjusted so as to keep the rates of switching between the stable states equal. Adapted from \cite{Aldridge2005}.}
\label{fig:cleland_switching}
\end{figure}

The overall change of the effective activation energies of interstate switching with the varying frequency and amplitude of the driving resonant force for a nonlinear nanomechanical beam is shown in Fig.~\ref{fig:cleland_switching}~(a).
Besides the bifurcation points discussed above, the switching rates display a universal dependence on the parameters near the critical point $F_c$ in Fig.~\ref{fig:kozinsky_hysteresis}~(b) where the two stable vibrational  states merge. One can tune the drive amplitude and the frequency in such a way that the rates of switching between the states are equal, see Appendix~\ref{subsubsec:bif_fluctuations}. In this case the activation energy $R$ is expected to depend on the distance to the critical frequency  as $[\omega_F - (\omega_F)_c]^2$ \cite{Dykman1979b,Dykman1980}. This scaling behavior of $\log W_\mathrm{sw}$ was clearly  demonstrated in the experiment, with $R$ varying over more than two decades, see Fig.~\ref{fig:cleland_switching}~(b).

\subsubsection{A kinetic phase transition}
\label{subsubsec:kinetic_phase}

Bistable vibrational systems allow one to study another fairly general group of nonequilibrium phenomena. They occur in
the parameter range where the  rates $W_{12}$ and $W_{21}$ of switching between the stable states $1\to 2$ and $2\to 1$ are close to each other; for concreteness, we use 1 and 2 to label the states of a resonantly driven mode with larger and smaller vibration amplitude, respectively. For weak noise, this happens where the effective activation energies $R_1$ and $R_2$ for switching $1\to 2$ and $2\to 1$ are equal or almost equal [we use that $W_{ij}\propto \exp(-R_i/D)$]. In this range the stationary populations $w_{1,2}$ of the states are also close to each other, as seen from the balance equation $w_1/w_2= W_{21} /W_{12}$. As the system parameters move away from this range, $R_1$ and $R_2$ becomes different, and the populations   quickly become exponentially different. It is this range that was studied by \textcite{Aldridge2005} close to the critical point.

The range where $R_1\approx R_2$  is similar to a smeared first order phase transition in an equilibrium system. Indeed, at such phase transition the free energies of the phases are equal, according to the Ehernfest classification, and the phases (for example, liquid and vapor) have comparable volumes (populations). One can conditionally associate the  large- and small-amplitude vibrational states with different phases of matter. In this context, the analogs of pressure and temperature in the liquid-vapor transition, i.e., the control parameters, are the amplitude $F$ and frequency $\omega_F$ of the driving field. For a certain relation between $F$ and $\omega_F$ we have $R_1=R_2$, which can be called a kinetic phase transition. The data in Fig.~\ref{fig:cleland_switching}(b) are obtained by moving along the corresponding line on the $(F,\omega_F)$ plane. The critical point $[F_c,(\omega_F)_c]$  (see Figs.~\ref{fig:kozinsky_hysteresis} and \ref{fig:cleland_switching}) is a counterpart of the critical point on the phase transition line.

The kinetic phase transition in a driven mode is accompanied by the onset of extremely narrow peaks in the power spectrum and the spectrum of the response to a probe field \cite{Dykman1979b,Dykman1994c}. The peaks are located at the driving frequency.  They result from the change of the state populations $w_1$ and $w_2$ induced by fluctuations or by the probe field. Such a change leads to the change of the vibration amplitude and phase between their values in the large- and small-amplitude states. The rates at which the populations change, and thus the widths of the peaks, are $\sim W_{12}\approx W_{21}$. They are exponentially smaller then the mode decay rate. Therefore the spectra are extremely sensitive to the parameters of the system, similar to the parameter sensitivity at the phase transition in an extended system.

A number of effects related to the kinetic phase transition, including the extremely narrow spectral peaks, have been observed in different types of resonantly driven micro- and nanomechanical systems \cite{Stambaugh2006a,Chan2006,Almog2007,Venstra2013,Chowdhury2017,Dolleman2019,Huber2020}.  Early on it was shown \cite{Cleland2005} that these effects enable a 100-fold improvement in frequency resolution compared to a conventional resonant-response based measurement.

\begin{figure}
\includegraphics{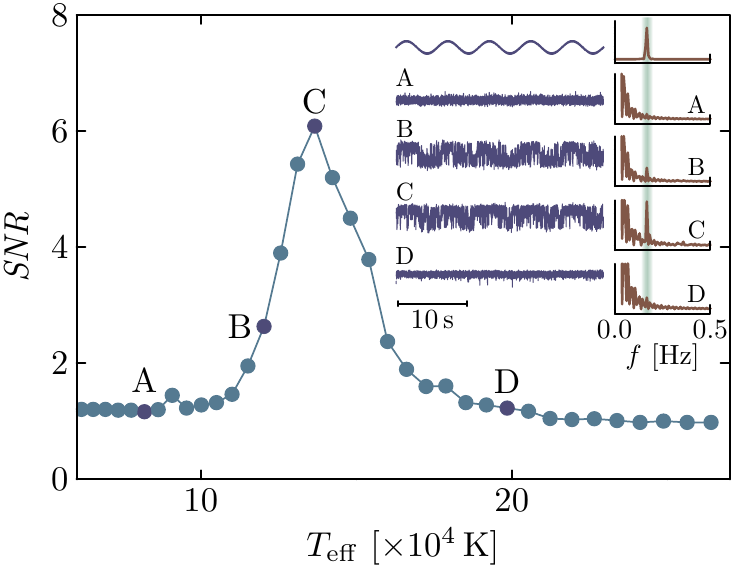}
\caption{Noise-enhanced measurement with a nanomechanical cantilever. The cantilever with geometric and inertial nonlinearity is driven into bistability of forced vibrations by a comparatively strong resonant drive. In addition, there is applied  a broad-band noise that emulates thermal noise. The drive is tuned to the kinetic phase transition. The amplitude of the drive is then slowly modulated, mimicking a signal at frequencies equal to the sum and difference of the modulation and drive frequencies. Main figure: the signal-to-noise ratio (SNR) as a function of the effective noise temperature. Inset: top row shows the excitation signal in time (left) and frequency (right) domains. The cantilever response at increasing noise intensity is shown in panels (A–D). Adapted from \cite{Venstra2013}.}
\label{fig:high_frequency_SR}
\end{figure}

We now discuss the fluctuation-induced response to the probe field and its relation to stochastic resonance. If the probe-field frequency $\omega_\mathrm{pr}$ is very close to the strong-drive frequency $\omega_F$, the probe field can be thought of as a modulation of the strong field amplitude and, hence, of the activation energies $R_{1,2}$, at frequency $|\omega_\mathrm{pr}-\omega_F|$.
As seen from Eq.~(\ref{eq:activation_general}),  modulation of $R_i$ results in an exponentially enhanced modulation of $W_{ij}$, for weak noise. However, if it is too fast, the populations of the states $w_{1,2}$  cannot ``adjust'' to the probe field. Therefore if  the noise intensity $D$ is very small, so that $W_{ij}\ll |\omega_\mathrm{pr}-\omega_F|$, the response is weak. It exponentially strongly increases with the increasing $D$ as  the rates $W_{ij}$ approach $\sim|\omega_\mathrm{pr}-\omega_F|$. For large $D$, on the other hand the sensitivity to the modulation of $R_{1,2}$ falls off. Thus the response displays a peak as a function of $D$. This is reminiscent of stochastic resonance in systems fluctuating in a static double-well potential. Such high-frequency stochastic resonance was observed in micro- and nanomechanical systems, as Fig.~\ref{fig:high_frequency_SR} illustrates, and several applications of this effect for sensing have been discussed by \textcite{Chan2006,Almog2007,Venstra2013}.

In the case of resonant parametric excitation at frequency $\omega_p\approx 2\omega_0$, a mode is ``automatically tuned'' into the kinetic phase transition. Indeed, the period-2 vibrational states have the same amplitude and differ just by phase. Noise-induced interstate transitions are phase-flip transitions, and the stationary populations of the states are equal by symmetry. However, an additional resonant additive drive at frequency $\omega_p/2$ lifts this symmetry. As a result, the stationary populations become different. They depend on the phase of the extra drive with respect to the phase of the strong parametric modulation \cite{Ryvkine2006a}. This happens already for a weak extra drive, where the both states are stable. Such symmetry lifting was observed  by \textcite{Mahboob2010} in a GaAs/AlGaAs based micromechanical resonator. The corresponding symmetry-breaking detector can resolve frequency shifts $\delta\omega_0/\omega_0\sim 10^{-7}$ in a single-shot measurement.

The fluctuation-mediated symmetry lifting has important consequences for coupled parametrically modulated modes. If  a mode vibrates at frequency $\omega_p/2$ and is in a state with a given phase, it exerts a symmetry-lifting force on the mode (or modes) it is coupled to. As a result, depending on the coupling, this second mode  is biased toward the state with the same or the opposite phase as the primary mode. Of course, the modes are on equal footing, they affect each other so as to have the same or opposite phases, and the effect comes through the fluctuations.

The effect of phase ordering in coupled parametrically excited modes was demonstrated by \textcite{Karabalin2011}. In the experiment two almost identical gated nanobeams were driven across the parametric instability. Depending on the sign of the coupling, the mode that experienced the instability later had predominantly the same or the opposite phase as the one that went through the instability first. The coupling could be controlled by the gate voltage, which provided a highly sensitive way of detecting this voltage.

The phase correlations between coupled parametric oscillators mediated by noise was used by \textcite{Mahboob2016a} to mimic the Ising dynamics of  coupled spins with two modes. The two different phases of a parametrically pumped micromechanical mode were associated with two spin projections. An implementation of quantum computation \cite{Goto2016a,Goto2019} and Boltzmann sampling \cite{Goto2018} with coupled nonlinear parametrically excited modes has been also discussed. Of significant importance in this respect is that switching between the period-2 states of a parametrically excited mode can be induced not only by classical, but also by quantum fluctuations even for $T=0$ \cite{Marthaler2006}. For modulated coupled nanomechanical modes this opens a way of  studying quantum phase transitions far from thermal equilibrium, in particular a transition to a Floquet time-crystalline state \cite{Dykman2018}, in a well-characterized environment.

\section{Resonant mode coupling}
\label{sec:coupling_modes}

A mechanical resonator has a large number of mechanical eigenmodes. These eigenmodes can couple with each other. This leads to a rich variety of different phenomena. We discussed in Sec.~\ref{sec:nonlinearity_general} some of these phenomena, in particular, those related to the dispersive coupling. Such coupling is important where the modes are far from resonance, but the frequency of a mode depends on the vibration amplitudes of other modes.
Here, we will discuss phenomena originating from resonant linear and nonlinear coupling as well as some resonant multi-mode effects of an external drive.

\subsection{Linear resonant  coupling}
\label{subsec:resonant_coupling}

A hallmark of nanoscale mechanical resonators is the wide tunability of their resonance frequencies by electrostatic means. This allows one to bring nanomechanical vibrational modes in and out of linear resonance. Strictly speaking, eigenmodes are defined by diagonalizing the quadratic in the displacements and momenta part of the  Hamiltonian  of the system. Where the resonator modes have very different eigenfrequencies,  one can speak of approximate modes, disregarding a quadratic in the displacements part of the potential energy that couples them. However, when the eigenfrequencies are tuned close to resonance, this part become substantial.

\begin{figure}[h]/
\includegraphics[scale=0.8]{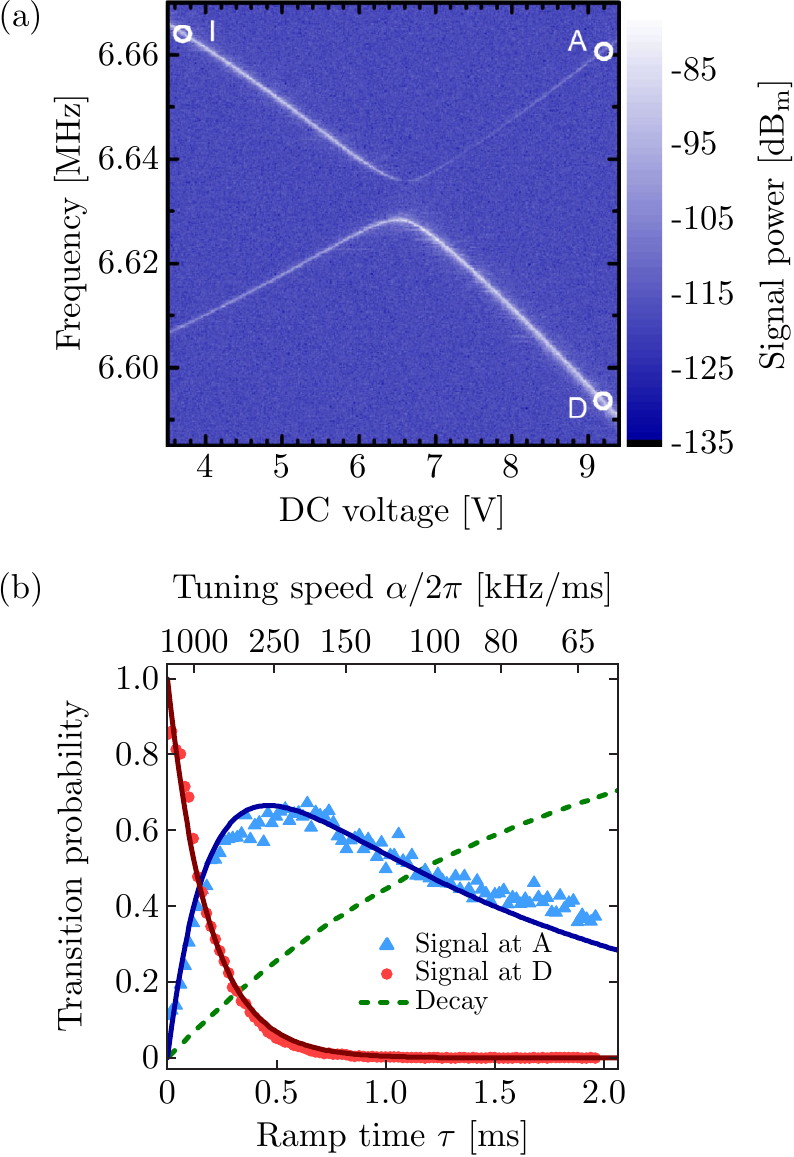}
\caption{Resonant mode coupling. (a) Resonant frequencies of two nanomechanical modes as  functions of the gate voltage. The two modes hybridize at the anticrossing. (b) Classical analogue of the Landau-Zener transition. The high-frequency mode is initially excited (point $I$ in panel (a)), and then the modes are swept through resonance over time $\tau$, which leads to energy exchange. The data (triangles and dots) show the signal power at points $A$ and $D$ in panel (a), respectively, as a function of $\tau$. The dashed line represents the decay probability of the total energy \cite{Faust2012a}.}
\label{fig:LandauZener}
\end{figure}

In the harmonic approximation, the potential energy of two modes can then be written as

\begin{equation}
U_\mathrm{12}(q_1,q_2)=\frac{1}{2}M_1\omega_1^2(t)q_1^2+\frac{1}{2}M_2\omega_2^2(t)q_2^2+M\Delta_{12}q_\mathrm{1}q_\mathrm{2},
\label{eq:potentialresonant}
\end{equation}
where $q_i$ is the displacement of the mode $i$ ($i=1,2$), $M_i$ is its effective mass, $\omega_i$ is its eigenfrequency, and $M\Delta_\mathrm{12}$ characterizes the coupling strength.
If the mode frequencies $\omega_i(t)$ vary in time in such a way that they go through resonance, there occurs anticrossing shown in Fig.~\ref{fig:LandauZener}(a). It can be observed while measuring the resonant frequencies of the resonator as functions of the gate voltage that tunes the frequencies~\cite{Faust2012a,Deng2016}.
This anticrossing is similar to the anticrossing of energy levels of a quantum system driven through resonance, even though the considered system is purely classical.

\textcite{Faust2012a} also reported observations of a classical analogue of the Landau-Zener transition (Fig.~\ref{fig:LandauZener}~b). The two-mode system is first prepared by detuning the mode eigenfrequencies by a large amount compared to the coupling. Energy is injected in mode 1, point I in Fig.~\ref{fig:LandauZener}(a). The system is then swept with the gate voltage to the final state where the modes are again practically independent from each other, points $A$ and $D$. As the modes go through resonance, the energy initially injected in mode 1 gets distributed between the two modes. The distribution depends on the ramp time of the gate voltage (Fig.~\ref{fig:LandauZener}~(b)).

 An important problem in the Landau-Zener tunneling is the effect of dissipation and noise on the transition, cf. \cite{Ao1989,Quintana2013,Malla2017} and references therein. The results by \textcite{Faust2012a} suggest that  nanomechanical modes can be used to study these effect in a well-controlled experimental setting.

In other experiments with nanomechanical modes, the response of coupled modes to pulses of resonant  drive were studied \cite{Faust2013,Okamoto2013}. The results allowed emulating Rabi oscillations in  a classical system.

The reason of the similarity of the classical dynamics of nanomechanical modes and the quantum dynamics of a two-level system is that the interference of linear modes has much in common with the interference of wave functions. We emphasize, however, that nanoresonators and quantum two-state systems, for example, qubits, are entirely different by nature as seen from the dimensions of their Hilbert spaces, in particular. Coupled classical linear modes may not be used as qubits, since it is the strong nonlinearity of qubits that underlies their application in quantum computing.

Nanomechanical resonators with a symmetric cross-section, like cylindrical wires, have degenerate flexural eigenmodes, which are orthogonal. Such modes are highly sensitive to perturbations that cause symmetry lifting. Consider a resonator with degenerate modes polarized in perpendicular directions ${\bf \hat{r}_1}$ and ${\bf \hat{r}_2}$. We now place the resonator in an inhomogeneous  force field with components $F_i$ ($i=1,2$). The partial derivatives of the force components $\partial F_i/\partial r_i$ shift the resonance frequencies of the eigenmodes. In contrast, the cross-derivatives $\partial F_i/\partial r_j$ with $i\neq j$ couple the eigenmodes. As a result, the new eigenmodes have different frequencies and are polarized in different directions. Such eigenmode rotation  has been directly measured~\cite{Gloppe2014,Lepinay2016,Rossi2016}.

The force that drives a nanowire can be non-potential, $\partial F_1/\partial r_2\neq \partial F_2/\partial r_1$. Such a force can come if a laser-driven nanowire is placed away from the waist of a focused laser beam. Then the eigenmodes are no longer pointing in orthogonal directions. It has been experimentally shown that the angle between them can be reduced down to zero \cite{Lepinay2018}. The  power spectra of the modes were found  to deviate from the conventional Lorentzian form.

Examples of other effects of resonant coupling of a few NVS modes include generation of circularly polarized mechanical oscillations using modes polarized in perpendicular directions \cite{Perisanu2010,Conley2008} and resonant coupling of an optomechanical resonator and a single bacterium~\cite{Gil-Santos2020}. A significant and important group of resonant-coupling effects concerns one-  and two-dimensional arrays of resonators; in such arrays many modes are brought in resonance at a time. Arrays of NVSs can display various types of topological effects ~\cite{Peano2015,Cha2018,Ren2020,Lin2021,Yamaguchi2021} as well as dynamical phase transitions and symmetry breaking effects \cite{Dykman2018,Matheny2019,Heugel2019a}, among others, and can be used as tunable phononic waveguides \cite{Hatanaka2019,Kirchhof2021}.


\subsection{Nonlinear resonant coupling}
\label{sec:nonlinear_resonant_coupling}

The study of nonlinear resonance has been attracting much attention and has a long history in quantum and classical mechanics, which goes back at least to Laplace and Poincar\'e on the classical side and to the Fermi resonance on the quantum side~\cite{Arnold1989,Fermi1931}. For two modes, nonlinear resonant mode coupling occurs where the ratio of their eigenfrequencies $\omega_2/\omega_1$ is close to a rational number $n/m$. In nonlinear resonance, modes can exchange energy with each other, similar to linear resonance. The coupling that leads to this exchange comes from the modes nonlinearity and therefore is unavoidable. Generally it falls off with the increasing order of the resonance, i.e., for two modes, with the increasing $n$ and $m$ in the ratio $n/m$.
Classically, multiple nonlinear resonances pave the road to chaos in conservative systems, a profound understanding that emerged in the 20th century.

A system of coupled nanomechanical modes provides a playground for studying nonlinear resonant effects. One can study them in a well-controlled setting in the regimes of strong to weak dissipation and explore a broad range of phenomena, from the aforementioned resonant energy exchange to various types of dynamical bifurcations, different scenarios of the onset of dynamical chaos, and resonant nonlinear friction to mention but a few.

For a two-mode system with the resonant condition $n\omega_1 \approx m\omega_2$, the simplest term in the potential energy that directly accounts for the resonant energy exchange in nonlinear resonance has the form
\begin{equation}
U_\mathrm{nl}^\mathrm{res}=M\Delta_{12}^{(nm)} q_1^{n}q_{2}^{m}
\label{eq:nonl_res}
\end{equation}
where $\Delta_{12}^{(nm)}$ is the coupling parameter. The importance of the coupling $U_\mathrm{nl}^\mathrm{res}$ is clear from the following argument. If $q_1$ and $q_2$ oscillate at frequencies $\omega_1$ and $\omega_2$, respectively, $U_\mathrm{nl}^\mathrm{res}$ contains a non-oscillating part. This is the ``normal form'' term \cite{Guckenheimer1997}: it is of the lowest order in nonlinearity that has  a non-oscillating part, drawing the similarity with the harmonic part of the Hamiltonian, which is independent of time.

\begin{figure}[h]
\includegraphics[scale=0.8]{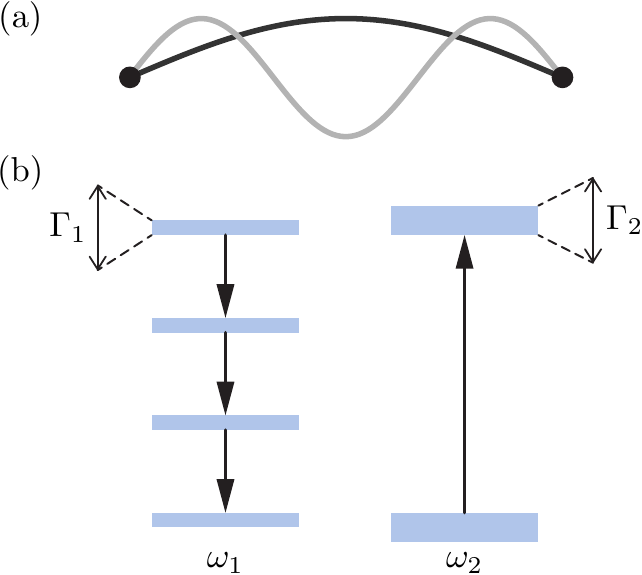}
\caption{Nonlinear resonant coupling.  (a) Spatial profile of two eigenmodes with resonance frequencies $\omega_\mathrm{1,2}$. (b) Energy diagram showing an energy exchange process between two harmonic oscillators with $\omega_2/\omega_1 \simeq 3$. The process simultaneously annihilates $n=3$ quanta in mode~1 and creates $m=1$ quantum in mode~2.}
\label{fig:Nonlinearesonantcoupling}
\end{figure}

The effect of the coupling (\ref{eq:nonl_res}) on the energy exchange between the modes is easy to understand in quantum terms. We write the displacement operator of an $i$th mode ($i=1,2$) in Eq.~(\ref{eq:nonl_res})  as $q_{i} = (\hbar/2M\omega_i)^{1/2}(a_i+a_i^\dagger)$, where  $a_i^\dagger$ and $a_i$ are  the raising and lowering operators. Therefore $U_\mathrm{nl}^\mathrm{res}$ contains terms $(a_1^\dagger)^n a_2^m + (a_2^\dagger)^m a_1^n$. They describe processes in which mode 1 goes up by $n$ energy levels  while mode 2 goes down by $m$ levels, or vice versa, mode 1 goes down by $n$ energy levels while mode 2 goes up by $m$ levels. For exact nonlinear resonance, in such processes the modes exchange energy $\hbar n \omega_1 =\hbar m\omega_2$, but the total energy of the modes is not changed. This is illustrated in Fig.~\ref{fig:Nonlinearesonantcoupling} for $n=3$ and $m=1$. If one of the modes is excited, the energy exchange happens periodically in time, as in the case of linear resonance. The energy exchange frequency is $\propto|\Delta_{12}^{(nm)}|$ and depends on the mode amplitudes [in fact, one has to take into account that, because of the mode nonlinearity, the frequencies $\omega_i$ depend on the mode amplitudes $A_i$, and a more accurate form of the resonance condition is $n\omega_1(A_1) = m\omega_2(A_2)$]. In quantum terms, the resonating energy levels are split, with the splitting  $\hbar\delta\omega_\mathrm{nl}$ that depends on the level numbers.

The onset of resonant nonlinear effects does not require an exact resonance. It suffices for the frequency detuning $|n\omega_1 - m\omega_2|$ to be smaller or on the order of the sum of the decay rates of the modes or the properly scaled maximal energy-exchange frequency. We note also that the symmetry of the modes may impose restrictions on the coupling: for example, in a uniform straight nanobeam the fundamental flexural mode is not coupled to odd powers of the first excited flexural mode. Selection rules apply also to other types of modes, and therefore in a number of cases special design was implemented to observe particular types of nonlinear resonance \cite{Asadi2021}.

An important asset for studying various aspects of nonlinear resonant effects with the NVSs is their tunability. The modes can be tuned in and out of resonance by sweeping their eigenfrequencies with the voltage applied to the gate electrode~\cite{Eichler2012} or by dynamically shifting the frequency of one of the modes by driving it into the nonlinear Duffing regime, where the vibration frequency depends on the vibration amplitude, cf. Eq.~(\ref{eq:Duffing_frequency_shift}), see  \cite{Antonio2012,Samanta2015,Mangussi2016,Chen2017,Hajjaj2018,Asadi2018,Luo2021,Arora2021,Houri2019a,Asadi2021,Shoshani2021} and references therein. The latter dependence also leads to the possibility of going through resonance during decay of the mode amplitude after the mode was initially excited.

In nanomechanical systems the nonlinear mode coupling is comparatively weak. It becomes stronger with the increasing mode amplitude, as seen from Eq.~(\ref{eq:nonl_res}). Therefore almost all observations in the classical domain refer to the case where one of the resonating vibrational modes is sufficiently strongly driven or is allowed to decay after being sufficiently strongly excited.

The amplitude dependence of both the mode frequencies and the effective coupling strength makes the non-linear resonance dynamics
much richer than in the case of linear resonance. Several types of ensuing behaviors have been seen in micro- and nanomechanical systems. Most observations refer to the cases where a low-frequency mode was driven close to resonance with a high-frequency mode with the frequency ratio $2:1$ or $3:1$.

\begin{figure*}[t]
\includegraphics[scale=0.8]{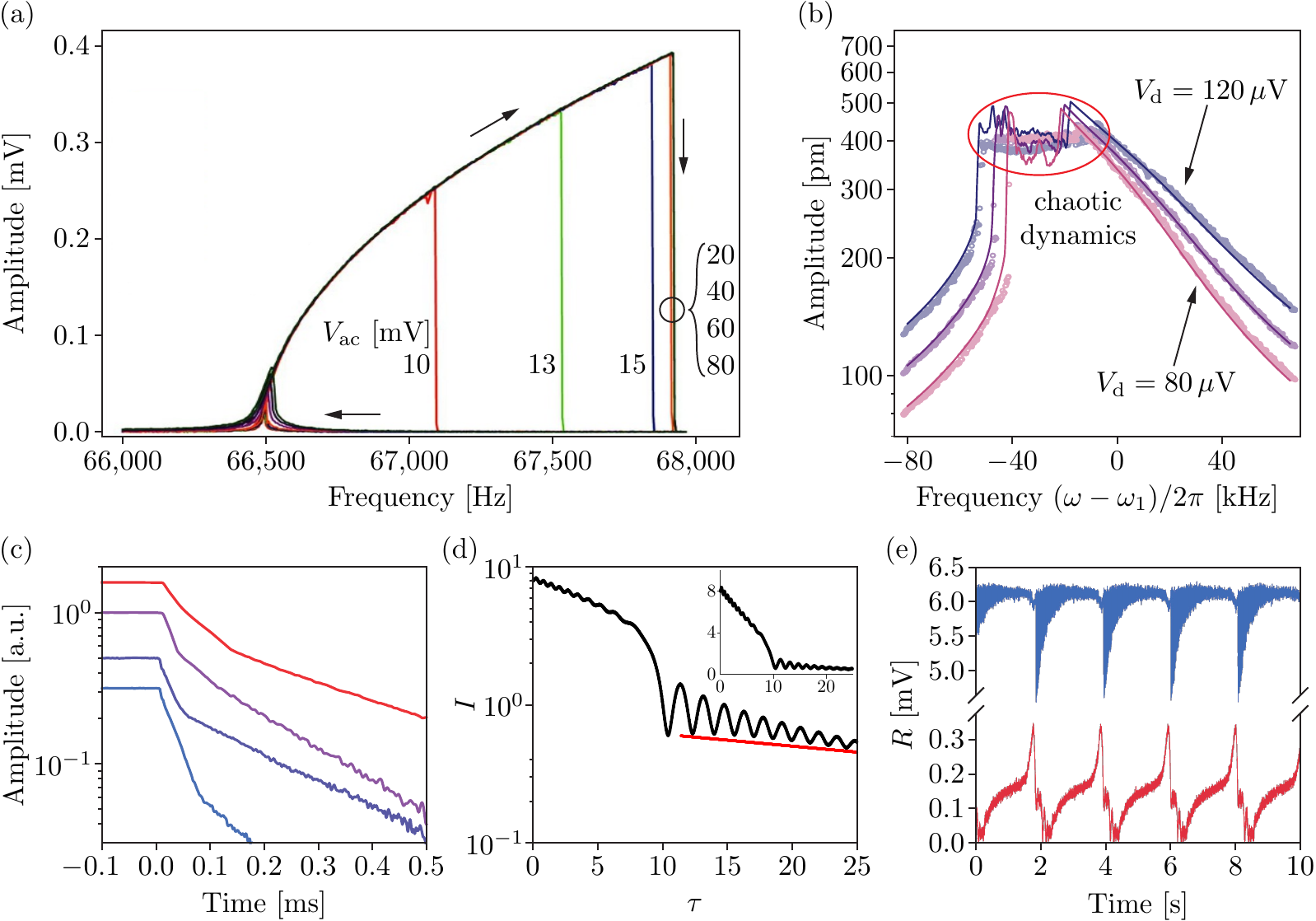}
\caption{Experiments and theoretical predictions showing the effects of nonlinear resonant coupling for $1:3$ resonance. (a) Nonlinear response of a resonantly driven lower-frequency micromechanical mode as a function of the drive frequency.  The mode switches from the large- to the small-amplitude branch at the frequencies that depend on the value of the drive amplitude indicated next to the vertical lines. Adapted from \cite{Antonio2012}.
(b) Nonlinear response of a multi-layer graphene nanoresonator where the nonlinear resonance with a higher-frequency mode results in a plateau. (c) Non-exponential energy decay in the same system after the drive is switched off at $t=0$. Adapted from \cite{Guttinger2017}. (d) Squared amplitude oscillations during decay as the lower-frequency mode goes through nonlinear resonance with a higher-frequency mode for the coupling larger than the decay rates. Adapted from \cite{Shoshani2017}. (e) Temporal amplitude of the lower-frequency (blue) and higher-frequency (red) responses of the micromechanical resonator showing the
bursting behavior as the system is driven close to the saddle-node bifurcation on the invariant cycle. Adapted from \cite{Czaplewski2018}.
}
\label{fig:Nonlin_res_observed}
\end{figure*}

For a micromechanical system that displays a $3:1$ resonance, it was found \cite{Antonio2012} that, for the low-frequency mode, the Duffing response curve shown in Fig.~\ref{fig:kozinsky_hysteresis}~(a)  changes due to the nonlinear resonance. As seen in Fig.~\ref{fig:Nonlin_res_observed}~(a), starting with a sufficiently strong driving amplitude, the frequency at which the mode switches from the large- to the small-amplitude branch becomes independent of the drive amplitude. This frequency is determined by the coupling to a mode with a three-times higher frequency. Such coupling opens a new channel of energy relaxation, limiting the increase of the amplitude. On a similar device it was found \cite{Czaplewski2018} that, in the comparatively strong drive regime the system may display very slow and strongly nonsinusoidal vibrations, producing a frequency comb with the spectral line spacing smaller than the mode eigenfrequency by a factor approaching $10^5$ and spanning a bandwidth larger then the mode decay rate by a factor $\sim 10^2$, see Fig.~\ref{fig:Nonlin_res_observed}~(e). This effect was related to a special type of the saddle-node bifurcation  (cf. Appendix~\ref{subsubsec:bif_fluctuations}), which in this case occurs on an invariant circle \cite{Shoshani2021}. When the system is near such a bifurcation point, its trajectory in the rotating frame is  burst-type.

In a nanomechanical resonator studied by \textcite{Guettinger2017} it was possible to tune the modes into a $3:1$ resonance not only by varying the parameters of the drive, but also by changing the eigenfrequencies by a gate voltage. It was observed that, when the low-frequency mode was driven sufficiently strongly, the dependence of its amplitude on the drive frequency displayed a plateau-like region, see Fig.~\ref{fig:Nonlin_res_observed}~(b).   Numerical simulations  suggest that in this region the dynamics in the rotating frame becomes chaotic.   The simulations took into account the nonlinear resonant coupling of the form (\ref{eq:nonl_res}) and successfully described switching between different vibrational branches of the coupled driven nonlinear modes.
Limit cycles and period doubling  for $3:1$ resonance in a MEMS resonator were observed  by \textcite{Houri2019a}.

The results \cite{Guettinger2017,Chen2017} also revealed a strongly nonexponential decay of the resonating modes after the low-frequency mode was excited to a large amplitude and the driving was switched off. This can be understood since, during decay, the modes exchange energy not only with the thermal reservoir, but also with each other. Such exchange is efficient while the modes are in resonance and the effective coupling strength is larger than the decay rates. This leads to a peculiar form of the time dependence of the amplitude shown in Fig.~\ref{fig:Nonlin_res_observed}~(c), which was in good agreement with the simulations \cite{Guettinger2017}.

Analytically, the time dependence of the mode amplitude has been described in two limiting cases. In the first case, the high-frequency mode has a large decay rate compared to the decay rate of the low-frequency mode and the appropriately scaled coupling, whereas in the second case the decay rates of the both modes are small on the coupling scale \cite{Shoshani2017}. In the first case, the high-frequency mode serves as a thermal reservoir that is switched on and off as the amplitude of the low-frequency mode changes, a direct analog of the nonlinear friction discussed in Sec.~\ref{subsec:nonlin_friction_mechanisms}, cf. Fig.~\ref{fig:Nonlinearfriction21}. In the second case, on the other hand, the mode decay is accompanied by strong amplitude oscillations. Interestingly, such oscillations are followed by a steep drop when the system  goes through a saddle point of the conservative motion and comes out of resonance,  as seen in Fig.~\ref{fig:Nonlin_res_observed}~(d).  In both limiting cases, for small amplitudes, where the modes are away from the resonance, the decay becomes exponential.

Besides nonlinear resonance of two modes, nano- and micromechanical systems allow studying multiple-mode resonance \cite{Mahboob2013,Luo2021}. \cite{Mahboob2013} observed in a nanomechanical system a resonance of 3 modes, in which $\omega_1 +\omega_2\approx \omega_3$ and $\omega_1\ll\omega_{2,3}$. The relevant resonant nonlinear coupling in this case has the form $U_\mathrm{nl}^\mathrm{res} = M\Delta_{123}q_1q_2q_3$. The resonance was achieved by tuning mode 1 with a piezoelectric transducer. It was shown that resonant driving of the high-frequency mode leads to excitation of coherent self-sustained vibrations of the modes 1 and 2, the effect called phonon lasing. A multiple-mode resonance was also studied in a cascade of beams with the frequencies of the successive beams decreasing by a factor of two \cite{Qalandar2014}. In this system there was demonstrated the energy transfer to a mode with frequency smaller by a factor of 4 than the excitation frequency.


\subsection{Parametrically-induced resonant coupling}
\label{sec:parametric_coupling}
It is a remarkable feature of mechanical resonators that it is possible to resonantly couple two vibrational modes using a parametric drive without any restriction on the ratio of their resonance frequencies $\omega_2/\omega_1$. A simple way to achieve such coupling is based on driving a resonator at the frequency $\omega_p$ equal to either the sum or the difference of the mode eigenfrequencies, $\omega_1+\omega_2$ or $\lvert \omega_1 - \omega_2 \rvert$, so that the drive resonantly modulates the coupling strength $\Delta_{12}^\mathrm{pump}(t)$ in the potential energy
\begin{equation}
U_{12}^\mathrm{pump}=M\Delta_{12}^\mathrm{pump}(t) q_1q_{2}.
\label{eq:parametric_mode_mode}
\end{equation}
The parametric drive is often implemented by modulating the stress in the resonator by  an electrostatic or a piezoelectric force. The coupling (\ref{eq:parametric_mode_mode}) may result also from driving one of the modes at the combination frequency $|\omega_1\pm \omega_2|$ in the presence of nonlinear mode-mode coupling \cite{Dykman1978,Sun2016}. The effect of the coupling (\ref{eq:parametric_mode_mode}) is similar to what happens in optomechanical systems~\cite{Aspelmeyer2014a} where the parametric coupling is used to cool and heat  mechanical vibrations, realize optomechanically-induced transparency in photon cavities, and hybridize mechanical vibrations with the optical field.

\begin{figure}[t]
\includegraphics[width=8.5cm]{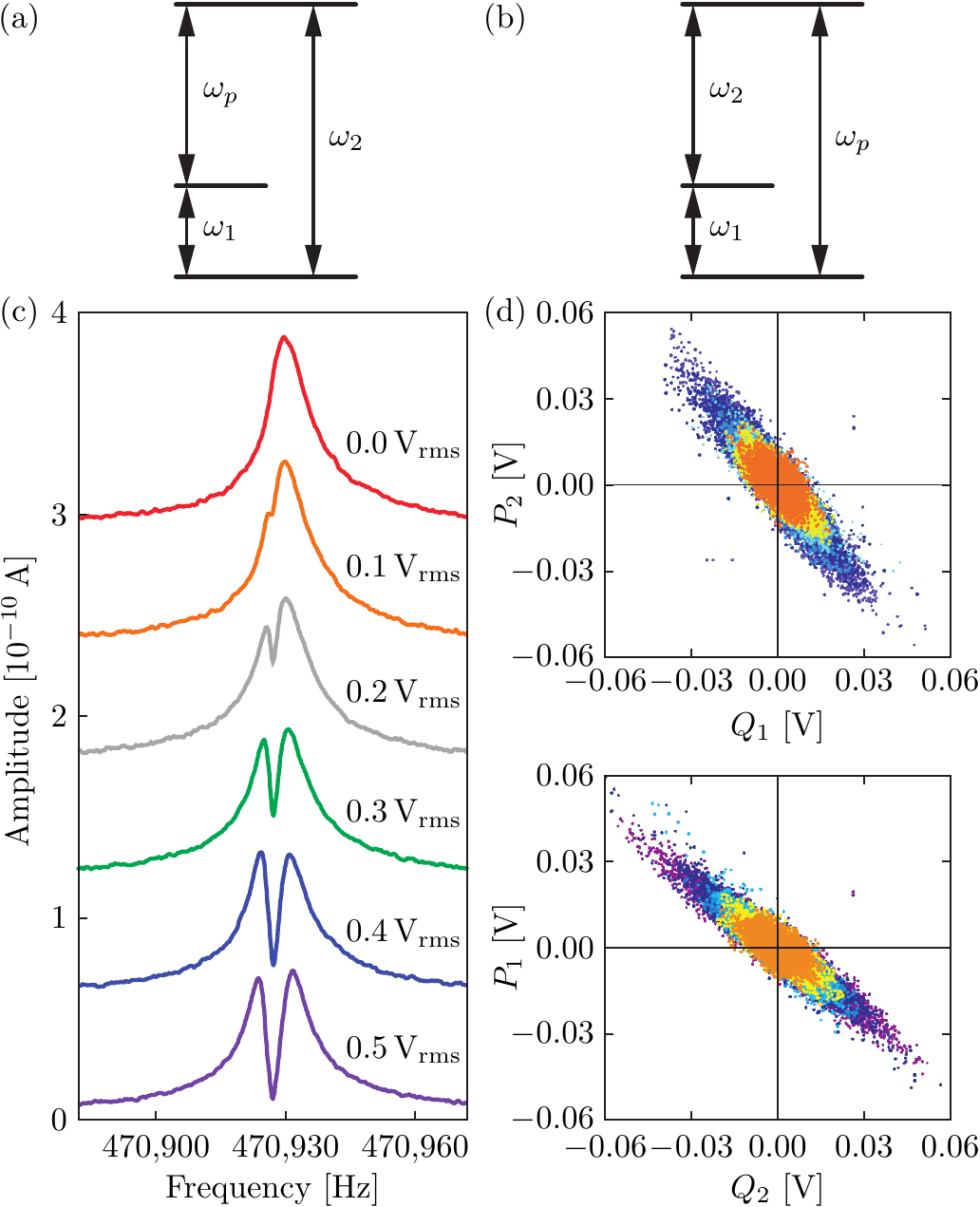}
\caption{Resonant mode coupling by a parametric drive. (a,b) Energy levels when the parametric drive frequency $\omega_\mathrm{p}$ is set at $\omega_2-\omega_1$ in (a) and $\omega_1+\omega_2$ in (b). (c) Driven response of the displacement of mode~2 to a weak probe field for different parametric drive powers with $\omega_\mathrm{p}=\omega_2-\omega_1$. The splitting of the peak at high parametric drive power indicates that modes~1 and 2 are significantly hybridized. (d) Correlations between modes~1 and 2 in the $(Q_1,P_2)$ and $(Q_2,P_1)$ phase spaces for $\omega_\mathrm{p}=\omega_2+\omega_1$ for several parametric drive amplitudes. Panels (c) and (d) adapted from \cite{Mahboob2012,Mahboob2014}.
}
\label{fig:modulated_coupling}
\end{figure}

If the drive frequency is  $\omega_p=\lvert \omega_1 - \omega_2 \rvert$, the driving leads to energy exchange between the modes. A simple way to understand this is suggested by Fig.~\ref{fig:modulated_coupling}(a) if one thinks of the driving as an electromagnetic field. In these terms, the interaction (\ref{eq:parametric_mode_mode}) describes a process in which a photon with energy $\hbar\omega_p$ and a quantum of the lower-frequency mode [mode 1 in Fig.~\ref{fig:modulated_coupling}(a)] are annihilated and a quantum of the higher-frequency mode is created, or vice versa, a quantum of a higher-frequency mode decays into a photon and a quantum of the lower-frequency mode. For $\omega_1<\omega_2$ the energy conservation condition is $\hbar\omega_p+ \hbar\omega_1 = \hbar\omega_2$.

Alternatively, if one notices that the modes coordinates $q_1$ and $q_2$ oscillate at frequencies $\omega_1$ and $\omega_2$, respectively, whereas $\Delta_{12}^\mathrm{pump}(t)$ oscillates at frequency $\omega_p=|\omega_1-\omega_2|$, in  $U_{12}^\mathrm{pump}$ there is a term that is independent of time. In the rotating wave approximation it has the same form as the static linear resonant mode-mode coupling discussed in Sec.~\ref{subsec:resonant_coupling}. In terms of the modes ladder operators the coupling energy is $\propto a_1^\dagger a_2 + a_2^\dagger a_1$. Therefore the energy exchange is similar to what was considered in Sec.~\ref{subsec:resonant_coupling}.

An important feature of the drive at $\lvert \omega_1 - \omega_2 \rvert$ is that it makes it possible to cool down the low-frequency mode [mode 1 in Fig.~\ref{fig:modulated_coupling}(a)] given that its relaxation rate is smaller than the relaxation rate of the high-frequency mode \cite{Dykman1978}.  The driving ``extracts'' the energy from the low-frequency mode and ``dumps'' it into the high-frequency mode which then quickly further dumps  it into the thermal reservoir coupled to this mode. Thus the high-frequency mode in this case serves as an effective  thermal reservoir for the low-frequency mode.

If the relaxation rate of the low-frequency mode is dominated by the energy exchange with the high-frequency mode, in the stationary state the populations of the excited states  of the modes should be equal. The occupation of the first excited energy level of the high-frequency mode (mode 2) is $\propto \exp(-\hbar\omega_2/k_BT)$. The above argument suggests that the occupation of the first excited state of the low-frequency mode (mode 1) should be $\propto \exp(-\hbar\omega_1/k_BT_\mathrm{eff})$ with $T_\mathrm{eff}= T\omega_1/\omega_2$. This means that the effective temperature of mode 1 is significantly lower than $T$. Surprisingly, the whole distribution over the excited states of mode 1 is of the Boltzmann form with  temperature $T_\mathrm{eff}$ in the absence of nonlinear friction \cite{Dykman1978}.  In the opposite case where the relaxation rate of mode 1 is much higher than that of mode 2, the effective temperature of mode 2 becomes $T\omega_2/\omega_1$, i.e., there occurs mode heating, see Appendix~\ref{sec:heating}.

Mode cooling in coupled mechanical modes has been experimentally demonstrated by \textcite{Mahboob2012,DeAlba2016}. The cooling is modest compared to what is achieved with optomechanical devices \cite{Chan2011,Teufel2011,Verhagen2012}, where the ratio of the mode frequencies can be much larger.

A profound effect of the driving-induced resonant coupling is seen in the response to a weak probe drive. This response displays Fano resonance. For a not too strong driving-induced coupling  and for the driving frequency
$\omega_p\approx|\omega_2-\omega_1|$, the amplitude of the response of a faster-decaying mode [mode 2 in Fig.~\ref{fig:modulated_coupling}(c)] to a probe field displays a narrow dip at frequency $\approx \omega_p+\omega_1$. The dip results from the interference of the direct resonant response of mode 2 and of the response of mode 1 ``uplifted'' by the drive (\ref{eq:parametric_mode_mode}) to the frequency $\omega_p+\omega_1 \approx \omega_2$. This is an analog of the optomechanically-induced transparency in photon cavities \cite{Weis2010,Safavi-Naeini2011,Qu2013}.

For a stronger driving $\Delta_{12}^\mathrm{pump}(t)$, Eq.~(\ref{eq:parametric_mode_mode}), where the rate of the driving-induced energy exchange between the modes becomes larger than their relaxation rates, the behavior of the modes is reminiscent of that at the mode anticrossing discussed in Sec.~\ref{subsec:resonant_coupling}. The resonance now is between $\omega_2$ and $\omega_1+\omega_p$. As mentioned above, the coupling Hamiltonians have the same form in the rotating wave approximation. The modes are strongly hybridized in this regime.

Experimentally, the Fano resonance and the mode hybridization in micro/nanomechanics were first demonstrated in GaAs-based semiconductor resonators~\cite{Mahboob2012,Okamoto2013}. In the response of the mode to a weak probe field shown in Fig.~\ref{fig:modulated_coupling}(c), the dip associated with the Fano resonance becomes more prominent with the increasing drive $\Delta_{12}^\mathrm{pump}(t)$. The response for the largest drive is nearly split in two peaks, the analog of anticrossing in Fig.~\ref{fig:LandauZener}(a),  indicating a significant mode hybridization. It is comparatively simpler to reach this hybridization regime -- also called the strong coupling regime -- in resonators based on nanoscale materials, such as graphene and nanotubes, because the stress can be modulated by a larger amount~\cite{Liu2015,Mathew2016,DeAlba2016,Zhu2017,Luo2018,Prasad2019}.

Parametric drive at frequency  $\omega_p\approx \omega_1+\omega_2$ leads to heating of the Brownian motion of both modes 1 and 2.  In the rotating wave approximation, in terms of the modes ladder operators the coupling energy is $ \propto  a_1^\dagger a_2^\dagger+a_1a_2$. The effect of the parametric drive can be thought of as a decay of a drive photon with creation of a quantum of mode~1 and a quantum of mode~2, cf. Fig.~\ref{fig:modulated_coupling}~(b). Such excitation corresponds to ``negative damping'' and is associated with a decrease of the dissipation rates of the modes. If in the absence of the driving one mode is decaying much faster than the other, the effect is described by the decrease of the linear friction coefficient and the increase of the effective temperature of the slower-decaying mode. When the dissipation rate is positive, this mode (mode 1, for concreteness) can amplify an externally applied weak drive at frequency $\omega_p-\omega_1$. When its dissipation rate goes through zero, the mode switches to the regime of self-sustained vibrations \cite{Dykman1978}.

Amplification of a weak radiation by a nanomechanical mode coupled to an optical cavity mode was observed by \textcite{Massel2011}. For coupled micromechanical modes, both the resonant parametric heating and the onset of oscillations   have been observed by~\textcite{Mahboob2014a}. In the parametric heating regime, a correlation in the displacement noise of the two modes was measured and a two-mode squeezing was found, see Fig.~\ref{fig:modulated_coupling}(d).

An interesting regime arises where both modes are pumped into the regime of self-sustained vibrations \cite{Sun2016}. The measurements show that the phase fluctuations of the two modes feature near-perfect anti-correlation, so that the sum of the phases  $\phi_1(t)+\phi_2(t)$ remains nearly constant. Such anti-correlation is a consequence of the discrete time-translation symmetry imposed by the periodic drive. This regime has not been accessed with the optomechanical systems fabricated thus far, since the dissipation rate of the optical cavity could not be driven to zero.

\begin{figure}[h]
\includegraphics{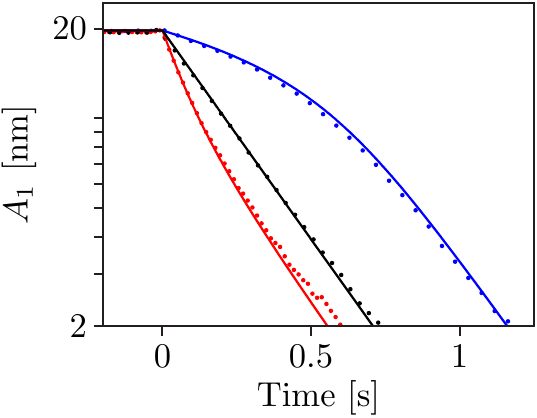}
\caption{Tailoring nonlinear friction with parametrically-modulated mode coupling. The vibration amplitude of mode~1 is shown as a function of time in a ring-down measurement. The black curve corresponds to linear friction. The red and blue curves correspond to positive and negative nonlinear fiction when the device is pumped at $\omega_2\mp 2\omega_1$, respectively. Adapted from \cite{Dong2018}}
\label{fig:Nonlineardecaydata}
\end{figure}

Parametric drive of the mode coupling can be also used to generate nonlinear friction in a controlled way. Pumping at $\omega_2-2\omega_1$ leads to positive nonlinear friction of mode~1,
if the damping rate of mode~2 is large enough so that any additional energy arriving from mode~1 is rapidly transferred to the environment. This results in a relaxation process where two quanta of mode~1 are simultaneously extracted and transferred to mode 2 along with a drive photon, in contrast to the discussed earlier linear friction that involves a transfer of one quantum of mode 1. Energy decay measurements of this nonlinear friction show that the vibrational amplitude decreases fastest at high amplitude (red data in Fig.~\ref{fig:Nonlineardecaydata})~\cite{Dong2018}. Pumping at $\omega_2+2\omega_1$ generates negative nonlinear friction, where the measured decay is slowest at high amplitude (blue data in Fig.~\ref{fig:Nonlineardecaydata}). Ultimately such negative nonlinear friction leads to the possibility of self-sustained vibrations in the system of coupled modes.

\section{Frequency fluctuations}
\label{sec:frequency_fluctuations}

Frequency fluctuations are one of the least understood chapters of the dynamics of nanomechanical systems. In a way, even the word ``frequency'' has been used somewhat ambivalently. Strictly speaking, the angular frequency is equal to $2\pi$ divided by the vibration period. It is thus associated with a discrete time interval. However,  in classical vibrational systems there is also considered an ``instantaneous''  frequency that continuously depends on time and is given by the derivative of the vibration phase $\varphi(t)$ over time. For perfect sinusoidal vibrations the two definitions of the frequency coincide, but in the presence of fluctuations they generally differ. Frequency fluctuations in NVSs have attracted much attention, as they often impose a limit on mass sensitivity \cite{Ekinci2005,Cleland2005,Yang2006a,Naik2009,Chaste2012} and other applications in sensing, like force and force gradient measurements
\cite{Weber2016,Braakman2019};
they are also a major limiting factor in the application of micromechanical systems as clocks, gyroscopes, and other devices \cite{Ng2013,Zaliasl2015,Miller2018}.

For an isolated linear mode, the displacement can be written as
\[q(t) = A\cos \varphi(t), \qquad \varphi(t)=\omega_0t+\phi.\]
Here $\omega_0$ is the mode eigenfrequency, i.e., a parameter of the system, whereas $\phi$ as well as the amplitude $A$ are  determined by the initial conditions. Coupling of the mode to the environment causes fluctuations of both $\phi$ and $\omega_0$, as well as amplitude fluctuations, making all these parameters time-dependent. The fluctuations of $\phi$ and $\omega_0$, which are of primary interest for this section, come from physically different sources and are called, respectively, phase and eigenfrequency fluctuations. Both of them contribute to fluctuations of the full vibrational phase, which can be now written as
\begin{align}
\label{eq:full_phase_formal}
\varphi(t) = \int_0^t\omega_0(t')dt' + \phi(t).
\end{align}
The time-dependent phase $\phi(t)$ is also often called the phase in the rotating frame. In high-$Q$ systems, of relevance are fluctuations of $\phi$ and $\omega_0$ that are slow on the time scale $\omega_0^{-1}$.

Fluctuations of $\phi(t)$ have been studied very broadly,  initially in various systems of self-sustained vibrations.
These studies can be traced back to 1930-1950s \cite{Berstein1938,Rytov1956,Rytov1956a}; they were later carried out for lasers, see \cite{Lax1967,Lax1968b} and references therein, and for time metrology \cite{Allan1966,Allan1988}.
A major feature of self-sustained vibrations is that the phase $\phi$ is arbitrary unless the vibrations are synchronized by an external source. Therefore phase fluctuations can accumulate in time. Generally, this leads to phase diffusion on a long time scale independent of the nature of the vibrational system.

\begin{figure}[h]
\includegraphics[scale=0.8]{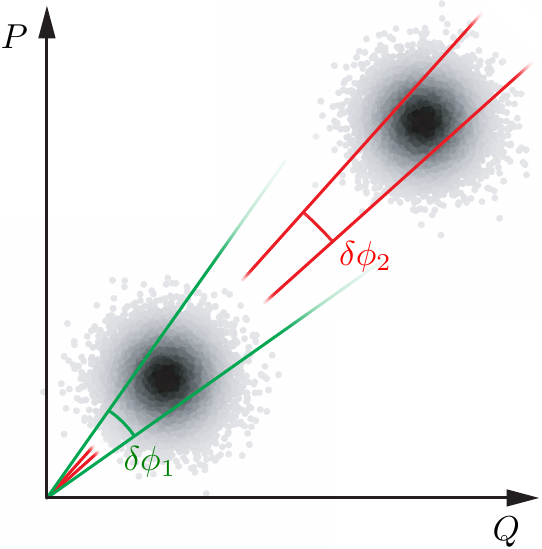}
\caption{
The effect of thermal phase noise. The noise is sketched as smearing of the values of the quadratures $Q,P$ of the mode. The uncertainty in the phase $\delta\varphi$ depends on the vibration amplitude. For a small amplitude its value $\delta\varphi_1$ is larger than the value $\delta\varphi_2$ for a larger amplitude.
}
\label{fig:phasenoiseschematic}
\end{figure}

In micro- and nanomechanical systems, an unavoidable source of  phase fluctuations  is the thermal (thermomechanical) noise. It comes along with friction from the coupling to a thermal reservoir and is described by the force $f_T(t)$ in Eq.~(\ref{eq:Brownian}) \cite{Cleland2002,Schmid2016}. It sets the so-called noise floor and thus imposes a fundamental limit on the precision with which the full phase $\varphi(t)$ and thus the frequency $\dot\varphi(t)$ can be determined, cf. Fig.~\ref{fig:phasenoiseschematic}.

A different source of fluctuations of the overall phase $\varphi(t)$  is fluctuations of the vibration eigenfrequency. A simple mechanism of such fluctuations is random attachment and detachment of molecules and the associated change of the mass and thus the eigenfrequency of a nanoresonator \cite{Yong1989,Cleland2002,Ekinci2004,Dykman2010a,Yang2011}.

Understanding frequency fluctuations requires separating the fluctuations of the eigenfrequency and of the slow part $\phi$ of the phase and characterizing both the spectra and the statistics of these fluctuations. In turn, such characterization provides a means for identifying the fluctuation sources. We will concentrate primarily on ``open-loop'' measurements done to this end. Such measurements are performed with no feedback loop. They are therefore free from the effect of the noise that comes from the feedback. In studying nano- and micromechanical systems, there are also often employed feedback-based methods, including those utilizing self-sustained vibrations \cite{Feng2008} as well as the phase-locked loop method \cite{Albrecht1991,Naik2009,Hanay2012,Olcum2015}, see \cite{Schmid2016} and \cite{Demir2021} and references therein.

We note that, in quantum terms, frequency fluctuations are usually associated with decoherence. A familiar source of decoherence in quantum systems is fluctuations of the level spacing, i.e., of the transition frequency, which corresponds to the eigenfrequency in the above picture. The thermomechanical noise described by Eq.~(\ref{eq:Brownian}) does not lead to fluctuations of the level spacing of a quantum oscillator in the absence of  nonlinearity. In this sense, there is a significant difference between the quantum and classical pictures of frequency fluctuations. The pictures can be reconciled, though, by realizing that, because of dissipation, for a nonzero temperature a quantum oscillator makes transitions  between its energy levels. The transitions happen at random, and this is a quantum analog of the effect of  thermal noise. The power spectra of quantum and classical linear oscillators have the same shape, see Sec.~\ref{subsec:quantum_aspects}.

\subsection{Allan variance}
\label{subsec:Allan}

The most broadly used method of characterizing frequency fluctuations is based on measuring the Allan variance. To find it, following the original approach \cite{Allan1966}, one has to measure the average frequency
${\bar f}(\tau)$ over time $\tau$ and to compare the values of this frequency obtained as the system evolves. Specifically, an $m$th value of the average frequency $\bar f_m\equiv \bar f_m(\tau)$ is determined by the  increment of the overall vibration phase $\varphi$ over the time from $t_m$ to $t_m+\tau$, and then $\bar f_m = [\varphi(t_m+\tau) - \varphi(t_m)]/2\pi\tau$.
If $f_0=\omega_0/2\pi$ is the mean value of $\bar f_m$, the Allan variance found from $N$ measurements is defined as
\begin{align}
\label{eq:Allan_defined}
\sigma_\mathrm{A}^2(\tau) = \frac{1}{2(N-1)f_0^2}\sum_{m=1}^{N-1}(\bar f_{m+1} - \bar f_m)^2.
\end{align}

The Allan variance can be expressed in terms of the power spectrum of the full phase $\varphi$. This gives $\sigma_\mathrm{A}^2(\tau)$ in a simple explicit form for several types of noise and for a different relation between $\tau$ and the mode relaxation time $\Gamma^{-1}$, in the closed-loop and open-loop measurements, see Appendix~\ref{sec:Allan_appendix}. The Allan variance as defined by Eq.~(\ref{eq:Allan_defined}) does not distinguish between fluctuations of the eigenfrequency and the rotating-frame phase $\phi$. Also, it does not provide information about the statistics of the fluctuations.

\begin{figure}[h]
\includegraphics{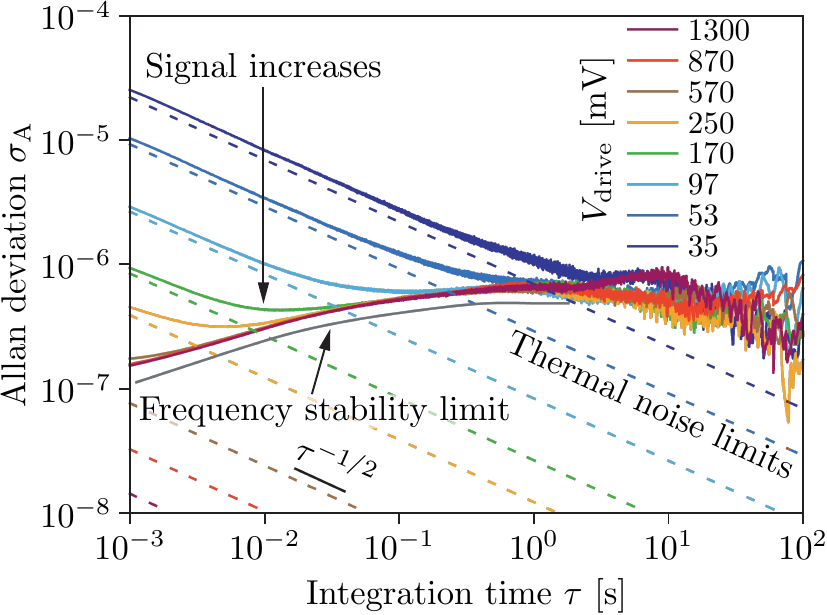}
\caption{Allan deviation as a function of integration time, from 1 ms to 100 s for different amplitudes of the resonant drive. Dashed lines indicate the expected stability from the output signal at each drive voltage and the total additive noise in the system, including the thermal and the measurement-related noises. The red line is a visual guide, highlighting the experimentally measured lower bound for frequency stability. Adapted from \cite{Sansa2016}.}
\label{fig:Hentz16}
\end{figure}

Figure \ref{fig:Hentz16} shows the measurements of the Allan deviation $\sigma_\mathrm{A}$ in a single-crystal Si nanoresonator based on the described approach \cite{Sansa2016}. The studied nanoresonators had $Q$ in the range $(5\div 7)\times 10^3$ and the experiments were done at room temperature. It is seen that the experimentally observed noise can be several orders of magnitude higher than the one expected from thermal fluctuations of the phase $\phi$ and described by equation
\begin{align}
\label{eq:repeat_H3}
\sigma_A^2(\tau) = (2\Gamma k_BT/\omega_0^4A^2)\tau^{-1},
\end{align}
cf. Eq.~(\ref{eq:Allan_thermal}). A significantly larger Allan deviation than what would be expected from thermal noise has been reported essentially for all nanomechanical systems studied thus far. These observations suggest that a major contribution to $\sigma_\mathrm{A}$ comes from other noise sources. Of particular importance in this respect are fluctuations of the mode eigenfrequency.

\subsection{Eigenfrequency fluctuations}
\label{subsec:eigenfrequency_fluctuations}

Several mechanisms of eigenfrequency fluctuations of nanoresonators have been discussed in the literature. Besides the fluctuations due to the mode nonlinearity and nonlinear mode coupling discussed in Sec.~\ref{subsubsec:nonlinearity_mode} they include the aforementioned noise due to random attachment and detachment of molecules, molecule diffusion along the resonator \cite{Yang2011,Atalaya2011,Schwender2018}, tension fluctuations due to temperature fluctuations, defect motion,  transitions between the states of two-level systems within a material or on the surface \cite{Fong2012,Hamoumi2018,MacCabe2020}, and local charge fluctuations \cite{Siria2012,Yazdanian2009,Miao2014,Dash2021}.

Different fluctuation mechanisms lead to eigenfrequency fluctuations with different time scales, i.e., with different correlation times $t_c$. The fluctuation statistics is also different. Often several fluctuation mechanisms with different $t_c$ and different statistics jointly affect the mode dynamics.

The presence of eigenfrequency fluctuations can be revealed by comparing the power spectrum of a mode or the spectrum of its response to a resonant drive with the results of a ringdown measurement where there is studied the decay of initially excited vibrations. In the absence of nonlinear friction the decay is exponential in time with the decrement given by the friction coefficient $\Gamma$, as seen from Eq.~(\ref{eq:eom_classical}). If there are no eigenfrequency fluctuations, $\Gamma$ is also the halfwidth of the power spectrum $S{}(\omega)$, cf. Eq.~(\ref{eq:Lorentz_classical}). However, often the shape of the spectrum deviates from the Lorentzian and the halfwidth $\Delta\omega$ exceeds $\Gamma$ even where the vibrations are linear \cite{Schneider2014,Guttinger2017,MacCabe2020}. This is a consequence of eigenfrequency fluctuations.

An advantageous approach to separating and characterizing eigenfrequency fluctuations is based on studying the mode dynamics in the presence of a close to resonance drive \cite{Maizelis2011,Fong2012,Gavartin2013,Zhang2014,Sun2015,Zhang2015a,Kalaee2019}.  The drive breaks the time translation symmetry of the system, cf. Sec.~\ref{subsec:driven_fluctuations}. As a result, fluctuations of the in-phase and quadrature vibration components  become different, which leads to several observable consequences.

One of these consequences is pronounced in  the correlators  $\langle u(t_1)...u(t_n)\rangle$ of the complex amplitude of the driven mode
\[u(t) = (2M\omega_F)^{-1}[M\omega_F q(t)-ip(t)]\exp(-i\omega_Ft),\]
where $\omega_F$ is the drive frequency, cf. Eq.~(\ref{eq:driven_linear}).
These correlators are nonzero only because of the broken time-translation symmetry, and it follows from the time-symmetry arguments that, for a linear mode, they do not depend on thermal noise and the corresponding phase fluctuations. However, they explicitly depend on fluctuations $\delta\omega_0(t)$ of the eigenfrequency. Studying these correlators provides a direct way to characterize the spectrum and statistics of the eigenfrequency fluctuations and enables measuring the correlators $\langle\delta\omega_0(t_1)...\delta\omega_0(t_n)\rangle$ \cite{Maizelis2011}. The ratio $\langle u^2\rangle/\langle u\rangle^2$ was used by \textcite{Gavartin2013} to characterize eigenfrequency fluctuations in a nanomechnical beam; it was also shown in this paper that the fluctuations can be suppressed with a feedback using a second mechanical mode as a frequency noise detector. The experiment on a micromechanical resonator in which the eigenfrequency was modulated by a telegraph noise \cite{Sun2015} demonstrated the possibility to reveal the noise statistics by measuring the moments $\langle u^n\rangle$.
\begin{figure}[h]
\includegraphics[scale=0.8]{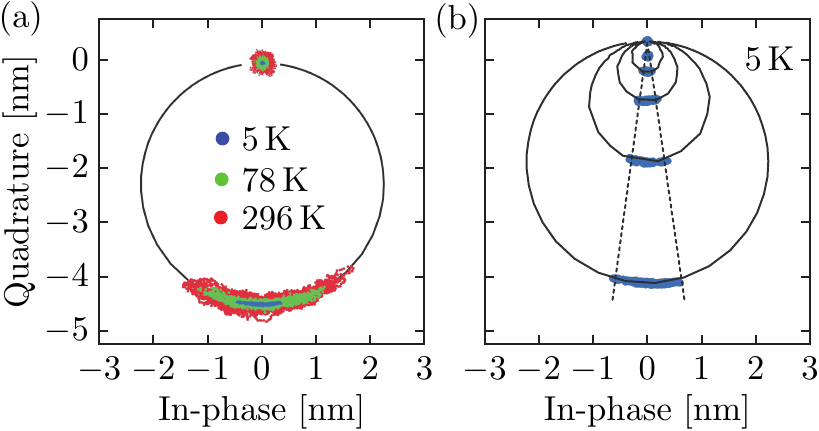}
\caption{Fluctuations of the in-phase and quadrature components of a resonantly driven SiN nanoresonator. The black circles show the driven response when the driving frequency $\omega_F$ is swept across the resonance. Panels (a) and (b) show the  increase of the fluctuations with the increasing temperature and the dependence on the driving amplitude; with no driving the mode fluctuates about zero (the data points at the top). Adapted from \cite{Fong2012}.}
\label{fig:hong_freq_fluct}
\end{figure}

The role of eigenfrequency fluctuations is clearly seen from Fig.~\ref{fig:hong_freq_fluct}, which shows fluctuations of the quadrature and in-phase components of a driven high-Q nanomechanical resonator \cite{Fong2012}. Whereas the fluctuational spread of the quadrature component is essentially independent of the amplitude of forced vibrations $A$, the spread of the in-phase component increases with the increasing amplitude. The observed spread of the  vibration phase counted off from $\omega_Ft$ ($\omega_F$ is the drive frequency), $\varphi(t) - \omega_Ft$, is practically independent of $A$ for low temperatures. For the thermomechanical noise, on the other hand it would fall off with the increasing $A$, $\sigma_\mathrm{A}^2\propto A^{-2}$, cf. Eq.~(\ref{eq:repeat_H3}) and Appendix~\ref{sec:Allan_appendix}. The observed phase spread is a direct consequence of the fluctuations of  $\omega_0$, with the phase deviation
\[\delta\phi(t)\approx \int_{-\infty}^t dt'\exp[-\Gamma(t-t')]\delta\omega_0(t')\]
for a weak eigenfrequency noise. The data allowed \cite{Fong2012} to study the low-frequency part of the spectrum of the fluctuations $\delta\omega_0(t)$. It was found to be of the $1/f$ type and was related to reorienting two-state elastic dipoles.

The power spectrum of the eigenfrequency fluctuations in a broad frequency range can be extracted directly from the power spectrum of a driven mode $S{}(\omega)$ \cite{Zhang2014,Zhang2015a}.  For a linear mode this can be qualitatively explained as follows. With no noise, a resonant force $F\cos\omega_Ft$ leads to the mode displacement
\begin{align}
\label{eq:naive_freq_fluct}
q(t) = \frac{F}{2M\omega_F}\,\mathrm{Re}\frac{i\exp(-i\omega_Ft)}{\Gamma -i(\omega_F-\omega_0)},
\end{align}
cf. Sec.~\ref{subsec:bistability}. If in this expression $\omega_0$ is fluctuating, that is, if $\omega_0$ is formally replaced by $ \omega_0+\delta\omega_0(t)$, the displacement $q(t)$ is also fluctuating. This should lead to an extra peak in the power spectrum of the mode. The form of this peak strongly depends on the interrelation between the correlation time $t_c$ of the fluctuations $\delta\omega_0(t)$ and the relaxation time of the mode $\Gamma^{-1}$.  Since the power spectrum is quadratic in $q(t)$, it is clear from Eq.~(\ref{eq:naive_freq_fluct}) that the peak is proportional to $F^2$, which allows one to identify it and separate it from other spectral features.

The replacement of $\omega_0$ with $\omega_0+\delta\omega_0(t)$ in Eq.~(\ref{eq:naive_freq_fluct}) is applicable if the correlation time of the eigenfrequency fluctuations $t_c$ is large compared to the mode relaxation time $\Gamma^{-1}$, so that the mode adiabatically follows these fluctuations. The fluctuation-induced slow time variation of the amplitude and phase of the forced vibrations at frequency $\omega_F$ lead to a narrow spectral peak, which is  centered at $\omega_F$. The width of this peak is $\sim t_c^{-1}$. For small $|\delta\omega_0(t)|$ the peak is proportional to the power spectral density $S_{\delta\omega_0}(\omega-\omega_F)$ of  $\delta\omega_0(t)$. Therefore the shape of the peak allows reading off this spectral density directly. In the experiment on a carbon nanotube resonator \cite{Zhang2014} carried out at $T=1.2$~K it was found that the spectral density of the eigenfrequency fluctuations is $S_{\delta\omega_0}(\omega)\propto \omega^{-\alpha}$ for small $\omega$, with $\alpha\approx 0.5$. A similar measurement was done for a silicon nanobeam, where such $1/f$-type scaling was also observed, with the exponent $\alpha\approx 0.7$ \cite{Sun2016}.

Slow eigenfrequency fluctuations determine the long-term stability of devices based on nano- and microresonators, including clocks. However, they don't lead to a broadening of the spectral response in the absence of the drive if the response is measured over time small compared to $ t_c$. The position of the spectral peak shifts from measurement to measurement in this case. If the duration of a measurement is $\gtrsim t_c$ and $\langle \delta\omega_0^2\rangle \gtrsim \Gamma^2$, the spectrum is broadened. This is an analog of the inhomogeneous spectral broadening. The effect was observed for a nanotube resonator \cite{Moser2014} and for a breathing mode in a phononic crystal \cite{MacCabe2020}.
\begin{figure}[t]
\centering
\includegraphics[scale=0.8]{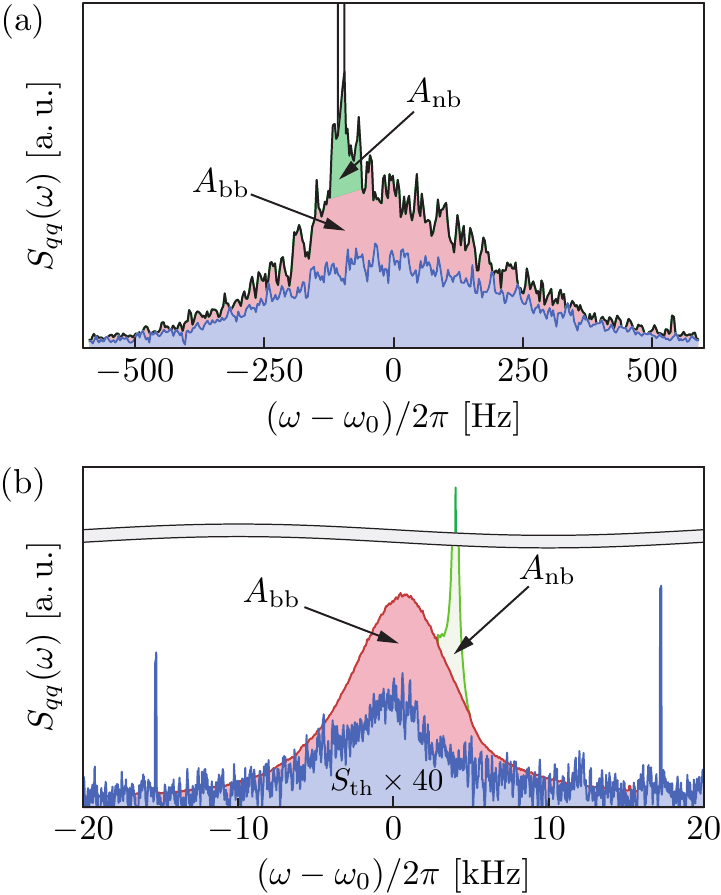}
\caption{Power spectra of driven modes with fluctuating eigenfrequencies. Panels (a) and (b) refer to a flexural mode of a carbon nanotube with the eigenfrequency $\omega_0/2\pi = 6.3$~MHz \cite{Zhang2014} and a breathing mode of a nanobeam phononic crystal with the eigenfrequency $\omega_0/2\pi \approx 425 $~MHz \cite{Kalaee2019}. Blue curves show the thermal power spectra without drive [multiplied by 40 in (b)]. The narrow peaks are centered at the drive frequencies. The regions marked as $A_\mathrm{bb}$ and $A_\mathrm{nb}$ indicate the driving induced parts of the power spectrum, which are due to the fast (broad-band) and slow (narrow-band) eigenfrequency fluctuations.  }
\label{fig:freq_noise}
\end{figure}

The opposite limit of comparatively fast eigenfrequency fluctuations, $t_c \ll \Gamma^{-1}$, is not directly described by Eq.~(\ref{eq:naive_freq_fluct}). A qualitative picture is somewhat more involved \cite{Zhang2014}, but roughly, one can think that the major spectral effect of the drive in this case comes from the effective heating of the mode, with $\delta T\propto F^2$.

Fast fluctuations $\delta\omega_0(t)$ lead to broadening of the power spectrum of a mode $S(\omega)$ in the absence of driving. The power spectrum and the mode susceptibility have a Lorentzian shape with a halfwidth $\Delta\omega$, which is a sum of $\Gamma$ and the characteristic intensity of the eigenfrequency fluctuations. This feature holds independent of the statistics of the fluctuations provided $\Delta\omega\ll \omega_0$. It has been often used to describe experiments, cf. Eq.~(\ref{eq:simple_Lorentzian}).

Studying the spectrum in the presence of driving allows one to separate the contribution of fast frequency fluctuations. The corresponding driving-induced part of the spectrum $S_F(\omega)$ can be written as
\begin{align}
\label{eq:weak_noise}
&S_F(\omega)\approx C_FF^2 \frac{(\Delta\omega - \Gamma)/\Gamma}{(\Delta\omega^2 + (\omega_F-\omega_0)^2}\frac{S(\omega)}{k_BT}
\end{align}
here strictly speaking, $\omega_0$ is also renormalized by the frequency fluctuations if they are non-Gaussian; [the explicit form of $C_F$ depends on the statistics of the fluctuations $\delta\omega_0(t)$ \cite{Zhang2014}]. The spectrum $S_F(\omega)$ has the same shape $S(\omega)$ as in the absence of driving, but is proportional to the squared driving amplitude. It is also proportional to the difference between the fluctuation- and decay-induced broadening $\Delta\omega - \Gamma$.

In Fig.~\ref{fig:freq_noise} we show the results of two experiments \cite{Zhang2014,Kalaee2019}, which demonstrate how the interplay of the driving and the eigenfrequency fluctuations affects the power spectrum. Even though the studied systems were very different, the observations clearly show the occurrence of a very narrow peak centered at the drive frequency and a broad spectrum with the shape similar to that in the absence of the driving. Comparing the areas of the peaks with and without driving allowed estimating the contribution of the fast eigenfrequency fluctuations to the observed spectral broadening $\Delta\omega$. In both experiments it was $\gtrsim 50$\%.

\section{Outlook and challenges}
\label{sec:outlook}

The goal of writing this review was to demonstrate the nontrivial physics of nanomechanical systems  and the possibility to use them as a platform for studying a broad range of nonlinear and nonequilibrium phenomena in a controlled setting, and also to indicate their numerous applications. These intertwined  aspects of nanomechanics make it a fairly unique interdisciplinary area of research and underlie the growing interest in the NVSs. The interest is further stimulated by the rapid progress in nanotechnology, which allows improving the existing types of the NVSs and making qualitatively new NVSs.  In closing the review, we present some of the nascent directions of the research in the field. Unavoidably, such a list is incomplete, particularly given the high rate at which new results are obtained.

An important aspect of the mesoscopic dynamics of the NVSs are coherent effects.
Among them, of much current interest are the effects of  coupling nanomechanical modes to qubits \cite{Lee2017,Arrangoiz-Arriola2019}, see also Sec.~\ref{subsubsec:qubit_coupling}, as well as using qubits
to entangle different mesoscopic modes  \cite{Wollack2021a}.
Related is a significant effort on developing nanomechanical resonators with  high-$Q$ microwave-frequency modes \cite{MacCabe2020,Wollack2021a}. Such modes can be brought to the ground quantum state already for temperatures $\lesssim 0.1$~K without sideband or active  feedback cooling (that reduces the $Q$-factor) and can be in resonance with superconducting qubits \cite{Mirhosseini2020}. An important type of these modes are vibrations localized around engineered defects in phononic crystals with frequencies in the gap of the spectrum of propagating modes.

A promising direction in the context of quantum information and coherent effects in nanomechanics is developing qubits based on the NVSs. This requires vibration  nonlinearity that remains large compared to the decay rate down to the quantum regime. Then resonant driving at the frequency of the transition from the ground to the first excited vibrational state does not lead to transitions to higher-energy states, similar to superconducting qubits. The corresponding nonlinearity can be achieved by coupling mechanical vibrations to the charge states of a double quantum dot. For a qubit based on a carbon nanotube resonator, the coherence time is predicted to be remarkably long~\cite{Pistolesi2021}. A very strong nonlinearity can be achieved also near a bifurcation point where a nanotube or a nanomembrane is close to collapse due to the strong gate voltage and can tunnel into the collapsed state \cite{Sillanpaa2011}.

Topological effects form another group of coherent phenomena that attract current attention. A possibility to observe such effects in phononic crystals in an optomechanical setting was indicated by \textcite{Peano2015}, and thermal phonons traveling along a topological edge channel with weak backscattering have been observed in an array of over 800 submicron  silicon membranes \cite{Ren2020}.  A two-dimensional metamaterial made out of submicron SiN membranes, which has time-reversal symmetry, has been demonstrated to have pseudospin-type edge states, which are robust to waveguide distortion \cite{Cha2018}. The  possibility to control the phononic band structure electrostatically makes  nanoresonator-based metamaterials interesting even where they are topologically trivial. Such control can be used for dynamical tuning of acoustic transparency and waveform engineering in phononic waveguides \cite{Cha2018a,Kurosu2018,Hatanaka2019}.

Yet another group of coherent effects is related to the dynamics of NVSs in a driving field with frequency $\omega_F$ close to an overtone of the mode eigenfrequency, $\omega_F\approx n\omega_0$. Such drive can resonantly excite vibrational states at frequency $\omega_F/n$ with $n>1$. The case $n=2$ corresponds to the parametric excitation discussed for a single mode in Sec.~\ref{subsec:parametric_main}. For coupled NVSs, the interaction between driven modes may lead to formation of  time crystals where  the discrete time-translation symmetry imposed by the periodic drive is broken, i.e., phase-matched vibrations occur at frequency $\omega_F/n\approx \omega_0$ \cite{Dykman2018}. The properties of such dissipation-free Floquet time crystals are strongly affected by the nontrivial geometric phase \cite{Lorch2019} that emerges for $n\geq 3$ \cite{Guo2013a,Zhang2017c}. The resonant-driving induced time crystals are expected  to have an exponentially long time before they are heated up and ultimately melted by the drive.

The ``incoherent'' side of the dynamics of coupled NVSs, i.e., the dynamics in the presence of relaxation and thermal fluctuations, is closely connected to the ``coherent'' side. Arrays of dissipative resonantly driven NVSs can display  time-crystalline behavior \cite{Dykman2018,Heugel2019a}, which in this case can have features related to breaking of the detailed balance.
Such arrays should enable, in particular, studying the effects  of disorder in  the eigenfrequencies and the coupling on the quantum and classical time-symmetry breaking transitions. Coupled NVSs can also display topological solitons \cite{Yamaguchi2021,Lin2021}. Topologically nontrivial dissipative networks with interesting dynamics can be created also by driving coupled resonators by radiation modulated at the difference of their frequencies \cite{delPino2021}. Challenging observations of broken-symmetry states with complex dynamics have been reported for networks of NVSs that display self-sustained vibrations \cite{Matheny2019}. Overall, the dynamics of coupled NVSs, which has been attracting attention for a long time \cite{Buks2002a,Karabalin2009,Lifshitz2003,Lifshitz2008,Cross2009}, is significantly enriched by the topological and Floquet aspects.

NVSs are playing an increasingly important role in studying condensed-matter systems. One of the pursued directions is establishing microscopic mechanisms of energy relaxation of vibrational modes. A significant  effort has been put recently into identifying the Landau-Rumer and Akhiezer relaxation, with the experiments covering the range from ultra-low to room temperatures  \cite{Rodriguez2019,MacCabe2020,Tepsic2021}. Understanding nonlinear damping and its dependence on the material, geometry, and temperature is on the agenda \cite{Atalaya2016,Steeneken2021}. Another direction is thermal effects in nanostructures. Heat transfer has been measured in graphene and MoSe$_2$ monolayers down to cryogenic temperatures, and slow equilibration between different vibrational branches of graphene has been established using photothermal response~\cite{Morell2019,Dolleman2020}; see also \cite{Sullivan2017}.  The mechanical detection and control of magnetic states and magnetic phase transitions in two-dimensional layered antiferromagnetic CrI$_3$ and FePS$_3$ materials have been demonstrated down to two layers~\cite{Jiang2020a,Siskins2020}. Charge density wave transition has also been measured in 2H-TaS$_2$ and 2H-TaSe$_2$ layered materials~\cite{Lee2021}. On the side of the electron-vibrational coupling, it has been shown using carbon nanotubes that, where the coupling is sufficiently strong, it results in controlled vibration  cooling or excitation of self-sustained vibrations in response to source-drain voltage \cite{Urgell2020,Wen2020}.

Much work is being done on characterizing two-level fluctuators with the NVSs. A promising direction is opened by passively cooling NVSs down to very low temperatures \cite{MacCabe2020,Maillet2020,Kamppinen2021,Gisler2021}, including sub-millikelvin temperatures, where even MHz-range modes are close to the quantum ground state \cite{Cattiaux2021}. The characteristic temperature dependencies of the decay rate and the frequency shift of the NVS modes  found by the ring-down measurements and by the measurements of the response and fluctuation spectra provide an insight into the relaxation mechanisms of two-level systems in nanostructures.

NVSs have demonstrated superior sensitivity and spatial resolution in studying superfluid $^3$He and $^4$He. They have been used to detect Cooper pair breaking in $^3$He \cite{Defoort2016}, as well as measure helium viscosity, detect modulated ``phonon wind''  \cite{Guenault2019,Guenault2020}, trap a single vortex \cite{Guthrie2021}, and study new effects of quantum turbulence at ultra-low temperatures \cite{Barquist2020,Barquist2021} in $^4$He. A series of first-order layering transitions of liquid helium on a carbon nanotube observed by \textcite{Noury2019} suggests the possibility of new types of phase transitions on smooth defect-free cylindrical surfaces.

Nanomechanics has essentially opened the field of experimental studies of nonlinear dynamics of fluctuating vibrational systems. The field is vibrant. Fluctuations can now be measured in real time. For a carbon nanotube, such measurements have revealed a weakly chaotic regime in which, at room temperature, the energy concentrates in low-frequency modes, disperses into higher-frequency modes, and then returns \cite{Barnard2019}, reminiscent of the  Fermi-Pasta-Ulam-Tsingou behavior. The interplay of fluctuations and nonlinearity leads to rich dynamics in systems with a few or even one vibrational mode away from thermal equilibrium, which is related to breaking of the detailed balance \cite{Roberts2021}, see Sec.~\ref{subsec:driven_fluctuations}. This includes scaling behavior of fluctuations near various types of bifurcation points \cite{Tadokoro2020,Jessop2020}, the onset of chaos in the rotating frame \cite{Guttinger2017,Houri2020}, and  new types of fluctuation squeezing in driven systems in the nonlinear regime \cite{Huber2020,Yang2021b}. Noise squeezing in a nonlinear regime may improve the measurement sensitivity, in particular for the phase-based measurements [see Sec.~\ref{sec:frequency_fluctuations}], by reducing the detrimental effect of thermal noise. Reducing measurement noise is also on the agenda, and to this end there are explored new detection methods, such as focused electron beams \cite{Pairis2019}.

Among various nonlinear resonant phenomena that can be accessed with the NVSs, of an increasing interest are nanomechanical frequency combs.
Such combs have been generated in coupled modes that display nonlinear resonance or in a single mode using feedback control \cite{Houri2019,Houri2019a,Singh2020,Houri2021}. A large number of spectral lines has been observed in nanomechanical systems with coupled modes displaying a SNIC bifurcation \cite{Czaplewski2018}, see Sec.~\ref{sec:nonlinear_resonant_coupling}, or by parametrically inducing resonant mode coupling \cite{Chiout2021}. However, driven nonlinear NVSs are expected to display a multiple-line comb even where there is just a single mode involved, but the dissipation is nonlinear \cite{Dykman2019}.

The high sensitivity of the NVSs provides a means for addressing fundamental physics problems. A part of them is related to the Casimir force at small distances and its dependence on the material properties and the geometry, as well as thermal and nonequilibrium  effects \cite{Tang2017,Wang2021a,Gong2021,Liu2021b}. The possibility of studying the interplay of quantum mechanics and gravity \cite{Schmole2016,Liu2021c} is being explored. A study of non-Newtonian gravity  and even the physics beyond the standard model, particularly with levitated  particles, which can be now cooled down to their ground quantum state, is also being discussed \cite{Moore2021,Gonzalez2021}. Very slowly decaying vibrations (with decay rate $<100~\mu$Hz) of levitated nanoparticles are being considered in a somewhat exotic context of the wave-function collapse \cite{Pontin2020}.

Much attention is attracted to  various coherent quantum effects in the coupled NVSs  \cite{Ockeloen-Korppi2018,Kotler2021,Lepinay2021}. These and a number of other quantum effects, such as cooling the vibrations to their ground state by coupling them to an electromagnetic cavity (see Sec.~\ref{sec:parametric_coupling}), using NVSs to convert microwave-frequency excitation of a superconducting qubit into an optical photon \cite{Mirhosseini2020}, or optically reading out a  transmon qubit  \cite{Delaney2021},  are often studied in the context of optomechanics, a  burgeoning area born out of nanomechanics \cite{Aspelmeyer2014a}. Recent improvements of nanofabricated NVSs~\cite{MacCabe2020,Beccari2021a,Seis2021} will lead to further advances in optomechanics.

One of the most important applications of NVSs is the emerging technology of single-molecule mass spectrometry with potentially high throughput. It will take advantage of the NVSs based inertial imaging \cite{Hanay2015,Sage2018}. Nanowires and nanotubes hold promise as cantilevers for the next-generation scanning probe microscopes.  By utilizing the fundamental modes polarized in perpendicular directions, such cantilevers
enable direct imaging of the components of the force fields and establishing whether the field is potential
\cite{Lepinay2016,Rossi2016}. Cantilevers functionalized with a magnetic material at their free end~\cite{Rossi2019} hold promise for imaging a large range of physical phenomena, such as skyrmions, superconducting vortices, and current-carrying edge states in two-dimensional systems~\cite{Braakman2019}. Magnetic resonance force microscopy with single nuclear spin sensitivity is another direction of great interest~\cite{Rose2018,Grob2019}.

\acknowledgments
A.B. acknowledges ERC Advanced Grant No. 692876, AGAUR (Grant No. 2017SGR1664), MICINN Grant No. RTI2018-097953-B-I00, the Fondo Europeo de Desarrollo, the Spanish Ministry of Economy and Competitiveness through Quantum CCAA and CEX2019-000910-S [MCIN/ AEI/10.13039/501100011033], Fundacio Cellex, Fundacio Mir-Puig, and Generalitat de Catalunya through CERCA. J.M. was supported by the National Natural Science Foundation of China through Grants No. 62150710547 and No. 62074107.  M.D. was supported in part by the National Science Foundation through Grants No. DMR-1806473 and No. CMMI 1661618.


\appendix
\section{Method of averaging: weak nonlinearity and weak damping}
\label{sec:averaging}


\subsection{Nonlinear vibrations with no damping}
\label{subsec:nonlin_averaging}

The Bogoliubov-Krylov method of averaging  used to derive the equation of motion for the complex amplitude of the mode is similar to the rotating wave approximation (RWA) in quantum mechanics. One thinks of the mode dynamics as vibrations at the mode frequency $\omega_0$ with the amplitude $A(t)$ and phase $\phi(t)$ that slowly vary on the time scale of the vibration period $2\pi/\omega_0$, that is, the mode coordinate is $q(t)=A(t)\cos[\omega_0 t+\phi(t)]$. We note that $\phi(t)$ here is the ``reduced'' phase, it does not contain the term $\omega_0t$. The complex slow variable $u(t)$ (the complex amplitude) defined by Eq.~(\ref{eq:u(t)}) is simply related to $A$ and $\phi$,
\begin{align}
\label{eq:u_t_Appendix}
u(t)=\frac{1}{2}\left(q-i\frac{p}{M\omega_0}\right)e^{-i\omega_0t}=\frac{1}{2}A(t)\exp[i\phi(t)].
\end{align}
For a harmonic mode that is not coupled to a bath,  $A$ and $\phi$ are independent of time and $u(t)=$~const.  A more general form of the averaging method is discussed in Appendix~\ref{subsec:action_angle}.

In this section we show how the dynamics is modified by the weak nonlinearity of the mode. To simplify the reading we repeat some equations from the main text.  We illustrate the Bogoliubov-Krylov method  by applying it to the Duffing oscillator. The goal is to describe the dynamics on times that largely exceed the vibration period $2\pi/\omega_0$. A direct perturbation theory in the nonlinearity does not apply, as it  leads to a secular ($\propto t$) correction to the oscillator displacement. Instead one should use the asymptotic perturbation theory.

The Hamiltonian of a nonlinear mode is
\begin{align}
\label{eq:H_0_general}
H_0=\frac{1}{2M}p^2 + U(q).
\end{align}
For the Duffing model the potential energy is
\[U(q)=\frac{1}{2}M\omega_0^2q^2 + \frac{1}{4}M\gamma q^4,\]
see Eq.~(\ref{eq:Duffing_potential}).

With the account taken of the relation $p = M\dot q$, we obtain from the definition of $u(t)$, Eq.~(\ref{eq:u_t_Appendix}), $\dot u\exp(i\omega_0 t) + \dot u^* \exp(-i\omega_0 t)=0$. Therefore
\begin{align}
\label{eq:u_substitution_Appendix}
\ddot q + \omega_0^2 q = 2i\omega_0 \dot u e^{i\omega_0 t} =  -2i\omega_0 \dot u{}^* e^{-i\omega_0 t},
\end{align}
and the Hamiltonian equation of motion for $u(t)$ reads
\begin{align}
\label{eq:Duffing_eom}
\dot u=i\frac{\gamma}{2\omega_0}\left(ue^{i\omega_0t} + u^* e^{-i\omega_0 t}\right)^3 e^{-i\omega_0 t}.
\end{align}

The time scale on which $u(t)$ varies because of the nonlinearity is seen from Eq.~(\ref{eq:Duffing_eom}) to be $\sim \omega_0/|\gamma|\gg \omega_0^{-1}$.  The right-hand side of Eq.~(\ref{eq:Duffing_eom})  contains the smooth term $\propto u|u|^2$, which is a constant for a time $\delta t\ll \omega_0/|\gamma|$ (since practically $u(t)$ does not change over this time), and the terms that oscillate as $\exp(\pm 2i\omega_0t), \, \exp(-4i\omega_0t)$. All these terms are of the same order of magnitude. However, if we now integrate them over time $\delta t \gg \omega_0^{-1}$, the contribution of the smooth term will be $\propto \delta t$, whereas the contribution of the fast-oscillating terms will be $\propto \omega_0^{-1}\ll \delta t$. Therefore to describe the dynamics of the oscillator on times $\gg \omega_0^{-1}$ the fast-oscillating terms can be disregarded and the equation of motion becomes
\begin{align}
\label{eq:Duffing_av}
\dot u \approx 3i\gamma u|u|^2/2\omega_0,\qquad u(t) \approx u(0) e^{3i\gamma |u|^2 t/2\omega_0}.
\end{align}

In this approximation $|u(t)|$ does not change in time, i.e., the vibration amplitude $A \approx (4|u|^2)^{1/2}$ does not change. However, the vibration phase acquires an extra term $3\gamma |u|^2 t/2\omega_0$. Comparing Eq.~(\ref{eq:Duffing_av}) with the expression $q(t)= u(t)\exp(i\omega_0t) + \mathrm{c.c.}$ [cf. Eq.~(\ref{eq:u_t_Appendix})], one can see that it corresponds to the change of the oscillator frequency (\ref{eq:Duffing_frequency_shift}),
\[\omega_0 \to \omega_0+ 3\gamma|u|^2/2\omega_0 = \omega_0+ 3\gamma A^2/8\omega_0.\]


\subsection{Effect of the coupling to a bath}
\label{subsec:averagin_cplng}

We now extend the Bogoliubov-Krylov method to describe the dynamics of the mode where it is weakly coupled to a thermal reservoir. There are two parts to this description, which are closely intertwined. One part is the evaluation of the average reaction force from the bath in slow time compared to $\omega_0^{-1}$. The second part refers to the random part of the force from the bath, the noise in the equations of motion for $u(t)$, and its properties  in slow time. In the ensuing approximation, the mode dynamics in {\it slow time} is Markovian.

For the coupling to the bath of the form $H_i=qh_\mathrm{b}$, the part of the force from the bath that describes the reaction of the bath to the mode (the backaction) is
\begin{align}
\label{eq:backaction_Appendix}
&F_\mathrm{b}^{\rm (r){}}(t) = -\delta h_\mathrm{b}(t) \approx   -\langle \delta h_{\rm b}(t)\rangle\nonumber \\
&\approx \int_0^\infty \mathcal{X}_\mathrm{b}(t')[ u(t-t')e^{i\omega_0(t-t')} + {\rm c.c.}]
\end{align}
cf. Eq.~(\ref{eq:bath_suscept2}). Here $\mathcal{X}_\mathrm{b}(t)$ is the time-dependent bath susceptibility. We remind that this expression is an approximation, as we have replaced the full reaction force $\delta h_\mathrm{b}(t)$ with its ensemble-averaged value $\langle \delta h_{\rm b}(t)\rangle$ and kept in the latter only the lowest-order term, which describes the linear response of the bath to the bath-mode coupling.

We now expand $u(t-t')=u(t)-t'\dot u(t) + \ldots$ and keep only the first term in this expansion \cite{Dykman1971}, relying on the smoothness of $u(t)$ (see also below). Then, using the definition of the Fourier transform of the bath susceptibility
\begin{align}
\label{eq:bath_suscept_defined}
\chi_\mathrm{b}(\omega) = \int_0^\infty dt \exp(i\omega t)\mathcal{X}_\mathrm{b}(t)
\end{align}
and  taking into account that $\chi_{\rm b}{}(\omega)=\chi_{\rm b}^*{}(-\omega)$ \cite{Landau1980}, we obtain for the reaction force
\begin{align}
\label{eq:coupling_expansion}
F_\mathrm{b}^{\rm (r){}}(t) \approx \chi_{\rm b}^*{}(\omega_0) u(t)e^{i\omega_0t} + {\rm c.c.}
\end{align}
This force is determined by the instantaneous value of $u(t)$ rather than the evolution of $u(t')$ for $t'\leq t$.

Substituting the force (\ref{eq:coupling_expansion}) into the full equation of motion for the mode coordinate $q(t)$ and using Eq.~(\ref{eq:u_substitution_Appendix}),  we obtain the equation of motion (\ref{eq:eom_classical}) for $u(t)$, which we reproduce here for completeness:
\begin{align}
\label{eq:Gamma_defined}
&\dot u = -(\Gamma - iP_{})u + \xi(t),  \quad \Gamma ={\rm Im}\,\chi_{\rm b}{}(\omega_0)/ 2M{}\omega_0,\nonumber\\
&P_{}=-{\rm Re}\,\chi_{\rm b}{}(\omega_0)/2M{}\omega_0
\end{align}
[the term $\xi(t)$ describes the noise; it does not come from $F_\mathrm{b}^{(\mathrm{r})}$ and is  discussed below]. In Eq.~(\ref{eq:Gamma_defined}) we assumed that $\Gamma\ll \omega_0$ and, in the spirit of the averaging method, disregarded the fast-oscillating term $\propto \Gamma u^*\exp(-2i\omega_0t)$ compared with $\Gamma u$.

In deriving Eqs.~(\ref{eq:coupling_expansion}) and (\ref{eq:Gamma_defined}) we further assumed that the bath susceptibility $\chi_{\rm b}{}(\omega) $ weakly varies with $\omega$ in a band of width $\sim \Gamma, |P|\ll \omega_0$ centered at $\omega_0$. It is this assumption that justifies disregarding the term $t'\dot u(t)$ and higher-order derivatives of $u(t)$ in the expansion of $u(t-t')$ in Eq.~(\ref{eq:backaction_Appendix}). In particular, the term $t'\dot u$ gives a correction
\[\sim \Bigl|(d\chi_{\rm b}/d\omega)_{\omega = \omega_0} \dot u\Bigr|\sim \Gamma\Bigl|u (d\chi_{\rm b}/d\omega)_{\omega = \omega_0}\Bigr|,\]
which is assumed small compared with the term $\sim |\chi_{\rm b}(\omega_0) u|$ kept in Eq.~(\ref{eq:Gamma_defined}). The assumption holds for $|d\log \chi_{\rm b}/d\omega|\ll 1$. It is  assumed that the higher derivatives of $\chi_{\rm b}$ are small  near $\omega_0$ as well. The typical frequency on which $\chi_{\rm b}(\omega)$ changes provides the other reciprocal ``fast'' time of the mode+bath system, in addition to $\omega_0^{-1}$. It is sometimes called the correlation time of the thermal reservoir $t_\mathrm{corr}$; note, however, that $\chi_{\rm b}(\omega)$ characterizes not just the reservoir, but also the coupling of the oscillator to the reservoir.

The approximation (\ref{eq:Gamma_defined}) is a Markovian approximation in slow time. It  holds if the response of the bath to the oscillator remains essentially unchanged where the oscillator frequency is changed not only by $ \Gamma$, but also by the polaronic frequency shift $P$. In the analysis of the nonlinear oscillator we will further assume that the response of the bath does not change due to the change of the oscillator frequency caused by the dependence of this frequency on the vibration amplitude.

The Bogoliubov-Krylov method of averaging can be applied also to the analysis of the effect of the thermal noise on the slow variables. The noise in Eq.~(\ref{eq:Gamma_defined})  is
\[\xi(t)= (-i/2M\omega_0)h_\mathrm{b}^{(0)}(t)\exp(-i\omega_0t),\]
where $h_\mathrm{b}^{(0)}(t)$ is the force on the oscillator from the bath calculated by disregarding the reaction of the bath to the oscillator. The noise correlator is simply expressed in terms of the bath power spectrum
\begin{align}
\label{eq:bath_FDT_again}
S_\mathrm{b}(\omega)=\int_{-\infty}^\infty dt e^{i\omega t}\langle h_\mathrm{b}^{(0)}(t) h_\mathrm{b}^{(0)}(0)\rangle.
\end{align}
Clearly,   $\langle \xi^*(t)\xi(t')\rangle \propto \int d\omega S_{\rm b}{}(\omega)\exp[-i(\omega-\omega_0)(t-t')]$.
We note that $S_\mathrm{b}(\omega)$ also defines the bath susceptibility, and thus the mode decay rate, via the fluctuation-dissipation theorem (cf. Sec.~\ref{sec:FDT}),
\[\mathrm{Im}\chi_\mathrm{b}(\omega) = S_\mathrm{b}(\omega)/2\hbar[\bar n(\omega)+1].\]

If we  replace $S_{\rm b}{}(\omega)\to S_{\rm b}{}(\omega_0)$ and use the relation (\ref{eq:Gamma_defined}) between $\Gamma$ and $\chi_\mathrm{b}(\omega_0)$, we obtain
\[
\langle \xi^*(t)\xi(t')\rangle =(\Gamma k_BT/M\omega_0^2) \delta(t-t').\]
As indicated in Sec.~\ref{subsubsec:slow_variables}, the $\delta$-function here is not a true $\delta$-function but, effectively,  a $\delta$-function on the time scale $\gg \omega_0^{-1}, t_\mathrm{corr}$. By expanding $S_{\rm b}(\omega)$ in a series about $\omega_0$, we find that $\langle \xi^*(t)\xi(t')\rangle$ has a peak at $t=t'$ with width $|t-t'|\lesssim |S_{\rm b}^{-1}d^2S_{\rm b}/d\omega^2|^{1/2}=t_\mathrm{corr}$, where the derivative of $S_{\rm b}$ is calculated for $\omega=\omega_0$ (the above expression may be considered a definition of $t_\mathrm{corr}$). The width $t_\mathrm{corr}$ is much smaller then $\Gamma^{-1}$ for $S_{\rm b}(\omega)$ smooth near $\omega_0$. The argument here coincides with the argument used in disregarding delay  in Eq.~(\ref{eq:coupling_expansion}).

The correlator $\langle \xi^*(t)\xi^*(t')\rangle$ has an extra factor $\exp[i\omega_0(t+t')]$ compared to the correlator $\langle \xi^*(t)\xi(t')\rangle$. This fast-oscillating factor averages out to zero on the time scale large compared to $\omega_0^{-1}$. Therefore in the analysis of the evolution of the slow variables $u(t), u^*(t)$ one can set $\langle \xi^*(t)\xi^*(t')\rangle=0$. The analysis of the higher-order correlators shows that the noise $\xi(t)$ is approximately Gaussian on the time scale large compared to $\omega_0^{-1}, t_\mathrm{corr}$. For a bath modeled  by a set of harmonic oscillators, with the coupling $h_\mathrm{b}$ nonlinear in the coordinates of these oscillators,  this was shown by \textcite{Dykman1971,Dykman1973a}. The analysis applies to the both classical and quantum cases. We note that a similar analysis has to be carried out to justify the expression for the reaction force in terms of the linear response of the bath to the mode.


\subsection{Ohmic dissipation}
\label{subsec:Ohmic_dissipation}

Here for completeness we describe the effect of coupling to the bath in a more restrictive but important case where the power spectrum $S_\mathrm{b}(\omega)$ is flat in a broad range from $\omega \ll \omega_0$ to $\omega\gg \omega_0$. The scale of the flatness is now $\omega_0$, not $\Gamma, |P|$. For a flat $S_\mathrm{b}(\omega)$, in the classical limit we can approximate the correlation function $s_\mathrm{b}(t)$ by a $\delta$-function,
\begin{align}
\label{eq:Ohmic_bath_class}
s_{\rm b}(t)\equiv  \langle h_\mathrm{b}^{(0)}(t) h_\mathrm{b}^{(0)}(0)\rangle =4M{}\Gamma k_BT\delta(t).
\end{align}
Equation~(\ref{eq:Ohmic_bath_class}) is essentially the definition of the parameter $\Gamma$ in terms of the correlator $s_\mathrm{b}(t)$, temperature, and the mode mass $M$ for the case considered in this section.

Taking into account that, for a classical bath, ${\cal X}_{\rm b}{} (t) = -(k_BT)^{-1}ds_{\rm b}{}/dt$, one obtains from Eqs.~(\ref{eq:backaction_Appendix})  the reaction force in the form
\begin{align}
\label{eq:classical_broadband}
F_\mathrm{b}^{\rm (r){}}(t) = \frac{s_{\rm b}(0)}{k_BT}q(t) - \int_0^\infty dt'\frac{s_{\rm b}(t')}{k_BT}\frac{dq(t-t')}{dt}.
\end{align}
For the $\delta$-correlated noise (\ref{eq:Ohmic_bath_class}), the last term in Eq.~(\ref{eq:classical_broadband}) gives the friction force $-2M{}\Gamma \dot q$ in the equation of motion (\ref{eq:Brownian}). Thus the reaction of the bath leads to viscous friction, with the friction force proportional to the mode velocity. With the account taken of Eq.~(\ref{eq:Ohmic_bath_class}), the overall dynamics of the mode is mapped on Brownian motion.

The first term in the right-hand side of  Eq.~(\ref{eq:classical_broadband}) renormalizes the mode frequency, $\omega_0^2 \to \omega_0^2 - s_{\rm b}(0)/M{}k_BT$. This is a classical polaronic effect. In calculating $s_{\rm b}(0)$ one should keep in mind that the power spectrum $S_{\rm b}(\omega)$ falls off for high frequencies, which makes $s_{\rm b}(0)$ finite.

To the best of our knowledge, for a classical oscillator
the frequency shift was first found by \textcite{Bogolyubov1945} for the model where the bath is a set of harmonic oscillators and $h_{\rm b}$ is linear in the coordinates $q_k$ of these oscillators. The corresponding Hamiltonian of the bath $H_\mathrm{b}$ reads
\begin{align}
\label{eq:CL_Hamiltonian}
H_b=\frac{1}{2}\sum_k(p_k^2 + \omega_k^2 q_k^2)
\end{align}
whereas the coupling Hamiltonian is $qh_\mathrm{b}=\sum_k\epsilon_k q q_k$. The coupling was weak and the dynamics was Markovian only in the rotating frame.

In the quantum theory, the constraint on the coupling parameters in the model (\ref{eq:CL_Hamiltonian}) that leads to a viscous friction force in the laboratory frame was found by \textcite{Caldeira1981}. The expression for the friction coefficient $\Gamma$  comes out if one assumes that the density of states of the bath weighted with the interaction has the form
\[\sum_k(\epsilon_k^2/\omega_k)\delta(\omega - \omega_k)= (4/\pi)M\Gamma  \omega.\]


\section{Oscillator decay rate in the Born approximation and the quantum kinetic equation}
\label{subsubsec:Born}

Equation (\ref{eq:Gamma_defined}) for the oscillator decay rate $\Gamma$   can be easily obtained also from a slightly different point of view. We first recall that the coordinate and momentum of the oscillator are expressed in terms of the ladder operators $a$ and $a^\dagger$ as $q=(\hbar/2M\omega_0)^{1/2}(a+ a^\dagger)$ and $p= -i (\hbar M\omega_0/2)^{1/2}(a - a^\dagger)$. In the analysis of the quantum dynamics it is convenient to use  the eigenfunctions $|k\rangle$ of the occupation number operator $a^\dagger a$; the energy of the isolated harmonic oscillator in a state $|k\rangle$ is $\hbar\omega_0(k+1/2)$.

\begin{figure}[h]
\includegraphics[width=6.25cm]{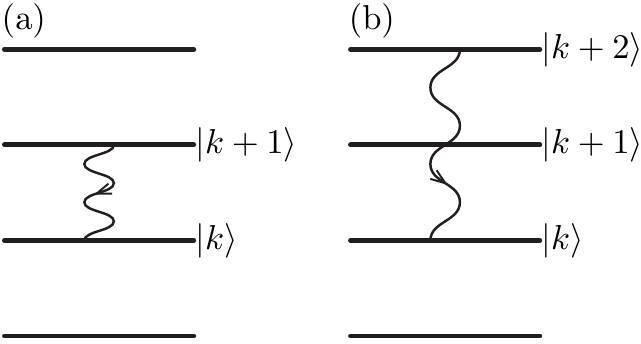}
\caption{Transitions between the oscillator energy levels with emission of excitations into a thermal bath. (a) Transitions between nearest levels lead, in the classical description, to a linear friction force proportional to the oscillator velocity. (b) Transitions between next-nearest levels lead, in the classical description,  to a nonlinear friction force proportional to the oscillator velocity multiplied by the squared coordinate (in the van der Pol model) or by the squared velocity (in the Rayleigh model). }
\label{fig:interlevel_decay}
\end{figure}

The coupling $q h_\mathrm{b}$ of the quantum oscillator to a thermal bath leads to transitions between the nearest oscillator energy levels in Fig.~\ref{fig:interlevel_decay}(a). The matrix elements of the coordinate $q$ are $\langle k|q|k-n\rangle = [\hbar k/2M{}\omega_0]^{1/2}\delta_{n,1}$ for $n\geq 0$. Therefore, to the leading order, the linear in $q$ coupling to the bath leads to transitions only between the neighboring levels. From the Fermi golden rule, the rate $W_{k+1\to k}$ of the transition $|k+1\rangle \to |k\rangle$ averaged over the states of the thermal bath is
\begin{align}
\label{eq:golden_rule}
&W_{k+1\to k} = (2\pi/\hbar)\,[\hbar(k+1)/2M{}\omega_0]\nonumber\\
&\times \bigl\langle\sum_{\nu_\mathrm{b}}|\langle \nu_\mathrm{b}|h_{\rm b}|\mu_\mathrm{b}\rangle|^2\delta (E_{\mu_\mathrm{b}}-E_{\nu_\mathrm{b}}+\hbar\omega_0)\bigr\rangle_{\mu_\mathrm{b}}
\end{align}
where $\mu_\mathrm{b}$ and $\nu_\mathrm{b}$ enumerate the bath states, $E_{\mu_\mathrm{b}}$ and $ E_{\nu_\mathrm{b}}$ are the energies of these states, and $\langle \cdot\rangle_{\mu_\mathrm{b}}$ indicates thermal averaging over the states $\mu_\mathrm{b}$, i.e., summation with the weight $\propto \exp(-E_{\mu_\mathrm{b}}/k_BT)$.

We now relate Eq.~(\ref{eq:golden_rule}) to the power spectrum $S_\mathrm{b}(\omega)$ of the operator $h_\mathrm{b}$, which determines the coupling of the bath to the oscillator,
\begin{align}
\label{eq:bath_spectrum_again}
&S_\mathrm{b}(\omega) = \int_{-\infty}^{\infty} dt e^{i\omega t} \langle h_\mathrm{b}(t) h_\mathrm{b}(0)\rangle = 2\pi\hbar Z_\mathrm{b}^{-1}\nonumber\\
&\times\sum_{\nu_\mathrm{b}}|\langle \nu_\mathrm{b}|h_{\rm b}|\mu_\mathrm{b}\rangle|^2\delta (E_{\mu_\mathrm{b}}-E_{\nu_\mathrm{b}}+\hbar\omega)e^{-E_{\mu_\mathrm{b}}/k_BT}
\end{align}
where $Z_\mathrm{b}$ is the bath partition function; in Eq.~(\ref{eq:bath_FDT_again}) and in the main text we used $h_\mathrm{b}^{(0)}$ instead of $h_\mathrm{b}$  in the definition of $S_\mathrm{b}(\omega)$ to emphasize that we are calculating the power spectrum in the absence of coupling to the oscillator.

Combining Eqs.~(\ref{eq:Gamma_defined}) and (\ref{eq:bath_spectrum_again}), we obtain
\begin{align}
\label{eq:Gamma_quantum_Appendix}
&W_{k+1\to k} = 2\Gamma(k+1)(\bar n+1),\nonumber\\
&\Gamma = S_\mathrm{b}(\omega_0)/4\hbar M\omega_0(\bar n+1)
\end{align}
[we recall that $\bar n \equiv \bar n(\omega_0)$]. This shows that $\Gamma$ determines the rate of the transitions between the states of a quantum oscillator due to its coupling to a thermal bath. In particular, $\Gamma = [W_{1\to 0}/2]_{\bar n=0}$. The transition rates (\ref{eq:golden_rule}) linearly increase with the level number.

\subsection{Master equation}
\label{subsec:master_equation}

In slow time compared to $\omega_0^{-1}, t_\mathrm{corr}$, quantum dynamics of the oscillator coupled to a bath can be described by a master equation for the oscillator density matrix in the rotating frame $\rho= U_0^\dagger (t)\rho_0 U_0(t)$, where $\rho_0$ is the density matrix in the laboratory frame and $U_0(t)=\exp(-i a^\dagger a\omega_0  t)$. For a linear oscillator with the coupling to the bath of the form of $H_i=qh_\mathrm{b}$, this equation was derived from the microscopic theory and carefully discussed by \textcite{Schwinger1961}; see also \cite{Senitzky1961}. An extension to weakly nonlinear oscillators was done by \textcite{Dykman1973a}. Where there hold the conditions of the Markovian approximation discussed in Sec.~\ref{subsec:averagin_cplng}, the master equation is Markovian in slow time and can be written in the form of a Lindblad equation,
\begin{align}
\label{eq:master_Lindblad}
\dot\rho = 2\Gamma \left[(\bar n+1)\mathcal{D}[a]\rho + \bar n\mathcal{D}[a^\dagger]\rho\right]\nonumber\\
 - iP[a^\dagger a,\rho] -i\hbar^{-1}[H_\mathrm{nl},\rho]
\end{align}
where
\begin{align}
\label{eq:Lindblad_operators}
\mathcal{D}[L]\rho = L\rho\L^\dagger -(L^\dagger L\rho + \rho L^\dagger L )/2
\end{align}
and $H_\mathrm{nl}$ describes nonlinear terms of the oscillator Hamiltonian; for example for the Duffing nonlinearity of the oscillator potential energy, which in the coordinate representation is described by $M\gamma q^4/4$ [cf. Eq.~(\ref{eq:Duffing_potential})], we have
\[H_\mathrm{nl} = 3\hbar^2 \gamma a^\dagger a (a^\dagger a +1 )/8M\omega_0^2.\]
The term $\propto P$ describes the polaronic effect of the shift of the oscillator frequency due to the coupling to a thermal bath.

Equation (\ref{eq:master_Lindblad}) corresponds to a linear friction force $-2M\Gamma\dot q$ in the phenomenological theory, as discussed in Sec.~\ref{sec:averaging}, and in the microscopic theory comes from the bath-induced transitions between neighboring energy levels of the oscillator. In contrast, the phenomenological nonlinear friction force
\[f_{\rm VdP} = -4M{}\Gamma\nl (q/q_0)^2\dot q, \quad q_0=(\hbar/2M{}\omega_0)^{1/2},\]
 corresponds, in the microscopic theory, to the bath-induced transitions over two energy levels of the oscillator, see Fig.~\ref{fig:interlevel_decay}(b). It comes from the interaction with the bath with energy $q^2h_\mathrm{b}^\mathrm{(nl)}$, see Eq.~(\ref{eq:H_nonlin_friction}), and specifically from the terms $a^2, a^\dagger{}^2$ in $q^2$. In the master equation the nonlinear friction is described by the term \cite{Dykman1975a}
\begin{align}
\label{eq:nonlin_fric_appendix}
(\dot\rho)_\mathrm{nl} =& 2\Gamma^\mathrm{(nl)} \left\{[\bar n(2\omega_0) +1]\mathcal{D}[a^2]\rho \right.\nonumber\\
&\left.+\bar n(2\omega_0)\mathcal{D}[a^\dagger{}^2]\rho\right\}.
\end{align}
The nonlinear friction coefficient $\Gamma^\mathrm{(nl)}$  is [cf. Eq.~(\ref{eq:coeff_nonlin_fric})]
\begin{align}
\label{eq:nonlin_fric_Appendix}
&\Gamma\nl =\frac{q_0^4}{2\hbar^2[\bar n(2\omega_0)+1]} S_\mathrm{b}^\mathrm{(nl)}(2\omega_0), \quad
\nonumber\\
&S_{\rm b}\nl(\omega)=\int_{-\infty}^\infty dt e^{i\omega t} \langle h_{\rm b}^{\rm (nl)}(t)h_{\rm b}^{\rm (nl)}(0)\rangle.
\end{align}

In the classical limit $k_BT\gg \hbar\omega_0$ Eq.~(\ref{eq:master_Lindblad}) goes into the Fokker-Planck equation for the probability distribution of a nonlinear oscillator, which corresponds to the stochastic classical equation of motion (\ref{eq:eom_nonlin}). Among other things, Eqs.~(\ref{eq:master_Lindblad}) and (\ref{eq:nonlin_fric_appendix}) allow one to calculate the power spectrum of a nonlinear oscillator.   The results on the spectra are discussed in the main text and in Appendix~\ref{sec:Duffing_spectra_Appendix}.

The master equation is easily extended to describe resonant and parametric driving. It allows studying the stationary probability distribution of a driven nonlinear oscillator as well as various transient quantum phenomena.


\section{Driving-induced cooling and heating for coupled modes}
\label{sec:heating}

The possibility to cool a mechanical mode by a driving that resonantly couples it to a high-frequency optical mode in a cavity plays a fundamental role in optomechanics \cite{Aspelmeyer2014a}. However, a mechanical mode can also be cooled down or heated up by coupling it to another mechanical mode or a mode of a different physical nature or just a thermal reservoir. An interesting and not a priori obvious part of the effect is that the stationary distribution of the driven mode is of the Boltzmann form with an effective temperature. This happens if there is no nonlinear friction and the relaxation rate of the considered mode is much smaller than the relaxation rate of the mode it is coupled to \cite{Dykman1978}.

Formally,  we consider two modes with eigenfrequencies $\omega_1$  and $\omega_2$ and the pumping (modulation) of the form of Eq.~(\ref{eq:parametric_mode_mode}), with energy
\[U_{12}^\mathrm{pump} = M q_1q_2 \bar\Delta_{12}\cos\omega_pt\]
[this corresponds to setting in Eq.~(\ref{eq:parametric_mode_mode})  $\Delta_{12}^\mathrm{pump}(t)  =\bar\Delta_{12}\cos\omega_pt$]. For brevity we set the effective masses of the modes to be equal to the same value $M$. The modulation frequency $\omega_p$ is either close to $|\omega_1-\omega_2|$ or $\omega_1+\omega_2$. In what follows we assume for concreteness that $\omega_1 < \omega_2$. We introduce $\epsilon_p=\pm 1$ such that $\omega_p$ is close to $\omega_2 -\epsilon_p\omega_1$, i.e., either to $\omega_2 - \omega_1$ or to $\omega_2 + \omega_1$.

In writing down the master equation for the modes we will assume that each mode is coupled to its thermal reservoir and, in the absence of the driving, their decay rates are $\Gamma_1$ and $\Gamma_2$. We will further assume that the decay rates do not change if the mode eigenfrequencies are slightly changed. For example, if $\omega_2$ is changed to $\omega_p+\epsilon_p\omega_1$; for the considered resonant pumping  $|\omega_p+\epsilon_p\omega_1-\omega_2|\ll \omega_{1,2}$.
We will write the master equation using the ladder operators $a_1, a_1^\dagger$ and $a_2, a_2^\dagger$ for the modes 1 and 2, similar to how it was done in Section~\ref{subsubsec:Born} for a single mode. We will also switch to the rotating frame and use the rotating wave approximation. A unitary transformation to the rotating frame is
\[U(t) = \exp[-i\omega_1 ta_1^\dagger a_1 -i(\omega_p + \epsilon_p\omega_1)t a_2^\dagger a_2].\]

In the rotating wave approximation, the master equation for the density matrix $\rho$ of the coupled modes reads
\begin{align}
\label{eq:QKE_parametric_coupling}
&\dot \rho = \sum_j \hat\Gamma_j\rho  + i\,\delta\omega_p[a_2^\dagger a_2,\rho]-i[\hat h_{12},\rho],\nonumber\\
&\hat\Gamma_j\rho = 2\Gamma_j \left[(\bar n_j+1)\mathcal{D}[a_j]\rho + \bar n_j\mathcal{D}[a_j^\dagger]\rho\right]\nonumber\\
&\delta\omega_p = \omega_p +\epsilon_p\omega_1 -\omega_2
\end{align}
where $\bar n_k=\bar n(\omega_k)$ ($k=1,2$). The Lindblad superoperators $\mathcal{D}[L]$ are defined in  Eq.~(\ref{eq:Lindblad_operators}).  Compared to Eq.~(\ref{eq:master_Lindblad}), in Eq.~(\ref{eq:QKE_parametric_coupling}) we have disregarded the nonlinearity of the modes and their eigenfrequency shifts due to the coupling to the thermal reservoirs.

The operator $\hat h_{12}$ describes the resonantly induced mode coupling,
\begin{align}
\label{eq:pumped_coupling_operator}
&\hat h_{12}= \Delta_{12}(a_1^\dagger a_2 + a_2^\dagger a_1), \quad \omega_p\approx \omega_2 - \omega_1\nonumber\\
&\hat h_{12}= \Delta_{12}(a_1 a_2 + a_1^\dagger a_2^\dagger), \quad \omega_p\approx \omega_2 + \omega_1\
\end{align}
where $\Delta_{12} = \bar\Delta_{12}/4\sqrt{\omega_1\omega_2}$. This parameter is of central importance, as it characterizes the coupling strength.

The physical picture of the mode dynamics is simplified in the case where the relaxation rates $\Gamma_1$ and $\Gamma_2$ are very different. For concreteness, we will assume that
\[\Gamma_2 \gg \Gamma_1.\]
In this case mode 2 adiabatically follows mode 1. If the coupling is sufficiently weak, $|\Delta_{12}| \ll \Gamma_2$, one can think of the linear response of mode 2 to the state of mode 1. This response is formed over time $\sim 1/\Gamma_2$ whereas the state of mode 1 varies over a significantly longer time.

In the adiabatic approximation the dynamics of mode 1 can be described by tracing out  mode 2. We introduce the density matrix of mode 1, $\rho_1 = \mathrm{Tr}_2 \rho$, where $\mathrm{Tr}_2$ denotes the trace over the states of mode 2. Similarly, $\langle a_2\rangle_2 = \mathrm{Tr}_2(a_2\rho)$ and  $\langle a_2^\dagger\rangle_2 = \mathrm{Tr}_2(a_2^\dagger\rho)$; we emphasize that these averages over the states of mode 2 are operators with respect to mode 1.

By taking  trace over mode 2 in Eq.~(\ref{eq:QKE_parametric_coupling}) we obtain
\begin{align}
\label{eq:QKE_mode1_coupled}
\dot\rho_1 = \hat\Gamma_1\rho_1 -i\Delta_{12}\bigl([a_1^\dagger, \langle a_2\rangle_2] +  [a_1, \langle a_2^\dagger\rangle_2]\bigr).
\end{align}
The equation for $\langle a_2\rangle_2$ has the form
\begin{align}
\label{eq:mean_a_2}
&\frac{d}{dt}\langle a_2\rangle_2 = \hat\Gamma_1\langle a_2\rangle_2 - (\Gamma_2 -i\delta\omega_p)\langle a_2\rangle_2\nonumber\\
& - i\mathrm{Tr}_2 (a_2[\hat h_{12},\rho]).
\end{align}
In the considered regime of fast relaxation rate $\Gamma_2$ we can look for the quasistationary solution of this equation. Respectively, we will disregard $d\langle a_2\rangle_2/dt$. We will also disregard the term $\propto \Gamma_1$ compared to the term $\propto \Gamma_2$. To describe the linear response of mode 2 to the coupling, we will calculate the last term in Eq.~(\ref{eq:mean_a_2}) to the lowest order in the coupling, i.e., we will set $\langle a_2^\dagger a_2\rangle_2\approx \bar n_2\rho_1$, whereas the term $\langle a_2^2\rangle_2$ will be disregarded. This gives
\[\langle a_2\rangle_2\approx -i\frac{\Delta_{12}}{\Gamma_2 - i\delta\omega_p}[(\bar n_2+1)a_1\rho_1 - \bar n_2 \rho_1 a_1].\]
Substituting this expression and the similar expression for $\langle a_2^\dagger\rangle_2$ into Eq.~(\ref{eq:QKE_mode1_coupled}), we obtain
\begin{align}
\label{eq:heating_cooling}
&\dot\rho_1 = -\Gamma_\mathrm{eff}\left[(\bar n_\mathrm{eff}+1)\mathcal{D}[a_1]\rho_1 + \bar n_\mathrm{eff}\mathcal{D}[a_1^\dagger]\rho_1\right]\nonumber\\
&-iP_\mathrm{eff}[a_1^\dagger a_1,\rho_1],
\end{align}
where
\begin{align}
\label{eq:effective_Gamma}
&\Gamma_\mathrm{eff} = \Gamma_1 + \epsilon_p\Gamma_2\frac{ \Delta_{12}^2}{\Gamma_2^2 + \delta\omega_p^2}, \nonumber\\
&\bar n_\mathrm{eff}=\Gamma_\mathrm{eff}^{-1}\left[\Gamma_1 \bar n_1 + \Gamma_2\frac{ \Delta_{12}^2}{\Gamma_2^2 + \delta\omega_p^2}\left(\bar n_2 -\frac{\epsilon_p-1}{2}\right)\right]
\end{align}
and $P_\mathrm{eff}= \delta\omega_p\,\Delta_{12}^2/ (\Gamma_2^2 + \delta\omega_p^2)$.

Remarkably, Eq.~(\ref{eq:heating_cooling}) maps the dynamics of the driven slowly decaying mode (mode 1) onto the dynamics of an undriven mode with an effective decay rate $\Gamma_\mathrm{eff}$ and an effective mean occupation number $\bar n_\mathrm{eff} $ \cite{Dykman1978}. As seen from Eq.~(\ref{eq:heating_cooling}), $\Gamma_\mathrm{eff}$ exceeds $\Gamma_1$ for $\omega_2\approx \omega_p+\omega_1$ and is smaller than $\Gamma_1$  for $\omega_2+\omega_1\approx \omega_p$. At the same time, the effective occupation number $\bar n_\mathrm{eff}$ is smaller or larger than $\bar n_1$. Respectively, in the stationary regime the effective temperature of the mode $k_BT_\mathrm{eff}= \hbar\omega_1/\ln[(\bar n_\mathrm{eff}+1)/\bar n_\mathrm{eff}]$
is lower or higher than the temperature of the thermal reservoir. This describes the sideband cooling for $\omega_2\approx \omega_p+\omega_1$ by ``superposing'' onto a low-frequency mode the distribution over the states of a higher-frequency mode. On the other hand, a sufficiently strong drive with $\omega_p\approx \omega_2+\omega_1$ leads to $\Gamma_\mathrm{eff}$ becoming equal to zero, which manifests an instability of the system.



\section{Forced vibrations}
\label{sec:Appendix_forced_vibrations}
\subsection{Resonant driving}
\label{subsec:resonant_forced}

A classical Duffing oscillator driven by a resonant force $F\cos\omega_Ft$ experiences the same friction force and the same noise from the thermal bath as in the absence of driving, provided the detuning of the drive frequency from the eigenfrequency $|\omega_F-\omega_0|$ is small compared to the reciprocal correlation time of the bath $t_\mathrm{corr}^{-1}$. We assume that  the driving is not extremely strong so that the amplitudes of the vibration overtones remain small compared to the amplitude of the main tone. It is then convenient to describe the dynamics by switching to the rotating frame and using the real variables $Q, P$, which are related to the coordinate and momentum of the driven oscillator by the expression
\begin{align}
\label{eq:quadratures_resonant_drive}
 Q+iP = [q + i(p/M\omega_F)]\exp(i\omega_F t)
 \end{align}
[cf.  Eq.~(\ref{eq:quadratures})]. These variables are similar to (twice)  the real and imaginary parts of the complex amplitude in the absence of driving $u^*(t)$, Eq.~(\ref{eq:u_t_Appendix}), except that for a driven oscillator  it is more convenient to change to the frame oscillating at frequency $\omega_F$ rather than $\omega_0$.

The equations for $Q,P$ are derived similarly to the equation for $u(t)$.  Disregarding small corrections to $Q,P$ that oscillate at frequency $\omega_F$ and its overtones, we obtain
\begin{align}
\label{eq:eom_quadratures}
&\dot  Q = \partial_Pg_r(Q,P) -\Gamma Q + {}\xi_Q(t)\nonumber\\
&\dot P = -\partial_Qg_r(Q,P) -\Gamma P +{}\xi_P(t),
\end{align}
where
\begin{align}
\label{eq:g_r}
&g_r= \frac{3\gamma}{32\omega_F}(Q^2+P^2)^2 -\frac{1}{2}\delta\omega(Q^2+P^2) -\frac{F}{2M\omega_F}Q,\nonumber\\
&\delta\omega=\omega_F-\omega_0,   \qquad |\delta\omega|\ll \omega_F
\end{align}
(the subscript $r$ here stands for ``resonant''). The major difference from Eq.~(\ref{eq:Gamma_defined}) is that Eq.~(\ref{eq:eom_quadratures}) is written in real variables and includes the Duffing nonlinearity $\propto \gamma$ and the driving force $\propto F$; the bath-induced frequency shift has been incorporated into $\omega_0$.

The variables $Q$ and $P$ and the time can be rescaled, so that the equations of motion contain only two parameters, in the absence of noise \cite{Dykman2012a}. The noise components ${}\xi_Q(t)$ and ${}\xi_P(t)$ are independent $\delta$-correlated Gaussian noises with the same intensity as in the absence of the driving
\begin{align}
\label{eq:noise_rotate_frame}
\langle {}\xi_Q(t) {}\xi_Q(0)\rangle = \langle {}\xi_P(t){}\xi_P(0)\rangle = (2\Gamma k_BT/M\omega_0^2)\delta(t).
\end{align}

\begin{figure}
\includegraphics{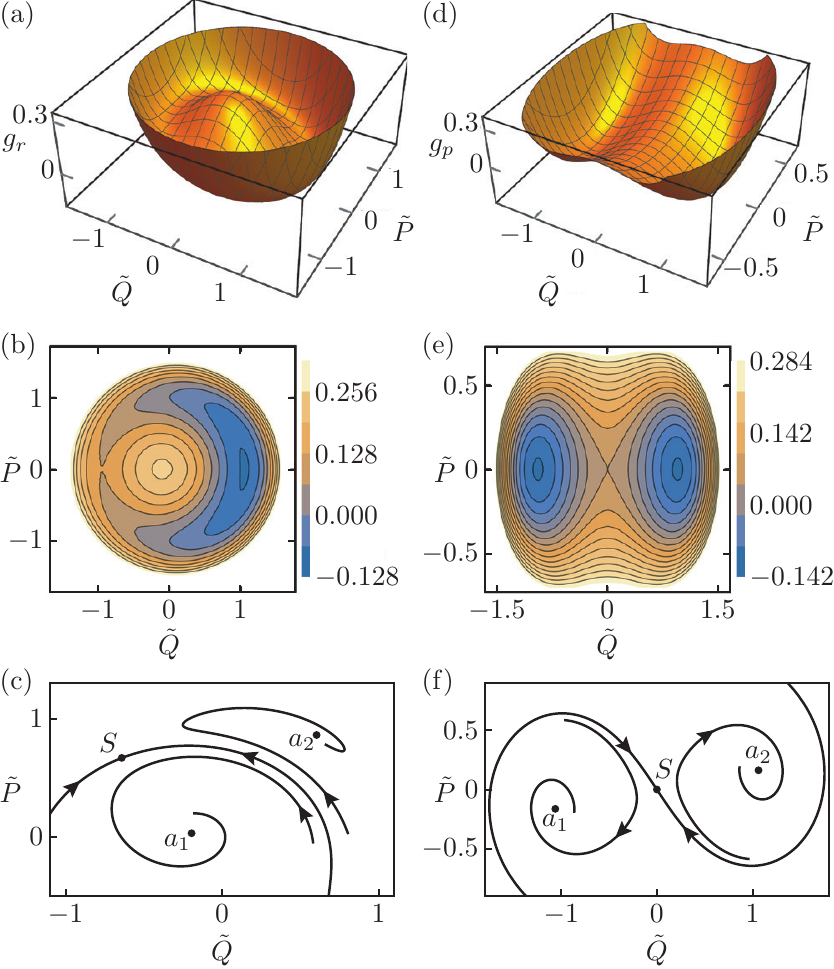}
\caption{The left and right columns refer to the resonantly driven and parametrically driven modes in the range of their bistability. The variables $\tilde Q$ and $\tilde P$ for the resonantly driven mode are, respectively,  the in-phase and the quadrature components $Q$ and $P$ multiplied by $[8\omega_F(\omega_F-\omega_0)/3\gamma]^{1/2}$. For the parametrically driven mode, $\tilde Q, \tilde P$ are the quadratures $Q$ and $P$ scaled by the factor $|2F_p/3\gamma|^{1.2}$. Panels  (a) and (b) show the effective Hamiltonian in the rotating frame for the resonantly driven mode $g_r$, Eq.~(\ref{eq:g_r}), and its cross-sections (the phase trajectories in the absence of decay)  for the scaled driving field intensity $\beta = 3\gamma F^2/32M^2\omega_F^3(\delta\omega)^3 = 0.01$. Panel (c) shows the phase portrait in the presence of dissipation for $\beta = 1/27$ and the scaled decay rate $\Gamma/|\delta\omega|=0.15$. The phase plane is separated into two parts by the separatrix, which  goes through the saddle points $S$. The trajectories on the opposite sides of the separatrices approach the stable states $a_1$ and $a_2$. As the decay rate goes to zero, the stable states $a_2$ and $a_1$ move toward the minimum and the maximum of $g_r$, respectively. Panels (d) and (e)   show, respectively, the effective Hamiltonian in the rotating frame $g_p$, Eq.~(\ref{eq:g_p}), and its cross-sections (phase trajectories in the absence of dissipation) for the scaled frequency detuning $\mu_p= -0.1$  defined in Eq.~(\ref{eq:parameter_mu}). Panel (f) shows the phase portrait in the presence of dissipation for $\mu_p=0.2, 2M\omega_p\Gamma/F_p=0.3$.  As the decay rate goes to zero, the stable states $a_1$ and $a_2$ move toward the minima of $g_p$.
}
\label{fig:phase_portraits}
\end{figure}

If the decay and the noise are disregarded, Eqs.~(\ref{eq:eom_quadratures}) become Hamiltonian equations for the coordinate $Q$ and momentum $P$ of the oscillator in the rotating frame (the in-phase and quadrature components). The function $g_r(Q,P)$ is the Floquet Hamiltonian. In the parameter range where the oscillator is bistable (in the presence of weak dissipation) it has the form of a tilted Mexican hat, see Fig.~\ref{fig:phase_portraits}(a). The cross-sections of the surface $g_r(Q,P)$ in Fig.~\ref{fig:phase_portraits}(b) show the oscillator trajectories (\ref{eq:eom_quadratures}) in the rotating frame in the limit of zero dissipation. Note the strong asymmetry (a horseshoe-like shape) of the trajectories that go around the minimum of $g_r(Q,P)$.

In the absence of noise, Eq.~(\ref{eq:eom_quadratures}) has 3 stationary solutions in the region inside the ``curvilinear triangle'' on the $(F,\omega_F)$-plane in Fig.~\ref{fig:kozinsky_hysteresis}(b). The solutions with the largest and smallest values of $Q^2+P^2$ are stable, and the solution with the intermediate $Q^2+ P^2$ is the saddle point. In the limit $\Gamma\to 0$ these states correspond, respectively, to the local minimum and maximum of the function $g_r(Q,P)$ and to its saddle point, see Fig.~\ref{fig:phase_portraits}(a). Near the saddle point $g_r(Q,P)$ has the shape of a hyperboloid. Through this point there goes the separatrix that separates the basins of attraction of the stable states for a finite damping. The phase portrait of the system in the range of bistability is shown in Fig.~\ref{fig:phase_portraits}(c), which complements  Fig.~\ref{fig:kozinsky_hysteresis}(c).

The boundaries of the range of the bistability, i.e., the sides of the ``curvilinear triangle'' in Fig.~\ref{fig:kozinsky_hysteresis}(b), are the bifurcation lines.  On the line with smaller $F$ the stable state with the larger amplitude merges with the saddle point, whereas on the  line with larger $F$ it is the smaller-amplitude state that merges with the saddle point.   The bifurcational values $F_{B1,2}\equiv F_{B1,2}(\Omega_r)$ of the drive amplitude as a function of the drive frequency are given by the expression
\begin{align}
\label{eq:bif_resonant}
&\tilde F_{B1,2}^2 = \frac{2}{27}\left[\Omega_r^3 +9\Omega_r \mp (\Omega_r^2-3)^{3/2}\right],\nonumber\\
&\tilde F= (3\gamma /32\omega_F^3\Gamma^3)^{1/2} F,\qquad \Omega_r = (\omega_F-\omega_0)/\Gamma.
\end{align}
Equation (\ref{eq:bif_resonant}) is written for the Duffing nonlinearity parameter $\gamma>0$; for $\gamma<0$ one should replace $\gamma\to |\gamma|$ in the expression for $\tilde F_{B1,2}$ and $\delta\omega\equiv \omega_F-\omega_0\to -\delta\omega$ in the expression for the scaled frequency detuning $\Omega_r$.

\subsubsection{Universality of fluctuations near a bifurcation point}
\label{subsubsec:bif_fluctuations}

The dynamics and fluctuations display universal features near bifurcation points. These features are characteristic, in particular, for merging of a stable state with a saddle point [called the saddle-node bifurcation \cite{Guckenheimer1997}]. The equations of motion (\ref{eq:eom_quadratures}) are simplified where the states are close to each other in phase space. The rates at which the dynamical variables $Q$ and $P$ approach their stable values become very different. The relaxation rate of the in-phase component $Q$ is $\approx 2\Gamma$, whereas the relaxation rate of the quadrature component $P$ goes to zero at the bifurcation parameter value, so that $P$ is a slow variable near the bifurcation point. This variable is an analog of a soft mode  in the phase transition theory. Its fluctuations can be analyzed in the adiabatic approximation, assuming that the component $Q(t)$ adiabatically follows $P(t)$. As a result,  the equation for the time evolution of $P(t)$ takes the form
\begin{align}
\label{eq:near_bifurcation}
&\dot P = -\partial_PU_B(P) + {}\xi_P(t), \\
 &U_B(P) = -\alpha_{B1} [F-F_B(\Omega_r)](P-P_B) + \alpha_{B2}(P-P_B)^3.\nonumber
\end{align}
The explicit form of the parameters $\alpha_{B1,2}$ and the value $P_B$ of the quadrature at the bifurcation point depend on $\Omega_r$ \cite{Dykman1980}.

For $\alpha_{B1}\alpha_{B2}(F-F_B)>0$ the potential $U_B$ has a minimum and a maximum, with the minimum corresponding to the stable state of the mode and the maximum corresponding to the saddle point. The relaxation rate near the stable state scales with the distance to the bifurcation point $F-F_B(\Omega_r)$ as $[12\alpha_{B1}\alpha_{B2}(F-F_B(\Omega_r)]^{1/2}$.

The ``softening'' of the potential $U_B(P)$ near the bifurcation point leads to an increase in fluctuations. Of particular interest is the effect of the fluctuation-induced switching  from the dynamically stable vibrational state to the coexisting state with a strongly different amplitude. Equation (\ref{eq:near_bifurcation}) reduces the problem of the switching rate to the problem of escape from a potential well for a static cubic-parabola potential $U_B(P)$ in a system with no inertia. The rate of escape  for this problem is well-known \cite{Kramers1940},
\[W_\mathrm{sw}\propto \exp(-M\omega_0^2\Delta U_B/2\Gamma k_BT),\]
where $\Delta U$ is the height of the potential barrier around the stable state,
\begin{align}
\label{eq:bif_barrier_height}
\Delta U_B = 4|\alpha_{B2}|\,[\alpha_{B1}(F-F_B)/3\alpha_{B2}]^{3/2}.
\end{align}
This barrier height scales as the distance to the bifurcation point to the power $3/2$. Such scaling has been seen in micro- and nanomechanical vibrational systems as well as in Josephson junction based systems \cite{Chan2007,Siddiqi2006a,Vijay2009,Defoort2015}. We note that, since $F_B\equiv F_B(\Omega_r)$, the bifurcation point can be approached by varying the drive amplitude or the drive frequency or both.

The values of the driving amplitude and frequency where $F_{B1} = F_{B2}$ give the critical point in Fig.~\ref{fig:kozinsky_hysteresis}. Near this point the variables $Q$ and $P$ also separate into comparatively fast and slow, with the equation for the slow variable $P$ having the form (\ref{eq:near_bifurcation}), except that the effective potential $U_B(P)$ has to be replaced with the potential $U_c(P)$, with
\begin{align}
\label{eq:U_c}
&U_c(P) = \alpha_{c1}(P-P_c)^4 -\alpha_{c2}[\omega_F - (\omega_F)_c](P-P_c)^2 \nonumber\\
&+
 \{\alpha'_{c3}[\omega_F-(\omega_F)_c] + \alpha''_{c3}(F-F_c)\}(P-P_c)
\end{align}
where $P_c, (\omega_F)_c$, and $F_c$ are the values of $P$, $\omega_F$, and $F$ at the critical point. These values as well as the parameters $\alpha_{c1,c2,c3}$ are easy to find from Eqs.~(\ref{eq:eom_quadratures}) and (\ref{eq:g_r}) \cite{Dykman1980}.

The critical point on the $(F,\omega_F)$-plane is reminiscent of the critical point on the line of the first-order phase transition. Fluctuations become strong and their correlation time diverges as the oscillator  approaches this point.   In the range of bistability, as determined by the interrelation between $F-F_c$ and $\omega_F - (\omega_F)_c$, the potential $U_c(P)$ has two minima. It becomes symmetric on the line on the $(F,\omega_F)$ plane where the coefficient at the linear in $P-P_c$ term is zero. On this line the switching rates between the two minima are equal to each other and the barrier height between the minima is  $\alpha_{c2}^2[\omega_F-(\omega_F)_c]^2/4\alpha_{c1}$. This barrier height as determined from the switching rate allows finding the eigenfrequency and the nonlinearity parameter of a nanomechanical mode with extremely high precision \cite{Aldridge2005}.



\subsection{Parametrically excited vibrations}
\label{subsec:parametric}

Parametric modulation of an oscillator can be described by incorporating into its Hamiltonian the term $-(F_pq^2/2) \cos\omega_p t$. The phenomenological equation of motion then takes the form (\ref{eq:driven_parametric}).  For the modulation at frequency $\omega_p$ close to $2\omega_0$, it is convenient to analyze the dynamics by switching to  the  quadratures $Q$ and $P$ that remain almost constant on the time scale $1/\omega_p$. The transformation is similar to that for the resonant drive, Eq.~(\ref{eq:quadratures_resonant_drive}),
\begin{align}
\label{eq:parametric_slow}
 Q+iP = \left(q + i\frac{p}{M\omega_p/2}\right)\exp(i\omega_p t/2)
 \end{align}
[as in Eq.~(\ref{eq:quadratures_resonant_drive}), $Q$ and $P$ are real]. In the rotating wave approximation the equations for $Q$ and $P$ have the same form as Eq.~(\ref{eq:eom_quadratures}), but now the function $g_r$ has to be replaced with the function $g_p$,
\begin{align}
\label{eq:quadratures_parametric}
&\dot  Q = \partial_Pg_p(Q,P) -\Gamma Q + {}\xi_Q(t),\nonumber\\
&\dot P = -\partial_Qg_p(Q,P) -\Gamma P +{}\xi_P(t),
\end{align}
where
\begin{align}
\label{eq:g_p}
&g_p= \frac{3\gamma}{16\omega_p}(Q^2+P^2)^2 -\frac{1}{2}\delta\omega_p(Q^2+P^2) ,\nonumber\\
&+\frac{F_p}{4M\omega_p}(P^2-Q^2), \qquad \delta\omega_p=\frac{\omega_p}{2} -\omega_0.
\end{align}
This equation applies provided $|\delta\omega_p|\ll \omega_p$.

If the decay and the noise are disregarded, Eqs.~(\ref{eq:quadratures_parametric}) become Hamiltonian equations for the coordinate $Q$ and momentum $P$ in the rotating frame. The function $g_p(Q,P)$ is the Hamiltonian (we note that this is not a Floquet Hamiltonian; this is the Hamiltonian in the frame oscillating at frequency $\omega_p/2$). In the parameter range where the oscillator has two stable states (in the presence of weak dissipation) it has the form of a symmetric double-well surface, see Fig.~\ref{fig:phase_portraits}(d). The cross-sections of these surface shown in Fig.~\ref{fig:phase_portraits}(e) illustrate the phase trajectories in the rotating frame in the limit of zero dissipation.

The symmetry is a feature of the parametric resonance. Indeed, incrementing the time by half of the modulation period $2\pi/\omega_p$ does not change the equation of motion in the laboratory frame, Eq.~(\ref{eq:driven_parametric}). Yet, as seen from Eq.~(\ref{eq:parametric_slow}), it leads to the change $Q\to -Q, \,P\to -P$.

The phase portrait in the presence of dissipation is shown in Fig.~\ref{fig:phase_portraits}(f). This figure refers to the parameter range where only two vibrational states are stable. As expected from the above arguments, the phase portrait has inversion symmetry. Similar to a resonantly driven mode, the regions of attraction to the stable states $a_1,a_2$ are separated by the separatrix that goes through the saddle point $\mathcal{S}$.

The variables $Q,P$ and the time can be rescaled, so that, in the absence of noise, the dynamics is described by two dimensionless parameters, $\mu_p$ and $f_p$,
\begin{align}
\label{eq:parameter_mu}
\mu_p =(\delta\omega_p/\Gamma) \sgn\gamma, \qquad f_p= F_p/2M\Gamma\omega_p.
\end{align}
Figure~\ref{fig:parametric_bifurcation}(a) shows the regions of the $(f_p,\mu_p)$-plane where there exist different numbers  of vibrational and steady states in the absence of nonlinear friction. The bifurcation lines $\mu_{B1,2}$ (the bifurcational values of $\mu_p$ as functions of $f_p$) are given by the equation
\begin{align}
\label{eq:mu_bifurcational}
\mu_{B1,2} = \mp (f_p^2 - 1)^{1/2}.
\end{align}
For weak modulation or large $-\mu_p$ the mode is not excited, the vibration amplitude is zero. At $\mu_p=\mu_{B1}$ the zero-amplitude state becomes unstable, and in the range $\mu_{B2} > \mu_p > \mu_{B1}$ the system has two stable vibrational states (these are period-2 states with the opposite phases). For $\mu_p>\mu_{B2}$ and $f_p>1$ the zero-amplitude state is also stable, the mode has 3 stable states and also 2 unstable period-2 states. On the line $f_p=1, \mu_p>0$ the stable period-2 states merge with the unstable period-2 states and disappear.
At the critical point $\mu_p=0, f_p=1$ all five stationary states merge.

Near the bifurcation lines (\ref{eq:mu_bifurcational}) the dynamics and fluctuations of the nascent states are controlled by a ``slow'' dynamical variable, similar to the case of a resonantly driven mode. This variable is a linear combination of $Q$ and $P$. A theory of fluctuations and the scaling of the rates of interstate switching in this parameter range was discussed by \textcite{Dykman1998,Lin2015}.

The nonlinear friction significantly modifies the bifurcation diagram, as seen from Fig.~\ref{fig:parametric_bifurcation}~(b). The line on which the stable and unstable period-2 states merge is tilted and the critical point shifts. A profound consequence of this change is the hysteresis with the varying modulation frequency, as described in Sec.~\ref{subsec:parametric_main}.


\subsubsection{Fluctuation squeezing in the linear regime}

Parametric modulation is often used as a way to squeeze fluctuations of one of the quadratures. The squeezing does not require exciting period-2 vibrations, it occurs already for a weak modulation. This can be seen from Eqs.~(\ref{eq:quadratures_parametric}) and (\ref{eq:g_p}) if one sets $\gamma=0$. In the absence of modulation, $F_p=0$, one has from these equations $\langle Q^2\rangle = \langle P^2\rangle =  k_BT/M\omega_0^2$ in the case where the noise ${}\xi_Q(t), {}\xi_P(t)$ is thermal. The stationary probability distribution $\rho(Q,P)$ is Gaussian, $\rho(Q,P) = Z^{-1}\exp[-M\omega_0^2(Q^2+P^2)/2k_BT]$; it is a Boltzmann distribution of a harmonic oscillator, $Z= M\omega_0^2/2\pi k_BT$.

In the presence of the modulation but below the excitation threshold, $f_p^2<1+\mu_p^2$,  the stationary probability distribution in the rotating frame $\rho(Q,P)$ is still Gaussian, if one disregards the nonlinearity, $\rho(Q,P) = Z^{-1}\exp(-\sum A_{ij}x_ix_j/2)$. Here $i,j=1,2$ and we use $x_1\equiv Q, x_2\equiv P$; the normalization factor is  $Z=2\pi/(\det \hat A)^{1/2}$. The matrix $\hat A$ can be easily found from the Fokker-Planck equation that corresponds to Eq.~(\ref{eq:quadratures_parametric}) [in terms of the theory of stochastic processes, the latter is the Langevin equation for the fluctuating variables $Q(t), P(t)$, see \cite{Risken1996}]. We can make a unitary transformation from $Q,P$ to $Q',P'$  so as to diagonalize the matrix $\hat A$. The variances $\langle Q'{}^2\rangle$ and  $\langle P'{}^2\rangle$ are given, respectively, by $A_+^{-1}$ and $ A_-^{-1}$, where $A_+^{-1}$ and $A_-^{-1}$ are the largest and the smallest eigenvalues of $\hat A^{-1}$,
\begin{align}
\label{eq:squeezed_param}
A_{\pm}^{-1} = \frac{ k_BT}{M\omega_0^2} \frac{1+\mu_p^2\pm |f_p|(1+\mu_p^2)^{1/2}}
{1+\mu_p^2 -f_p^2}.
\end{align}

One can easily see that $A_-^{-1} < k_BT/M\omega_0^2$ for $f_p^2< 1+\mu_p^2$, which shows that the variance $\langle P'{}^2\rangle = A_-^{-1}$ is smaller than the variance of the quadratures in the absence of driving. This demonstrates
squeezing of classical fluctuations. The squeezing becomes more pronounced as the scaled modulation amplitude $|f_p|$ approaches the critical value $(1+\mu_p^2)^{1/2}$ where period-two vibrations are excited. Close to the critical $|f_p|$, the eigenvalue $A_-^{-1}$ is 1/2 of its value in the absence of the modulation. This is known as the $3~$dB limit of squeezing. While $A_-$ decreases,  fluctuations of the other quadrature increase, $\langle Q'{}^2\rangle=A_+^{-1} > k_BT/M\omega_0^2$. The difference between the variances was clearly demonstrated already in the first experiment on squeezing in  nanomechanical systems \cite{Rugar1991}.


\subsubsection{Squeezing of fluctuations about the state of forced vibrations}
\label{subsubsec:fluctuations_driven}

Here we expand the discussion in Sec.~\ref{subsec:driven_fluctuations} to describe what underlies the power spectrum-based detection of fluctuation squeezing in driven underdamped nonlinear modes. The detection exploits the fact that the spectrum of fluctuations about a stable state of forced vibrations of a nonlinear mode is double-peaked. The peaks are resolved for sufficiently weak damping.  Their occurrence can be understood from the equations of motion for the quadratures of a driven mode (\ref{eq:eom_quadratures}) and (\ref{eq:quadratures_parametric}).

In the limit of zero damping and in the absence of noise, the stationary states of the mode in the rotating frame lie at $\partial_Pg = \partial_Q g=0$ where $g=g_r$ and $g=g_p$ for the resonant and parametric modulation, respectively. The functions $g_r, g_p$ are effective Hamiltonians in the rotating frame, and their extrema play the same role in the dynamics as the minima of the Hamiltonian function $(P^2/2M) +U(Q)$ of a classical particle with coordinate $Q$ and momentum $P$ in a potential $U(Q)$, except that $g_r,g_p$ do not have the form of a sum of the kinetic and potential energies. An important characteristic of the motion near an extremum of $g_r,g_p$ is the frequency
\begin{align}
\label{eq:rotating_frame_frequency}
\omega_\mathrm{rot} =  (\partial^2_Pg \partial^2_Qg)^{1/2}.
\end{align}
For a particle with the Hamiltonian $(P^2/2M) +U(Q)$ this expression goes into the familiar expression for the vibration frequency near a potential minimum $(\partial^2_Q U/M)^{1/2}$.

In the presence of weak damping and weak noise, after a transient the periodically driven mode approaches one of the stable states, depending on where it was initially prepared (the stable states are slightly shifted from the extrema of $g_r, g_p$ for weak damping). This is again similar to a particle in a potential well, including the case of a double-well potential. The mode then fluctuates about this state for a long time compared to the relaxation time $\sim 1/\Gamma$ (see Section~\ref{subsec:rare_events}). These fluctuations correspond to random vibrations of $Q(t)$ and $P(t)$ at frequency $\omega_\mathrm{rot}$, as seen by linearizing Eqs.~(\ref{eq:eom_quadratures}) and (\ref{eq:quadratures_parametric}) about a stable state. Again, the random vibrations of $Q(t), P(t)$  are similar to thermal vibrations of a particle in a potential well. However, they occur in the rotating frame.

As seen from Eqs.~(\ref{eq:quadratures_resonant_drive}) and (\ref{eq:parametric_slow}), vibrations of $Q,P$ modulate  the forced vibrations of  the coordinate and momentum of the driven mode in the laboratory frame $q(t), p(t)$. Therefore the power spectrum of fluctuations of the mode measured in the laboratory frame has peaks at frequencies $\omega_F \pm \omega_\mathrm{rot}$ or $(\omega_p/2)\pm\omega_\mathrm{rot}$ for resonant and parametric modulation, respectively \cite{Drummond1980c,Dykman1994c,Dykman2011}.

On the experimental side, the double-peak spectrum of fluctuations about a stable vibrational state of a resonantly driven MEMS was observed by \textcite{Stambaugh2006a}, but in this experiment the spectral peaks significantly overlapped. In the experiment \cite{Huber2020} the damping was small and the peaks were well-resolved.

As seen from Fig.~\ref{fig:phase_portraits}(b) and (e), the orbits of motion on the $(Q,P)$-plane with a given $g$ are strongly non-circular  both for resonant and parametric driving. Therefore the variances of the quadratures $Q$ and $P$ are different, which means squeezing. If we disregard dissipation,
\begin{align}
\label{eq: squeez_angle}
&\langle (\delta Q)^2\rangle = \frac{k_BT}{2M\omega_0^2}(1+ e^{-4\varphi_*}),
&
\langle P^2\rangle = \frac{k_BT}{2M\omega_0^2}(1+ e^{4\varphi_*}),\nonumber\\
&\exp (2\varphi_*) = |\partial^2_Q g|^{1/2}/ |\partial^2_P g|^{1/2}.
\end{align}
Here $\delta Q=Q-Q_0$, where $Q_0$ is the value of $Q$ at the considered extremum of $g$, and the second derivatives of $g$ are calculated at the extremum,  with $g$ being $g_r$ or $g_p$ for the resonant and parametric modulation, respectively.

 Along with the squeezing comes the difference in the areas of the spectral peaks at the frequencies $\omega_F+\omega_\mathrm{rot}$ and $\omega_F-\omega_\mathrm{rot}$ for a resonantly driven mode, as well as the peaks at  $\frac{1}{2}\omega_p+\omega_\mathrm{rot}$ and $\frac{1}{2}\omega_p-\omega_\mathrm{rot}$ for a parametrically modulated mode. The difference in the areas is directly related to the squeezing parameter \cite{Dykman2012a,Huber2020}. The ratio $\RR_\mathrm{peaks}$ of the areas of the peaks is
 \[\RR_\mathrm{peak} = \tanh^2\varphi_*.\]
 Depending on the parameters and on whether the mode is in the larger- or smaller-amplitude state, for resonant driving, the larger-area peak is on the higher or lower-frequency side of $\omega_F$ or $\omega_p/2$.

The expression for $\RR_\mathrm{peak}$ applies also to the ratio of the areas of the peaks in the imaginary parts of the susceptibility of a strongly driven strongly underdamped mode. Such susceptibility describes the response to a weak probe force at frequency $\omega_\mathrm{pr}$ close to resonance. The peak that corresponds to resonant amplification of the force always has a smaller area than the one that corresponds to the absorption. We note that the  expression for the ratio of the areas of the susceptibility peaks applies also in the quantum regime.


\section{Spectra of nonlinear underdamped vibrational modes: quantum and classical}
\label{sec:Duffing_spectra_Appendix}

The spectra of nonlinear modes (oscillators) are determined by two processes. One is decay of the vibration amplitude. Such decay makes the vibrations nonsinusoidal and thus leads to a frequency
``uncertainty'' and to a spectral broadening.  The other is frequency fluctuations. Here we will consider the frequency fluctuations that come from the interplay of the dependence of the vibration frequency on the amplitude and the amplitude fluctuations due to thermal noise or a broad-band noise from other sources. The two mechanisms of the spectral broadening are not simply superposed, but compete, in some sense, because the decay rate of the amplitude $\Gamma$  is also the reciprocal correlation time of the frequency fluctuations, as explained in Sec.~\ref{subsec:spectra_nonlinear_conservative}. Therefore the shape of the spectrum is determined by the ratio of the fluctuational frequency spread $\overline{\delta\omega_0}$ to $\Gamma$.

The broadening of the spectrum of an oscillator due to the nonlinearity was first discussed for a quantum oscillator \cite{Ivanov1965}. The analysis was done for the limiting cases $\overline{\delta\omega_0}\ll \Gamma$ and $\Gamma\to 0$. A complete solution of the problem that showed the evolution of the spectrum with the varying  $\overline{\delta\omega_0}/ \Gamma$ was obtained first in the classical theory \cite{Dykman1971} and then in the quantum theory \cite{Dykman1973a}. It described the interplay of the nonlinearity  and decay and offered an insight into the paradox of the harmonic oscillator (see below).

\begin{figure}[h]
\includegraphics{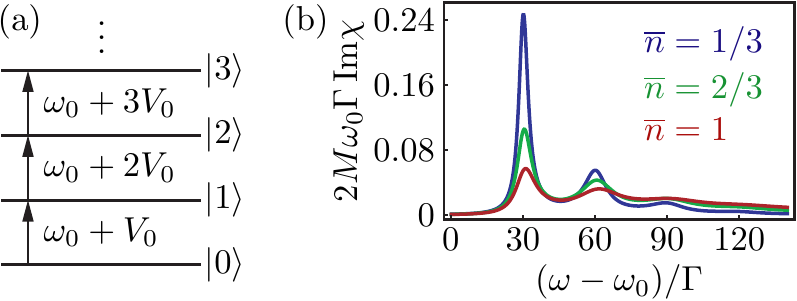}
\caption{(a) Sketch of the transitions between the energy levels of the Duffing oscillator. The difference between the neighboring transitions frequencies is $V_0=3\hbar\gamma/4M\omega_0^2$. (b) Fine structure of the imaginary part of the oscillator susceptibility [the power spectrum is  $S_{}(\omega)\propto (\bar n +1)\mathrm{Im}~\chi_{}(\omega)$].  The plot refers to the ratio of the nonlinearity parameter $V_0$ to the decay rate $V_0/\Gamma=30$. The curves from top to bottom at the first maximum, $\omega-\omega_0\approx V_0$, correspond to $\bar n = 1/3, 2/3$ and 1. For $\bar n\to 0$ the spectrum becomes a Lorentzian curve, $2M\omega_0\mathrm{Im}~\chi_{}(\omega) =\Gamma/[\Gamma^2 + (\omega - \omega_0)^2]$.  }
\label{fig:fine_structure_Duffing}
\end{figure}

In the quantum analysis, it is necessary to take into consideration that the energy levels of a nonlinear oscillator are nonequidistant.  In the Duffing model (\ref{eq:Duffing_potential}), the energy of a $k$th level is
\[E_k=\hbar k[\omega_0 + V_0(k+1)/2], \qquad V_0=3\hbar\gamma/4M\omega_0^2,\]
for $|V_0| k\ll \omega_0$. The transition frequencies $(E_k-E_{k-1})/\hbar$ are shown in Fig.~\ref{fig:fine_structure_Duffing}. They depend on the level number $k$, that is, on the energy $E_k$. This is the quantum analog of the energy dependence of the oscillator vibration frequency in the classical limit. The parameter $V_0$ is proportional to the Duffing nonlinearity parameter $\gamma$. It is the discreteness of the transition frequencies that determines the quantum effects of the nonlinearity on the susceptibility and the power spectrum.

So far, in nanomechanical systems studied in the quantum regime \cite{O'Connell2010,Satzinger2018,Chu2018,Wang2019,Arrangoiz-Arriola2019, Wollack2021a,MacCabe2020,Cattiaux2021}, including the systems studied in quantum optomechanics \cite{Aspelmeyer2014, Aspelmeyer2014a,Lepinay2021,Kotler2021}, the Duffing nonlinearity $|V_0|$ was small compared to the decay rate $\Gamma$. This impeded an observation of quantum effects of this nonlinearity in the spectra. However, quantum effects can be pronounced for electromagnetic modes in nonlinear microwave cavities and for Josephson junction based systems, cf. \cite{Schuster2007,Bertet2012}. They underlie the operation of the transmon qubit \cite{Koch2007}, which is the basic element of the nowadays superconducting quantum computers.

A quantum picture of the nonlinearity-induced spectral change is in some sense more intuitive than the classical, and the classical results follow from the quantum results as a limiting case. Therefore we present this picture first.

\subsection{``Paradox'' of the quantum harmonic oscillator}
\label{subsec:paradox}

The oscillator susceptibility $\chi_{}(\omega)$ near resonance, $\omega\approx \omega_0$, is formed by the transitions between neighboring levels. A naive way to describe it is to think of  the oscillator as a set of two-level systems formed by the pairs of neighboring states $\Ket{k-1}, \Ket{k}$ with $k=1,2,...$. Each such system makes a partial contribution to the resonant susceptibility, which is described by a function $\phi_{}(k,\omega)$. The overall susceptibility of the oscillator can be then sought in the form of a sum of such partial susceptibilities,
\begin{align}
\label{eq:susc_as_sum}
\chi_{}(\omega) = (2 M\omega_0)^{-1}\sum_{k=1} \phi_{}(k,\omega).
\end{align}
One might further assume that a partial susceptibility $\phi_{}(k,\omega)$ is given by the familiar expression for the susceptibility of a two-level system \cite{Weisskopf1930a}
\begin{align}
\label{eq:two_level}
&\phi_\mathrm{W}(k,\omega) =\frac{k}{\bar n +1}\,\frac{\rho_{k-1}}{\Gamma_\mathrm{W}(k)-i(\omega-\omega_0-V_0k)},\nonumber\\
&\Gamma_W(k) = \Gamma[2k(2\bar n +1)-1].
\end{align}
Here $\omega_0+V_0k$  is the transition frequency of the $\ket{k-1}\to\ket{k}$ transition, as seen from Fig.~\ref{fig:fine_structure_Duffing}(a). The parameter $\Gamma_W(k)$ is given by the half-sum of the reciprocal lifetimes of the states $\Ket{k-1}$ and $\Ket{k}$.  In calculating it we used that  the reciprocal lifetime of a state $\ket{k}$ is $W_{k\to k+1} + W_{k\to k-1}$, where the transition rates $W_{k\to k\pm 1}$ are given by Eq.~(\ref{eq:golden_rule}), and we also took into account that $W_{k\to k+1}/W_{k+1\to k} = \bar n/(\bar n+1)$ [the Einstein relation \cite{Landau1980}], with $\bar n = [\exp(\hbar\omega_0/k_BT)-1]^{-1}$. Further along the lines of the Weisskopf-Wigner theory, in Eq.~(\ref{eq:two_level}) the coefficient $\rho_{k-1}$ is the population of the state $k-1$ from which the system makes a transition, $\rho_k= \exp(-\hbar k\omega_0/k_BT)/(\bar n +1)$. The function Im~$\phi_\mathrm{W}(k,\omega)$ has the familiar form of a Lorentzian centered at the transition frequency $\omega_0+V_0k$ and having halfwidth $\Gamma_\mathrm{W}(k)$.

An obvious flaw of this picture, which was noticed already by \textcite{Weisskopf1930}, is that it does not describe the susceptibility of a harmonic oscillator in the limit $V_0=0$.  In this limit the functions Im~$\phi_\mathrm{W}(k,\omega)$ are Lorentzians centered at the same frequency, but their half-widths $\Gamma_W(k)$ are different, so that the whole spectrum is not Lorentzian.   This has become known as the paradox of the harmonic oscillator \cite{Belavin1969,Zeldovich1969}. \textcite{Weisskopf1930} studied the effect for a three-state system with equal transition frequencies and showed that, indeed, such a system is not described by a set of two independent two-level systems.

The breakdown of the approximation (\ref{eq:two_level}) with decreasing $|V_0|/\Gamma$ is a characteristic quantum effect. The transition  frequencies $\omega_0+V_0k$ with different $k$ are close to each other, and to distinguish the partial spectra $\phi_\mathrm{W}(k,\omega)$ one has to wait for  a time $t\gg |V_0|^{-1}$. However, because of the coupling to a thermal bath, the oscillator stays in a state $k$ for a time $\sim \Gamma_\mathrm{W}^{-1}(k)$. If this time is less than $|V_0|^{-1}$, the partial spectra may not be distinguished.

The time $\Gamma_\mathrm{W}^{-1}(k)$ can be thought of as the time it takes to ``switch'' from one two-level system to another. The switching couples the partial spectra with different $k$ to each other. This coupling is described by a system of linear equations \cite{Dykman1984}, which  can be obtained from the quantum master equation for the oscillator density matrix $\rho$ (\ref{eq:master_Lindblad}).

Following the standard Kubo formula we relate the susceptibility of the oscillator to the Fourier transform of the correlator $\langle a(t) a^\dagger(0)\rangle = \mathrm{Tr} [a \exp(-iHt)a^\dagger \rho_\mathrm{full}\exp(i Ht)]$ where $\rho_\mathrm{full}$ is the density matrix of the system and the bath. Tracing out the bath and switching to the rotating frame, we reduce the trace to that over the states of the oscillator, with the density matrix satisfying Eq.~(\ref{eq:master_Lindblad}) with the initial condition
\[\rho(0) = a^\dagger \exp(-\hbar\omega_0 a^\dagger a/k_BT)/(\bar n +1).\]
The Fourier transform of the trace over the oscillator states $\ket{k}$ is a sum over $k$ of the Fourier transforms of the corresponding matrix elements
\[\phi(k,\omega) = \int_0^\infty dt\exp[i\omega-\omega_0)t]k^{1/2}\bra{k}\rho(t)\ket{k-1}.\]
From Eq.~(\ref{eq:master_Lindblad}) we obtained a system of equations for $\phi(k,\omega)$. It reads
\begin{align}
\label{eq:coupled_spectra}
&[\Gamma_W(k)-i(\omega-\omega_0-V_0k)] \phi_{}(k,\omega) \nonumber\\
&-2\Gamma k[\bar n \phi_{}(k-1,\omega)+(\bar n +1)\phi_{}(k+1,\omega)]\nonumber\\
 &= k\rho_{k-1}/(\bar n +1).
\end{align}
It is seen from this equation that, for $|V_0|\gg \Gamma_W(k)$, the partial spectra $\phi_{}(k,\omega)$ near their maxima are indeed given by Eq.~(\ref{eq:two_level}). However, one can easily check that, for a harmonic oscillator, $V_0=0$, the solution is
\[ \phi_{}(k,\omega)= \frac{k}{\bar n +1}\,\frac{\rho_{k-1}}{\Gamma - i(\omega-\omega_0)} \quad (V_0=0).\]
In other words, all ``partial spectra'' have the same spectral shape, a profound quantum effect. With the account taken of Eq.~(\ref{eq:susc_as_sum}), this expression gives the familiar expression (\ref{eq:Lorentz_classical}) for the susceptibility of a harmonic oscillator.

\subsection{The susceptibility in the explicit form }
\label{subsec:suscept_explicit}
Equation (\ref{eq:coupled_spectra}) can be solved and the expression for the susceptibility $\chi_{}(\omega)$ can be obtained in the form of an integral of an elementary function \cite{Dykman1973a},
\begin{widetext}
\begin{align}
\label{eq:suscept_time_explicit}
&\chi_{}(\omega) = \int_0^\infty dt e^{i\omega t}\mathcal{X}_{}(t), \qquad \mathcal{X}_{}(t) =\frac{ i}{2M\omega_0} e^{-i\omega_0t} e^{\Gamma t}\psi_{00}^{-2}(t), \qquad \psi_{00}(t) = \cosh(\aleph_0\Gamma t) +  \Lambda_0\sinh(\aleph_0\Gamma t),\nonumber\\
&\aleph_0=\left[1+i\frac{V_0}{\Gamma}(2\bar n +1)-\frac{V_0^2}{4\Gamma^2}\right]^{1/2} \quad (\mathrm{Re}~\aleph_0>0), \qquad
\Lambda_0=\aleph_0^{-1}\left[1+i\frac{V_0}{2\Gamma}(2\bar n +1)\right].
\end{align}
\end{widetext}

It is seen from the explicit expression (\ref{eq:suscept_time_explicit}) that the susceptibility of a quantum Duffing oscillator is determined by two dimensionless parameters: the ratio $V_0/\Gamma$ of the difference of the transition frequencies (cf. Fig.~\ref{fig:fine_structure_Duffing}a) to the decay rate and the oscillator thermal occupation number $\bar n$. The power spectrum $S_{}(\omega)\propto \mathrm{Im}~\chi_{}(\omega)$ has a peak near $\omega=\omega_0$, which is not only non-Lorentzian, but is actually asymmetric, in contrast to the case of a harmonic oscillator. For $|V_0|\gg \Gamma(2\bar n +1)$ the power spectrum can have a fine structure, with the spectral peaks centered at frequencies $\approx \omega_0+kV_0$, cf. Eq.~(\ref{eq:two_level}). This fine structure is illustrated in Fig.~\ref{fig:fine_structure_Duffing}(b).  It exists in a temperature range limited from above and below. On the one hand, the temperature should be sufficiently high so that the excited states of the oscillator are populated. On the other hand, the fine structure smears out with the increasing temperature, as the linewidths $2\Gamma_W(k)$ increase and the condition $|V_0|\gg \Gamma_W(k)$ ceases to hold for smaller and smaller $k$.

\subsection{Dispersively coupled vibrational modes}
\label{subsec:coupled_modes}

A similar effect on the susceptibility of the considered mode $n=0$  comes from its dispersive coupling to other modes. The  energy of the dispersive coupling is
\[U_\mathrm{disp}=\frac{3}{4}M\sum_{n_1\neq n_2} \gamma_{n_1n_1 n_2n_2}
q^2_{n_1}q^2_{n_2}\]
where the subscripts $n_{1,2}$ enumerate the modes. This is a part of the total energy of the nonlinear mode coupling described by Eq.~(\ref{eq:nonlin_multimode}).

Because of the dispersive coupling, the frequency $\omega_0$ of the mode $n=0$ becomes dependent on the states $\ket{k_n}$ of other modes,
\begin{align}
\label{eq:dispersive_frequency}
&\omega_0 \to \omega_0\{k_n\}=\omega_0+\sum_{n>0} V_n\left[k_n+(1/2)\right],\nonumber\\
& V_n=
\frac{3\hbar}{2M_n}\frac{\gamma_{00nn}}{\omega_0\omega_n} .
\end{align}
Here we assume that the coupling parameters $\gamma_{n_1n_1n_2n_2}$ have been renormalized to allow for the cubic in the mode coordinates terms in the potential energy of the modes, cf. Eq.~(\ref{eq:nonlin_multimode}); $M_n$ is the effective mass of mode $n>0$ ($M_0\equiv M$). Equation (\ref{eq:dispersive_frequency}) is the quantum analog of Eq.~(\ref{eq:dispersive_shift}) for the frequency shift in terms of the mode amplitudes.

The frequencies $\omega_0\{k_n\}$ form a ladder for each $n$, similar to the frequency ladder in Fig.~\ref{fig:fine_structure_Duffing}. As in the case of the internal mode nonlinearity, the susceptibility of the mode $n=0$ is affected by the coupling of the transition amplitudes for different $k_n$. The overall expression for the susceptibility of the mode $n=0$ can be written in the same form as Eq.~(\ref{eq:suscept_time_explicit}) provided one replaces \cite{Dykman1973a}
\begin{align}
\label{eq:susceptibility_multimode}
&e^{\Gamma t}\psi_{00}^{-2}(t)\to \psi_{00}^{-1}(t)\prod_n e^{\Gamma_n t}\psi_{0n}^{-1}(t),\nonumber\\
&\psi_{0n}(t)= \cosh (\aleph_n\Gamma_nt) + \Lambda_n\sinh (\aleph_n\Gamma_nt).
\end{align}
Here, the parameters $\aleph_n$ and $ \Lambda_n$ are again given by Eq.~(\ref{eq:suscept_time_explicit}) with the replacement
\begin{align}
\label{eq:cpld_modes_parameters}
&V_0\to V_n,\quad \Gamma\to \Gamma_n, \quad \bar n \to \bar n_n=\left[\exp(\hbar\omega_n/k_BT)-1\right]^{-1},\end{align}
In the above expressions, $\Gamma_n$ is the decay rate of mode $n$ and $\bar n_n$ is its thermal occupation number; we also use $\Gamma_0\equiv \Gamma$ and $\bar n_0\equiv \bar n$.

Where the difference in the frequencies of transitions with different $k_n$ is large compared to the decay rates of the involved modes, $|V_n|\gg \Gamma_n(2\bar n_n+1)$, the susceptibility given by Eqs.~(\ref{eq:suscept_time_explicit})- (\ref{eq:cpld_modes_parameters}) becomes a sum of partial susceptibilities for the transitions where modes $n$ are in different Fock states $\Ket{k_n}$, similar to Eq.~(\ref{eq:two_level}). On the other hand, the  modes with $|V_n|(2\bar n_n+1)\ll \Gamma_n$, i.e., the modes that are weakly dispersively coupled to the considered mode $n=0$ compared to their decay rates, only slightly perturb $\chi_{}(\omega)$.

\subsection{Classical limit}
\label{subsec:classical_spectrum}

The expressions for the susceptibility simplify in the classical limit. In this limit, the thermal occupation numbers of the modes are $\bar n_n\approx k_BT/\hbar\omega_n$. At the same time, the nonlinearity parameters $V_n$, which are explicitly related to the discreteness of the modes energy spectra, are $\propto \hbar$. Therefore they can enter only in combination with $\bar n_n$. Respectively, in the classical limit we have in Eqs.~(\ref{eq:suscept_time_explicit}) -- (\ref{eq:cpld_modes_parameters})
\begin{align}
\label{eq:classical_spectrum}
&\aleph_n\to (1+4i\alpha_n)^{1/2}, \quad \Lambda_n\to \aleph_n^{-1}(1+2i\alpha_n)\nonumber,\\
&\alpha_n=3\gamma_{00nn}k_BT(2-\delta_{n,0})/8M_n\omega_0\omega_n^2\Gamma_n.
\end{align}

In the classical limit, the susceptibility does not have fine structure. However, it is very different from the simple Lorentzian limit (\ref{eq:simple_Lorentzian}). The power spectrum $S_{}(\omega)\propto \mathrm{Im}~\chi_{}(\omega)$ becomes asymmetric with the increasing $|\alpha_n|$, both because of the internal nonlinearity and of the dispersive coupling to fluctuating modes \cite{Dykman1971}.

\begin{figure}[h]
\includegraphics{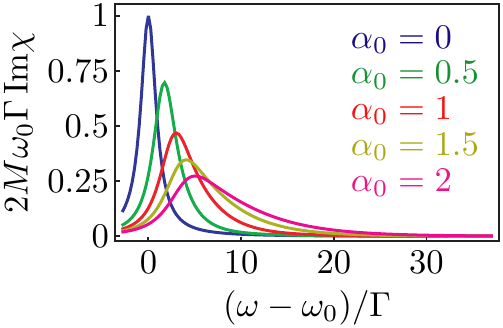}
\caption{Imaginary part of the susceptibility of the classical Duffing oscillator Im~$\chi(\omega)\propto S(\omega)$. The shape of the spectrum is determined by the single parameter $\alpha_0$, Eq.~(\ref{eq:classical_spectrum}). The curves from top to bottom at the  maximum refer to $\alpha_0=0$ (a Lorentzian spectrum with halfwidt $\Gamma$), $\alpha_0=0.5, 1, 1.5, 2$.  }
\label{fig:classic_Duffing}
\end{figure}

\begin{figure}[h]
\includegraphics{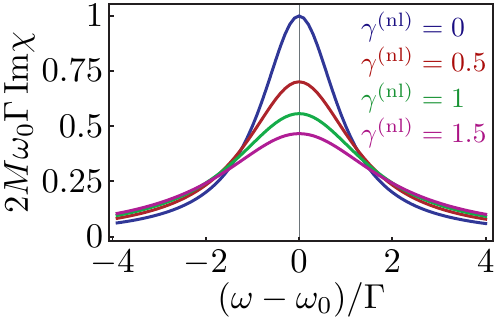}
\caption{Imaginary part of the susceptibility of an oscillator with nonlinear friction in the absence of the dependence of the vibration frequency on the amplitude. The dynamics is described by Eqs.~(\ref{eq:master_Lindblad}) -- (\ref{eq:nonlin_fric_Appendix}) in the classical limit [cf. also Eq.~(\ref{eq:eom_nonlin}) of the main text] with the  Duffing nonlinearity parameter $\gamma=0$. The spectra from top to bottom correspond to the scaled nonlinear friction coefficient $\gamma^\mathrm{(nl)}\equiv 2\Gamma^\mathrm{(nl)} k_BT/\hbar\omega_0\Gamma = 0, 0.5,1$, and 1.5. The nonlinear friction parameter $\Gamma^\mathrm{(nl)}$ is defined in Eq.~(\ref{eq:nonlin_fric_Appendix}).
}
\label{fig:nonlin_fric_spectrum}
\end{figure}

The evolution of the spectrum of a single mode $n=0$ with the varying single parameter of the internal nonlinearity $\alpha_0$ is shown in Fig.~\ref{fig:classic_Duffing}.  The width of the spectrum increases with the increasing $\alpha_0$; we note that $\alpha_0\propto T$ if the decay rate is independent of temperature. For $\alpha_0=0$ the spectrum is Lorentzian, whereas for $|\alpha_0|\gg 1$ the spectrum near the maximum has the form $S_{}(\omega)\propto |\omega-\omega_0|\exp[-(\omega-\omega_0)/2\alpha_0\Gamma]$ for $\alpha_0(\omega-\omega_0)>0$. The expression for the power spectrum of a classical oscillator that coincides with the presented result  was given by \textcite{Renz1985}.

For a comparison, we show in Fig.~\ref{fig:nonlin_fric_spectrum} the effect of nonlinear friction on the susceptibility spectrum in the absence of conservative nonlinearity. The spectrum remains symmetric in this case, but is strongly non-Lorentzian. The deviation from the Lorentzian form is a signature of the vibration nonlinearity,   while the symmetry of the spectrum enables distinguishing the effects of conservative and dissipative nonlinearity.

An interesting behavior occurs where the number of the modes $N$ dispersively coupled to the considered mode is large even though the coupling to each mode is small. One would expect some kind of the central limit theorem to apply in this case, leading to a Gaussian power spectrum $S_{}(\omega)$ \cite{Barnard2012}. This is indeed the case \cite{Zhang2015a}.  A Gaussian spectrum emerges if $|\alpha_n|\ll 1$ for all $n$, but $\sum_n\alpha_n^2\gg1$ and $\sum_n\alpha_n^2\Gamma_n^2 \gg \Gamma_m^2 (\forall m)$. In this case $S_{}(\omega)\propto \exp[-(\omega-\tilde\omega_0)^2/2\sigma^2]$ with $\tilde\omega_0 =\omega_0 + 2\sum_n\alpha_n\Gamma_n$ and $\sigma^2 = 4\sum_n\alpha_n^2\Gamma_n^2$.

In conclusion of this section, we note that the coupling of interstate transitions and the related distortion of the spectral lines is a generic property of systems with close transition frequencies. Such coupling occurs in different types of systems. Besides various vibrational systems, like Josephson junctions, microwave cavities, and NVSs, examples range from the cyclotron resonance in semiconductors to electron spin resonance in strong magnetic fields in systems with $S>1/2$.  The coupling of transitions is important also for classical vibrational systems with fluctuating frequency. An example is provided by nano- and micromechanical resonators with a fluctuating number and/or positions of attached molecules \cite{Vig1999,Yang2011}. The spectra of such systems can also be asymmetric and display a fine structure \cite{Dykman2010a}. On the formal side, for different physical mechanisms the full spectra can be often described by linear equations for coupled partial spectra. These equations are convenient for a numerical analysis.


\section{The action-angle variables}
\label{subsec:action_angle}

The Duffing model has been very successful in describing many observations of  nanomechanical systems, and in the majority of cases the analysis was based on the Bogoliubov-Krylov method of averaging outlined in Appendix~\ref{subsec:nonlin_averaging}. As indicated above, this method is similar to the rotating wave approximation (RWA), and it is used  throughout the present paper. However, we have to indicate that it may become inapplicable even where the nonlinearity is still comparatively weak, that is, the nonparabolic in $q$  terms in the potential energy of a vibrational mode $U(q)$ are still small compared to $M\omega_0^2q^2/2$. A simple example is provided by a mode with a  broken inversion symmetry. In this case, in the nonlinear part of the potential one has to keep the cubic in $q$ term,
\[U(q) = \frac{1}{2}M\omega_0^2q^2 + \frac{1}{3}M\beta q^3 + \frac{1}{4}M\gamma q^4\]
[cf. also Eq.~(\ref{eq:nonlin_multimode})]. Such modes have been extensively studied in the literature, see \cite{Kozinsky2006,Chan2008a,Eichler2011a,Meerwaldt2012,Eichler2013,Huang2019,Ochs2021}. In fact the lack of inversion symmetry is fairly generic for flexural nanomechanical modes, as it comes, for example, whenever a gate voltage is applied and a nanoresonator is bent or just from the capacitive part of the potential energy $\propto (\partial^3 C_\mathrm{g}/\partial q^3)(V_\mathrm{g}^\mathrm{DC})^2$, cf. Eq.~(\ref{eq:capacitivesoftening}).

Cubic nonlinearity leads to several effects, including vibrations at the second overtone of the eigenfrequency, i.e., at the frequency $\approx 2\omega_0$, and the change of the dependence of the vibrations frequency on the amplitude. In the RWA, the latter is described by the renormalization of the Duffing parameter \cite{Landau2004a}
\begin{align}
\label{eq:renormalized_gamma}
\gamma\to \gamma_\mathrm{eff} =\gamma-\frac{10\beta^2}{9\omega_0^2}
\end{align}
(see \cite{Eichler2013,Huang2019} for some other effects).

It is immediately seen from Eq.~(\ref{eq:renormalized_gamma}) that the term $\propto \beta^2$ can significantly change the character of the amplitude dependence of the mode frequency (\ref{eq:Duffing_frequency_shift}). Indeed, if $\gamma>0$, but $\gamma_\mathrm{eff}<0$, even the sign of the  slope $d\omega/dA^2$ of the frequency dependence on the amplitude changes. However, it is clear that for large amplitudes the term $\propto q^4$ in $U(q)$ becomes more important than the term $\propto q^3$. Simple dimensional arguments show that for the amplitudes $ A^2 \gtrsim \omega_0^2\gamma_\mathrm{eff}/\gamma^2$, the RWA approximation (\ref{eq:renormalized_gamma}) becomes inapplicable. For small $\gamma_\mathrm{eff}/\gamma$ this happens where the nonlinear part of the energy $\sim M\gamma A^4$ is still small compared to the harmonic part $\sim M\omega_0^2A^2$. Therefore it is necessary to find an alternative approach that would not rely on the RWA.

The analysis of the mode dynamics beyond the RWA is simplified in the case of weak damping, where the decay rate $\Gamma\ll \omega_0$. Here it is convenient to use the method of averaging in the form developed in the dynamics of Hamiltonian systems \cite{Arnold1989}. In this method one changes from the coordinate and momentum of the mode to its action-angle variables. This is a canonical transformation of variables. The coordinate and momentum are functions of the action $I$ and the phase (angle) $\varphi$ and are periodic in  $\varphi$,
\[q(I, \varphi+2\pi) = q(I,\varphi), \quad p(I, \varphi+2\pi) = p(I,\varphi). \]
The action and phase variables of the Hamiltonian system are defined as
\[I=(2\pi)^{-1}\oint p\,dq ,\qquad \varphi = \frac{\partial}{\partial I}\int p\,dq\]
\cite{Landau2004a}. The vibration frequency of the mode is a function of $I$ or, equivalently, of the mode energy $E$,
\[\omega(I)=(\partial I/\partial E)^{-1},\]
and $\omega(I)\partial_\varphi q = p/M$, whereas $\omega(I)\partial_\varphi p = -\partial_qU(q)$.

To describe the mode dynamics in the presence of a friction force $-2\Gamma p$ and a driving force $F\cos\omega_F t$, one can change from the variables $(q,p)$ to $(I,\varphi)$. The resulting equations for $I$ and $\varphi$ read
\begin{align}
\label{eq:eom_I_phi_full}
&\dot I = R\partial_\varphi q, \quad \dot\varphi = \omega(I) -R\partial_Iq, \nonumber\\
&R=-2\Gamma p + F\cos\omega_Ft.
\end{align}
The key observation that underlies the averaging principle is that the action varies in time only because of the friction and the driving, which are assumed small. In contrast, the phase accumulates at frequency $\omega(I)$, which is assumed large. Therefore the time evolution of $q$ and $p$ is fast oscillations with a slowly varying in time action $I$. The contribution of fast oscillations to $\dot I$  does not accumulate in time. Therefore on the time scale large compared to $\omega_F^{-1}, \omega^{-1}(I)$ the motion can be described by averaging over the fast oscillations for a given action.

The effect of the driving is most pronounced (see Appendix~\ref{subsec:resonant_forced}) where the driving is resonant. This means that the vibrations occur at frequencies $\omega(I)$ close to $\omega_F$, that is, $|\omega(I)-\omega_F|\ll \omega_F$ (the analysis of parametric driving can be done similarly).  It is then convenient to write
\begin{align}
\label{eq:varphi_0}
\varphi = \omega_F t + \varphi_0(t),\quad \dot\varphi_0 = \omega(I)-\omega_F -R\partial_I q.
\end{align}
The phase $\varphi_0$ is also a slow variable. For a given $\varphi_0$ the functions  $q(I,\varphi), p(I,\varphi)$ are periodic in $\omega_Ft$.

It follows from the above arguments that the equations for $I,\varphi_0$ can be obtained by averaging the full equations (\ref{eq:eom_I_phi_full}) over fast oscillations,
\begin{align}
\label{eq:averaging_action}
\dot I = \overline{R\partial_\varphi q }, \qquad \dot\varphi_0 = \omega(I) -\omega_F-\overline{R\partial_Iq}.
\end{align}
Here
\[\overline{L(I,\varphi)} = (2\pi)^{-1}\int_0^{2\pi}d\theta L(I,\theta+\varphi_0)
\]
where $L$ is an arbitrary function of $I,\varphi$, which is periodic in $\varphi$ (in $R$ we replace $F\cos\omega_Ft$ with $F\cos\theta$; in fact the averaging is done over the period $2\pi/\omega_F$ for fixed $I,\varphi_0$).

Stationary states of forced vibrations are given by Eq.~(\ref{eq:averaging_action}) in which one sets $\dot I=\dot\varphi_0 =0$. One can see, in particular, that in the case of a comparatively weak Duffing nonlinearity the resulting equations give the same result for the vibration amplitude and phase as Eqs.~({\ref{eq:eom_quadratures}) and (\ref{eq:g_r}) or, equivalently, Eq.~(\ref{eq:bistability_naive}). However, the method allows one to go beyond the weak-nonlinearity limit and study effects that are not described by the RWA, cf. \cite{Dykman1990d,Soskin2003,Shoshani2017,Huang2019,Miller2021,Ochs2021}. We note that if the involved characteristic frequencies $\omega(I)$ become significantly different from $\omega_0$, the approximation of the frequency-independent coefficients of linear and nonlinear friction may become inapplicable. A more general approach to describing dissipation may be necessary in this case.


\section{Thermoelastic and Akhiezer relaxation}
\label{sec:app_Akhiezer}

In this section we consider the mechanisms of Akhiezer and thermoelastic relaxation of NVS modes. These mechanisms have common origin and can be described within the same general framework, as indicated in Sec.~\ref{sec:FDT}. In both mechanisms, relaxation comes from  inelastic scattering of thermal phonons off the low-frequency  NVS mode. The process is sketched in Fig.~\ref{fig:scattering_process_with_phonon_scattering}. The mechanisms are particularly important  if the mode eigenfrequency $\omega_0$ is small compared to the temperature in frequency units,  $\hbar\omega_0\ll k_BT$. To find the mode decay rate $\Gamma$ in this case it is usually necessary to take into consideration that thermal phonons are scattered off each other, and their scattering rate can be comparable to $\omega_0$.

\begin{figure}[h]
\includegraphics{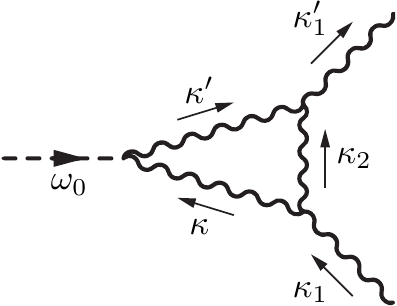}
\caption{Feynman diagram showing scattering of phonon $\kappa$ off the NVS mode into phonon $\kappa'$. The phonons $\kappa, \kappa'$  themselves are scattered off other phonons, and the diagram provides an example of such scattering. The resulting lifetime of the involved phonons can be smaller than the  eigenfrequency  $\omega_0$ of the NVS mode.}
\label{fig:scattering_process_with_phonon_scattering}
\end{figure}

The coupling Hamiltonian that describes the scattering of thermal phonons off  the low-frequency NVS mode has the form
\begin{align}
\label{eq:cplng_scattering}
H_i = qh_\mathrm{b}^\mathrm{scatt}, \quad h_\mathrm{b}^\mathrm{scatt}= \sumprime{\kappa,\kappa'}V_{\kappa\kappa'}b_\kappa^\dagger b_{\kappa'},
\end{align}
where $b_\kappa$ and $ b_\kappa^\dagger$ are annihilation and creation operators of the vibrational modes coupled to the NVS, cf. Eq.~(\ref{eq:anharmonic_linear}). These modes provide a thermal reservoir. We call them phonons and assume that their frequencies $\omega_\kappa$  have a (quasi) continuous spectrum.

It follows from the results of Sec.~\ref{sec:FDT}, see also   Eqs.~(\ref{eq:Gamma_defined}), (\ref{eq:bath_FDT_again}),
that, to the leading order in $H_i$,  the decay rate of the considered low-frequency mode can be expressed in terms of the power spectrum of the operator $h_\mathrm{b}^\mathrm{scatt}$. It is convenient to write this expression as
\begin{align}
\label{eq:G_two_phonon_general}
&\Gamma = -\frac{\hbar}{2M{}k_BT}\,{\rm Im}\,\sum_{\kappa,\kappa'}V_{\kappa'\kappa}^*\int_0^\infty dt e^{i\omega_0 t-\ep t}\phi_{\kappa\kappa'}(t),\nonumber\\
&\phi_{\kappa\kappa'}(t)=\frac{-i}{\hbar}\sum_{\kappa_0,\kappa_0'}V_{\kappa_0'\kappa_0}
\langle b_{\kappa}^\dagger (t) b_{\kappa'}(t) b_{\kappa_0'}^\dagger (0) b_{\kappa_0}(0)\rangle
\end{align}
($\ep \to +0$.) Function $\phi_{\kappa\kappa'}(t) $ is a two-phonon correlation function.

Equation (\ref{eq:G_two_phonon_general}) immediately gives the Landau-Rumer-Krivoglaz type result (\ref{eq:Landau_Rumer}), if one disregards the interaction between the high-frequency phonons, in which case $\phi_{\kappa\kappa'}(t) = (-i/\hbar)V_{\kappa'\kappa}\bar n(\omega_\kappa) [\bar n(\omega_{\kappa'})+1]\exp[i(\omega_\kappa-\omega_{\kappa'})t]$.

The interaction between the phonons can be taken into account by deriving a quantum kinetic equation for  $\phi_{\kappa\kappa'}$. The modes $\kappa$ can be well-defined and the calculation can be done in a fairly general case where the resonator is spatially nonuniform, but the nonuniformity is smooth on the wavelength of thermal phonons $\lambda_T$ \cite{Atalaya2016}. This means, for example, that the size of the ripples on a nanomembrane or the scale of nanotube bending or twisting largely exceed $\lambda_T$.

Here we will outline the analysis of the phonon-phonon scattering for a simple case where the resonator is spatially uniform. In a uniform system, thermal phonons are characterized by their wave vector $\kb$ and the branch $\alpha$, i.e., the phonon label $\kappa$ is $\kappa\equiv (\kb,\alpha)$; for thin resonators, $\alpha$ includes the number of the quantized state of motion in the direction of the confinement. The strain of the considered low-frequency mode varies on the length $L$ that largely exceeds the thermal wavelength of phonons $\lambda_T$; for a flexural mode, $L$ is the length of the resonator. Therefore the modes $\kappa$ and $\kappa'$ coupled to it  have close wave vector, $|\kb \ -\kb'|\ll |\kb|\sim 1/\lambda_T$. For thermal modes to resonantly scatter off the low-frequency mode, their frequencies should be also close, $|\omega_\kappa - \omega_{\kappa'}|\sim \tau_\kappa^{-1}, \, \omega_0\ll \omega_\kappa$, where $\tau_\kappa$ is the relaxation time of mode $\kappa$. The conditions on the wave vectors and the mode frequencies are usually met in a sufficiently broad range of $\kb$, if the modes $\kappa,\kappa'$ belong to the same branch $\alpha$ [the situation may be more complicated in anisotropic systems \cite{Herring1954}]; we will consider coupling to the modes of the same branch.

Given the difference in the spatial scales $L$ and $\lambda_T$, it is convenient to switch from $\phi_{\kappa\kappa'}$ to its Wigner transform. For a spatially uniform system [$\kappa\equiv (\kb,\alpha), \kappa'\equiv (\kb'\alpha)$] it has the form
\begin{align}
\label{eq:Wigner_phonons}
\Phi_\alpha(\rb{},\kb{},t) =&\frac{\mathbb V}{(2\pi)^d}\int d\kb_1\,d\kb'_1 \,e^{i(\kb'_1-\kb_1)\rb{}}\phi_{(\kb_1\alpha) (\kb_1'\alpha)}(t) \nonumber\\
&\times \delta\left(\frac{\kb_1 + \kb'_1}{2}-\kb{}\right),
\end{align}
where  $d$ is the dimensionality of thermal phonons and ${\mathbb V}$ is the volume, area, or length of the resonator, depending on the dimensionality. Function $\Phi_\alpha(\rb,\kb,t)$ is the two-phonon correlation function for the branch $\alpha$.

One can also introduce the coefficient $V_\alpha(\rb{},\kb{})$, which is given by the same expression as $\Phi_\alpha$, except that $\phi_{(\kb_1\alpha) (\kb_1'\alpha)}(t)$ is replaced by $V_{(\kb_1'\alpha) (\kb_1\alpha)}$. Then the decay rate (\ref{eq:G_two_phonon_general}) can be written as
\begin{align}
\label{eq:decay_in_terms_of_Wigner}
\Gamma =& -\frac{\hbar}{2M{}k_BT}\,{\rm Im}\,\int\frac{d\rb{}\,d\kb{}}{(2\pi)^d}\sum_\alpha V^*_\alpha(\rb{},\kb{})
\nonumber\\
&\times \int_0^\infty dt \, \Phi_\alpha(\rb{},\kb{},t)\exp(i\omega_0 t - \ep t).
\end{align}
This equation presents the decay rate as an integral  of the local (for given $\rb{}$) decay rate ``density''.

The parameters $V_\alpha(\rb{},\kb{})$ take a simple form for the deformation potential coupling of the considered mode to phonons \cite{Gurevich1988}. The deformation potential model assumes that the phonon wavelengths are much smaller than the length over which there varies the mode-induced displacement field $q\boldsymbol{\varphi}(\rb)$. In the model  the change of the phonon frequency $\delta\omega_\kappa$ is proportional to the divergence of the displacement field,
\begin{align}
\label{eq:Gruneisen_appendix}
\delta\omega_\kappa =- \omega_\kappa\gamma_\kappa ^{\rm (G)}q\n \boldsymbol{\varphi}(\rb),
\end{align}
cf. Eq.~(\ref{eq:Gruneisen}).  From Eq.~(\ref{eq:Gruneisen_appendix}) and from Eq.~(\ref{eq:Wigner_phonons}) written for  $V_\alpha(\rb,\kb)$ we can directly express the coupling parameters in terms of the Gr\"uneisen parameters $\gamma_{\kb \alpha}^\mathrm{(G)}$,
\begin{align}
\label{eq:coupling_V_deformation}
V_\alpha(\rb,\kb) = -\hbar\omega_{\kb\alpha}\gamma_{\kb\alpha}^{\rm (G)}\n\boldsymbol{\varphi}(\rb).
\end{align}

Equations (\ref{eq:decay_in_terms_of_Wigner}) and (\ref{eq:coupling_V_deformation}) reduce the problem of finding the decay rate of a low-frequency mode to calculating the correlation function $\Phi_\alpha$ of thermal phonons in the Wigner representation. We note that it is not assumed that the mode is described by a plane wave, as in the Akhiezer theory of ultrasound absorption \cite{Akhiezer1938}. The general analysis below does not use the model (\ref{eq:Gruneisen_appendix}).

\subsection{Kinetic equation}

Time evolution of the function $\Phi_\alpha(\rb{},\kb{},t)$ is determined by phonon-phonon scattering. If the phonon-phonon coupling and the disorder are weak, one can sum the perturbation series for the functions $\phi_{\kappa\kappa'}$ \cite{Atalaya2016} and obtain a Markov kinetic equation for  $\Phi_\alpha$ (strictly speaking, with a renormalized phonon spectrum),
\begin{align}
\label{eq:kinetic_Boltzmann}
\partial_t\Phi_\alpha  + \vb_{\kb{}\alpha}\partial_\rb{} \Phi_\alpha ={\rm St}[\Phi_\alpha], \quad \Phi_\alpha\equiv \Phi_\alpha(\rb{},\kb{},t).
\end{align}
Here $\vb_{\kb{}\alpha}$ is the group velocity of the phonon of the branch $\alpha$ with the wave vector $\kb{}$ and ${\rm St}$ is the collision integral. The initial condition follows from Eq.~(\ref{eq:G_two_phonon_general}),
\begin{align}
\label{eq:initial_Phi}
\Phi_\alpha(\rb{},\kb{},0) = -i\hbar^{-1}V_\alpha(\rb{},\kb{})\bar n_{\kb{}\alpha}(\bar n_{\kb{}\alpha}+1),
\end{align}
where $\bar n_{\kb{}\alpha} \equiv \bar n(\omega_{\kb{}\alpha})$ is the phonon thermal occupation number.

The typical momentum exchange in a phonon-phonon collision or a collision with a short-range scatterer is $\sim \hbar/\lambda_T$. Therefore the collision rate is independent of $\rb$ and the collision integral is local,
\begin{align}
\label{eq:bar_lambda_uniform}
{\rm St}[{}\Phi_\alpha({}\rb,\kb,t]=\frac{\mathbb V}{(2\pi)^d}\sum_{\alpha_0}\int d\kb_0 L{}_{\kb\alpha}^{\kb_0\alpha_0} {}\Phi_{\alpha_0}({}\rb,\kb_0,t).
\end{align}
For phonon-phonon scattering,  the coefficients $L{}_{\kb\alpha}^{\kb_0\alpha_0}$ are quadratic in the parameters of the cubic anharmonicity $V_{\kappa_1\kappa_2\kappa_3}$ introduced in Eq.~(\ref{eq:lattice_nonlinearity}); they are real and are given by the expression for $\tilde\Lambda_{\kb\alpha}^{\kb_0\alpha_0}$ in \cite{Atalaya2016}. The locality of the collision integral holds also in the presence of a smooth disorder, where thermal phonons are no longer plane waves.

The analysis of the dynamics of thermal phonons and ultimately of the decay of the low-frequency mode can be conveniently done in terms of the right and left eigenmodes of the collision integral $\psi_\nu(\kb,\alpha)$ and $\Psi_\nu(\kb,\alpha)$,
\begin{align}
\label{eq:stoss_eigenfunctions}
&{\rm St}[\psi_\nu(\kb,\alpha)] = -\ep_\nu \psi_\nu(\kb,\alpha),\nonumber\\
&[\mathbb V/(2\pi)^d]\sum_\alpha\int d\kb \Psi_{\nu'}(\kb,\alpha) \psi_\nu(\kb,\alpha) = \delta _{\nu,\nu'}.
\end{align}
Since the coefficients $L{}_{\kb\alpha}^{\kb_0\alpha_0}$ are real, the functions $\psi_\nu, \Psi_\nu$ and the eigenvalues $\ep_\nu$ are real or  form complex-conjugate pairs, with Re~$\ep_\nu \geq 0$. The real parts of the eigenvalues determine the decay rates of the two-phonon correlation functions. The zeroth eigenvalue, $\nu=0$, is $\ep_0=0$. It corresponds to the stationary value of the two-phonon correlator,
\[\psi_0(\kb,\alpha) = \hbar\omega_{\kb\alpha}\bar n_{\kb\alpha}(\bar n_{\kb\alpha}+1), \quad \Psi_0(\kb,\alpha)=\frac{\hbar\omega_{\kb\alpha}}{k_BT^2C\rho {\mathbb V}}.\]
The eigenfunctions and eigenvalues with $\nu>0$ can be found using the explicit form of the collision operator.

Except for special fine-tuned cases, the eigenfunctions $\psi_\nu$ form a complete set. One can then seek the solution of the kinetic equation in the form
\begin{align}
\label{eq:solution_eigenvectors}
{}\Phi_\alpha({}\rb,\kb,t) = \sum_\nu T_\nu({}\rb,t)\psi_\nu(\kb,\alpha).
\end{align}
Functions $T_\nu$ describe the spatial structure of the correlator $\Phi_\alpha(\rb,\kb,t)$. The equation for these functions reads
\begin{align}
\label{eq:R_dependent_functions}
&\partial_t T_\nu({}\rb,t) + \sum_{\nu'}\vb_{\nu\nu'}\partial_\rb T_{\nu'}({}\rb,t) = -\ep_\nu T_\nu ({}\rb,t),\nonumber\\
&\vb_{\nu\nu'}=\frac{\mathbb V}{(2\pi)^d}\sum_\alpha \int d\kb \Psi_{\nu}(\kb,\alpha)\vb_{\kb\alpha} \psi_{\nu'}(\kb,\alpha).
\end{align}
We will outline the solution of this equation in the limiting cases of the thermoelastic and Akhiezer relaxation.


\subsubsection{Thermoelastic relaxation}
\label{subsubsec:TER_Appendix}

The decay rate $\Gamma$ of the low-frequency mode is determined by the evolution of ${}\Phi_\alpha({}\rb,\kb,t)$ on the time scale $\lesssim \omega_0^{-1}$, as seen from Eq.~(\ref{eq:decay_in_terms_of_Wigner}). We will start with the case where $\omega_0^{-1}$ is large compared to the relaxation time of thermal phonons $\tau_{\rm ph}=\max [{\rm Re}~\ep_{\nu>0}^{-1}]$. As seen from Eq.~(\ref{eq:R_dependent_functions}), $\tau_{\rm ph}$ determines the  long-time decay of the functions $T_{\nu>0}$. The decay of $T_0$ can be still slower, and this is the case we now consider.

The slow evolution of $T_0({}\rb,t)$ can be described in the adiabatic approximation in which the functions $T_{\nu>0}$ adiabatically follow $T_0$. Then $T_\nu({}\rb,t)\approx -\ep_\nu^{-1}\vb_{\nu 0}\partial_\rb T_0({}\rb,t)$ for $\nu > 0$ and $t\gg \tau_{\rm ph}$. Equation~(\ref{eq:R_dependent_functions}) for $T_0$ then takes the form
\begin{align}
\label{eq:adiabatic_T_0}
&\partial_t T_0({}\rb,t) = \sum_{ij}D_{ij} \partial_{r_i} \partial_{r_j}T_0({}\rb,t), \nonumber\\
 &D_{ij}=\sum_{\nu>0}(\vb_{0\nu})_i(\vb_{\nu 0}))_j/\ep_\nu.
\end{align}
Using the explicit form of $\Phi_\alpha(\rb,\kb,t)$ and $\psi_0(\kb,\alpha)$ one can show that $iT_0({}\rb,t)$ can be interpreted as the scaled coordinate-dependent increment of the temperature of high-frequency phonons compared to the ambient temperature. Respectively, Eq.~(\ref{eq:adiabatic_T_0}) has the form of the standard equation of thermal diffusion.  Using the completeness of the set of the  eigenfunctions $\psi_\nu$, one can further show that the expression for  $D_{ij}$  coincides with the standard expression \cite{Lifshitz1981a} for thermal diffusivity. In an isotropic medium $D_{ij}=D\delta_{ij}$; in terms of the thermal conductivity and the specific heat,  $D=\kappa_T/C\rho$.

The boundary conditions for the function $T_0(\rb,t)$ follow from its proportionality to the temperature increment. At a free side of a nanoresonator there is no heat flux in the direction $\hat{\bf n}$ normal to the side, and then $(\hat{\bf n}\partial_\rb T_0)=0$. Such boundary condition on the temperature increment was used in the analysis of thermoelastic relaxation by \textcite{Lifshitz2000}. On the other hand, at the surfaces where the resonator is clamped the temperature may be equal to the ambient temperature, and then $T_0=0$ (but the clamping area can also have a thermal contact resistance).

A convenient strategy is to find the eigenvalues $\mu_n$ and eigenfunctions $T_{0n}(\rb)$ of the diffusion equation (\ref{eq:adiabatic_T_0}), express $T_0(\rb,t)$ as a sum of $T_{0n}\exp(-\mu_nt)$, and then find the decay rate of the low frequency mode from Eq.~(\ref{eq:decay_in_terms_of_Wigner}) and Eq.~(\ref{eq:solution_eigenvectors}) in which we keep only the term with $\nu=0$ \cite{Atalaya2016}. We illustrate this strategy for an important type of NVSs \cite{Zener1938,Landau1986,Lifshitz2000}, a long and thin rectangular nanobeam. We assume that the beam is clamped at $x=0$ and $x=L$, has width $W$ in the $y$-direction and thickness $l_\perp$ in the $z$-direction, with the length $L\gg W,l_\perp$ and with $W,l_\perp\gg l_T$.  The beam bends in the $z$-direction. Therefore the temperature is nonuniform in the $z$-direction, but it can be uniform in the $y$-direction. Since $l_\perp\ll L$, the low-lying eigenvalues of the diffusion equation (\ref{eq:adiabatic_T_0}) correspond to the eigenmode $\propto \sin(\pi z/l_\perp)$.  If we choose $T_0=0$ at $x=0$ and $x=L$, the eigenmodes of Eq.~(\ref{eq:adiabatic_T_0}) are $T_{0n}(\rb)=(2/\sqrt{V})\sin(\pi z/l_\perp)\sin(n\pi x/L)$, and the corresponding eigenvalues are $\mu_n=\pi^2D(l_\perp^{-2} + n^2L^{-2})$. For small $n$, where the term $\propto L^{-2}$ in $\mu_n$ can be disregarded, $\mu_n\approx \tau_Z^{-1}=D(\pi/l_\perp)^2$. This expression coincides with the Zener relaxation rate $\tau_Z^{-1}$ used in the equation for the thermoelastic decay rate (\ref{eq:TER}).

The phenomenological analysis of the thermoelastic relaxation corresponds to the assumption that the Gr\"uneisen parameter is the same for all phonons, $\gamma_{\kb\alpha}^{\rm (G)}\equiv \gamma^{(G)}= E\alpha_T/C\rho(1-2\nu_\mathrm{P})$, where $\alpha_T$ is the linear thermal expansion coefficient and $\nu_\mathrm{P}$ is the Poisson ratio. Then from Eqs.~(\ref{eq:coupling_V_deformation}),  (\ref{eq:initial_Phi}), and (\ref{eq:solution_eigenvectors}), with the account taken of the explicit form of $\psi_0(\kb,\alpha)$, we have
\[T_0(\rb,0) = i\hbar^{-1}\gamma^{(G)}\n\boldsymbol{\varphi}.\]
To find $T_0(\rb, t)$ one should expand $T_0(\rb,0)$ in the eigenmodes $T_{0n}(\rb)$. Then integration over time in Eq.~(\ref{eq:decay_in_terms_of_Wigner}) gives the decay rate $\Gamma$ of the NVS mode in the form
\begin{align}
\label{eq:TER_modes}
\Gamma=\frac{E^2\alpha_T^2T}{2MC\rho(1-2\nu_\mathrm{P})^2}\sum_n\frac{\mu_n^{-1}}{1+(\omega_0/\mu_n)^2}
\left|\int d\rb \n\boldsymbol{\varphi}T_{0n}(\rb)\right|^2.
\end{align}
The eigenfunctions of the thermal diffusion equation were also used by \textcite{Zener1938} in the analysis of the $Q$-factor of the beam vibrations based on the coupled equations of motion of a slow mode and temperature. These equations were derived from thermodynamic arguments and only diffusion transverse to the beam was considered.

For a  flexural mode we have
\[\n\boldsymbol{\varphi} = -(1-2\nu_\mathrm{P})L^{1/2}z\partial^2_x\zeta \,(\int \zeta^2 dx)^{-1/2},\]
where $\zeta(x)$ is the displacement of the central plane in the $z$-direction \cite{Landau1986}. Using the explicit form of $\zeta(x)$ for the lowest flexural mode and the expression for the mode eigenfrequency $\omega_0 \approx 6.5(l_\perp/L^2)(E/\rho)^{1/2}$, and setting  $\mu_n\approx \mu_0$,  from Eq.~(\ref{eq:TER_modes}) we obtain
\begin{align}
\label{eq:TER_summed}
\Gamma=\Gamma^{\rm TER} \approx 0.98 \frac{E\alpha_T^2T\omega_0}{2C\rho}\,\frac{\omega_0\tau_Z}
{1+(\omega_0\tau_Z)^2},
\end{align}
which essentially coincides with Eq.~(\ref{eq:TER}).

We note that  it is necessary to keep several terms in the sum over the eigenmodes $T_{0n}$ in Eq.~(\ref{eq:TER_modes}) [Eq.~(\ref{eq:TER_summed}) includes the whole sum]. If the aspect ratio $L/l_\perp$ is not very large, one should take into account the dependence of the relaxation times of the thermal modes $\mu_n^{-1}$ on the mode number. The described method applies to any geometry. It also allows taking into account the difference between the values of the Gr\"uneisen parameter  for different phonons, and moreover, going beyond the deformation potential approximation (\ref{eq:Gruneisen_appendix}) all together.


\subsubsection{Akhiezer relaxation}
\label{subsubsec:Akhiezer_Appendix}

When the mode eigenfrequency significantly exceeds the rate of thermal diffusion, $\omega_0\gg \tau_Z^{-1}$, one should take into account a finite time it takes for the phonons to locally equilibrate. The corresponding mechanism is the extension to a resonator mode of the Aihezer mechanism of decay of ultrasound \cite{Akhiezer1938}.

In terms of the formalism described here, the Akhiezer damping is determined by the evolution of the function $T_\nu(\rb,t)$ over the phonon relaxation time $\tau_{\rm ph}\ll \tau_Z$.
On this time scale one can disregard the drift term in Eq.~(\ref{eq:R_dependent_functions}). Indeed, this term comes from the spatial nonuniformity of the phonon distribution. The characteristic scale of the nonuniformity is the size of the system, which largely exceeds the phonon mean free path $\sim v_\mathrm{ph}\tau_\mathrm{ph}$ ($v_\mathrm{ph}$ is the characteristic phonon velocity), so that $|v_\mathrm{ph}\partial_\rb T_\nu|\ll |T_\nu|/\tau_\mathrm{ph}$. Then the functions $T_\nu(\rb,t)$ exponentially decay in time as $\exp(-\ep_\nu t)$ for $\nu>0$.

From Eqs.~(\ref{eq:decay_in_terms_of_Wigner}) and (\ref{eq:solution_eigenvectors}), the Akhiezer decay rate of the considered low-frequency mode $\Gamma \equiv \Gamma^{\rm Akh}$ is
\begin{align}
\label{eq:Gamma_Akh}
&\Gamma^{\rm Akh} = \frac{1}{2Mk_B T }\,\frac{\mathbb V}{(2\pi)^{2d}} {\rm Re}\sum_{\alpha,\alpha'}
\int d\rb d\kb d\kb'\; V^*_\alpha(\rb,\kb) \nonumber \\
& \times\sum_{\nu>0}\frac{\psi_\nu(\kb,\alpha)\Psi_{\nu}(\kb',\alpha')}{\epsilon_\nu-i\omega_0} V_{\alpha'}(\rb,\kb')\bar n_{\kb'\alpha'}(\bar n_{\kb'\alpha'} + 1).
\end{align}
The term $\nu=0$, which describes thermal diffusion, does not contribute to the Akhiezer relaxation.

In deriving Eq.~(\ref{eq:Gamma_Akh}) no assumptions have been made about the structure of the considered low-frequency mode and the symmetry of the medium. It is important, though, that if we describe the coupling to phonons by the deformation potential  (\ref{eq:coupling_V_deformation}) and assume that the coupling parameters $\gamma_{\kb \alpha}^{\rm (G)}$ are the same for all phonons, it follows from Eq.~(\ref{eq:stoss_eigenfunctions}) that $\Gamma^{\rm Akh}=0$. This means that to describe the Akhiezer relaxation it is necessary to allow for the dependence of $\gamma_{\kb \alpha}^{\rm (G)}$ on the phonon quantum numbers $\kb,\alpha$.

In the analysis of the Akhiezer relaxation the phonon decay rates Re~$\ep_\nu$  are often replaced by a characteristic parameter $\tau_{\rm ph}^{-1}$, see \cite{Maris1968,Iyer2016} and references therein. In this approximation we can simplify Eq.~(\ref{eq:Gamma_Akh}) using the completeness of the eigenfunctions $\psi_\nu$, $\sum_{\nu>0} \psi_\nu(\kb,\alpha)\Psi_{\nu}(\kb',\alpha') = [(2\pi)^d/{\mathbb V}]\delta (\kb-\kb')\delta_{\alpha\alpha'} - \psi_0(\kb,\alpha)\Psi_0(\kb',\alpha')$. If we denote the averaging over phonons  by an overline,
\begin{align}
\label{eq:kb_averaging}
\overline{ B_\alpha(\rb,\kb)} = &\frac{\hbar^2}{(2\pi)^d C\rho k_BT^2}\sum_\alpha\int d\kb B_\alpha(\rb,\kb)\nonumber\\
&\times\omega_{\kb\alpha}^2 \bar n_{\kb\alpha}(\bar n_{\kb\alpha} + 1)
\end{align}
[here $B_\alpha(\rb,\kb)$ is an arbitrary function of $\rb, \kb,\alpha$], we can rewrite Eq.~(\ref{eq:Gamma_Akh}) as
\begin{align}
\label{eq:Gamma_Akh_simple}
&\Gamma^{\rm Akh} = \frac{q_0^2}{\hbar }C\rho T   \frac{\omega_0\tau_{\rm ph}}{1+\omega_0^2\tau_{\rm ph}^2}
\nonumber\\
&\times\int d\rb  \left[\overline{|v_\alpha(\rb,\kb)|^2} - |\overline{v_\alpha(\rb,\kb)}|{}^{^2}\right]
\end{align}
where $v_\alpha(\rb,\kb) = V_\alpha(\rb,\kb)/\hbar\omega_{\kb\alpha}$ and $q_0=(\hbar/2M\omega_0)^{1/2}$.
Equation (\ref{eq:Gamma_Akh_simple}) explicitly shows that, to describe the Akhiezer relaxation, one has to take into account the dependence of the parameters $\gamma_{\kb\alpha}^{\rm (G)}$ on $\kb,\alpha$. From this point of view, it may be more appropriate to interpret the parameter $(\gamma^{\rm (G)})^2$ in Eq.~(\ref{eq:Akhiezer}) as the variance rather than the squared mean value of $\gamma_{\kb\alpha}^{\rm (G)}$.

An advantageous feature of the presented technique is that it allows one to consider an intermediate parameter range between the limits of the thermoelastic and Akhiezer relaxation. Also, it immediately applies to micro- and nanoresonators of an arbitrary geometry and various boundary conditions.

\section{Allan variance in the limiting cases}
\label{sec:Allan_appendix}

As indicated in Sec.~\ref{sec:frequency_fluctuations}, a most common way to characterize frequency fluctuations is based on the Allan variance $\sigma_\mathrm{A}^2(\tau)$. It is defined  in terms of the frequencies $\bar f_m$ measured from the increments of the full vibrational phase $\varphi$ over the time intervals $(t_m, t_m+\tau)$ as
\begin{align}
\label{eq:Allan_defined_appendix}
\sigma_\mathrm{A}^2(\tau) = \frac{1}{2(N-1)f_0^2}\sum_{m=1}^{N-1}(\bar f_{m+1} - \bar f_m)^2,
\end{align}
where $f_0$ is the mean value of $\bar f_m$,  cf. Eq.~(\ref{eq:Allan_defined}). We recall that $\varphi(t)$ determines  the displacement in the laboratory frame, which is $\propto \cos[\varphi(t)]$; it should not be confused with the phase $\phi(t)$ in the rotating frame, which is counted off from $2\pi f_0t$;

If the dead-time between the successive measurements is zero, $t_{m+1}-t_m=\tau$, the Allan variance can be simply expressed in terms of the power spectrum $S_\varphi(\omega)$ of the fluctuations of the full vibrational  phase,
\begin{align}
\label{eq:Allan_phase_spectrum}
\sigma_\mathrm{A}^2(\tau) = \frac{8}{\pi\omega_0^2\tau^2}\int_0^\infty d\omega \sin^4(\omega\tau/2) S_\varphi(\omega).
\end{align}

Allan variance is used particularly broadly to characterize noise of self-sustained vibrations in systems with feedback. In such systems the vibration amplitude $A$ is kept almost constant by the feedback loop, but the phase is not fixed (unless the vibrations are synchronized by an external source). Noise causes phase fluctuations, which accumulate in time. If the noise is thermal (thermomechanical), as in the equation of Brownian motion (\ref{eq:Brownian}) or in Eq.~(\ref{eq:Gamma_defined}), the phase is diffusing \cite{Berstein1938}. Then, from the above equations and Eq.~(\ref{eq:Allan_phase_spectrum}), $\sigma_\mathrm{A}^2(\tau) $ displays a characteristic dependence on $\tau$ and $A$,
\begin{align}
\label{eq:Allan_thermal}
\sigma_\mathrm{A}^2(\tau) = (2\Gamma k_BT/M\omega_0^4 A^2)\tau^{-1}.
\end{align}

In the range where $\tau$ is small compared to the decay time of the oscillator $\Gamma^{-1}$, this expression applies also  to a mode driven by a sufficiently strong resonant force $F\cos\omega_Ft$ with no feedback loop. For a strong drive the amplitude of forced vibrations largely exceeds the thermal displacement $(k_BT/M\omega_0^2)^{1/2}$, and thus amplitude fluctuations are relatively small.  Fluctuations in this case can be analyzed using the equation of motion for the complex amplitude $u(t)$ of a driven linear mode
\begin{align}
\label{eq:driven_linear}
&\dot u = -[\Gamma +i(\omega_F- \omega_0)] u(t) -i\frac{F}{4M\omega_F} + \xi(t),\nonumber\\
&u(t) = \frac{1}{2M\omega_F}[M\omega_F q(t) - ip(t)]\exp(-i\omega_Ft),
\end{align}
where $\xi(t)$ is thermal noise with the correlator (\ref{eq:noise_correlator}), cf. Sec.~\ref{subsec:resonant_forced}.

In  the absence of noise and in the stationary regime, the vibration amplitude  is $A_\mathrm{st}\approx 2|u_\mathrm{st}| =(F/2M\omega_F)[(\omega_F-\omega_0)^2 +\Gamma^2]^{-1/2}$. The vibration phase  $\phi_\mathrm{st}$ as counted off from the drive phase is
\begin{align}
\label{eq:phase_stationary}
\phi_\mathrm{st}=-\frac{1}{2}\pi - \arctan[(\omega_F-\omega_0)/\Gamma].
\end{align}
For weak thermal noise the vibration phase $\phi(t)= \varphi(t) - \omega_F t$ fluctuates about $\phi_\mathrm{st}$. This behavior is very different from the phase diffusion for self-sustained vibrations. Interestingly, it follows from  Eq.~(\ref{eq:driven_linear}) that the dependence $\sigma_\mathrm{A}^2\propto \tau^{-1}$ has the same form in the both cases provided $\Gamma\tau\ll 1$. It is this characteristic dependence that gives the so-called noise floor for thermal-noise dominated fluctuations in nanomechanical systems, cf. \cite{Cleland2002,Ekinci2004,Sansa2016,Sadeghi2020} and references therein.

For a resonantly driven mode  subject to thermal noise  and in the absence of a feedback loop, Eq.~(\ref{eq:driven_linear})  allows one also to find a simple expression for the Allan variance in the opposite limit of a long time $\tau$, where $\Gamma\tau\gg 1$,
\begin{align}
\label{eq:Allan_driven_long_tau}
\sigma_\mathrm{A}^2(\tau) = (3 k_BT/M\omega_0^4 A^2)\tau^{-2}
\end{align}

Of significant importance is a different regime where the Allan variance is dominated not by thermal fluctuations of the slow part of the phase, but rather by eigenfrequency fluctuations. These fluctuations often have $1/f$-type component (the flicker noise). In this case it follows from Eq.~(\ref{eq:full_phase_formal}) that  $S_\varphi(\nu)\propto \nu^{-3}$ for small $\nu$. Then from Eq.~(\ref{eq:Allan_phase_spectrum}) $\sigma_\mathrm{A}^2 \propto \tau^{-4}$. As mentioned in Sec.~\ref{sec:frequency_fluctuations}, the Allan variance does not distinguish between the eigenfrequency fluctuations and the fluctuations of the rotating-frame phase $\phi$.

A convenient approach to  an open-loop measurement of the Allan variance is based on measuring the ratio of the quadrature and in-phase components  of the vibrations of a driven mode. This ratio gives $\tan \phi(t)$.  In the measurement,  the drive frequency $\omega_F$ is often chosen maximally close to the measured mode eigenfrequency $\omega_0^\mathrm{meas} = 2\pi f_0$. The relation between the phase fluctuation $\Delta\phi(t)$ and the fluctuation $\Delta\omega_0^\mathrm{meas}$ in this case can be obtained from Eq.~(\ref{eq:phase_stationary}). To do this one should replace $\omega_0$ in this equation with $\omega_0^\mathrm{meas}$ and the stationary phase $\phi_\mathrm{st}$ with the time-dependent phase $\phi$. One then finds $\Delta\omega_0^\mathrm{meas} = \Gamma\Delta\phi$.  The relation applies in the adiabatic limit, where the change $\Delta\phi(t)$ is slow compared to the oscillator relaxation time $\Gamma^{-1}$.


%

\end{document}